\documentclass[prd,aps,twocolumn,a4paper,showkeys,nofootinbib]{revtex4-1}
\usepackage{amsmath}
\usepackage{amsfonts}
\usepackage{amssymb}	
\usepackage{graphicx}
\usepackage{bm}
\usepackage{color}
\usepackage[utf8]{inputenc}

\def\GMc2{G M_{\odot} c^{-2}}

\def\lm{{\ell m}}

\def\lm{{\ell m}}

\def\lm{{\ell m}}

\def\ii{{\rm i}}

\def\TEOBResumS{\texttt{TEOBResumS}}
\def\TEOBiResumSM{\texttt{TEOBiResumS\_SM}}
\def\TEOBiResumM{\texttt{TEOBiResumMultipoles}}

\def\SEOBNRvq{{\texttt{SEOBNRv4}}}

\newcommand{\be}{\begin{equation}}  
\newcommand{\ee}{\end{equation}}
\newcommand{\bea}{\begin{eqnarray}}           
\newcommand{\eea}{\end{eqnarray}} 
\newcommand{\beqn}{\begin{eqnarray*}}
\newcommand{\eeqn}{\end{eqnarray*}}
\begin{document}
\title{A multipolar effective one body waveform model for spin-aligned black hole binaries.}
\author{Alessandro \surname{Nagar}${}^{1,2}$}
\author{Gunnar \surname{Riemenschneider}${}^{1,3}$}
\author{Geraint \surname{Pratten}$^{4,5}$}
\author{Piero \surname{Rettegno}${}^{1,2}$}
\author{Francesco \surname{Messina}${}^{6,7}$}
\affiliation{${}^1$INFN Sezione di Torino, Via P. Giuria 1, 10125 Torino, Italy}
\affiliation{${}^2$Institut des Hautes Etudes Scientifiques, 91440 Bures-sur-Yvette, France}
\affiliation{${}^{3}$ Dipartimento di Fisica, Universit\`a di Torino, via P. Giuria 1, 10125 Torino, Italy}
\affiliation{${}^{4}$ School of Physics and Astronomy and Institute for Gravitational Wave Astronomy, University of Birmingham, Edgbaston, Birmingham, B15 9TT, United Kingdom}
\affiliation{${}^{5}$ Departament de Fisica, Universitat de les Illes Balears and Institut d'Estudis Espacials de Catalunya, Crta. Valldemossa km 7.5, E-07122 Palma, Spain}
\affiliation{${}^{6}$ Dipartimento di Fisica, Universit\`a degli studi di Milano Bicocca, Piazza della Scienza 3, 20126 Milano, Italy}
\affiliation{${}^{7}$ INFN, Sezione di Milano Bicocca, Piazza della Scienza 3, 20126 Milano, Italy}
\begin{abstract}
  We introduce \TEOBiResumSM{}, an improved version of the effective-one-body (EOB) 
  waveform model \TEOBResumS{} for spin-aligned, coalescing black hole binaries, 
  that includes subdominant gravitational waveform modes completed through merger and ringdown.
  Beyond the dominant $(\ell,|m|)=(2,2)$ one, the more robust multipoles all over 
  the parameter space are: $(2,1)$, $(3,3)$, $(3,2)$, $(4,4)$ and $(5,5)$. 
  Modes as $(3,1)$, $(4,3)$ and $(4,2)$ can also be generated, but are less
  robust. The multipolar ringdown EOB waveform stems from suitably fitting many numerical 
  relativity (NR) waveform data from the Simulating eXtreme Spacetimes (SXS) collaboration 
  together with test-mass waveform data. Mode-mixing effects are not incorporated. 
  The orbital (nonspinning) part of the multipolar waveform amplitudes includes
  test-mass results up to (relative) 6PN order and, for most modes,
  is Pad\'e resummed. The $m$=odd waveform multipoles (up to $\ell=5$) 
  incorporate most of the currently available spin-dependent analytical information. 
  Each multipolar amplitude is additionally orbital-factorized and resummed. 
  Improving on previous work, we confirm that certain
  $m=\text{odd}$ modes, e.g. the $(2,1)$, and even the $(3,1)$, may develop
  a zero (or a minimum) in the amplitude for nearly equal-mass binaries
  and for several combinations of the individual spins. 
  A remarkable EOB/NR agreement around such zero is found for these modes.
  The new waveform, and radiation reaction, prompts a new NR-calibration of
  the spinning sector of the model, done with only $32$ datasets.
  The maximum $(2,2)$ EOB/NR unfaithfulness $\bar{F}$ with Advanced LIGO noise against 
  the SXS catalog ($\sim 595$ datasets) is always below $0.5\%$ for binaries with total
  mass $M$ as $10M_\odot\leq M \leq 200M_\odot$, except for a single
  outlier with $\max{(\bar{F})}\sim 0.85\%$. When $(2,1)$, $(3,3)$ and $(4,4)$
  modes are included, one finds an excellent EOB/NR agreement up to $M\sim 120M_\odot$,
  above which the performance degrades slightly and moves above $3\%$
  We also point out that the EOB dynamics may develop unphysical features
  for large, anti-aligned, spins and this may impact the correct construction
  of the $(2,1)$ mode in some corners of the parameter space.
\end{abstract}

\maketitle

\section{Introduction}
The recent observation made by LIGO~\cite{TheLIGOScientific:2014jea} and Virgo~\cite{TheVirgo:2014hva} 
of gravitational wave (GW) signals from twelve coalescing compact binaries 
marked the beginning of the era of gravitational wave astronomy.
Of these detections, ten were associated to coalescing binary black holes
(BBHs)~\cite{Abbott:2016blz,Abbott:2016nmj,Abbott:2017vtc,Abbott:2017gyy,Abbott:2017oio,LIGOScientific:2018mvr}
and two to a binary neutron star (BNS)~\cite{TheLIGOScientific:2017qsa,Abbott:2020uma}.

Up to recent times, gravitational waveform models used on LIGO and Virgo data only
incorporated the dominant $(\ell=2, m=2)$ mode. This may be sufficient when the
binary system is highly symmetric (e.g nearly equal masses and nearly equal spins),
but for binaries when one object is more massive than the other, or when
the spins are very different, modeling the subdominant multipoles becomes an absolute
necessity to avoid potential biases in the parameters~\cite{OShaughnessy:2014shr,Varma:2016dnf}.
Similarly, at large inclinations, the modeling of gravitational wave modes
beyond the dominant mode becomes increasingly important as higher modes are
geometrically suppressed in the face-on/off limit.
For this reason, there were recent efforts in building waveform models that incorporate
the subdominant modes. This was the case for phenomenological models, both in the
spinning~\cite{London:2017bcn} or nonspinning case~\cite{Mehta:2017jpq},
or for effective-one-body (EOB) models~\cite{Cotesta:2018fcv} for spin aligned
black hole binaries. In addition, Ref.~\cite{Varma:2018mmi} took advantage of 
a huge number of high-quality numerical relativity simulations from the SXS
collaboration to construct a numerical relativity (NR) surrogate model with
as many modes as possible (also including the $m=0$ ones).

Within the effective-one-body framework~\cite{Buonanno:1998gg,Buonanno:2000ef,Damour:2000we,Damour:2001tu,Damour:2015isa}
for coalescing black-hole binaries, the {\tt SEOBNRv4HM} model introduced
in Ref.~\cite{Cotesta:2018fcv} is the higher-mode version of
the \SEOBNRvq{}~\cite{Bohe:2016gbl} spin-aligned model, calibrated
to NR simulations, while {\tt SEOBNRv4HMP} is its precessing version~\cite{Ossokine:2020kjp}
and represents current state of the art. Alternatively to \SEOBNRvq{}, a different spin-aligned
EOB model, informed by NR simulations, is \TEOBResumS{}.
This model was introduced in~\cite{Nagar:2018zoe},
and used to independently infer the parameters of GW150914~\cite{Abbott:2016blz}.
Although this waveform model is limited to the $\ell=m=2$ dominant mode,
is publicly available either as a stand-alone $C$ code based on the GSL library
or through the LIGO {\tt LALSuite}~\cite{lalsuite} library.
One of the advantages of this model is that it implements
the description of the inspiral dynamics based on the
(high-order) post-adiabatic (PA)
approximation~\cite{Nagar:2018gnk,Akcay:2018yyh,Nagar:2018plt}.
This allows one to generate long-inspiral waveforms so
efficiently to be of direct use for parameter estimation
purposes (see also Ref.~\cite{Rettegno:2019tzh} where the same
approach is applied to the {\tt SEOBNRv4} Hamiltonian).

Recently, in a companion paper~\cite{Nagar:2019wds},
hereafter Paper~I, the nonspinning sector of \TEOBResumS{},
was augmented with all subdominant waveform modes, completed
through merger and ringdown, up to $\ell=m=5$ included.
This defined the \TEOBiResumM{} model. In doing so, the EOB orbital
interaction potential was improved thanks to a more stringent
comparison with state-of-the-art NR simulations with small uncertainties.
This led us to construct a multipolar model with EOB/NR unfaithfulness
at most of the order of  $2\%$ (and typically well below $1\%$)
for total mass up to $200M_\odot$ and mass ratio in the range $1\leq q \leq 10$.
It was also possible to verify that the model performs excellently
up to $q=18$, that is the NR dataset with the largest
mass ratio currently available to us.

The purpose of the present work is to generalize the results of Paper~I
to the case of spin-aligned black-hole binaries. To improve the robustness
of the multipolar waveform amplitudes towards merger, we build
upon Refs.~\cite{Nagar:2016ayt,Messina:2018ghh}, implementing the corresponding
orbital factorization and resummation paradigm,
though limited to the $m=$odd waveform modes. Together with the changes
in the nonspinning part of the dynamics discussed in Paper~I, this
led us to a new determination of the next-to-next-to-next-to-leading (NNNLO) spin-orbit
effective parameter $c_3$ introduced long ago~\cite{Damour:2014sva,Nagar:2015xqa}.
The construction of the multipolar waveform around the amplitude peak of each
multipole (e.g. around merger), of the next-to-quasi-circular (NQC) corrections
and of the postpeak-ringdown phase follows the procedure discussed,
multipole by multipole, for the nonspinning case in Paper~I. The only
difference is that some of the NR-informed fits incorporate now a
suitable spin-dependence. The reader should be aware that this paper
stems from Refs.~\cite{Damour:2014sva,Damour:2014yha,Nagar:2015xqa,
  Nagar:2016ayt,Nagar:2016iwa,Nagar:2017jdw,Nagar:2018zoe} and it
is essentially the follow up of Refs.~\cite{Nagar:2018zoe,Nagar:2019wds}.
As such, it adopts the same notations and conventions. For this reason,
we shall assume the reader to be familiar with the notation and language
of those papers, that might not be reintroduced if not absolutely necessary.

The paper is organized as follows. In Sec.~\ref{sec:model} we review the elements
of the EOB dynamics that remained unchanged with respect to~\cite{Nagar:2018zoe}
and Paper~I; we discuss the structure of the new multipolar waveform and
the related new determination of $c_3$. Section~\ref{sec:NR_catalog} summarizes
describes in  details the numerical waveforms employed in this paper
(either to inform the model or to check it), focusing in particular on an estimate 
of their uncertainty. Section~\ref{sec:barF} 
probes the $(2,2)$ mode all over the current release of the SXS catalog~\cite{Boyle:2019kee}. 
Section~\ref{sec:HM} focuses on the behavior of higher multipolar modes, 
highlighting several aspects related to their accurate modelization. 
In particular, it is pointed out, and explained, the peculiar behavior of 
some $m=1$ modes. The important
EOB/NR unfaithfulness computations with higher modes are also performed
there. Our concluding remarks are then collected in Sec.~\ref{sec:conclusions}.
The bulk of the text is complemented by several Appendixes.
Appendix~\ref{sec:nospin_lim} discusses in detail the nonspinning limit
of the model; Appendix~\ref{sec:NRsystematics} highlights a few 
systematics in the SXS waveform data that are relevant for higher modes;
Appendix~\ref{sec:app_fits} reports all the NR-informed fits
that are needed to accurately build the merger and ringdown part
of the multipolar waveform. 

If not otherwise specified, we use natural units with $c=G=1$.
Our notations are as follows: we denote with $(m_1,m_2)$ the individual masses,
while the mass ratio is $q\equiv m_1/m_2\geq 1$. The total mass and symmetric
mass ratio are then $M\equiv m_1+m_2$ and $\nu = m_1 m_2/M$.
We also use the mass fractions $X_{1,2}\equiv m_{1,2}/M$ and $X_{12}\equiv X_1-X_2=\sqrt{1-4\nu}$.
We address with $(S_1,S_2)$ the individual, dimensionful, spin components along the
direction of the orbital angular momentum. The dimensionless spin variables are
denoted as $\chi_{1,2}\equiv S_{1,2}/(m_{1,2})^2$. We also use $\tilde{a}_{1,2}\equiv X_{1,2}\chi_{1,2}$,
the effective spin $\tilde{a}_0=\tilde{a}_1-\tilde{a}_2$ and $\tilde{a}_{12}\equiv \tilde{a}_1-\tilde{a}_2$.
\begin{table*}[t]
\caption{\label{tab:PNchoices} Resummation choices used to build our multipolar EOB waveform.
  The bar denotes resummation using the inverse Taylor expansion, as described by Eq.~(5)
  of Ref.~\cite{Nagar:2016ayt}. The PN-order should be intended relative to the leading-order
  term and also indicates the order of the additional (spinning) test-particle terms.
  For example, 3.5PN means that we take a polynomial of the form
  $1+x^{3/2}+x^2+x^{5/2}+...+ x^{7/2}$, with the known $\nu$ dependence in the coefficients.
  Instead, $1.5^{+1}$PN means that we add to the $\nu$-dependent 1.5PN-accurate polynomial
  an additional term proportional to $x^{5/2}$ obtained by suitably incorporating spinning p
  article terms as illustrated in Sec.~VB of Ref.~\cite{Messina:2018ghh}.
  We denote Pad\'e resummation by $P^n_d$, where $N=n+d$ is the PN order.}
\begin{center}
\begin{ruledtabular}  
	\begin{tabular}{lccccc}
		\rule[0,6cm]{0cm}{0cm}
		$(\ell,m)$ &\multicolumn{2}{c}{Resummation choices}\hspace{2.5cm} &\multicolumn{2}{c}{Relative PN order}\hspace{1.5cm}\\
		\hfill & orbital & spin & orbital & spin\\
		\hline
		\hline
		\rule{0pt}{5ex}
		$(2,2)$ & $P^5_0[\rho_{22}^{\rm orb}]$ & $T[\rho_{22}^{\rm S}]$ & $3^{+2}$PN & 3.5PN without NNLO SO term\\
		\rule{0pt}{5ex}
		$(2,1)$ & $P_1^5[\rho_{21}]$ & $\hat{f}_{21}^{\rm S}= X_{12}\overline{\hat{f}^{\rm S_{(0)}}_{21}}-\dfrac{3}{2}\tilde{a}_{12}x^{1/2}\overline{\hat{f}_{21}^{\rm S_{(1)}}}$ & $3^{+3}$PN & 2.5PN\\
		\rule{0pt}{5ex}
		$(3,3)$ & $P_2^4[\rho_{33}]$ & $\hat{f}_{33}^{\rm S}= X_{12}\overline{\hat{f}^{\rm S_{(0)}}_{33}}+\left(-\dfrac{1}{4}+\dfrac{5}{2}\nu\right)\tilde{a}_{12}x^{3/2}\hat{f}_{33}^{\rm S_{(1)}}$ & $3^{+3}$PN & 2.5PN\\
		\rule{0pt}{5ex}
		$(3,2)$ & $P^4_2[\rho_{32}]$ & $T[\rho_{32}^{\rm S}]$  & $2^{+2}$PN & $1.5^{+1}$PN (SO only)\\
		\rule{0pt}{5ex}
		$(3,1)$ & $P_2^3[\rho_{31}^{\rm orb}]$ & $\hat{f}_{31}^{\rm S}= X_{12}\overline{\hat{f}^{\rm S_{(0)}}_{31}}+\left(-\dfrac{9}{4}+\dfrac{13}{2}\nu\right)\tilde{a}_{12}x^{3/2}\hat{f}_{31}^{\rm S_{(1)}}$ & $3^{+2}$PN & 2.5PN\\ 
		\rule{0pt}{5ex}
		$(4,4)$ & $P^6_0[\rho_{44}^{\rm orb}]$ & $T[\rho_{44}^{\rm S}]$ & $2^{+4}$PN & $1.5^{+2}$PN (SO only)\\
		\rule{0pt}{5ex}
		$(4,3)$ & $P_2^4[\rho_{43}^{\rm orb}]$ & $\hat{f}_{43}^{\rm S}= X_{12}\hat{f}^{\rm S_{(0)}}_{43}-\dfrac{5}{4}\tilde{a}_{12}x^{1/2}$ & $1^{+5}$PN & 0.5PN (SO only)\\
		\rule{0pt}{5ex}
		$(4,2)$ & $P^6_0[\rho_{42}^{\rm orb}]$ & $T[\rho_{42}^{\rm S}]$ & $2^{+4}$PN & $1.5^{+3}$PN (SO only)\\
		\rule{0pt}{5ex}
		$(4,1)$ & $P_2^4[\rho_{41}^{\rm orb}]$ & $\hat{f}_{41}^{\rm S}= X_{12}\hat{f}^{\rm S_{(0)}}_{41}-\dfrac{5}{4}\tilde{a}_{12}x^{1/2}$ & $1^{+5}$PN & 0.5PN (SO only)\\
                \rule{0pt}{5ex}
                $(5,5)$ & $P_0^6[\rho_{55}^{\rm orb}]$  & $\hat{f}_{55}^{\rm S}=X_{12}\overline{\hat{f}_{55}^{{\rm S}_{(0)}}} + 10\nu\dfrac{(1-3\nu)}{3-6\nu}\tilde{a}_{12}x^{3/2}$& $1^{+5}$PN & 2PN
	\end{tabular}
\end{ruledtabular}
\end{center}
\label{tab:pn_order}
\end{table*}

\section{The model: Relative dynamics and multipolar waveform}
\label{sec:model}
In this Section we collect the analytical elements of \TEOBiResumSM{}
that change with respect to the original implementation of \TEOBResumS{}
of~\cite{Nagar:2018zoe} or that stem from results of Paper~I.
The modifications regard all building blocks of the model: the Hamiltonian,
the inspiral, EOB-resummed, waveform as well as the merger-ringdown part.
However, the structure of the Hamiltonian is precisely the
same of \TEOBResumS{}: there is thus no need to describe it here in detail
and we address the reader to Sec.~II of Ref.~\cite{Nagar:2018zoe}.
The modifications are limited to the NR-informed effective 5PN coefficient
$a_6^c(\nu)$ (that coincides with the function determined in Paper~I)
as well as the effective NNNLO spin-orbit parameter $c_3(\nu,\tilde{a}_1,\tilde{a}_2)$.
This one needs to be redetermined, by phasing comparison with NR simulations,
because of both the new $a_6^c(\nu)$, that has changed with respect to Ref.~\cite{Nagar:2018zoe},
and the new analytical choice for the factorized
(and resummed) multipolar waveform taken from Ref.~\cite{Messina:2018ghh}.
In addition, we also present here a new, spin-dependent, description
of the multipolar merger and ringdown waveform, that is based on
fits informed by NR simulations. These fits incorporate some,
but not all, spin dependence for all modes up to $\ell=m=5$,
as we detail in Appendix~\ref{sec:app_fits}.
We start by discussing the structure of the resummed waveform.

\subsection{Inspiral multipolar waveform}
\label{sec:waveform}
The waveform amplitudes we use here incorporate several factorization and resummation
procedures that have been discussed in previous
literature~\cite{Damour:2008gu,Pan:2010hz,Nagar:2016ayt,Messina:2018ghh}.
One should be warned that there are not ubiquitous recipes for what concerns
the choice of resummation and/or the multipolar order to use: each multipolar
amplitude can, in principle, be treated separately from the others.
In practice, following Paper~I, we attempt to comply at the idea of
using 6PN-accurate hybrid orbital (i.e. nonspinning) amplitudes that are,
whenever possible, resummed using Pad\'e approximants. By ``hybrid'' we mean that
the $\nu$-dependent terms, analytically known up to 3PN accuracy,
are augmented by test-particle terms up to getting a relative 6PN order
in all the residual waveform amplitudes. The spin sector takes advantage of some
of, but not all, the new PN information at next-to-next-to-leading-order (NNLO) that
was recently presented in Ref.~\cite{Cotesta:2018fcv} adapting (yet unpublished)
results of S.~Marsat and A.~Boh\'e. Practically all the structure of the waveform
was discussed in Sec.~IIIB,~IIIC and~IIID and  of Paper~I. Since we are adopting
the same notation and nomenclature introduced there, it is not worth to repeat it
here. We only recall that the acronym NQC stands for ``next-to-quasi-circular''
and that $f_\lm$'s or $\rho_\lm\equiv (f_\lm)^{1/\ell}$ functions are the residual
waveform amplitudes. For resumming the $m=\text{odd}$ mode waveform amplitudes
we implement the orbital-factorization and resummation scheme
of Ref.~\cite{Messina:2018ghh}. In brief, following the notation of this latter
reference, our analytical choices for the waveform amplitudes are listed in
Table~\ref{tab:PNchoices}. We give below more details, discussing explicitly,
and separately, the orbital and spin sectors.

\subsubsection{Orbital sector} All $\nu$-dependent terms in the multipolar amplitudes
up to $\ell=6$ are augmented with test-particle terms up to relative (hybrid) order 6PN
{\it except} for the $(2,2)$ and $(3,1)$ modes, that rely on $3^{+2}$~PN information,
consistently with previous work. For most of the modes, such 6PN-accurate,
hybridized, amplitudes are additionally Pad\'e resummed consistently with the
choice made in the extreme-mass-ratio limit in Ref.~\cite{Messina:2018ghh}.
Note however that some multipoles actually behave {\it better} (when compared
with test-mass numerical data) when they are left in nonresummed form.
Table~\ref{tab:pn_order} lists, in the second column, the analytical representation
chosen for the orbital factors up to $\ell=m=5$.
We address the Pad\'e approximant of order $(n,d)$ with the usual notation
$P^n_d$, where $n$ is the polynomial order of the numerator and $d$ the one
of the denominator. For notational consistency, we also indicate
with $P^n_0$ the Taylor-expanded form of the functions. The subdominant modes 
that do not contain spin information are not reported in the table. 
The $(5,1)$, $(6,1)$, $\ell=7$ and $\ell=8$ modes are kept in Taylor-expanded
form at (global) $3^{+2}$~PN order for simplicity, consistently with previous work.
All other $\rho_\lm^{\rm orb}$'s with $\ell=5$ and $\ell=6$ are resummed
as $P^4_2\left(\rho_\lm^{\rm orb}\right)$ approximants. 

\subsubsection{Spin sector} The spin-dependent terms in the waveform amplitudes are 
incorporated only in those multipoles where the $\nu$-dependence beyond the leading order
is analytically known, i.e. up to $\ell=m=5$, as illustrated in Table~\ref{tab:pn_order}.
For some modes, the $\nu$-dependent information is augmented with spinning-particle
terms, according to the hybridization procedure discussed in Ref.~\cite{Messina:2018ghh}.
Note that the analytical resummation of the residual waveform amplitudes to improve their
robustness in the strong-field, fast-velocity regime when $m=\text{even}$ is not the
same as when $m=\text{odd}$.  
  For the $m=\text{even}$ modes, the residual amplitudes are written as
  \be
  P^n_d\left[\rho^{\rm orb}_\lm\right] + \rho_\lm^{\rm S},
  \ee
  where we explicitly indicate the fact that the orbital part is Pad\'e resummed
  (including in this nomenclature also the plain Taylor-expansion) according to
  Table~\ref{tab:pn_order}. By contrast, the spin-dependent part is kept in
  Taylor-expanded form, with the (relative) PN order given in Table~\ref{tab:pn_order}.
  Here, the notation $T[\rho_{\lm}^{\rm S}]$ is an explicit reminder that we are
  using the $\rho_\lm^{\rm S}$ in Taylor-expanded form.
  The amount of analytical information used in each mode is listed in the fifth
  column of the table. First of all, note that we {\it do not} include the
  NNLO spin-orbit term in $\rho_{22}^{\rm S}$ that was recently computed and is part of 
  either {\tt SEOBNRv4}~\cite{Bohe:2016gbl} and {\tt SEOBNRv4HM}~\cite{Cotesta:2018fcv}.
  As it was pointed out already in Ref.~\cite{Nagar:2016ayt}, this term has 
  a large impact on the EOB waveform towards merger for large, positive,
  spins, so that the EOB/NR difference is {\it larger} with this term than
  without it (see Fig.~6 of~\cite{Nagar:2016ayt}). By contrast, the NLO-accurate 
  amplitude alone already delivers an excellent representation of the corresponding 
  NR amplitude and thus gives a more robust starting point for the action
  of the NQC factor. We do, however, include the LO cubic-in-spin term in $\rho_{22}^{\rm S}$.
  Browsing the fifth column of Table~\ref{tab:pn_order} the notation adopted
  indicates that the $\nu$-dependent terms in $(\rho_{32}^{\rm S},\rho_{44}^{\rm S},\rho_{42}^{\rm S})$ 
  were {\it hybridized} with some of the higher-order, spin-orbit, terms obtained 
  in the limit of a spinning particle on a Schwarzschild black hole in Ref.~\cite{Nagar:2019wrt}.  
  The rational behind such hybridization procedure is discussed in Sec.~VB of
  Ref.~\cite{Messina:2018ghh} and allows one to incorporate some of the leading-in-$\nu$-dependence
  by suitably ``dressing'' the $\nu=0$ information. One finds that the additional terms are such
  to increase the EOB/NR waveform ampltiude agreement towards merger in a natural way.
  To be explicit, we have 
  \be
  \rho_{32}^{\rm S}=c^{{\rm SO}_{\rm lo}}x^{1/2}+c^{\rm SO_{\rm nlo}}_{32}x^{3/2}+c^{{\rm SO}_{\rm nnlo}}_{32}x^{5/2} ,
  \ee
  where $(c^{{\rm SO}_{\rm lo}},c^{{\rm SO}_{\rm nlo}})$ are the usual known terms with the full
  $\nu$ dependence (see e.g.~\cite{Messina:2018ghh} for their explicit form), while
  \be
  c^{{\rm SO}_{\rm nnlo}}_{32}= -\dfrac{2571199}{1924560}\tilde{a}_0 - \dfrac{1844993}{1924560} \tilde{a}_{12}X_{12},
  \ee
  that reduces to the known spinning test-particle terms when $\nu\to 0$.
  Similarly, $\rho_{44}^{\rm S}$ reads
  \be
  \rho_{44}^{\rm S}=c^{\rm SO_{\rm lo}}_{44}x^{3/2}+c^{\rm SO_{\rm nlo}}_{44}x^{5/2}+c^{\rm SO_{\rm nnlo}}_{44}x^{7/2},
  \ee
  where
  \begin{align}
    c^{\rm SO_{\rm nlo}}_{44}&=-\dfrac{199}{550}\tilde{a}_0 - \dfrac{491}{550}\tilde{a}_{12}X_{12},\\
    c^{\rm SO_{\rm nnlo}}_{44}&=\dfrac{527001653}{264264000}\tilde{a}_0 + \dfrac{3208967}{264264000}\tilde{a}_{12}X_{12},
  \end{align}
  For $\rho_{42}^{\rm S}$ we have
  \begin{align}
  \rho_{42}^{\rm S}=c^{\rm SO_{\rm lo}}_{42}x^{3/2}+c^{\rm SO_{\rm nlo}}_{42}x^{5/2}+c^{\rm SO_{\rm nnlo}}_{42}x^{7/2}+c^{\rm SO_{\rm nnnlo}}_{42}x^{9/2},
  \end{align}
  where the $\nu$-dressed spinning particle coefficients read
  \begin{align}
    c^{\rm SO_{\rm nlo}}_{42} &=-\dfrac{219}{550}\tilde{a}_0 + \dfrac{92}{275}\tilde{a}_{12}X_{12},\\
    c^{\rm SO_{\rm nnlo}}_{42}&=-\dfrac{329051729}{264264000}\tilde{a}_0 + \dfrac{169512229}{264264000}\tilde{a}_{12}X_{12},\\
    c^{\rm SO_{\rm nnnlo}}_{42}&=-\left(\dfrac{32079746680643}{16482145680000} + \dfrac{17581}{51975}{\rm eulerlog}(x,2)\right)\tilde{a}_0 \nonumber\\
              &\!\!\!\!\!\!\!\!\!\!\!\!-\left(\dfrac{28943192016227}{16482145680000} - \dfrac{10697}{51975}{\rm eulerlog}(x,2)\right)\tilde{a}_{12}X_{12}.
  \end{align}
For the $m=\text{odd}$ modes, we apply in full the factorization of the orbital term
and subsequent resummation of the spin factor with its inverse Taylor representation
as illustrated in Ref.~\cite{Messina:2018ghh}. Recalling the notation therein,
each $m$-odd waveform mode is written as
\be
h_\lm^{(\epsilon)}=h_\lm^{N,(\epsilon)'}\tilde{h}^{(\epsilon)}_\lm ,
\ee
where $h_\lm^{N,(\epsilon)'}$ is the usual Newtonian prefactor~\cite{Damour:2008gu}
with the overall factor $X_{12}$ factorized out, while
\be
\label{eq:hodd}
\tilde{h}^{(\epsilon)}_\lm\equiv X_{12}\hat{h}^{(\epsilon)}_\lm,
\ee
and $\hat{h}^{(\epsilon)}_\lm$ is the usual relativistic correction~\cite{Damour:2008gu}.
The $m$-odd relativistic waveform correction is then factorized as
\begin{equation}
  \tilde{h}^{(\epsilon)}_\lm =\hat{S}^{(\epsilon)}_{\rm eff}\hat{h}_\lm^{\rm tail}e^{{\rm i}\delta_\lm}
  \left[P^n_d\left(\rho_\lm^{\rm orb}\right)\right]^\ell \hat{f}_\lm^{\rm S},
\end{equation}
where $(\hat{h}^{\rm tail}_\lm,\delta_\lm)$ are the usual, well known, tail factor and
residual phase correction~\cite{Damour:2008gu}. 
The spin-dependent $\hat{f}^{\rm S}_\lm$ functions that we use are 
summarized in Table~\ref{tab:pn_order}.
The same table also lists the Pad\'e approximants $P^n_d[\rho_\lm^{\rm orb}]$ adopted
for the orbital factors. For the spin factors, we take advantage of the new NNLO
results of Ref.~\cite{Cotesta:2018fcv}, in particular
those concerning the $\ell=m=5$ mode. This multipole is also resummed
consistently with the others. In particular, it also includes the 2PN-accurate
(or relative LO) spin-square term. The inverse-resummed
factor $\overline{\hat{f}^{S_{(0)}}_{55}}$ explicitly reads
\be
\overline{\hat{f}^{S_{(0)}}_{55}}= \left(1+\dfrac{10}{3}\tilde{a}_0x^{3/2}-\dfrac{5}{2}\tilde{a}_0^2 x^2\right)^{-1}. 
\ee
The global structure of the spin factors is illustrated in Table~\ref{tab:pn_order}
and we do not discuss here any further as it is a straightforward application of
the procedure of Ref.~\cite{Messina:2018ghh} once modified with the new PN
terms published in Ref.~\cite{Cotesta:2018fcv} and the spinning-particle terms
of Ref.~\cite{Nagar:2019wrt}.
\subsubsection{Residual phase corrections $\delta_\lm$}
Let us finally detail the expression of the $\delta_\lm$ we use. 
Following Ref.~\cite{Damour:2012ky}, we mostly use them in Pad\'e resummed
form, augmenting, for some modes, the 3.5PN, $\nu$-dependent terms with the
next, 4.5PN-accurate, contribution in the test-particle limit~\cite{Fujita:2014eta}.
In addition, we only rely on {\it nonspinning} information, although spin-dependent
terms are available~\cite{Cotesta:2018fcv}. Explicitly, the expressions we use read
\begin{widetext}
\begin{align}
\delta_{22} =&~ \frac{7}{3} y^{\frac{3}{2}}\frac{808920 \nu \pi \sqrt{y} + 137388 \pi^2 y + 35 \nu^2 \left(136080 + (154975 - 1359276 \nu) y\right)}{808920 \nu \pi \sqrt{y} + 137388 \pi^2 y + 35 \nu^2 \left(136080 + (154975 + 40404 \nu) y\right)}, \\
\delta_ {21} =&~ \frac {2} {3} y^{\frac {3} {2}}\frac {5992 \pi \sqrt{y} + 2465 \nu (28 - 493 \nu y)}{69020 \nu + 5992 \pi \sqrt{y}} , \\
\delta_ {33} =&~ \frac {13} {10} y^{\frac {3} {2}}\frac {1 + \frac{94770 \pi}{566279 \nu} \sqrt{y}} {1 + \frac{94770 \pi}{566279 \nu} \sqrt{y} + \frac{80897}{3159}\nu y}, \\
\delta_ {32} =&~  \frac {10 + 33 \nu} {15 (1 - 3 \nu)} y^{\frac {3} {2}} \frac{1}{1 -  
	\frac{260 (1 - 3 \nu)}{7(10 + 33\nu)}\pi y^{\frac{3}{2}}
	+\frac{1}{(10  + 33\nu)^2}\left(\frac{91120}{27} + \frac{9112}{9} \nu - \frac{100232}{3}\nu^2 + \frac{130000}{147} \pi^2 - \frac{412880}{49} \nu \pi^2 + \frac{848640}{49} \nu^2 \pi^2\right) y^3}, \\
\delta_ {31} =&~  \frac {13} {30} y^{\frac {3} {2}}\frac {4641 \nu + 1690 \pi \sqrt{y}} {4641 \nu + 1690 \pi \sqrt{y} + 18207 \nu^2 y}, \\
\delta_ {44} =&~  \frac {112 + 219 \nu} {120 (1 - 3 \nu)} y^{\frac{3}{2}}\frac {1}{1  - \frac{201088 (1  - 3  \nu)}{231(112  + 219  \nu)}\pi y^{\frac {3} {2}} -
	\frac{1  - 3  \nu}{(112  + 219  \nu)^2}\left( \frac{49409024}{25} + \frac{96612288}{25}\nu + \frac{8854306816}{17787}\pi^2 - \frac{
	49478908928}{17787} \nu \pi^2\right)y^3}, \\
\delta_ {43} =&~  \frac {486 + 4961 \nu} {810 (1 - 2 \nu)} y^{\frac {3} {2}} \Bigg[1  -\frac{254502 (1- 2\nu)}{77 (486  + 4961  \nu)} \pi y^{\frac {3} {2}} + 
	\frac{1}{(486  + 4961  \nu)^2} \Big(\frac{122106771}{5} + \frac{2004460533}{10} \nu  + \\ \nonumber
	&- \frac{2492887617}{5} \nu^2 + \frac{45723320316}{5929}\pi^2 - \frac{415427177628}{5929} \nu \pi^2 + \frac{647961073992}{5929} \nu^2 \pi^2\Big) y^3\Bigg]^{-1}, \\ 
\delta_ {42} =&~  \frac {7 (1 + 6 \nu)} {15 (1 - 3 \nu)} y^{\frac {3} {2}}\frac {1} {1  - 
	\frac{6284 (1  - 3  \nu)}{1617 (1  + 6  \nu)} \pi y^{\frac {3} {2}} + 
	\frac{1 - 3  \nu}{(1  + 6 \nu)^2} \left(\frac{6893}{175} + \frac{41358}{175}\nu + \frac{8646784}{871563}\pi^2 - \frac{22195088}{290521}\nu \pi^2\right) y^3}, \\
\delta_ {41} =&~  \frac {2 + 507 \nu} {10 (1 - 2 \nu)} y^{\frac {3} {2}} + \frac {1571} {3465}\pi^3 y^3, \\
\delta_ {55} =&~  \frac {96875 + 857528 \nu} {131250 (1 - 2 \nu)} y^{\frac {3} {2}},
\end{align}
\end{widetext}
where $y = \hat{H}_{\rm EOB} \Omega$, with $\hat{H}_{\rm EOB}$ and $\Omega$ being the energy and orbital frequency of the binary system respectively.
For completeness, let us also list the original Taylor expanded functions that are then resummed using the Pad\'e approximants explicitly written above.
\begin{align}
\delta_{22}^{\rm Taylor} &= \frac73 y^\frac32 - 24 \nu y^\frac52 + \frac{428}{105} \pi y^3 \nonumber\\
&+ \left(\frac{30995}{42} + \frac{962}{5}\nu\right)\frac{\nu}{27} y^\frac72, \\
\delta_{21}^{\rm Taylor} &= \frac23 y^\frac32 - \frac{493}{42} \nu y^\frac52 + \frac{107}{105} \pi y^3, \\
\delta_{33}^{\rm Taylor} &= \frac{13}{10} y^\frac32 - \frac{80897}{2430} \nu y^\frac52 + \frac{39}{7} \pi y^3\ , \\
\delta_{31}^{\rm Taylor} &= \frac{13}{30} y^\frac32 - \frac{17}{10} \nu y^\frac52 + \frac{13}{21} \pi y^3\ ,\\
\delta_{32}^{\rm Taylor}&= \dfrac{10+33\nu}{15(1-3\nu)}y^\frac32+ \frac{52}{21}\pi y^3 \nonumber\\
                                      &+\left(\dfrac{208}{63}\pi^2-\dfrac{9112}{405}\right)y^{9/2}\ ,\\
\delta_{44}^{\rm Taylor}&=\frac{112+219\nu}{120\left(1-3\nu\right)}y^\frac32 \nonumber\\
                                     &+\dfrac{25136}{3465}\pi y^3+\left(\dfrac{201088}{10395}\pi^2-\dfrac{55144}{375}\right)y^{9/2},\\
\delta_{43}^{\rm Taylor}&=\dfrac{486+4961\nu}{810(1-2\nu)}y^{3/2}+\dfrac{1571}{385}\pi y^3 \nonumber\\
                                     &+\left(-\dfrac{18611}{300}+\dfrac{3142}{385}\pi^2\right)y^{9/2},   \\                                  
\delta_{42}^{\rm Taylor}&=\dfrac{7(1+6\nu)}{15(1-3\nu)}y^{3/2}+\dfrac{6284}{3465}\pi y^3\nonumber\\
                                     &+\left(\dfrac{25136}{10395}\pi^2-\dfrac{6893}{375}\right)y^{9/2} \ .                                    
\end{align}
Comparing with Appendix~D of Ref.~\cite{Damour:2012ky}, we are here explicitly using
4.5PN terms in some of the higher modes, since we found that they improve the
EOB/NR frequency agreement close to merger. In practice, after factorizing
the leading contribution following~\cite{Damour:2012ky} , 
the approximants we use for each mode are: $\delta_{22}\to P^2_2$;
$\delta_{21}\to P^2_1$; $\delta_{33}\to P^1_2$; 
$\delta_{32}\to P^0_2$; $\delta_{31}\to P^1_2$;  $\delta_{44}\to P^0_2$ and $\delta_{43}\to P^0_2$.
\subsection{Multipolar peak, ringdown and next-to-quasi-circular corrections}
The modelization of the peak and postpeak waveform multipole by multipole
is done following precisely the same procedure adopted in the nonspinning case,
but incorporating spin dependence (whenever possible) in all fits.  As we
detail in Appendix~\ref{sec:app_fits}, in practice we include: (i) complete spin-dependence
for what concerns peak quantities and postpeak fits in all $\ell=m$ modes up
to $\ell=5$; (ii) modes like $(2,1)$, $(3,2)$, $(4,3)$ and $(4,2)$ include spin
dependence for peak frequency and amplitude, but they adopt the simpler nonspinning
fits for the parameters entering the postpeak waveform description;
(iii) the $(3,1)$ and $(4,1)$ mode only rely on nonspinning information. 
The values at the NQC determination points are either obtained with dedicated
fits of the corresponding NR quantities, or directly from the postpeak behavior.
All considered, this approach allows one to obtain a rather robust description
of the ringdown waveform all over the parameter space. 
   
\subsection{NR-informed EOB functions: $\bm{a_6^c}$ and $\bm{c_3}$}
\label{sec:params}
Finally, we discuss the NR-informed functions that enter the EOB dynamics.
For $a_6^c(\nu)$, we use the function determined in Paper~I. Note that this was obtained using the Pad\'e resummed $P^4_2[\rho_{22}^{\rm orb}]$
description of the residual $\ell=m=2$ waveform amplitude hybridized
with test-particle terms up to 6PN. For simplicity, we adopt it here
{\it even if} we are here using $\rho_{22}^{\rm orb}$ at $3^{+2}$~PN accuracy.
The differences in the dynamics, at the nonspinning level, are consistent
with the NR uncertainty, so it is not worth to proceed with a new, more
consistent, determination of this function. The expression adopted from
Paper~I is
\be
a_6^c=n_0\dfrac{1+n_1\nu + n_2\nu^2+n_3\nu^3}{1+d_1\nu},
\ee
where
\begin{align}
  n_0 &= 5.9951,\\
  n_1 &=-34.4844,\\
  n_2 &=-79.2997,\\
  n_3 &=713.4451,\\
  d_1 &=-3.167.
\end{align}
This, together with the new analytical description of the spin-sector
of the waveform (and radiation reaction) calls for a new determination of $c_3$.
This is obtained precisely following Sec.~IIB.2 of Ref.~\cite{Nagar:2018zoe},
i.e. by determining the good values of $c_3$ such that the EOB/NR dephasing is
within the nominal NR phase uncertainty at NR merger. This is done using 32 NR
datasets, 30 from SXS and 2 from the BAM code. The configurations used
are listed in Table~\ref{tab:c3}, together with the value of $c_3$ that assures
an EOB/NR phasing at merger that is smaller than (or comparable with) the
nominal numerical uncertainty (see~\cite{Nagar:2018zoe}. Note also that
these values are such to assure that the EOB frequency evolution towards
merger is correctly reproducing the corresponding NR one.
 \begin{table}[t]
   \caption{\label{tab:c3}Binary configurations, first-guess values of $c_3$
     used to inform the global interpolating fit given in Eq.~\eqref{eq:c3fit},
     and the corresponding $c_3^{\rm fit}$ values.}
   \begin{center}
 \begin{ruledtabular}
   \begin{tabular}{lllll}
     $\#$ & ID & $(q,\chi_1,\chi_2)$ & $c_3^{\rm first\;guess}$ & $c_3^{\rm fit}$\\
     \hline
    1 & SXS:BBH:0156 &$(1,-0.95,-0.95)$  & 88 & 87.87\\
    2 & SXS:BBH:0159 &$(1,-0.90,-0.90)$ & 85.5 & 85.54\\
    3 & SXS:BBH:0154 &$(1,-0.80,-0.80)$ & 81  & 80.90\\
    4 & SXS:BBH:0215 &$(1,-0.60,-0.60)$ & 71.5 & 71.72\\
    5 & SXS:BBH:0150 &$(1,+0.20,+0.20)$ & 38.0 & 36.92\\
    6 & SXS:BBH:0228 &$(1,+0.60,+0.60)$ & 22.0 & 21.94\\
    7 & SXS:BBH:0230 &$(1,+0.80,+0.80)$ & 15.5 & 16.25\\
    8 & SXS:BBH:0153 &$(1,+0.85,+0.85)$ & 14.5 & 15.25\\
    9 & SXS:BBH:0160 &$(1,+0.90,+0.90)$ & 14.9 & 14.53\\
    10 & SXS:BBH:0157 &$(1,+0.95,+0.95)$ & 14.3 & 14.20\\
    11 & SXS:BBH:0177 &$(1,+0.99,+0.99)$ & 14.2 & 14.32\\
    12 & SXS:BBH:0004 &$(1,-0.50,0)$ & 54.5 & 56.61 \\
    13 & SXS:BBH:0231 &$(1,+0.90,0)$ & 27.0 & 26.18  \\
    14 & SXS:BBH:0232 &$(1,+0.90,+0.50)$ & 19.0 & 18.38\\
    15 & SXS:BBH:0005 &$(1,+0.50,0)$ & 34.3 & 34.34\\
    16 & SXS:BBH:0016 &$(1.5,-0.50,0)$ & 57.0 & 58.19\\
    17 & SXS:BBH:0255 &$(2,+0.60,0)$ & 29.0 & 29.75\\
    18 & SXS:BBH:0256 &$(2,+0.60,+0.60)$ & 22.8 & 23.68\\
    19 & SXS:BBH:0257 &$(2,+0.85,+0.85)$ & 15.7 & 17.73\\ 
    20 & SXS:BBH:0036 &$(3,-0.50,0)$ &  60.0 & 60.39\\
    21 & SXS:BBH:0267 &$(3,-0.50,-0.50)$ & 69.5 & 65.28\\ 
    22 & SXS:BBH:0174 &$(3,+0.50,0)$ &  30.0 & 31.20\\
    23 & SXS:BBH:0286 &$(3,+0.50,+0.50)$ & 26.0 & 27.28\\
    24 & SXS:BBH:0291 &$(3,+0.60,+0.60)$ & 23.4 & 24.22\\
    25 & SXS:BBH:0293 &$(3,+0.85,+0.85)$ & 16.2 & 18.48\\
    26 & SXS:BBH:0060 &$(5,-0.50,0)$ & 62.0 & 61.91\\
    27 & SXS:BBH:0110 &$(5,+0.50,0)$ & 31.0 & 29.97\\
    28 & SXS:BBH:1375 &$(8,-0.90,0)$ & 64.0 & 78.27\\
    29 & SXS:BBH:0064 &$(8,-0.50,0)$ & 57.0 & 63.23\\
    30 & SXS:BBH:0065 &$(8,+0.50,0)$ & 28.5 & 28.86\\
    31 & BAM &$(8,+0.80,0)$ & 24.5 & 20.85\\
    32 & BAM &$(8,+0.85,+0.85)$ & 16.3 & 18.11 
 \end{tabular}
 \end{ruledtabular}
 \end{center}
 \end{table}
The data of Table~\ref{tab:c3} are fitted with a global function as
$c_3(\nu,\tilde{a}_0,\tilde{a}_{12})$ that is actually simplified with
respect to previous work. The fit template reads
\begin{align}
  \label{eq:c3fit}
& c_3(\tilde{a}_1,\tilde{a}_2,\nu)=
p_0\dfrac{1+n_1\tilde{a}_0+n_2\tilde{a}_0^2+n_3\tilde{a}_0^3+n_4\tilde{a}_0^4}{1+d_1\tilde{a}_0}\nonumber\\
&+ p_1 \tilde{a}_0\nu\sqrt{1-4\nu}+p_2\left(\tilde{a}_1-\tilde{a}_2\right)\nu^2,
\end{align}
where the parameters are
\begin{align}
p_0&=\;\;\;45.235903,\\
n_1&=-1.688708,\\
n_2&=\;\;\;0.787959,\\
n_3&=-0.018080,\\
n_4&=-0.001906,\\
d_1&=-0.751479,\\
p_1&= \;\;\;47.3756,\\
p_2&= -36.1964.
\end{align}
\begin{figure}[t]
\begin{center}
\includegraphics[width=0.45\textwidth]{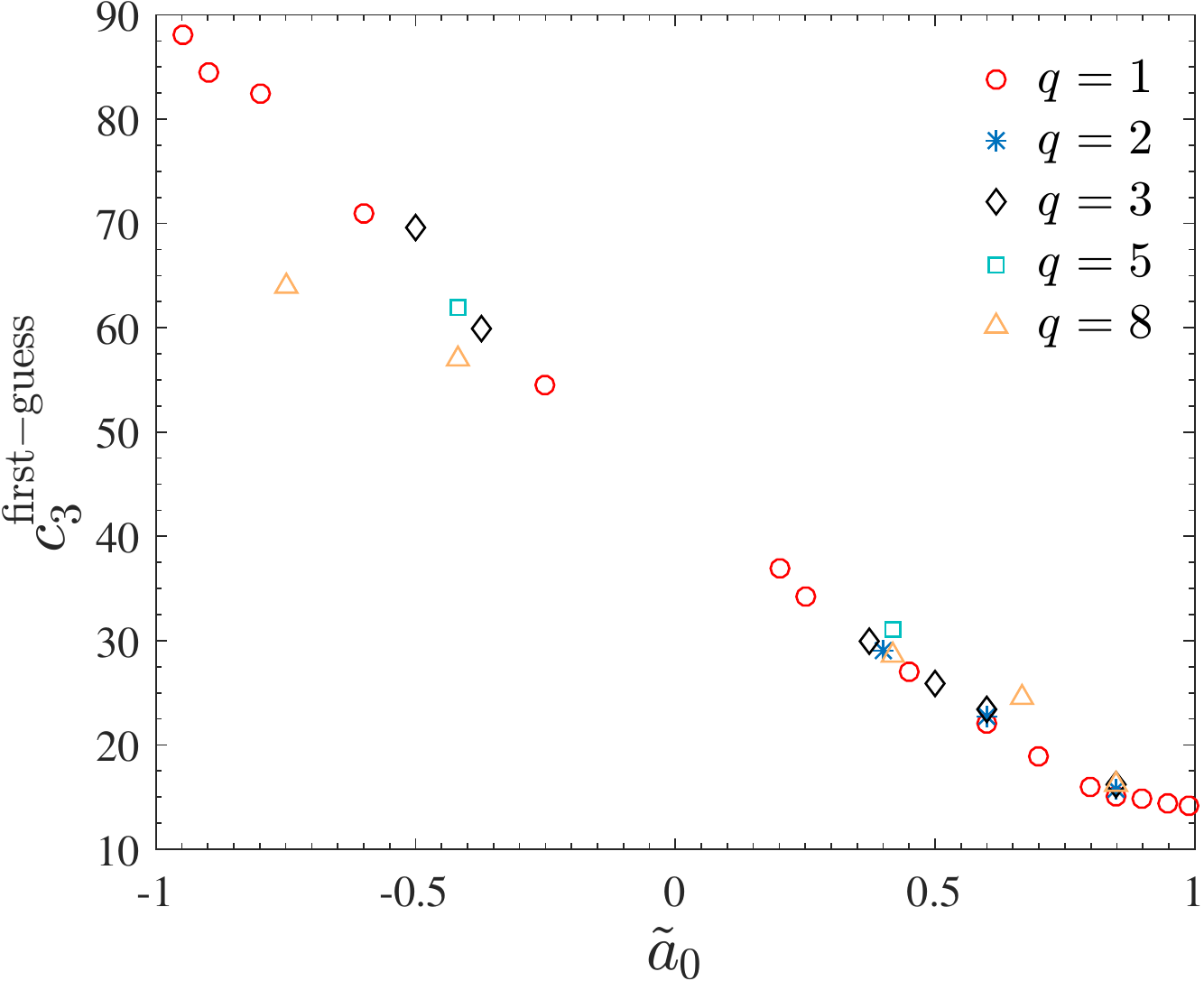}
\caption{The first-guess $c_3$ values of Table~\ref{tab:c3} versus
  the spin variable $\tilde{a}_0\equiv S_{1}/(m_1 M)+S_2/(m_2M)$.
  The unequal-spin and unequal-mass points can be essentially
  seen as a correction to the equal-mass, equal-spin values.}
\label{fig:c3}
\end{center}
\end{figure}
Figure~\ref{fig:c3} highlights that the span of the ``best'' (first-guess)
values of $c_3$ is rather limited (especially for positively aligned spins)
around the equal-mass, equal-spin case. At a practical level, this eases up
the fitting procedure, that, following Ref.~\cite{Nagar:2018zoe}, is performed
in two steps. First, one fits the equal-mass, equal-spin data with
a quasi-linear function of $\tilde{a}_0=\tilde{a}_1+\tilde{a}_2$
with $\tilde{a}_1=\tilde{a}_2$. This delivers the six
parameters $(p_0,n_1,n_2,n_3,n_4,d_1)$. Note that the
analytical structure of the fitting function was chosen in order
to accurately capture the nonlinear behavior of $c_3$ for $\tilde{a}_0\to 1$.
In the second step one subtracts this fit, computed for the unequal-mass,
unequal-spin data, from the corresponding $c_3^{\rm first-guess}$ values and
fits the residual. This gives the parameters $(p_1,p_2)$.
The novelty with respect to Ref.~\cite{Nagar:2018zoe} is that,
thanks to the new analytical improvements, one finds that the unequal-spin
and unequal-mass correction can be represented, in Eq.~\eqref{eq:c3fit},
with acceptable accuracy, only with the two parameters $(p_1,p_2)$,
as we shall illustrate quantitatively in Sec.~\ref{sec:barF}, after an 
assessment of the accuracy of the NR waveforms at our disposal. 
\begin{table*}[t]
	\begin{center}
		\begin{ruledtabular}
			\begin{tabular}{l c l | l c c| cc c| cc |lll}
				& & & \multicolumn{2}{c}{Parameter interval ranges} &&\multicolumn{2}{c}{Waveform count $\#$}& &  && \multicolumn{2}{l}{{$\bar{F}_{\rm LevH/LevM }$ } }& \\
				& & & $q\equiv m_1/m_2$ & $\chi_{1,2}$  & & total & with LevM		& & $\langle N_{\rm orb}\rangle$ && $\bar{F}^{\rm max}_{\rm NR/NR}$ & $ \langle \bar{F}^{\rm max}_{\rm NR/NR}\rangle$& \\
				\hline
				&  {\it Calibration set} & \\
				\hline\hline
				& \texttt{SXS} 	&& $[1.0,\ 10.0]$ 	&  $0$ 							&&	$19$ 	& $18$ & & $21.98$ &  & $0.075\%$ 	& $0.0092\%$ & \\
				& \texttt{SXS}	  	&& $[1.0,\ 1.0]$  	& $\left[-0.95, 0.9942\right]$  &&	$38$ 	& $37$ & & $22.77$ &  & $0.22\%$ 	& $0.020\%$ & \\
				&  \texttt{SXS}					&& $[1.3,\ 8.0]$  	& $\left[-0.9, 0.96\right]$  	&&	$78$ 	& $73$ & & $25.09$	&  & $0.11\%$ 	& $0.0088\%$ & \\\hline
				& \texttt{BAM} 					&& $[4.0,\ 18.0]$  	& $0$  							&&	$3$  	& $-$  & & $8.11$	&  & $<0.1\%$ & $<0.1\%$ & \\			
				&  								&& $[2.0,\ 18.0]$  	& $\left[-0.85, 0.85\right]$  	&&	$16$ 	& $-$  & & $11.13$ 	&  & $<0.1\%$ & $<0.1\%$ &\\
				\hline
				& {\it Validation set} & \\
				\hline\hline
				& \texttt{SXS} 		&& $[1.0,\ 10.0]$  	& $0$  							&&	$67$ 	& $52$ & & $24.98$ &  & $0.066\%$ & $0.0050\%$ &\\
				& \texttt{SXS}  				&& $[1.0,\ 1.16]$  	& $\left[-0.97, 0.998\right]$  	&&	$79$ 	& $77$ & & $20.29$ &  & $0.0093\%$ & $0.0029\%$ &\\
				& \texttt{SXS}  							&& $[1.17,\ 8.0]$  	& $\left[-0.9, 0.95\right]$  	&&	$309$ 	& $287$& & $20.74$ &  & $0.056\%$ & $0.0052\%$ &\\\hline
				& long \texttt{SXS}  	&& $[1.41,\ 1.83]$  & $\left[-0.5, 0.5\right]$  	&&	$5$ 	& $5$  & & $144.05$ &  & $1.52\%$ & $0.98\%$ &\\
			\end{tabular}
		\end{ruledtabular}
		\caption{\label{tab:NR_summary} Numerical Relativity datasets used in this work.  From left to right, the columns  report: catalog origin and use; 
		interval of parameters covered for the mass ratio $q$ and the spins $\chi_{1,2}$; total number of waveforms in the particular sub-catalog;
			the number of \texttt{SXS} data with a second resolution LevM available; the average waveform length expressed in number of orbits, 
			$\langle N_{\rm orb}\rangle$, counted here between the relaxation time  (i.e., after the junk radiation) and the waveform amplitude peak; 
			the absolute  maximum $\bar{F}^{\rm max}_{\rm NR/NR}$ and the average of the individual maxima $ \langle \bar{F}^{\rm max}_{\rm NR/NR}\rangle$ 
			of the unfaithfulness $\bar{F}_{\rm NR/NR}$ computed between the highest, LevH,  and second highest, LevM, resolutions. 
			The $\bar{F}_{\rm NR/NR}$-uncertainties versus total mass $M$ are depicted in Fig.~\ref{fig:barF_SXS_error} (spinning configurations)  
			and Fig.~\ref{fig:barF_SXS_nospin} (nonspinning configurations).}
	\end{center}
\end{table*}
\begin{figure*}[t]
	\begin{center}
		\includegraphics[width=0.45\textwidth]{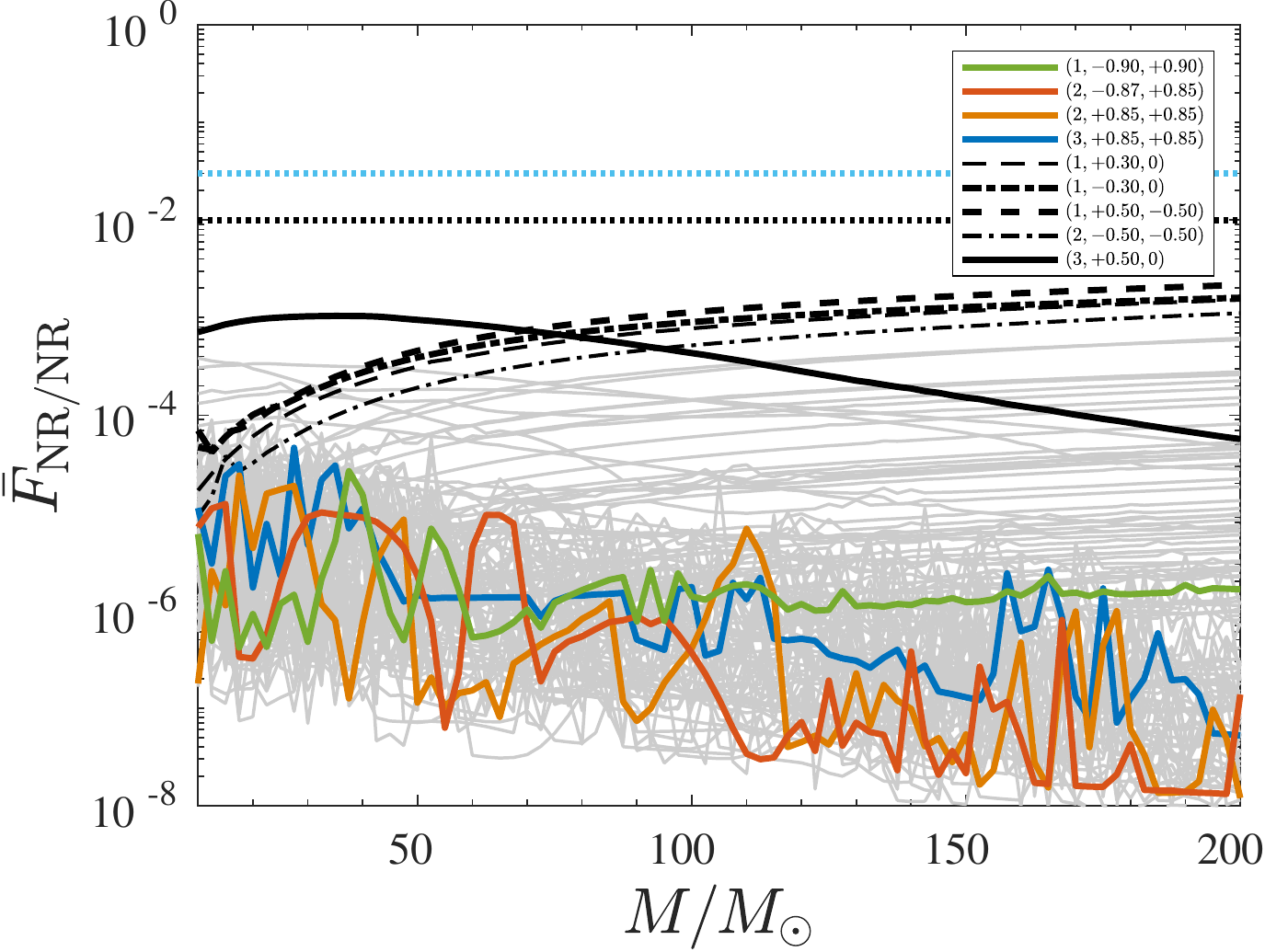}
		\includegraphics[width=0.45\textwidth]{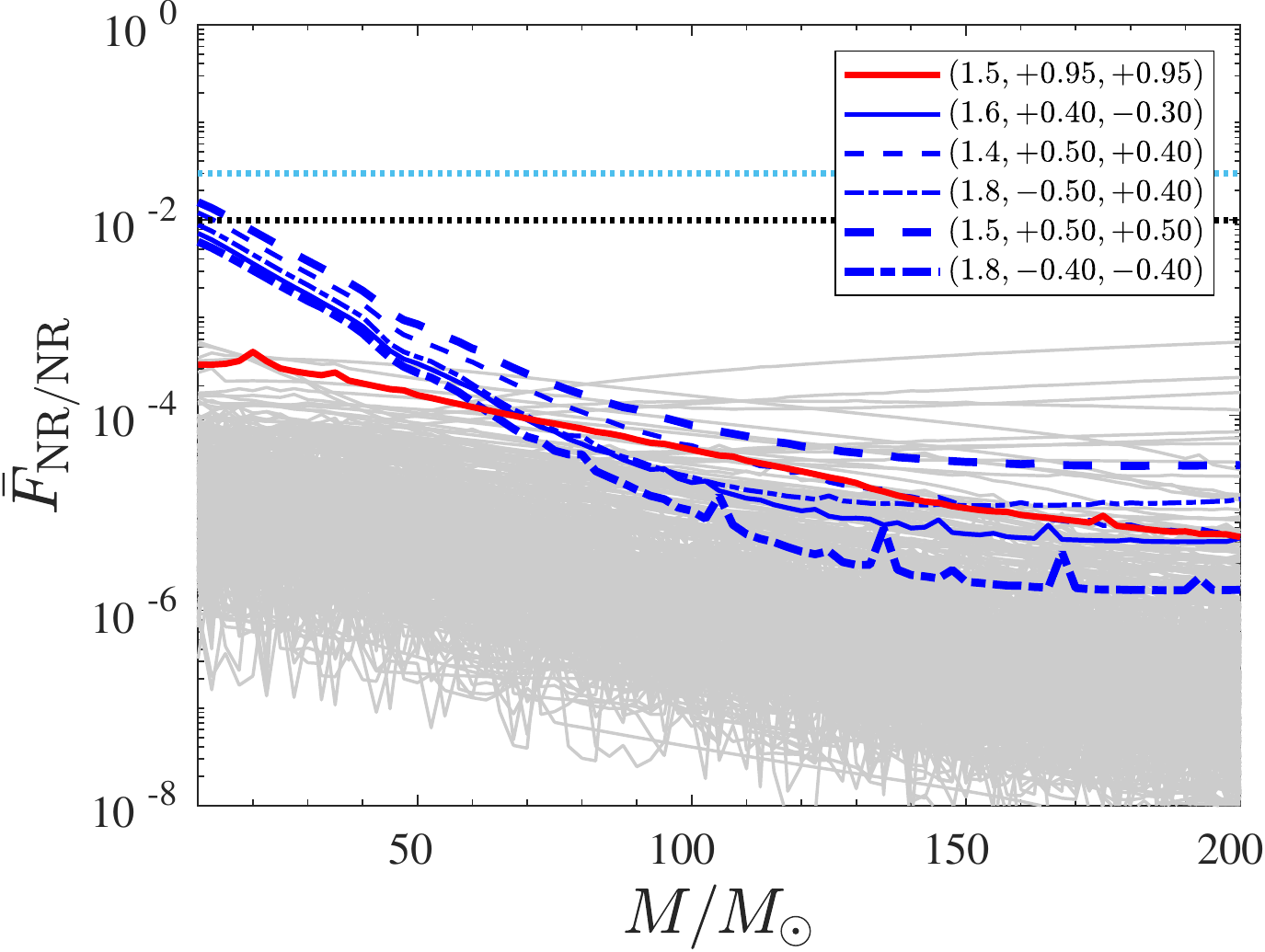}
		\caption{NR/NR unfaithfulness uncertainty computed  from Eq.~\eqref{eq:barF}
		        using the highest and next to highest resolution waveform for each SXS dataset.			
			The left panel  refers to  the 116 spinning waveforms used to globally 
			inform the model (either $c_3$ or ringdown) that were already available
			the the time  of Ref.~\cite{Nagar:2018zoe}. The right panel refers to the 
			393 spinning datasets recently released and discussed in Ref.~\cite{Boyle:2019kee}. 
			The same quantity for nonspinning waveforms is computed in  
			Appendix~\ref{sec:nospin_lim} and shown in Fig.~\ref{fig:barF_SXS_nospin}.}
		\label{fig:barF_SXS_error}
	\end{center}
\end{figure*}
\section{Numerical Relativity waveforms}
\label{sec:NR_catalog}
\subsection{Waveforms overview}
The NR data used here were separated into two categories (see Table~\ref{tab:NR_summary}).
On the one hand, a set of waveforms used for the \textit{calibration} of the postpeak and ringdown waveform;
on the other hand, a set used for the \textit{validation} of the full waveform model\footnote{With the exceptions 
of the fitting of a small set of problematic parameters. These parameters are subdominant 
and not well resolved in the NR data available in the \textit{calibration} set. 
See Appendix~\ref{sec:app_fits} for further details.}.
The \textit{postpeak-calibration} set consists of the following:
(i) we enlarge the set of 23 nonspinning waveforms (19 SXS, 3 BAM, 1 test-particle)
used in~\cite{Nagar:2019wds} ; (ii) we use 38 SXS, spin-aligned, 
equal-mass waveforms with spins between $-0.95\leq \chi_{1,2}\leq 0.9942$, see 
Table~\ref{tab:equal_SXS}; and (iii) 78 SXS spin-aligned, unequal-mass, waveforms 
going up to mass ratio $q=8$ with spins in the range $-0.9\leq \chi_{1,2}\leq 0.96$, 
see Tables~\ref{tab:q2_SXS_old}~--~\ref{tab:q3_SXS_old}, plus a single, high-quality, 
waveform with $(q,\chi_1,\chi_2)=(8,-0.9,0)$; (iv) 16 \texttt{BAM}, spin-aligned, 
unequal-mass waveforms with $2\leq q \leq 18$, encompassing
two $q=8$ waveforms with $\chi_1=\chi_2=\pm 0.85$ and 
two $q=18$ waveforms with $\chi_1=\pm0.8$ and $\chi_2=0$, see Table~\ref{tab:BAM}.
All this, is complemented (v) by a sample of waveforms for a test-particle 
inspiralling and plunging on a Kerr black hole~\cite{Harms:2014dqa} 
with dimensionless black hole spin $\hat{a}$ in the interval $-0.99\leq \hat{a}\leq 0.99$.
The 154 NR waveforms (excluding the test-particle waveforms)  in 
the \textit{Calibration} set contain an average length of $22.35$ orbits, 
while the eccentricity never exceeds $0.004$. 

The \textit{validation} set consists of 460 waveforms from the SXS catalog~\cite{SXS:catalog}. 
The waveforms span mass ratios up to $q=8$ and spins in the range $-0.97\leq\chi_{1,2}\leq 0.998$.
This set includes 5 long waveforms with an average length of $144.05$ orbits
between the relaxation time and the peak of the dominant mode. The average
length of the remaining waveforms is of   $21.29$ orbits. 
Eccentricity is limited to $0.001$. The waveforms are listed in
Tables~\ref{tab:SXS_new1}~--~\ref{tab:SXS_new8}. Further details on the 
SXS catalog can be found in Refs.~\cite{Boyle:2019kee,Buchman:2012dw,
Chu:2009md,Hemberger:2013hsa, Scheel:2014ina,Blackman:2015pia,
Lovelace:2011nu,Lovelace:2010ne,Lovelace:2014twa,Mroue:2013xna,
Kumar:2015tha,Chu:2015kft}.
The nonspinning datasets are listed in Tables~\ref{tab:nospin_1}-\ref{tab:nospin_2}.

\subsection{Estimating NR uncertainties for the $\ell=m=2$ mode}
\label{sec:NR_error}
The most recent update of the SXS catalog is detailed in Ref.~\cite{Boyle:2019kee}.
In particular, that reference gave an estimate of the NR uncertainty due to numerical truncation error 
on each waveform (either precessing or nonprecessing) by computing the maximal 
unfaithfulness (or mismatch, see below), in flat noise, between  the $\ell=m=2$ waveform computed at 
highest and second highest resolutions available. This is found to be $\sim 10^{-4}$, 
that is then taken as a reliable estimate of the NR error. To ease the reader, we perform 
again here this uncertainty computation, although we (i) restrict it only
to the case of nonprecessing waveform and (ii) we use the zero-detuned,
high-power noise spectral density of Advanced LIGO~\cite{aLIGODesign_PSD}. 
The uncertainty of the \texttt{BAM} waveforms was estimated in~\cite{Khan:2015jqa} 
and will be referenced and summarized for the practical purposes of this work. 
Considering two waveforms $(h_1,h_2)$, the unfaithfulness
is a function of the total mass $M$ of the binary and is defined as
\be
\label{eq:barF}
\bar{F}(M) \equiv 1-F=1 -\max_{t_0,\phi_0}\dfrac{\langle h_1,h_2\rangle}{||h_1||||h_2||},
\ee
where $(t_0,\phi_0)$ are the initial time and phase, $||h||\equiv \sqrt{\langle h,h\rangle}$,
and the inner product between two waveforms is defined as 
$\langle h_1,h_2\rangle\equiv 4\Re \int_{f_{\rm min}^{\rm NR}(M)}^\infty \tilde{h}_1(f)\tilde{h}_2^*(f)/S_n(f)\, df$,
where $\tilde{h}(f)$ denotes the Fourier transform of $h(t)$, $S_n(f)$ is the zero-detuned,
high-power noise spectral density of Advanced LIGO~\cite{aLIGODesign_PSD} and
$f_{\rm min}^{\rm NR}(M)=\hat{f}^{\rm NR}_{\rm min}/M$ is the initial frequency of the
NR waveform at highest resolution, i.e. the frequency measured after the junk-radiation
initial transient.
Waveforms are tapered in the time-domain so as to reduce high-frequency 
oscillations in the corresponding Fourier transforms.
Figure~\ref{fig:barF_SXS_error} illustrates the outcome of Eq.~\eqref{eq:barF} when
$(h_1,h_2)$ are the $\ell=m=2$ waveforms corresponding to the highest and 
second-highest resolution available for each SXS dataset.
The left panel of Fig.~\ref{fig:barF_SXS_error} displays $\bar{F}(M)$ 
for the 116 spinning waveforms of the \textit{Calibration} set; 
in the right panel, we have the  393 spinning waveforms 
in the \textit{Validation}. For almost all waveforms, the uncertainty  is below $0.5\%$,
except for the 5 long \texttt{SXS} (blue in the right panel of the figure) 
that will deserve a dedicate discussion in Sec.~\ref{sec:longNR} below.
As a very conservative, global, estimate of the NR uncertainty, we take it to 
be at the $0.5\%$ level. This choice is made to prevent over fitting of 
the NR-informed parameters, although we will see that very often a 
much better EOB/NR agreement arises naturally.
Finally, note that the analysis of the quality of the \texttt{SXS} data is here
limited to the uncertainties due to the numerical truncation error, because, as pointed out 
in Sec.~4 of Ref.~\cite{Boyle:2019kee}, is the largely dominant one.
The accuracy of the \texttt{BAM} waveforms was studied in Ref.~\cite{Khan:2015jqa},
considering several uncertainty sources. In particular, Figs.~2 and 3 
of Ref.~\cite{Khan:2015jqa}  illustrate that the NR uncertainty is $\bar{F}\approx 0.1\%$ 
or less. Similarly to the SXS case, and to avoid overfitting and be conservative,
we assume the uncertainties on alla BAM waveforms at the $0.3\%-0.5\%$  level
and use this as target for EOB/NR comparisons. 

\subsection{Long-inspiral Numerical Relativity waveforms}
\label{sec:longNR}
Let us comment on the 5, very-long, waveforms listed  in Table~\ref{tab:SXS_long}.
All these waveforms show an inspiral of over 100 orbits before a common horizon 
appears\footnote{ {\tt SXS:BBH:1110} is excluded from this analysis since the waveform 
needs additional post-processing.}.
The unfaithfulness between the two highest resolution levels is shown as blue lines in Fig.~\ref{fig:barF_SXS_error}.
All dataset show a rather large $\bar{F}$ for low masses, up to $1.5\%$ for {\tt SXS:BBH:1415}, 
that then decreases  to the average $\bar{F}$ around $60M_{\odot}$. 
This suggests that the long inspiral is more sensitive to resolution and/or 
other systematics effects, so that the numbers of Fig.~\ref{fig:barF_SXS_error} should 
be taken as a rather conservative uncertainty estimates. 

\begin{figure*}[t]
\begin{center}
\includegraphics[width=0.45\textwidth]{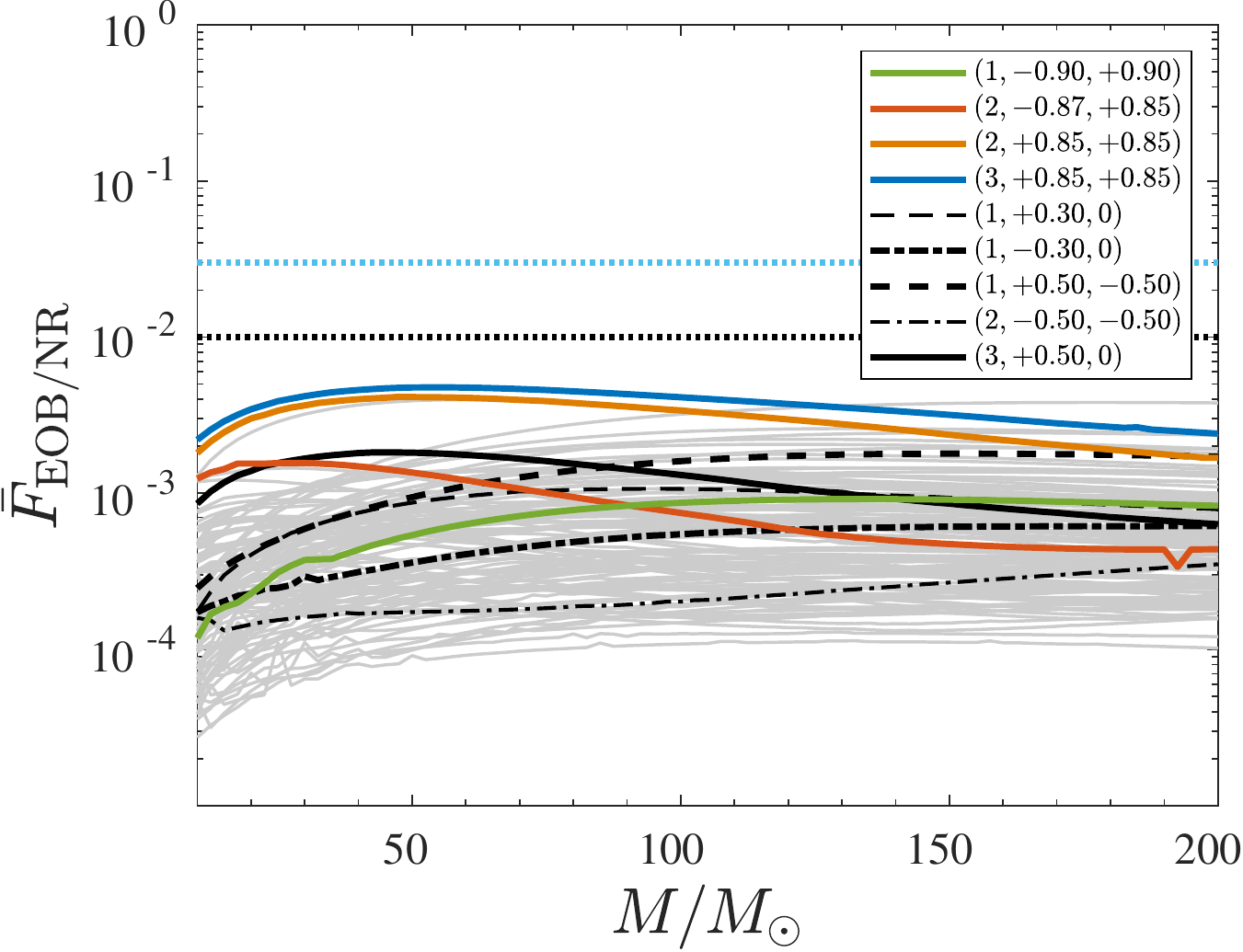}
\hspace{5mm}
\includegraphics[width=0.45\textwidth]{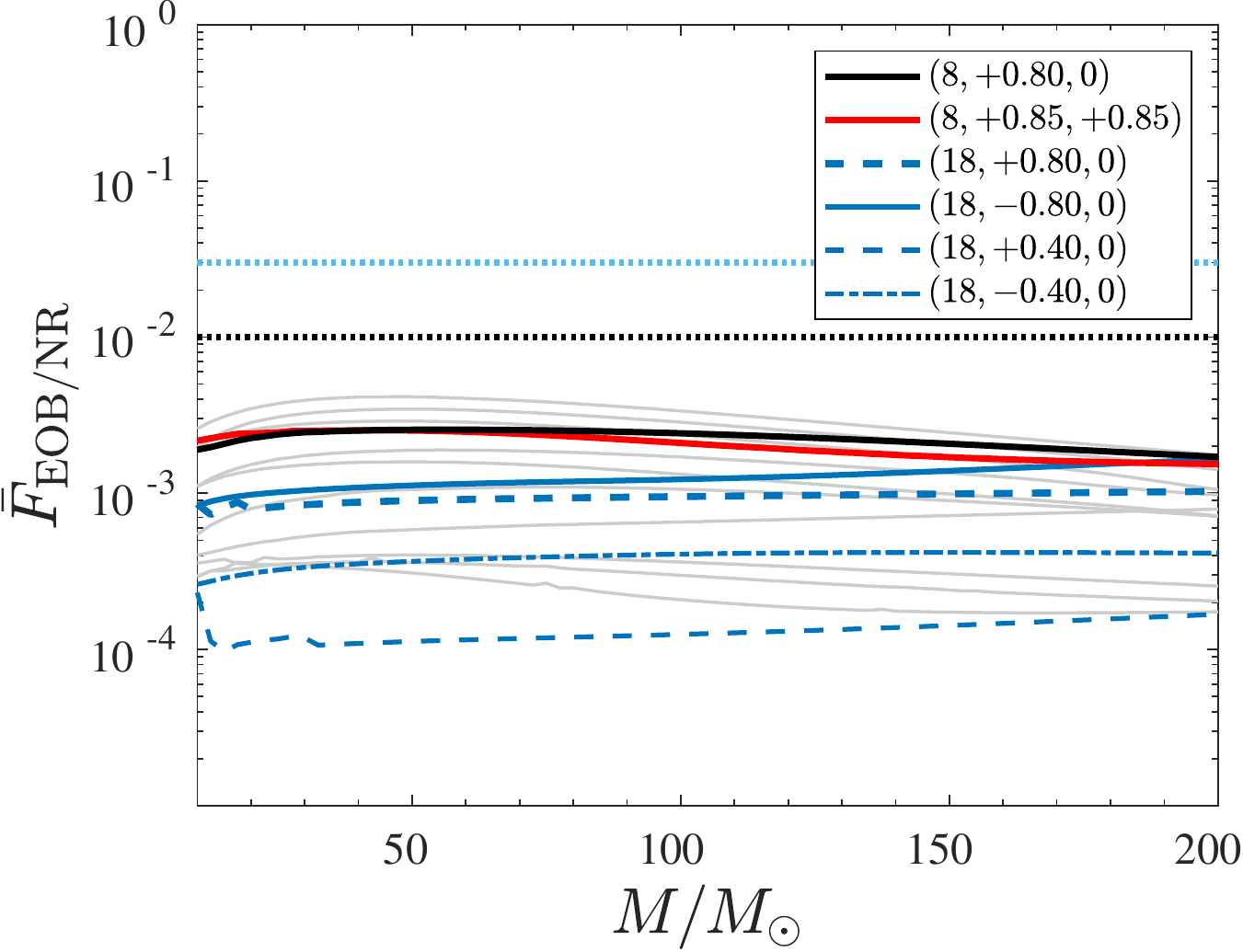}
\caption{EOB/NR unfaithfulness for the $\ell=m=2$ mode obtained from Eq.~\eqref{eq:barF}. 
Left panel: computation using SXS waveforms publicly released before February 3, 2019. 
Right panel: same computation done with  BAM waveform data. As explained in Sec.~\ref{sec:params}, 
a subset of all this data (see Table~\ref{tab:c3}) is used to inform the $c_3$ EOB function. 
Comparison with Figs.~1 and~3 of Ref.~\cite{Nagar:2018zoe} 
allows one to appreciate the improvement with respect to the 
original implementation\footnote{The reader should actually note 
that we changed from the, outdated, zero-detuned, high-power 
noise spectral density of Ref.~\cite{dcc:2974} used in Ref.~\cite{Nagar:2018zoe}, 
to its most recent realization, Ref.~\cite{aLIGODesign_PSD}.} 
of \TEOBResumS{}. Comparison with Fig.~\ref{fig:barF_SXS_error} 
highlights that the $\bar{F}_{\rm EOB/NR}$ 
is either of the order of, or larger than the NR/NR uncertainties.}
\label{fig:barF_SXSBAM}
\end{center}
\end{figure*}

\section{The $\ell=m=2$ mode: EOB/NR unfaithfulness}
\label{sec:barF}
\begin{figure}[t]
\begin{center}
\includegraphics[width=0.45\textwidth]{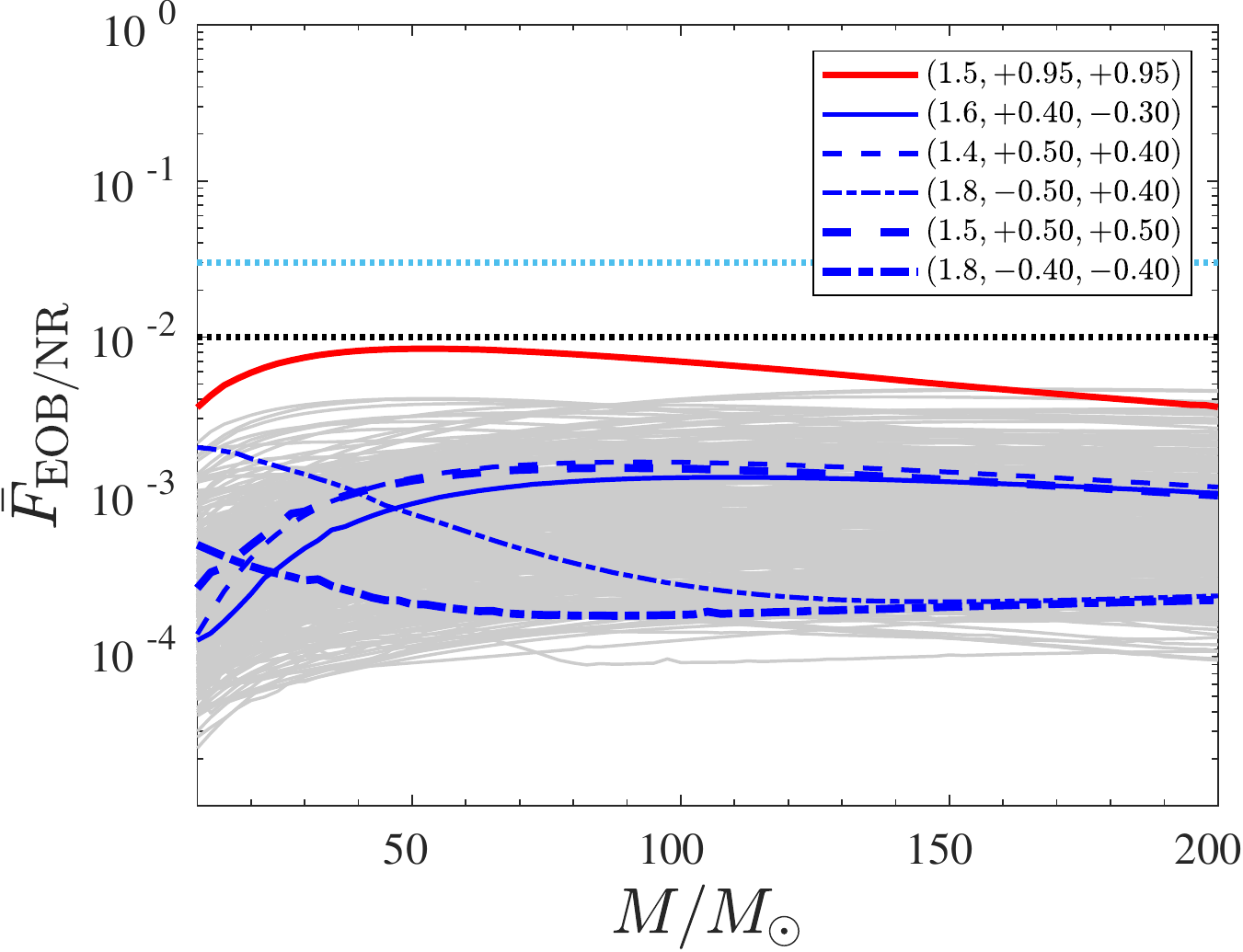}
\caption{\label{fig:barF_SXS_new}EOB/NR $\ell=m=2$ unfaithfulness computation 
with SXS waveform data publicly released after February 3, 2019.
None of these datasets was used to inform the model in the dynamical 
EOB functions $(a_6^c,c_3)$, although several were used for the 
postmerger waveform part. It is remarkable that $\bar{F}^{\rm max}_{\rm EOB/NR}$ is always 
below $0.4\%$ except for a single outlier, red online, that however 
never exceeds $0.85\%$. The plot includes five exceptionally long 
waveforms, each one developing more than 139 GW cycles before 
merger,  SXS:BBH:1412, 1413, 1414, 1415 and 1416 (blue online).}
\end{center}
\end{figure}

\begin{figure}[t]
\begin{center}
\includegraphics[width=0.45\textwidth]{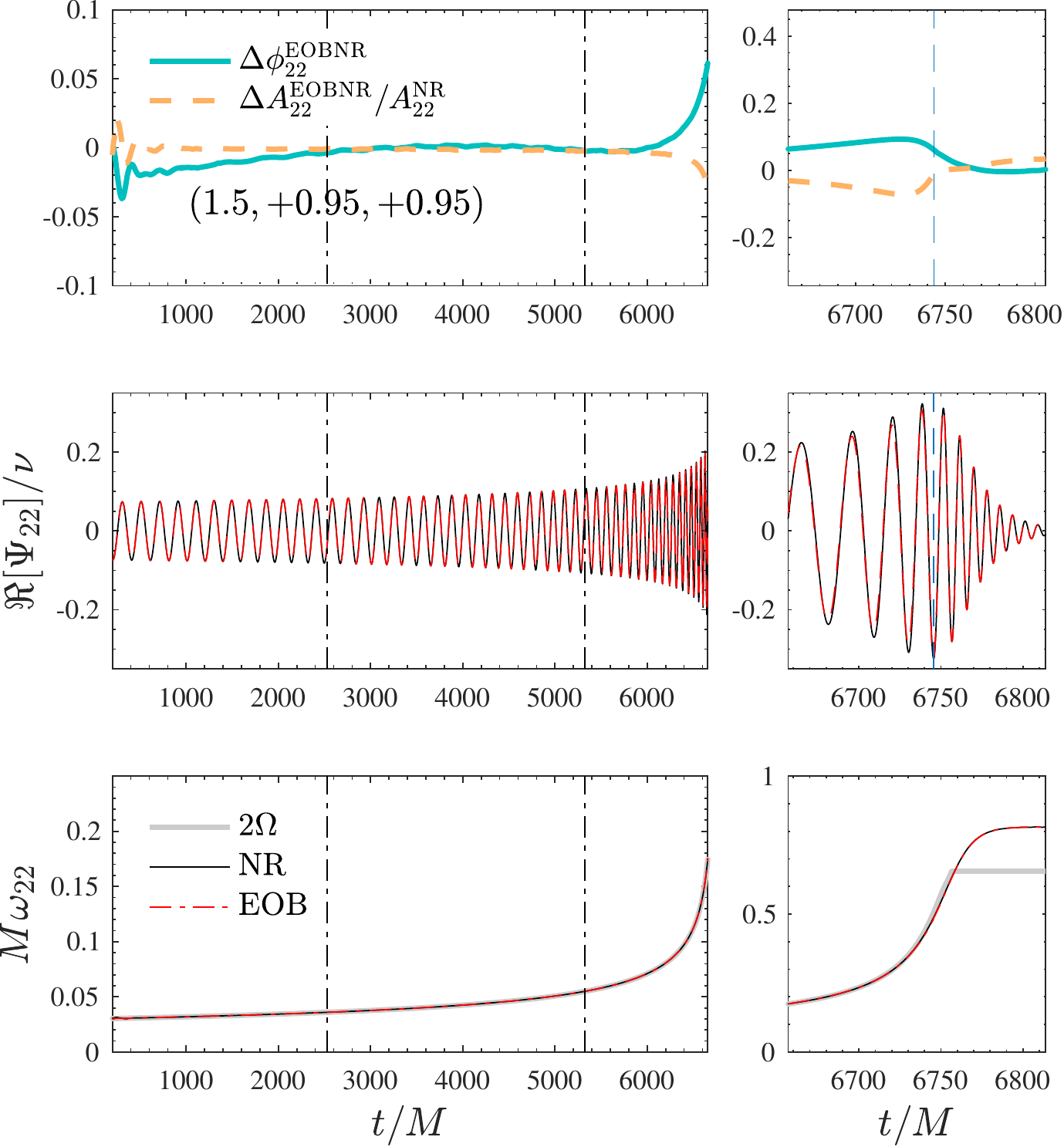}
\caption{\label{fig:phasing_1146} Improved EOB/NR phasing comparison for {\tt SXS:BBH:1146} when 
the value of $c_3^{\rm fit}=15.96$ used in Fig.~\ref{fig:barF_SXS_new} is lowered to $c_3=11.1$. 
Top panel: (relative) amplitude and phase differences. Middle panel: real part of the waveform. 
Bottom panel: gravitational frequencies. For convenience, also twice the EOB orbital frequency 
$2\Omega$ is shown on the plot. The dash-dotted vertical lines indicate the alignment frequency region,
while the dashed one the merger time. This comparison illustrates that {\tt SXS:BBH:1146} is 
an outlier in Fig.~\ref{fig:barF_SXS_new} only because  of the rather limited amount of NR waveforms used to 
inform $c_3^{\rm fit}$.}
\end{center}
\end{figure}

\begin{figure}[t]
\begin{center}
\includegraphics[width=0.45\textwidth]{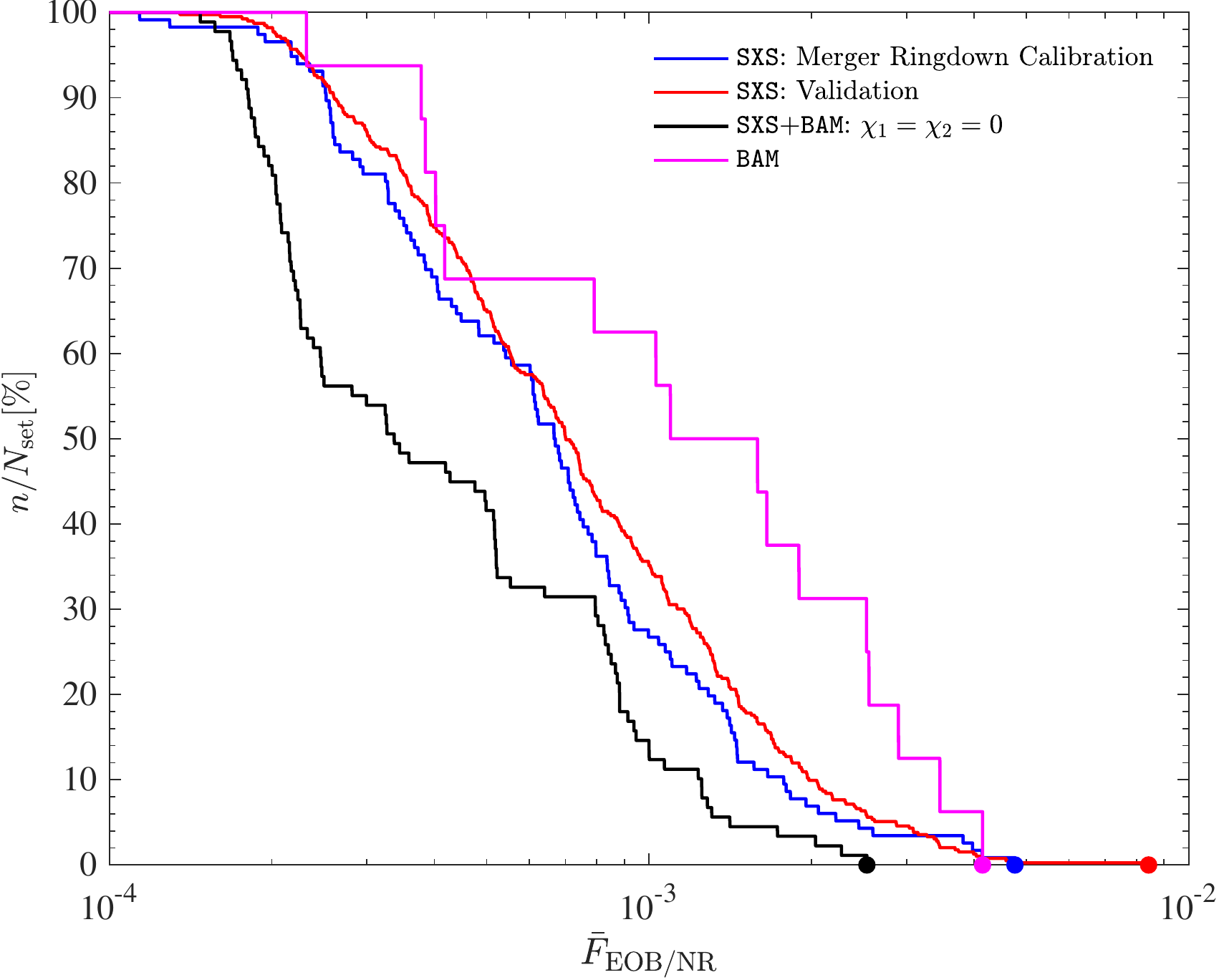}
\caption{Global representation of  $\bar{F}_{\rm EOB/NR}^{\rm max}$ 
all over the SXS (595)  and BAM (19) NR simulations.
The various SXS subsets, nonspinning (black online, 89 waveforms), merger-ringdown calibration 
(blue online, 116 spin-aligned waveforms)  and validation (red online, 388 spin-aligned waveform) 
discussed in the text are represented separately. The plot shows  the fraction (expressed in $\%$) 
$n/N_{\rm set}$, where $N_{\rm set}$ is the total number of waveforms in a given 
NR-waveform set and $n$ is the number of waveforms, in the same set, that, 
given a value $\bar{F}$, have  $\bar{F}^{\rm max}_{\rm EOB/NR}\geq \bar{F}$. 
The colored marker highlight the largest values in each NR dataset. 
Note that this plot incorporates  460 new SXS waveforms that were 
not included in Fig.~6 of~\cite{Nagar:2018zoe}.
}
\label{fig:histo}
\end{center}
\end{figure}

We start now discussing the performance of the analytical waveform 
model in terms of EOB/NR unfaithfulnesses plots for the $\ell=m=2$ 
mode, obtained computing Eq.~\eqref{eq:barF} between EOB and
NR waveforms. Both EOB and NR waveforms are tapered in the time-domain
so as to reduce high-frequency oscillations in the corresponding
Fourier transforms. Figure~\ref{fig:barF_SXSBAM} illustrates $\bar{F}$ versus $M$
evaluated over the same NR waveform data used in Ref.~\cite{Nagar:2018zoe}, with
the SXS data in the left panel and the BAM data in the right panel. As mentioned above,
a subset of this data, listed in Table~\ref{tab:c3}, (both SXS and BAM) was used
to inform the $c_3(\nu,\tilde{a}_1,\tilde{a}_2)$ function. The global performance of
the model is largely improved with respect to Ref.~\cite{Nagar:2018zoe},
see Fig.~1 there\footnote{In this respect, it is interesting to note
 that $\bar{F}$ for $(2,+0.85,+0.85)$ is now around the $10^{-3}$ level,
 while in Fig.~1 of~\cite{Nagar:2018zoe} is around $10^{-4}$.
 This happens because the difference between $c_3^{\rm fit}$ and
 $c_3^{\rm first-guess}$ is now larger than what it was in~\cite{Nagar:2018zoe},
 see Table~I there. A priori, a more flexible fitting function
 for $c_3$ would allow one to obtain even smaller values of $\bar{F}_{\rm EOB/NR}$.
 Since the EOB/NR performance of the model is already rather good,
 we content ourselves of the current, simple, analytical
 representation of $c_3$.}. Remarkably, the model performs excellently 
 also for large mass ratios and large spins, without any outlier above 
 the $1\%$ threshold, but $\bar{F}^{\rm max}_{\rm EOB/NR}\lesssim 0.5\%$ all over.

After February 3, the SXS collaboration publicly released another 455\footnote{The 5 very \textit{long} 
	($>100$ GW cycles) simulations are separately discussed in Sec.~\ref{sec:longNR_EOB} below.} 
	new simulations at an improved accuracy.  This part of the catalog 
mostly covers the same region of parameter space of the previous data, except for a few
waveforms spanning mass ratios between 4 and 8, with spins higher than
what considered before. The catalog also includes a few extremely
long waveforms, with more than 100 orbits. As an additional cross 
check of the robustness and accuracy of our model,
we compute $\bar{F}_{\rm EOB/NR}$ all over this new 
set of NR waveforms.
The result is displayed in Fig.~\ref{fig:barF_SXS_new}.
We find that $\bar{F}^{\rm max}_{\rm EOB/NR}$ always 
remains {\it below} $0.85\%$, a value reached only by one dataset, 
 $(1.5,+0.95,+0.95)$ {\tt SXS:BBH:1146}, while for all others we 
 have $\bar{F}^{\rm max}_{\rm EOB/NR}\lesssim 0.4\%$. 
This is not surprising since the set of NR waveforms used to inform $c_3$
does not cover, except for one single dataset with $(1.5,-0.5,0)$, the parameter 
space with $1<q<2$. In this respect, to better understand the behavior of this outlier 
in Fig.~\ref{fig:barF_SXS_new} we checked that $c_3^{\rm fit}(1.5,+0.95,+0.95)=15.96$ 
yields an accumulated EOB/NR phase difference  $\sim 4.7$~rad at merger once the two 
waveforms are aligned during the inspiral.
Interestingly, by {\it lowering} the value of $c_3$, and thus {\it increasing} the magnitude of the spin-orbit effective coupling and
thus making the EOB waveform longer, we can easily reconcile it with the NR data.
For convenience we illustrate this result in Fig.~\ref{fig:phasing_1146}, that is obtained
with $c_3=11.1$ (the two dash-dotted vertical lines indicate the alignment region). We also point the 
reader to Table~\ref{tab:SXS_new3}, where the 
NR uncertainty for this dataset is estimated to be $\bar{F}_{\rm NR/NR}=0.0446\%$.
On a different note, this suggests that the current model could be additionally, and easily,
improved by also considering {\tt SXS:BBH:1146} to inform $c_3^{\rm fit}$.
Yet, this results highlights the robustness of our model: without any additional input 
from NR simulations to determine $c_3$, it is able to deliver rather accurate waveforms 
even in a region of the parameter space 
previously not covered by NR data. The model performance is summarized 
in Fig.~\ref{fig:histo}. For each dataset considered above, the figure exhibits the 
fraction of waveform whose $\bar{F}_{\rm EOB/NR}^{\rm max}$ is larger or equal 
a given value $\bar{F}$. Thanks to the additional analytical information incorporated and to
the improved waveform resummation, \TEOBiResumSM{} is currently the EOB model 
that exhibits the lowest EOB/NR unfaithfulness for the $\ell=m=2$ mode.

\begin{figure}[t]
\begin{center}
\includegraphics[width=0.45\textwidth]{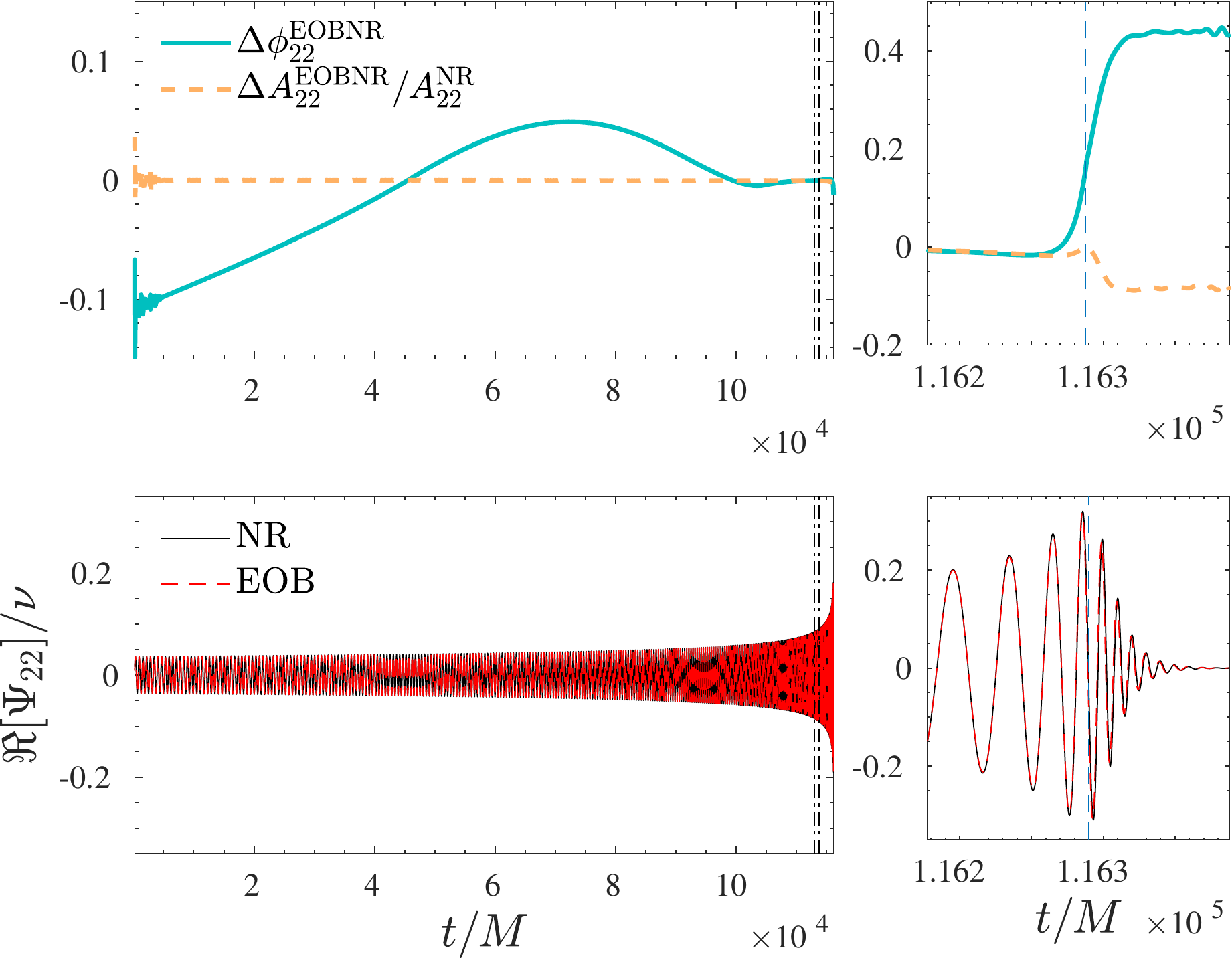}
\caption{EOB/NR phasing comparison for SXS:BBH:1415, $(1.5,+0.50,+0.50)$. Note that it
  doesn't seem possible to flatten the phase difference up to $t/M\simeq 1\times 10^5$.
  The vertical lines indicate the alignment frequency region $[M\omega_L,M\omega_R]=[0.038,0.042]$.
   See text for additional discussion.}
\label{fig:SXS_1415}
\end{center}
\end{figure}

\subsection{Long-inspiral Numerical Relativity waveforms}
\label{sec:longNR_EOB}
It is interesting to note that the 5, long, NR simulation exhibit an excellent agreement 
($\bar{F}_{\rm EOB/NR}\simeq 10^{-3}$, see Fig.~\ref{fig:barF_SXS_new})  with the 
analytical waveform , even during the long inspiral phase. Note that this is below the NR-uncertainty
estimate in the right panel of Fig.~\ref{fig:barF_SXS_error} without any input in the model coming from
this data. Despite such good agreement for the usual standard, an illustrative time-domain comparison
done for {\tt SXS:BBH:1415}, see Fig.~\ref{fig:SXS_1415}, highlights some features that is worth 
commenting on\footnote{See Paper~I and references therein for additional details concerning the alignment procedure.}.
At first glance,  the phase agreement is excellent for any standard quality assessment, always between
$\pm 0.1$~rad. However, contrary to our expectations, we didn't succeed in flattening the phase difference
by aligning the EOB and NR waveform during the early inspiral. This is usually achieved by 
narrowing the alignment frequency window $[M\omega_L,M\omega_R]$ and moving it to
early-inspiral frequency values. By contrast, to achieve a rather flat 
phase difference on a reasonably large time-interval we had to progressively displace interval the 
frequency window to {\it higher} frequencies, until hitting  $[M\omega_L,M\omega_R]=[0.038,0.042]$,
that corresponds to the two, dash-dotted, vertical lines in Fig.~\ref{fig:SXS_1415}. In view of the rather large 
uncertainty on this NR waveform,  we cannot really state whether this is due to
some systematics in the NR waveforms or in missing physics within the EOB model. 
Additional analyses done using more sophisticated phasing diagnostics, e.g. the gauge-invariant 
$Q_\omega\equiv \omega^2/\dot{\omega}$ function~\cite{Baiotti:2010xh,Baiotti:2011am,Damour:2012ky}, 
where $\omega\equiv \dot{\phi}$ is the gravitational wave frequency, might be necessary to better 
investigate the low-frequency consistency between the EOB and NR waveforms. 
An analogous behavior is shared also by the other long-term waveforms. However
 Fig.~\ref{fig:barF_SXS_new} highlights that the $\bar{F}_{\rm EOB/NR}$ for {\tt SXS:BBH:1414} 
 $(1.83,-0.5,+0.4)$ and {\tt SXS:BBH:1416} $(1.78,-0.4,-0.4)$ show a qualitatively different behavior, 
 with $\bar{F}_{\rm EOB/NR}$ that is starting at a slightly increased value for $M=10M_\odot$ and
 then is progressively decreasing with $M$. When aligning the waveforms in the same frequency interval $[0.038,0.042]$,
 so to obtain a quasi-flat phase difference also outside the alignment window, one finds that the
 EOB/NR phase difference grows linearly backwards for $10^5M$, to reach the $0.5$-$0.7$~rad at the
 beginning of the inspiral. Although this fact might explain the behavior of $\bar{F}_{\rm EOB/NR} $ 
 seen in Fig.~\ref{fig:barF_SXS_new}, conclusive NR-quality assessments require more 
 detailed investigations that are postponed to future work.
\begin{figure}[t]
\begin{center}
\includegraphics[width=\columnwidth]{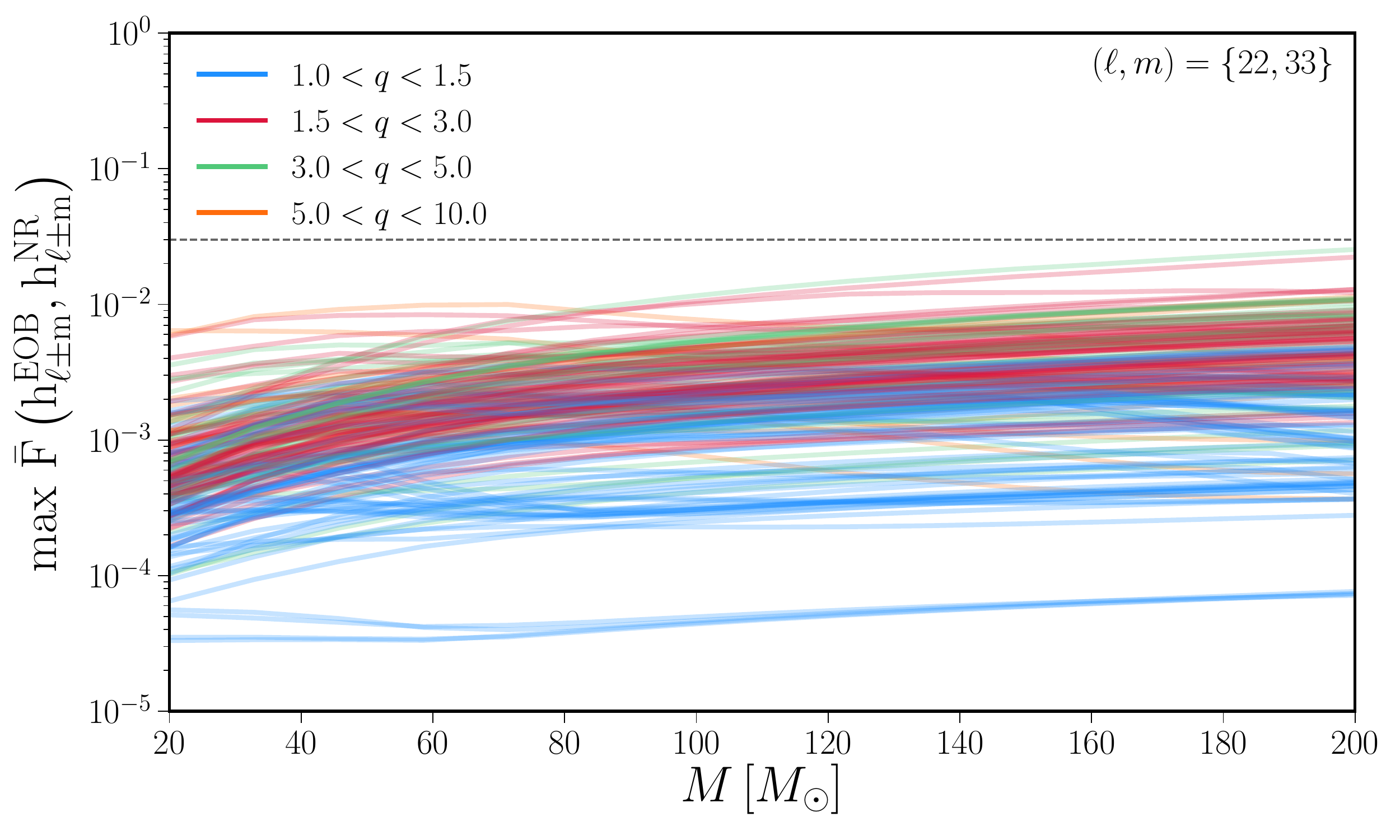}\\
\includegraphics[width=\columnwidth]{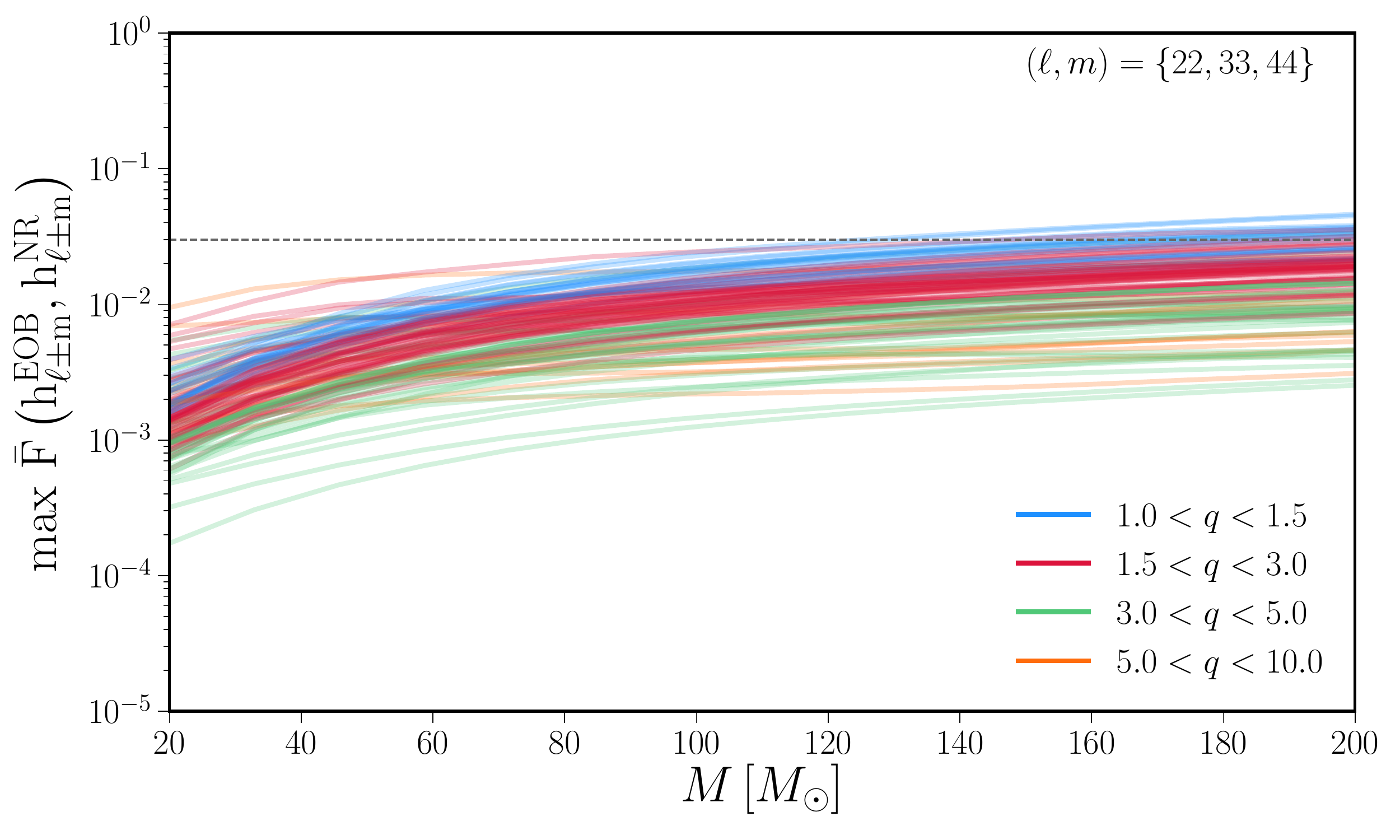}
\caption{EOB/NR unfaithfulness computation putting together all $\ell=m$ modes
up to $\ell=4$. Plotted is the worst-case performance maximing the unfaithfulness
over the sky, Eq.\eqref{eq:barF_max}. The worst-case mismatches arise from
near edge-on configurations, when the power emitted in the $(2,2)$ mode is minimized.}
\label{fig:mismatch_leqm}
\end{center}
\end{figure}

\section{Higher multipolar modes}
\label{sec:HM}
\begin{figure*}[t]
  \begin{center}
    (a)
  \includegraphics[width=0.35\textwidth]{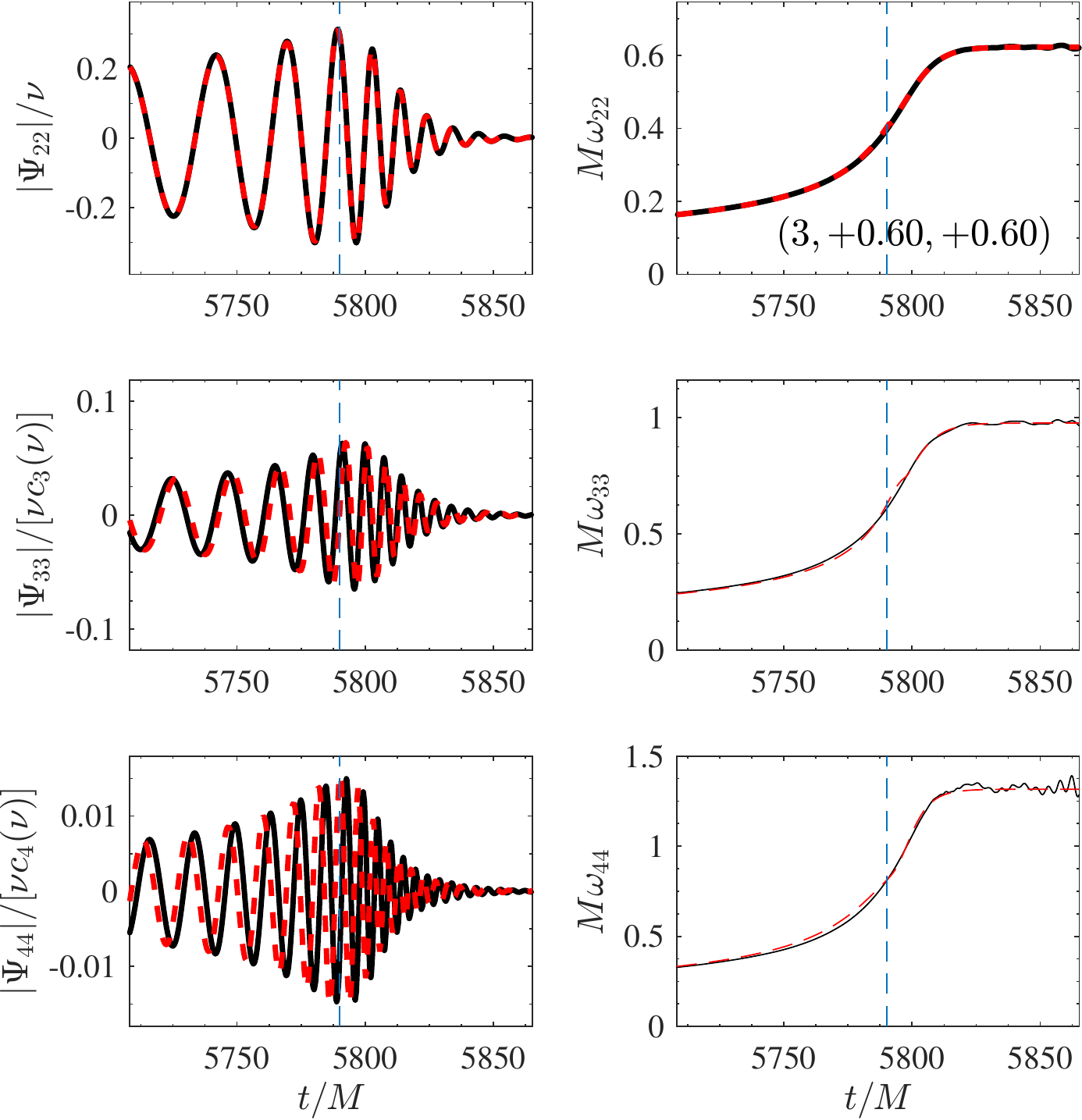}\qquad\qquad
  (b)
  \includegraphics[width=0.35\textwidth]{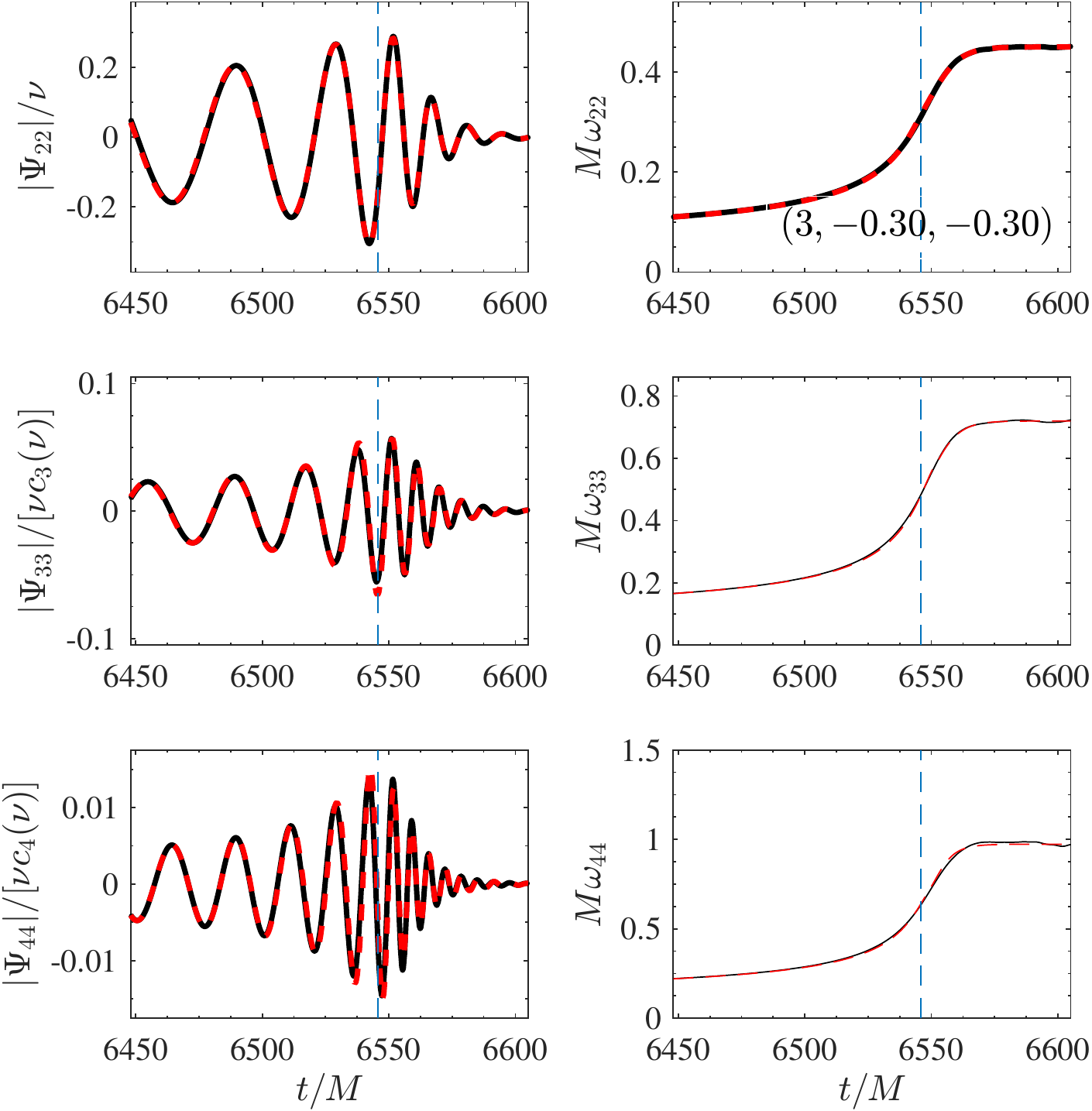}\\
  \vspace{0.5cm}
  (c)
  \includegraphics[width=0.35\textwidth]{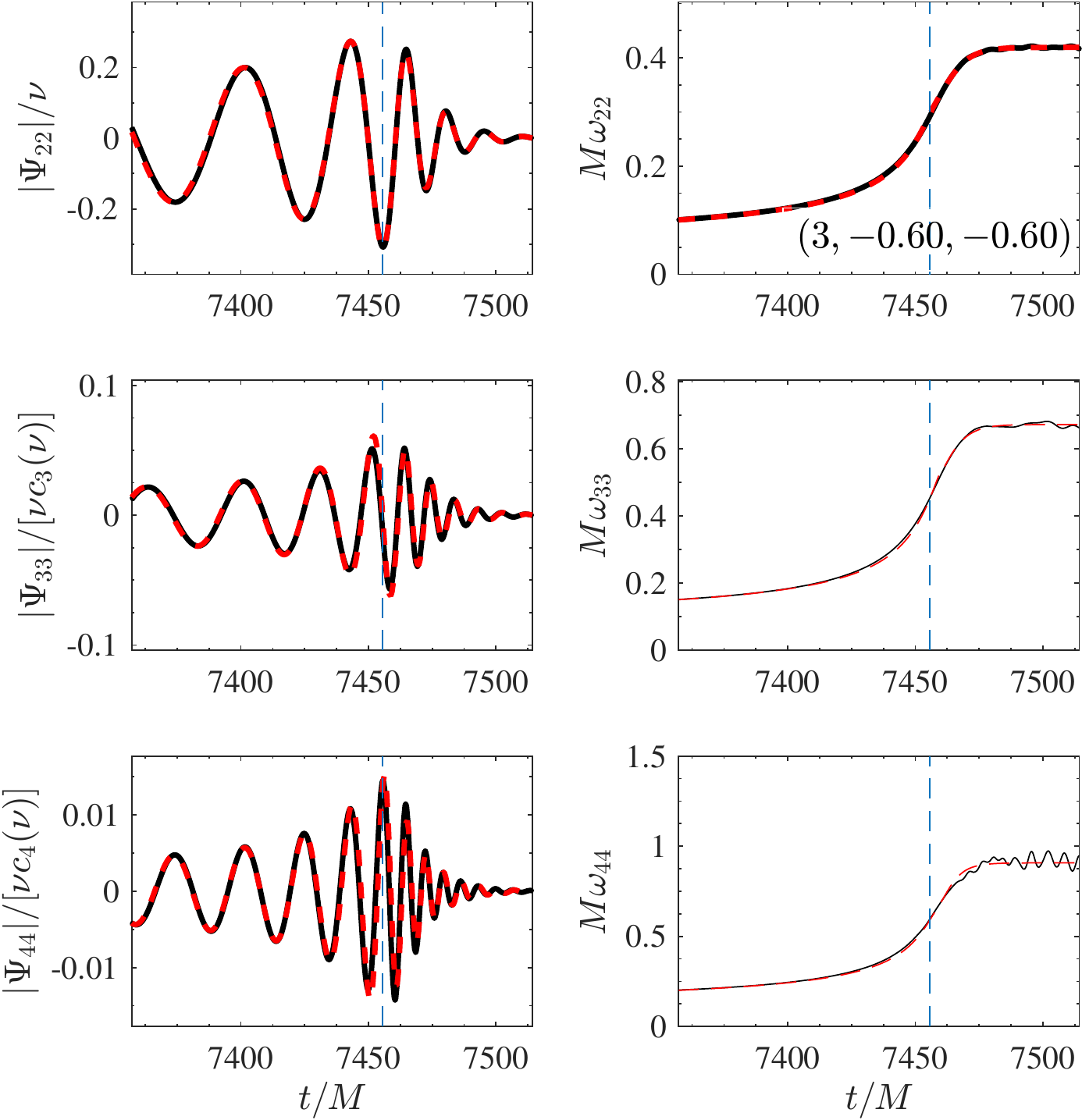}\qquad\qquad
  (d)
\includegraphics[width=0.35\textwidth]{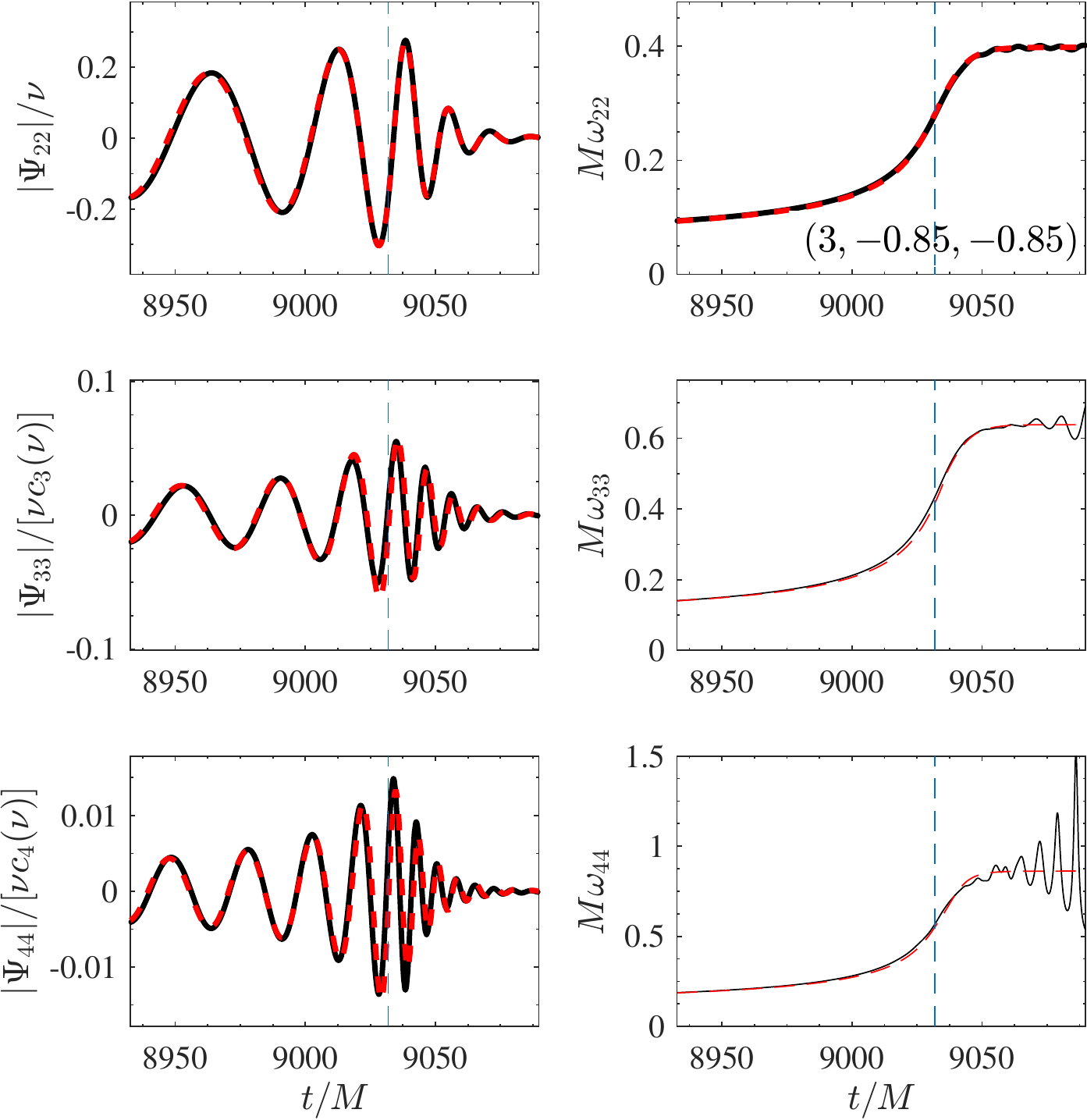}
\caption{Behavior of $(2,2)$, $(3,3)$ and $(4,4)$ modes for a few, illustrative, spin-aligned configurations with $q=3$: comparing
NR (black) with EOB (red) waveform. Each panel plots the real part (left columns) and the instantaneous frequency (right columns).}
\label{fig:ell_eq_emm}
\end{center}
\end{figure*}
\subsection{Multipoles $(2,2)$, $(3,3)$ and $(4,4)$}
Let us move now to discussing the quality of the higher modes. For illustrative
purposes, we consider explicitly four configurations with $q=3$, with
equal spins, both aligned or anti-aligned to the orbital angular momentum.
More precisely, we use $(3,-0.85,-0.85)$, $(3,-0.60,-0.60)$, $(3.-0.30,-0.30)$
and $(3,+0.60,+0.60)$. The qualitative (and quantitative) behavior discussed
here for this configuration is general enough to be considered paradigmatic
all over the SXS waveform catalog.
Figure~\ref{fig:ell_eq_emm} illustrates the behavior of the $(2,2)$, $(3,3)$ and
$(4,4)$ mode. For each multipoles, we show the real part of the EOB/NR
waveforms together with the instantaneous GW frequency $\omega_{\lm}$.
The EOB waveform is aligned to the NR one around merger, so to highlight
the excellent EOB/NR agreement there. 
The EOB/NR agreement is rather good either for spins both anti-aligned or 
aligned with the orbital angular momentum. We should, however, mention
that when the spins are large and aligned there is an increasing dephasing
accumulating between the EOB and NR $(4,4)$ mode, as one can see 
in Fig.~\ref{fig:ell_eq_emm} (a). 
As it was the case for the $\ell=m=2$ mode discussed above, a global
understanding of the actual performance of the model comes from
EOB/NR unfaithfulness computations. In addition to Eq.~\eqref{eq:barF}, 
due to the non-trivial angular dependence introduced  by the subdominant spherical harmonics, 
we consider the worst-case performance of the model by maximizing the unfaithfulness over the sky
\begin{align}
\label{eq:barF_max}
\max \; \bar{F}  (h_1 , h_2) \equiv \max_{\theta,\phi} \; \bar{F} (h_1 , h_2) .
\end{align}
In Fig. \ref{fig:mismatch_leqm}, we show the worst case performance for the $\ell = m$ modes 
up to $\ell = 4$, finding excellent agreement up to $\sim 120 M_{\odot}$ above which the model 
performance degrades slightly and moves above 3\%. In all cases, the worst case mismatches 
arise from near edge-on configurations, where the power in the (2,2)-mode is minimized. 
The worst mismatches occur for mass ratios $1\leq q \leq 1.5$ and equal-spin configurations, in which 
the approximate symmetry of the binary leads to a suppression of odd-$m$ modes. 
For these binaries, the degraded performance will be driven by the accuracy of
the $(4,4)$ mode in both the EOB model and the underlying NR data itself.
\subsection{Other subdominant multipoles}
\label{sec:largerspins}
\begin{figure*}[t]
  \begin{center}
    (a)
  \includegraphics[width=0.35\textwidth]{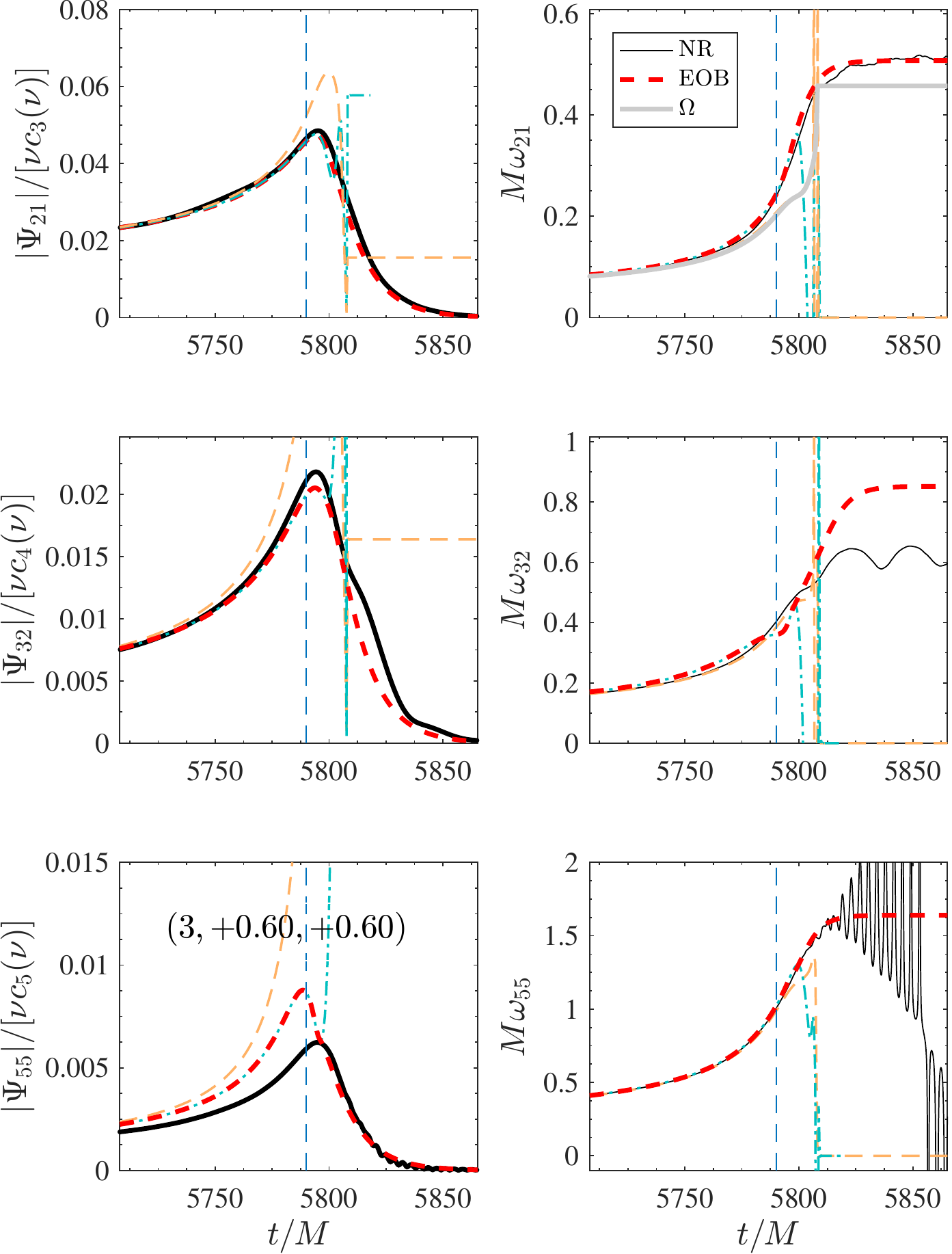}\qquad\qquad
  (b)
  \includegraphics[width=0.35\textwidth]{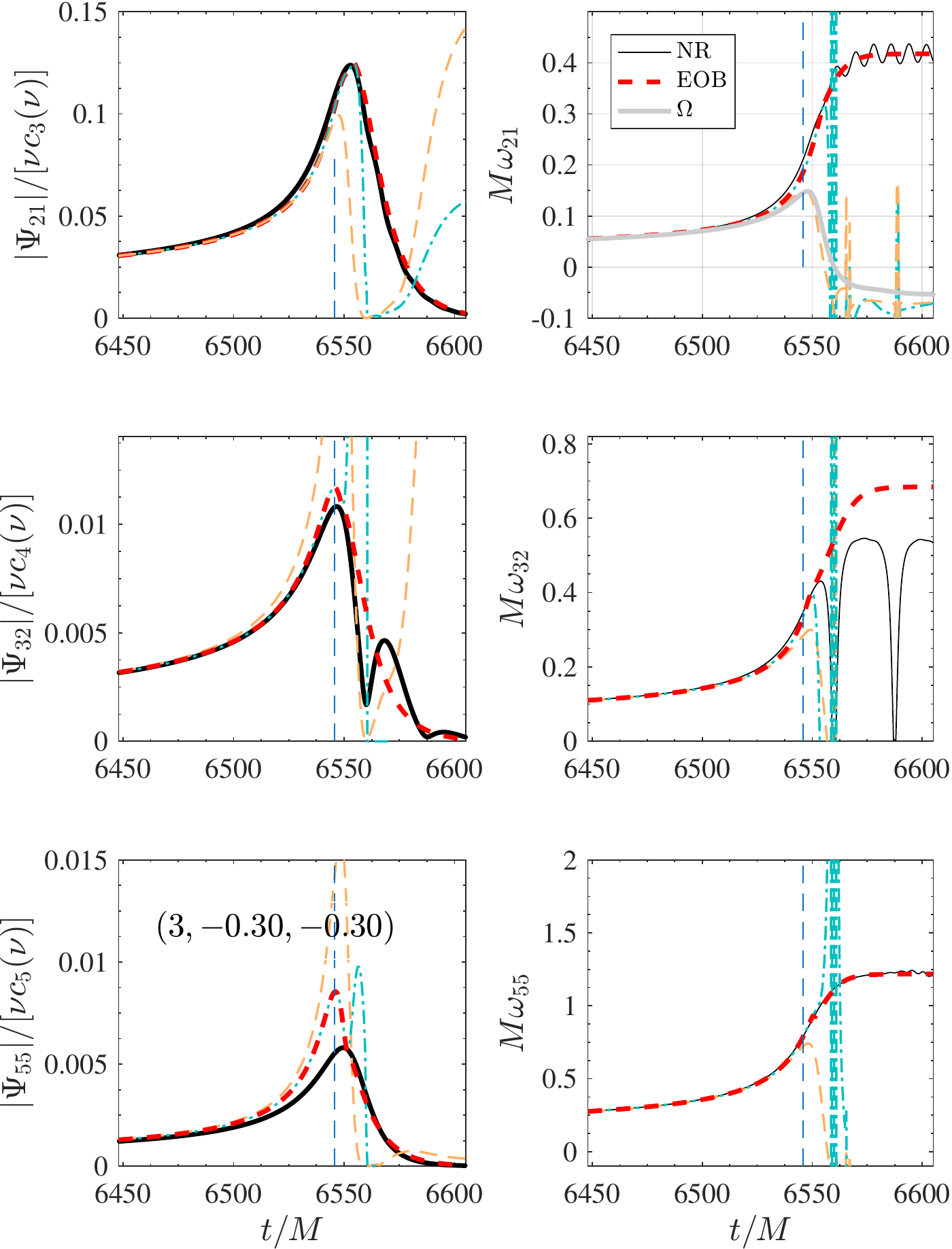}\\
  \vspace{0.5cm}
  (c)
  \includegraphics[width=0.35\textwidth]{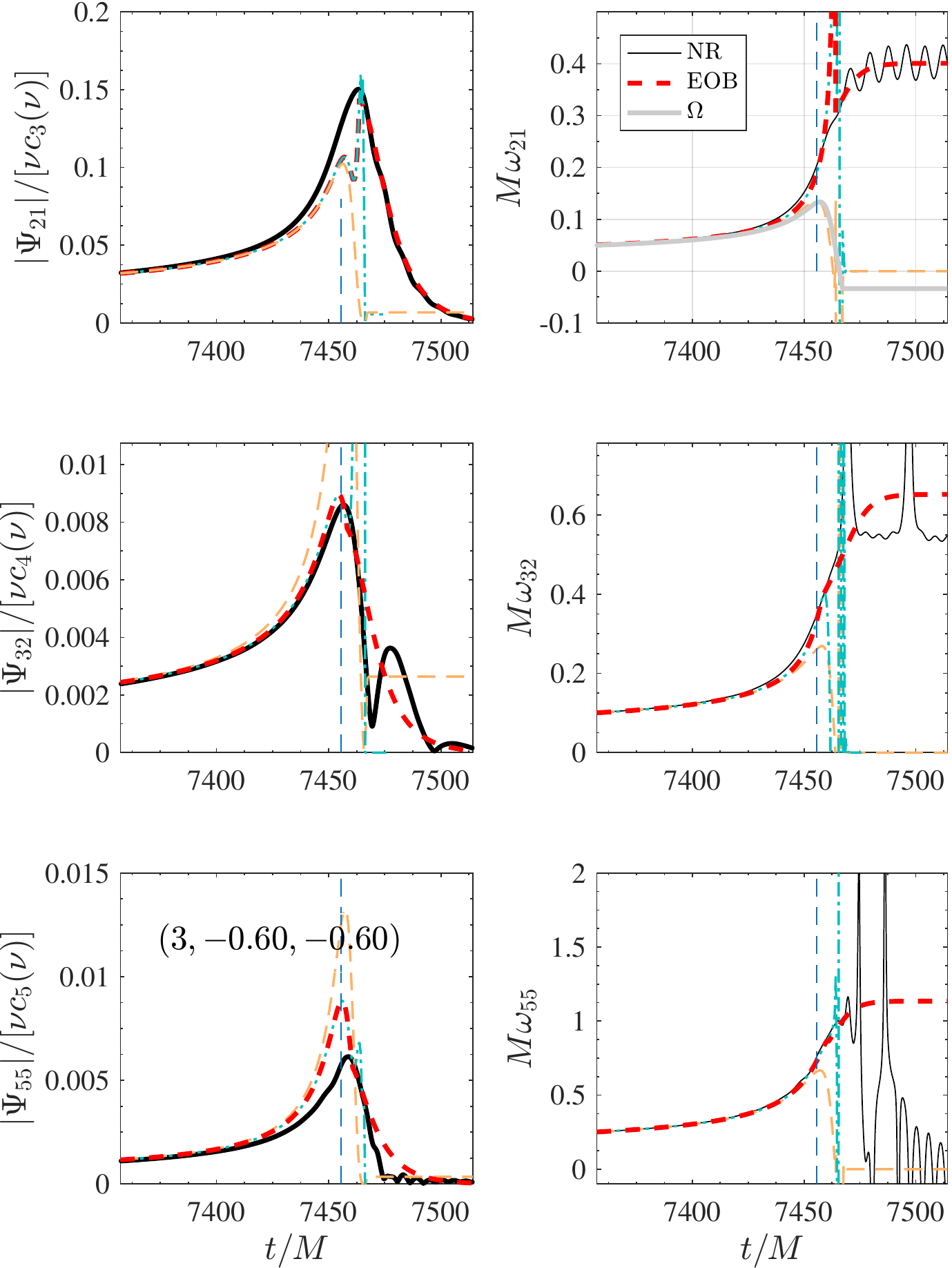}\qquad\qquad
  (d)
\includegraphics[width=0.35\textwidth]{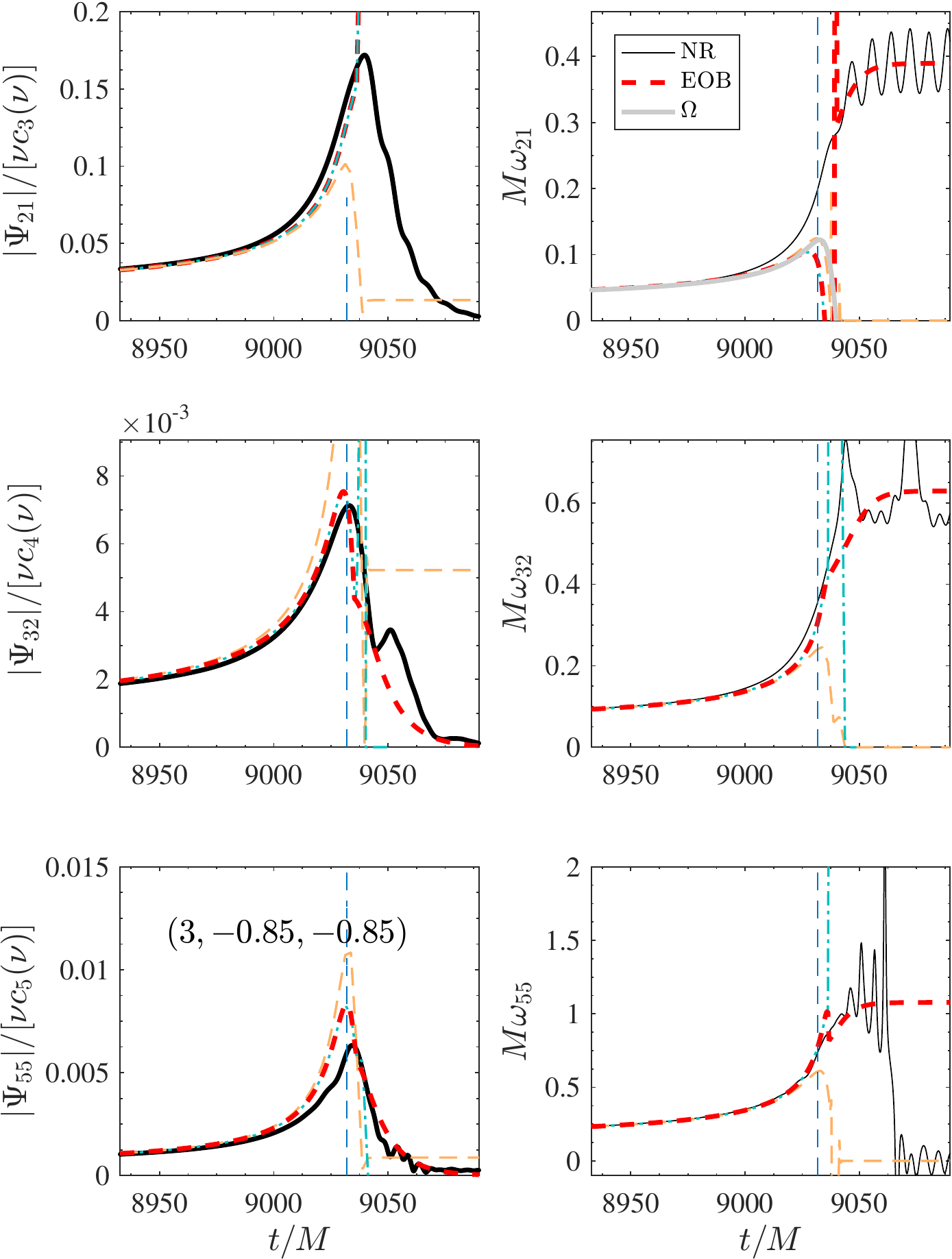}
\caption{Frequency and amplitude for the $(2,1)$, $(3,2)$ and $(5,5)$ modes for the same BBH configurations of Fig.~\ref{fig:ell_eq_emm}. On top
of the NR (black) and complete EOB curves (red, dashed), the plots also show: (i) the analytical EOB waveform, without NQC corrections and ringdown
(orange online) and (ii) the NQC-augmented EOB waveform (light-blue online). The dashed, vertical, line marks the merger location, i.e. the peak of 
the $\ell=m=2$ waveform amplitude. The $(2,1)$ frequency plots also incorporate the orbital frequency $\Omega$ (grey online). The construction of
the $(2,1)$ mode through merger and ringdown cannot be accomplished correctly for large values of the spins anti-aligned with the orbital 
angular momentum [see panel (c) and (d)].}
\label{fig:ell_neq_emm}
\end{center}
\end{figure*}
\begin{figure}[t]
\begin{center}
\includegraphics[width=\columnwidth]{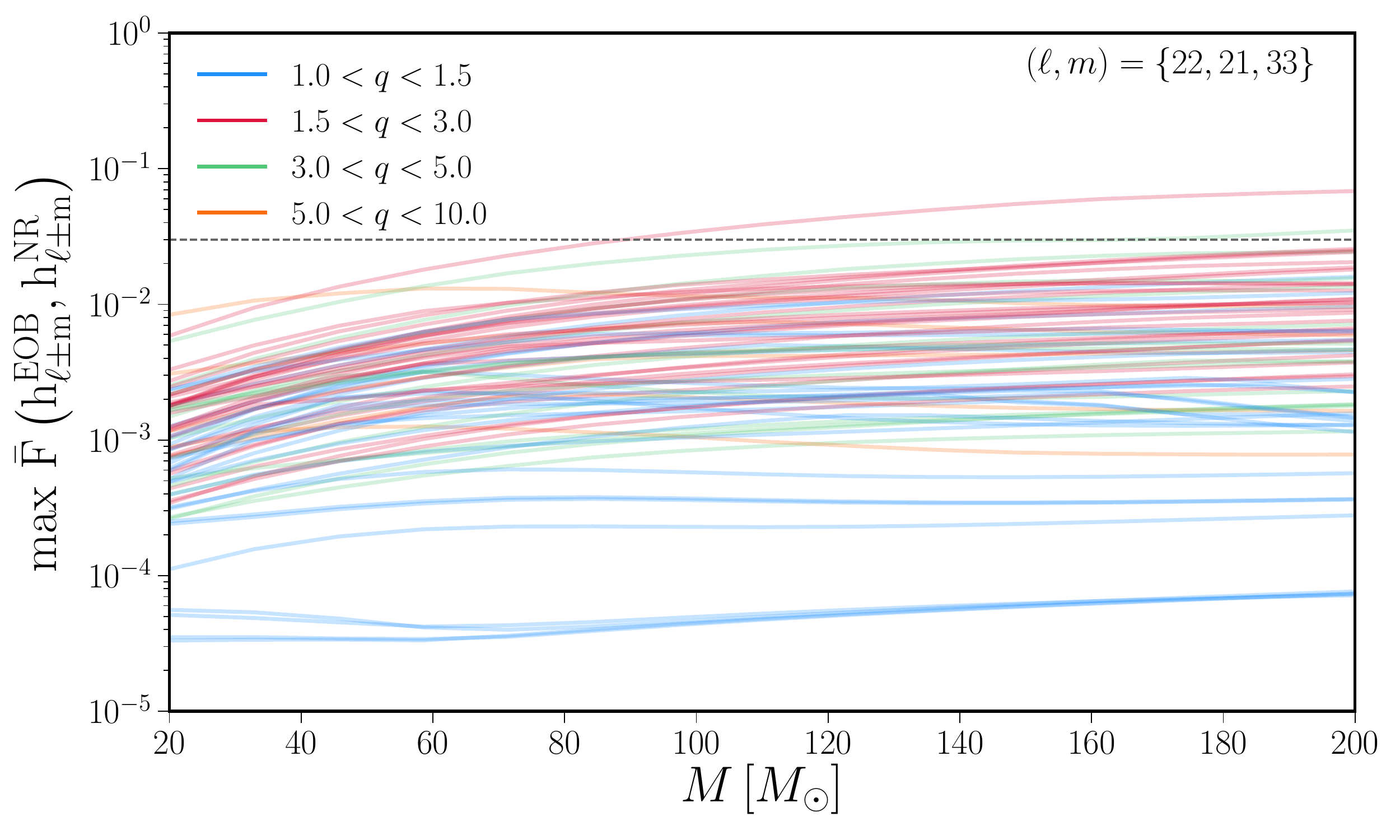}
\caption{EOB/NR unfaithfulness, maximized over the direction from the sky, when
  including $(2,2)$, $(2,1)$ and $(3,3)$ modes. Here we only consider a
  subset of the SXS waveforms with $\chi_i > -0.4$, where the $(2,1)$ EOB waveform mode
  does not present pathologies (see Fig.~\ref{fig:ell_neq_emm}).
  The worst case configuration is {\tt SXS:BBH:0239}, a binary of mass ratio
  and spins $(2.0,-0.37,+0.85)$.}
\label{fig:mismatch_22-21-33}
\end{center}
\end{figure}
\begin{figure}[t]
\begin{center}
\includegraphics[width=\columnwidth]{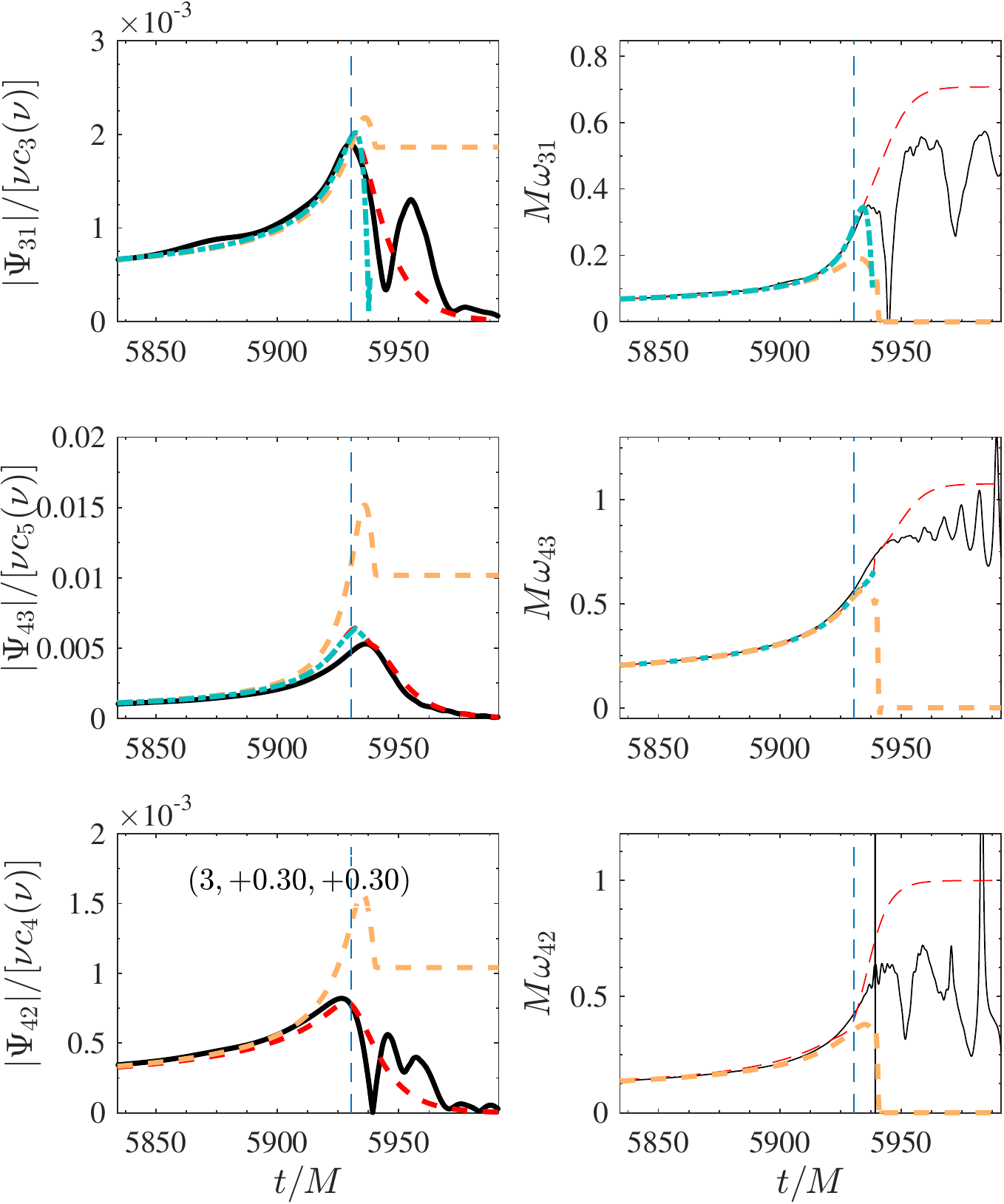}
\caption{Illustrative EOB/NR comparison for modes $(3,1)$, $(4,3)$ and $(4,2)$ for $(3,+0.3,+0.3)$.
  This behavior is analogous to the nonspinning case and is robust until the spins are mild.
  For larger spins, these modes may suffer the same problem related to the NQC factor
  discussed above for the $(2,1)$ mode.}
\label{fig:31etal}
\end{center}
\end{figure}
\subsubsection{Multipoles $(2,1)$, $(3,2)$ and $(5,5)$}
Let us discuss now modes $(2,1)$, $(3,2)$ and $(5,5)$, that
can be robustly constructed over most (but crucially not all) the parameter space.
To illustrate the typical behavior, we consider the same BBH configurations
show in Fig.~\ref{fig:ell_eq_emm}, but we focus now on amplitude 
and frequency. Each panel of the figure compares 
four curves: the NR one (black), the analytical EOB waveform (orange), 
the NQC-corrected EOB waveform (light-blue) and the complete
EOB waveform that includes the ringdown part. In addition, on the
$(2,1)$ frequency we also superpose the EOB orbital frequency,
as a grey line. The blue, dashed, vertical lines in the plot mark the
location of the merger point, i.e. the peak of the $(2,2)$ waveform
amplitude.
A few considerations first on the NR waveforms: during the ringdown,
one clearly sees in the $(2,1)$ and $(3,2)$ the effect of mode mixing,
that shows up as amplitude modulations and frequency oscillations.
The origin of these features has been explained in details in Ref.~\cite{Taracchini:2014zpa}.
By contrast, the $(5,5)$ mode shows features that clearly highlight
some lack of accuracy in the NR data. This is more evident in both
$(3,-0.60,-0.60)$ and $(3,-0.85,-0.85)$ configurations (see bottom rows
of the (c) and (d) panels of Fig.~\ref{fig:ell_neq_emm}.
Let us focus first on the $(3,2)$ mode. Despite the absence of mode-mixing,
the complete EOB waveform qualitatively reproduces the behavior
of the NR one around peak and postpeak, especially for what concerns
the amplitude. By contrast, the ringdown frequency, i.e. in the postpeak regime, 
is flat and {\it systematically larger} than the NR one because of lack of
the physical information in the ringdown modelization. It is however 
interesting to note that  the approximation is more reliable for 
large, anti-aligned, spins.  Similarly, the  shape of the waveform 
entailed by the action of the NQC is rather accurate and yields a 
reliable approximation of the frequency behavior up to merger.
By contrast, the situation is different for the $(2,1)$ mode.
When spins are aligned with the angular momentum, the standard
procedure for improving the behavior of the merger waveform via
NQC and the ringdown attachment works well, consistently with
the nonspinning case discussed in Ref.~\cite{Nagar:2019wds}.
This is clear for the case $(3,+0.60,+0.60)$ of Fig.~\ref{fig:ell_neq_emm}
and the procedure remains robust at least up to $(3,-0.30,-0.30)$ as
the figure illustrates. By contrast, as the magnitude of the
anti-aligned spins increase, the NQC correction becomes progressively
inaccurate and the resulting waveform becomes incompatible with the NR ones.
This is for example the case for $(3,-0.85,-0.85)$, where the NQC
correction is unable to act so as to smoothly connect
the inspiral, plunge and merger waveform to the ringdown (postmerger)
part. This latter is, by contrast, reliable, except for the
mode-mixing oscillation, that is missing by construction.
We tracked the reason of the unphysical behavior of the NQC correction
as follows. In our approach, that is the same of the nonspinning case, Paper~I, 
the NR information used to determine the NQC parameters is extracted $2M$
after the $(2,1)$ peak. As a consequence, for a successful implementation,
the NQC factor should be evaluated there. Unfortunately, the
EOB dynamics in this region, that is {\it after} merger time
(i.e. the peak of the $(2,2)$ mode), may develop unphysical features
depending on the values of the spins. 
The simplest way to explain  what is going on is by looking at the 
orbital frequency, $\Omega$. This is shown as a grey line in
the $(2,1)$ panels of Fig.~\ref{fig:ell_neq_emm}.
One sees that for both $(3,-0.60,-0.60)$ and $(3,-0.85,-0.85)$ $\Omega$
becomes very small around the peak of the $(2,1)$ mode until it crosses
zero and becomes negative. This is unexpected for this configuration and 
not what it is supposed to be. The unphysical character of this feature 
can be understood by qualitative comparison with the system made by
a point-particle inspiralling and plunging on a Kerr black hole.
In this case, the orbital frequency changes sign for configurations
where the spin of the black hole is antialigned with the orbital
angular momentum and large: the frame dragging exerted by the
black-hole space time on the particle is responsible of the sign
change in the frequency (see e.g. Ref.~\cite{Harms:2014dqa} ).
One should be aware that such dynamical behavior reflects on
the waveform, and in particular on the QNMs frequency excitations,
notably also at the level of the $(2,2)$
mode, that should have a zero at the time when the angular velocity
of the particle changes sign (i.e., from counterrotation with respect
to the black hole, to rotating with the black hole). Such qualitative
features are not present in the NR waveform, so we believe that the
EOB frequency behavior for this configuration is incorrect after merger
time. This suggests that the current Hamiltonian should be modified
so to avoid this feature. At a practical level, the fact that $\Omega$
crosses zero when the values of the relative separation $r$ is small, but
finite, implies that the NQC functions $n_4\equiv p_{r_*}/(r\Omega)$ 
and $n_5\equiv p_{r_*}/(r\Omega)\Omega^{2/3}$ (see Paper~I)
become very large and prevent the related NQC correction to the 
phase to act efficiently so to correctly modify the bare inspiral
waveform. This is is evident in panel (c) and (d) of Fig.~\ref{fig:ell_neq_emm}.
This problem affects the $(2,1)$ for any mass ratio when the anti-aligned
spin(s) are sufficiently large. For example, a similar behavior is found
also for $(8,-0.90,0)$. As a consequence, to use the current multipolar
model for actual parameter estimation studies, it will be necessary
to determine the precise region of the parameter space where the $(2,1)$
mode is reliable. Selecting only those datasets with $\chi_i>-0.4$,
Fig.~\ref{fig:mismatch_22-21-33} shows the EOB/NR unfaithfulness,
maximized over the sky, when including modes $(2,2)$, $(2,1)$ and $(3,3$.
Further improvement, as well as the determination of the precise range
of reliability of the $(2,1)$ mode through merger and ringdown, are
postponed to future work. Here we will just briefly
explore, in Sec.~\ref{sec:no_uc} below, a possible modification
to the current spin-orbit sector of the Hamiltonian that
may eventually improve the behavior of the $(2,1)$ mode
in the anti-aligned spin case.
\subsubsection{Multipoles $(3,1)$, $(4,3)$ and $(4,2)$}
From fits of the SXS waveforms we can also obtain a postmerger/ringdown description
of the $(3,1)$, $(4,2)$ and $(4,3)$ modes. For simplicity and robustness,
the $(3,1)$ ringdown relies on the nonspinning fits of Ref.~\cite{Nagar:2019wds},
while for $(4,3)$ and $(4,2)$ the relevant information is found in Appendix~\ref{sec:42}-\ref{sec:43}.
When the magnitude of the spins are  relatively mild, these modes can be 
modeled rather accurately (modulo mode mixing during ringdown) as in the
nonspinning case~\cite{Nagar:2019wds}. Figure~\ref{fig:31etal} illustrates this fact 
for $(3,+0.30,+0.30)$, with the usual EOB/NR comparison as we did above. 
For $(4,3)$ and $(4,2)$ modes one can appreciate the relevant action of
the NQC factor. When spins are larger (and notably anti-aligned) one can
have $\Omega$-driven pathological effects like the $(2,1)$ mode discussed
above. Seen also the (average) lower accuracy of the corresponding NR modes
all over the SXS catalog, we postpone a more detailed discussion (and possible
improvements) to future work.
\subsubsection{Improving the behavior of the $(2,1)$ multipole}
\label{sec:no_uc}
\begin{figure}[t]
\begin{center}
\includegraphics[width=\columnwidth]{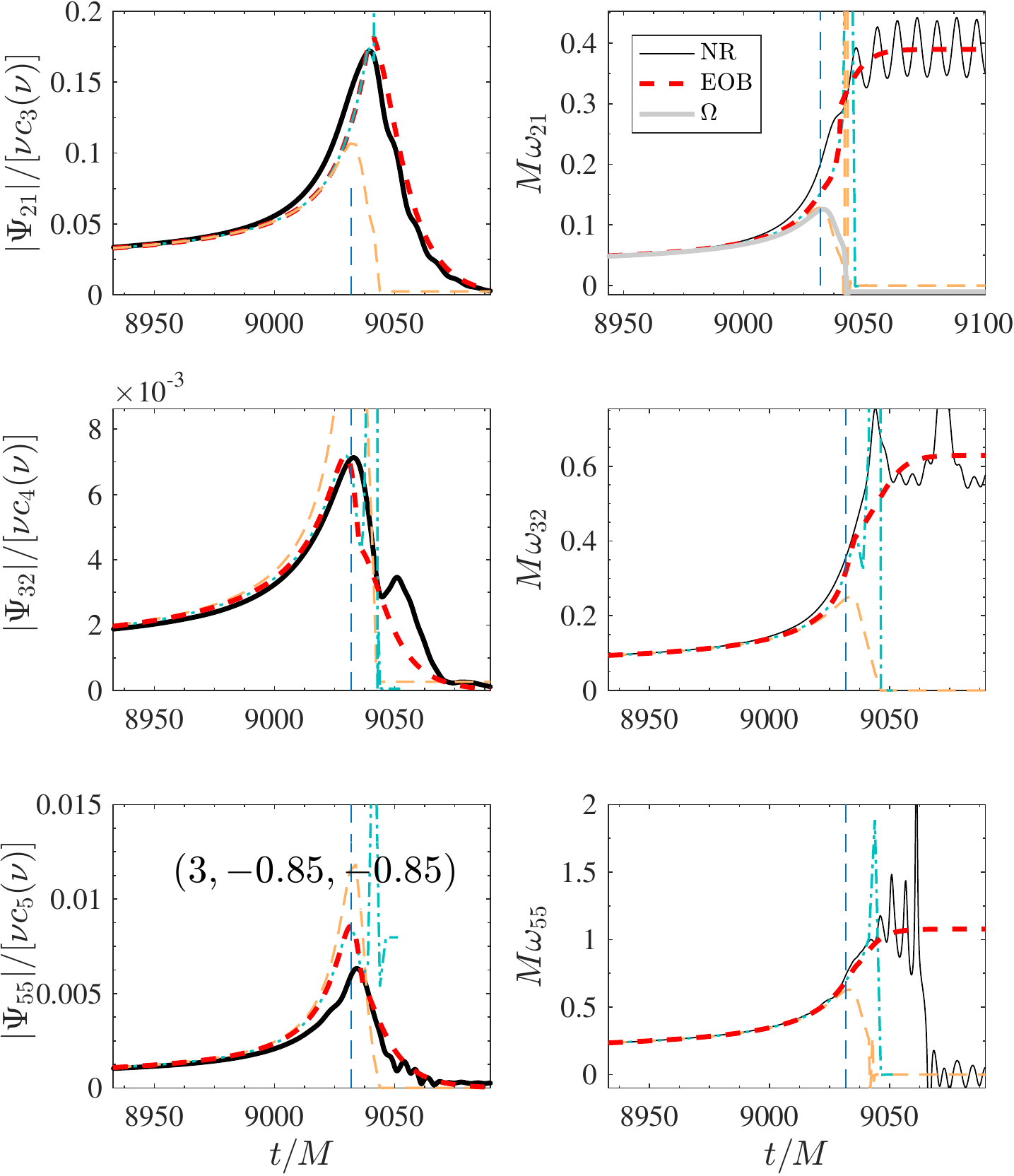}
\caption{Attempt of improving the behavior of the $(2,1)$ mode for $(3,-0.85,-0.85)$
  by modifying the spin-orbit sector of the EOB Hamiltonian. The related change
  in the EOB orbital frequency, $\Omega$, that is seen now to decrease more mildly
  after its peak than in Fig.~\ref{fig:ell_neq_emm} (d), is sufficient to improve
  the efficiency of the NQC correction, so to get a more acceptable frequency growth
  that can be smoothly connected with the ringdown. See text for additional details.}
\label{fig:no_uc}
\end{center}
\end{figure}
The correct behavior of the orbital frequency $\Omega$ in the strong-field
regime is determined by subtle compensation between the orbital and spin-orbit
part of the Hamiltonian. This is the region where our analytical understanding
is weaker, as we have to rely on resummed results that are analytically
incomplete. From the practical point of view, to NQC-complete
the inspiral $(2,1)$ mode following the current scheme it would be sufficient
the behavior of $\Omega$ be milder after the merger. In practice, we found that
this is possible by implementing a small modification to the resummed
$(G_S,G_{S_*})$ functions. The spin-orbit sector of the Hamiltonian is based on
Ref.~\cite{Damour:2014sva}, in particular the gyro-gravitomagnetic functions
are given by Eqs.~(38),~(39),~(41), and (42), where the inverse separation $u$
is replaced by the inverse centrifugal radius $u_c$. While $G_S^0=2 u u_c^2$,
Eq.~(38) of Ref.~\cite{Damour:2014sva}, has the structure of the Kerr gyro-gravitomagnetic
function, the dependence of $u_c$ introduced in the other functions, $G_{S_*}^0$,
$\hat{G}_S$ and $\hat{G}_{S_*}$ was an arbitrary choice. One finds that
replacing such $u_c$ dependence with the, more natural, $u$-dependence
is sufficient to provide small modifications in the behavior
of $\Omega$ that entail a far more robust behavior of the
NQC correction. In practice, we use
\begin{align}
  G_S   &= 2 u u_c^2\hat{G}_S(u),\\
  G_{S_*}&=\dfrac{3}{2}u^3\hat{G}_{S_*}(u),
\end{align}
where $(\hat{G}_S,\hat{G}_{S_*})$ are given by Eqs.~(41)-(42) of Ref.~\cite{Damour:2014sva}
where $u_c$ is replaced by $u$. The result of this change for $(3,-0.85,-0.85)$
is illustrated in Fig.~\ref{fig:no_uc}. Note that, since the dynamics has now changed,
to get a good $(2,2)$ EOB/NR phasing agreement we had to use $c_3=86.5$ instead
of $c_3^{\rm fit}=79.98$ from Eq.~\eqref{eq:c3fit}.
Comparing Fig.~\ref{fig:no_uc} with the panel (d) of Fig.~\ref{fig:ell_neq_emm} one immediately
notices the different behavior of the orbital frequency, whose peak is shallower
than before. The consequence of this behavior is that the action of the NQC factor
on both amplitude and frequency is more correct than before, though not yet fully
accurate for this latter. Although improvable, this proves that the scheme for
completing the EOB waveform through merger and ringdown for all modes that was
seen to be efficient in the nonspinning case~\cite{Nagar:2019wds} {\it can} be
straightforwardly generalized to the spinning case provided the dynamics, i.e.
the orbital frequency, behaves correctly. The result of Fig.~\ref{fig:no_uc}
gives us a handle to improve the description of spin-orbit effects within the
EOB Hamiltonian in future work.
\begin{table}[t]
  \caption{\label{tab:21freq} Frequency of the minimum of the $(2,1)$ amplitude for a few BBH configurations considered
    in Ref.~\cite{Cotesta:2018fcv} and not publicly available. $M\Omega_0$ is the (orbital) frequency corresponding to
    a minimum (or a zero) in the amplitude. Our EOB-predicted value, from the zero of $\hat{f}_{21}^{\rm S}$ in
    Table~\ref{tab:PNchoices}, is more consistent with the NR one than the straightforward PN value.}
\begin{center}
\begin{ruledtabular}  
	\begin{tabular}{llllllll}
		Name                     & $q$  & $\chi_1$ & $\chi_2$   &  $\hat{S}$ &$\!\!M\Omega_0^{\rm NR}$ & $\!\!\!M\Omega_0^{\rm EOB}$ & $M\Omega_0^{\rm PN}$ \\
		\hline
		SXS:BBH:0614      & 2    &  0.75    & $-0.5$  &   0.278 & 0.083     & 0.0968  &  0.057 \\
		SXS:BBH:0612      & 1.6  &   0.5    & $-0.5$  &   0.115  & 0.068    & 0.0712  & 0.047 \\
		SXS:BBH:1377      & 1.1  &  $-0.4$  & $-0.7$  &   $-0.268$ & 0.033  & 0.0330  & 0.029 \\
	\end{tabular}
\end{ruledtabular}
\end{center}
\end{table}
\begin{figure}[t]
\begin{center}
\includegraphics[width=0.4\textwidth]{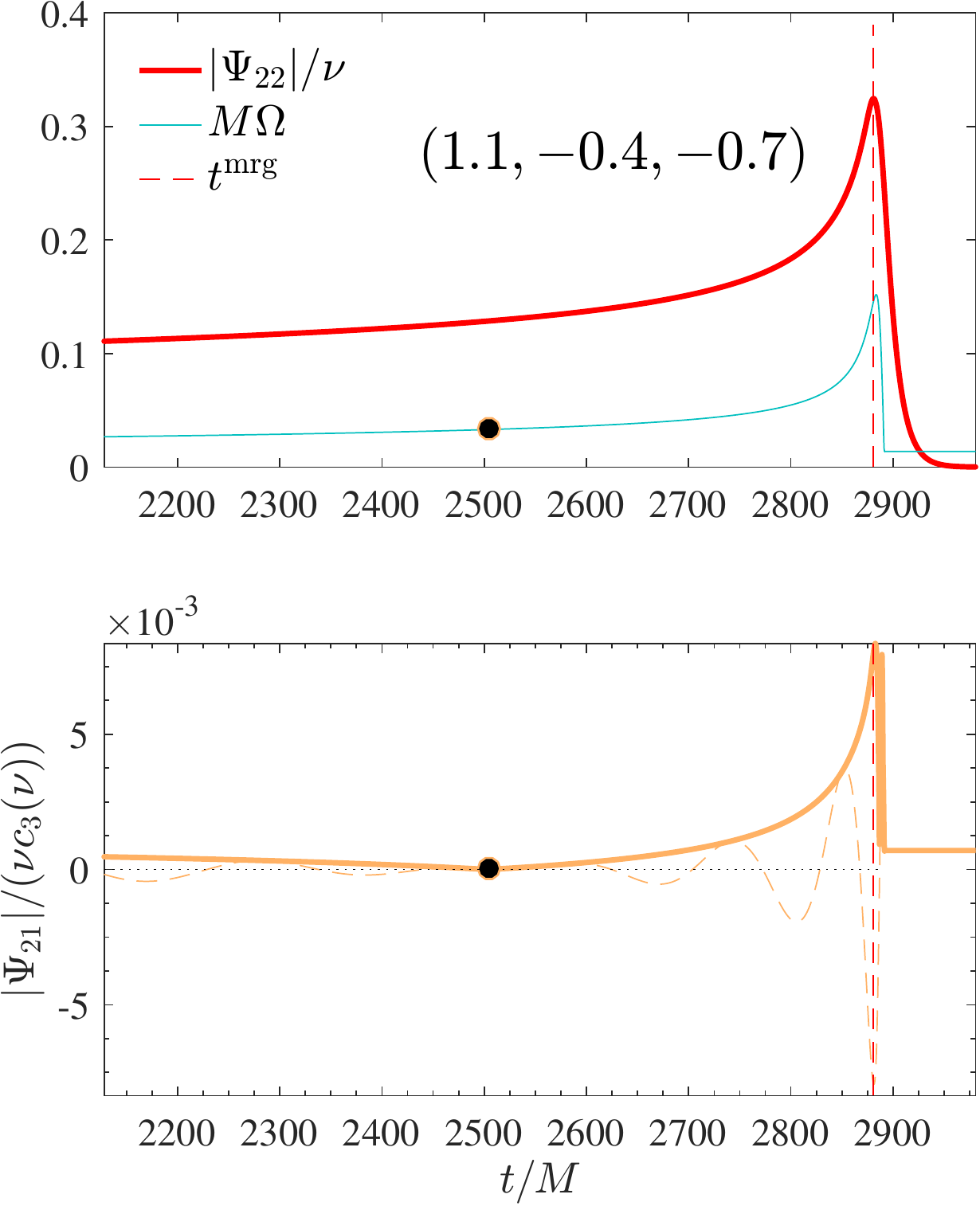}\\
\includegraphics[width=0.4\textwidth]{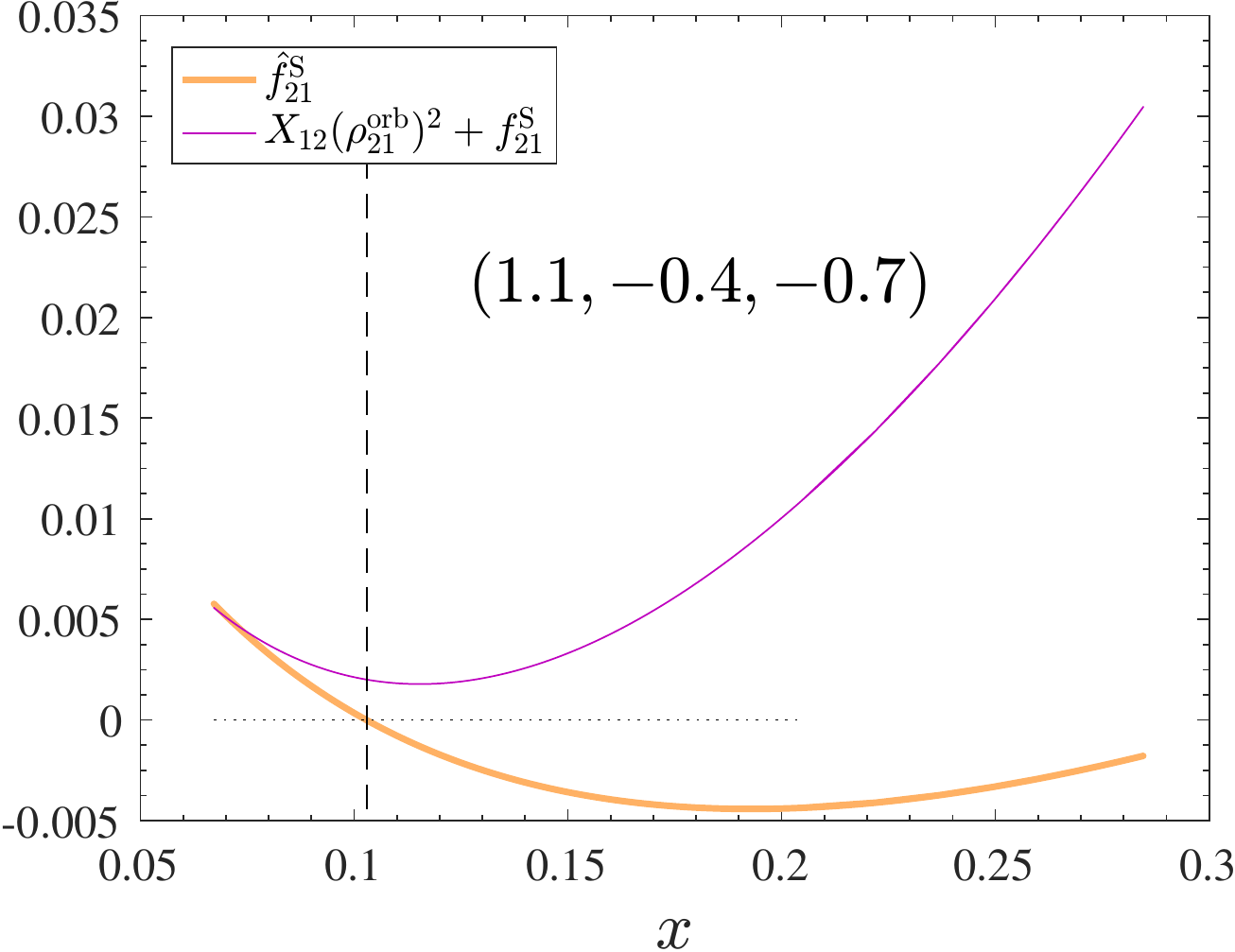}
\caption{\label{fig:h21_m04_m07}Top and medium panels: occurrence of a zero in the $(2,1)$ amplitude
  in configuration $(1.1,-0.4,-0.7)$, corresponding to NR dataset {\tt SXS:BBH:1377} analyzed in Ref.~\cite{Cotesta:2018fcv}.
  This dataset is not publicly available through the SXS catalog. The EOB-predicted value of the frequency
  is perfectly compatible with the NR value reported in Ref.~\cite{Cotesta:2018fcv} (see the last row of
  Table~\ref{tab:21freq}). The bottom panel compares the zero location of the resummed (orange) and nonresummed (magenta)
  amplitudes. See text for details.}
\end{center}
\end{figure}
\begin{figure}[t]
\begin{center}
  \includegraphics[width=0.42\textwidth]{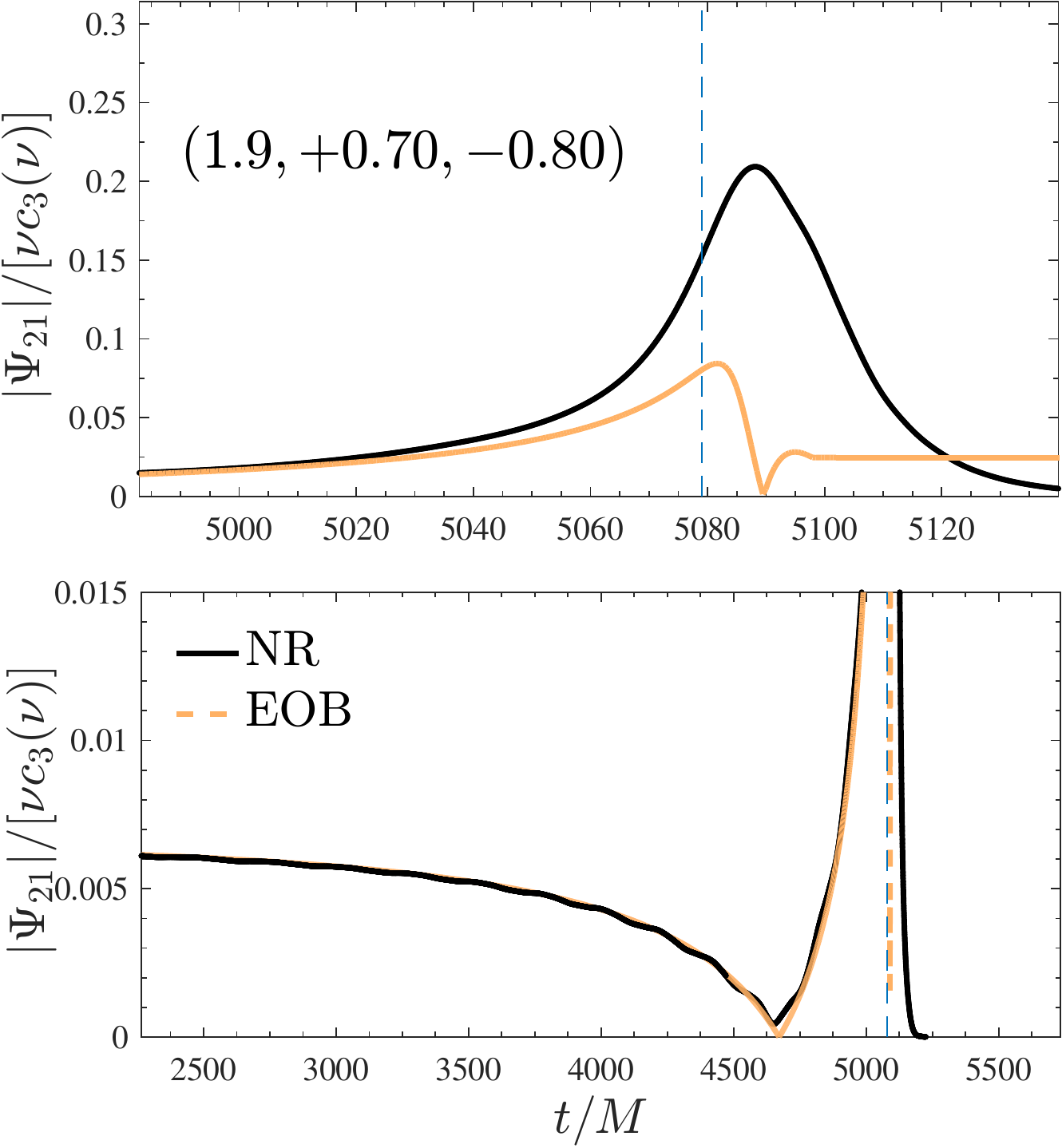}
\caption{Mode $(2,1)$: comparison between the EOB amplitude (orange) and the corresponding NR one from dataset {\tt SXS:BBH:1466}.
  The purely analytical EOB waveform multipole can accurately predict the location of the minimum (that analytically is a zero of the modulus)
  consistently with the one found in the NR data. The excellent agreement shown is obtained {\it naturally}, 
  without the need of calibrating any additional parameter entering the waveform amplitude. The dashed vertical line
  corresponds to merger time, i.e. the peak of the $\ell=m=2$ waveform. The cusp in the analytical amplitude occurs because of 
  a zero in $\hat{f}_{21}^{\rm S}$ as illustrated in Fig.~\ref{fig:f21S_1p9}.}
\label{fig:h21_q1p9}
\end{center}
\end{figure}

\begin{figure}[t]
\begin{center}
  \includegraphics[width=0.42\textwidth]{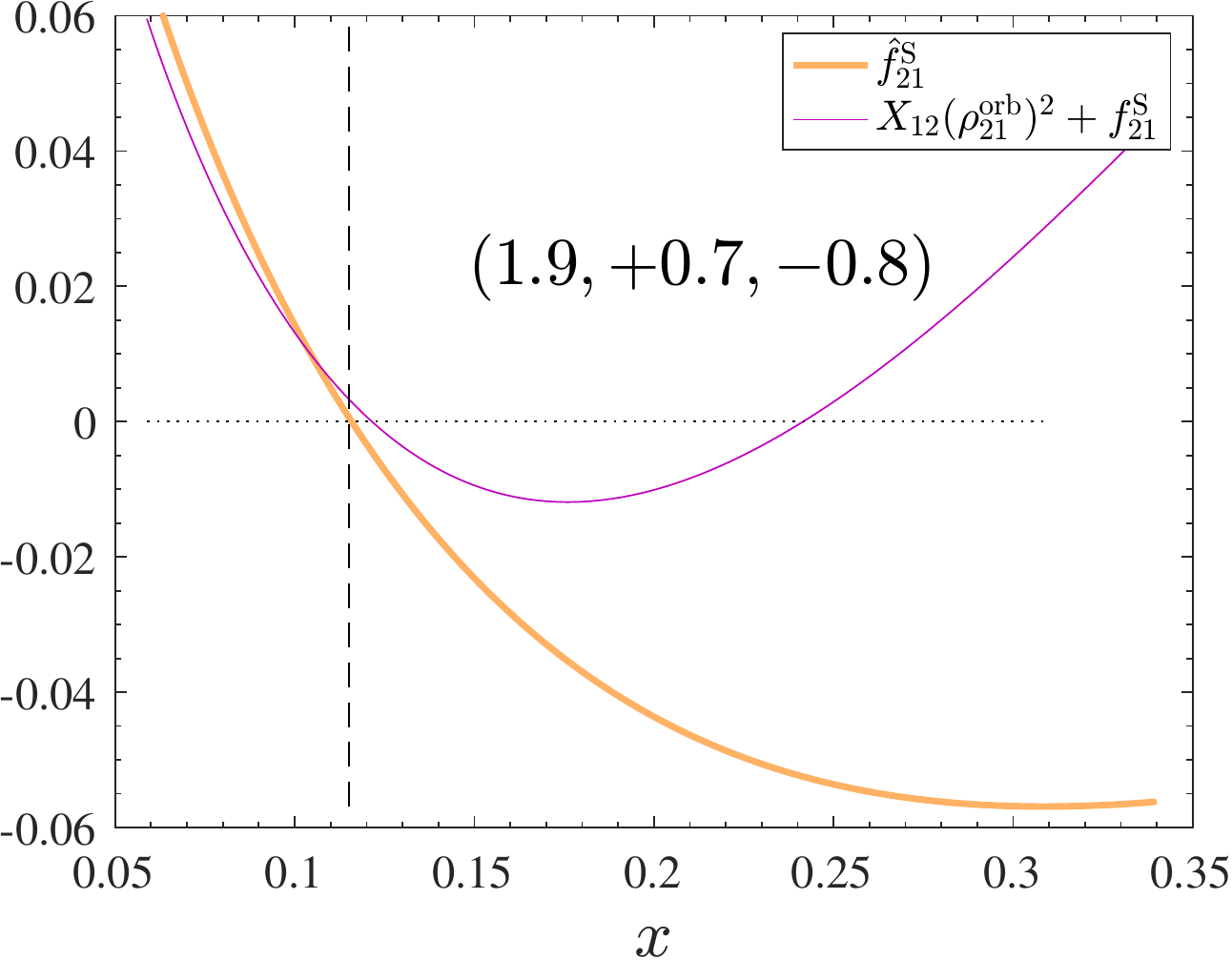}
\caption{\label{fig:f21S_1p9} Complementing Fig.~\ref{fig:h21_q1p9}: the behavior of the resummed versus
nonresummed amplitude versus $x=\Omega^{2/3}$. }
\end{center}
\end{figure}
\begin{figure}[t]
\begin{center}
  \includegraphics[width=\columnwidth]{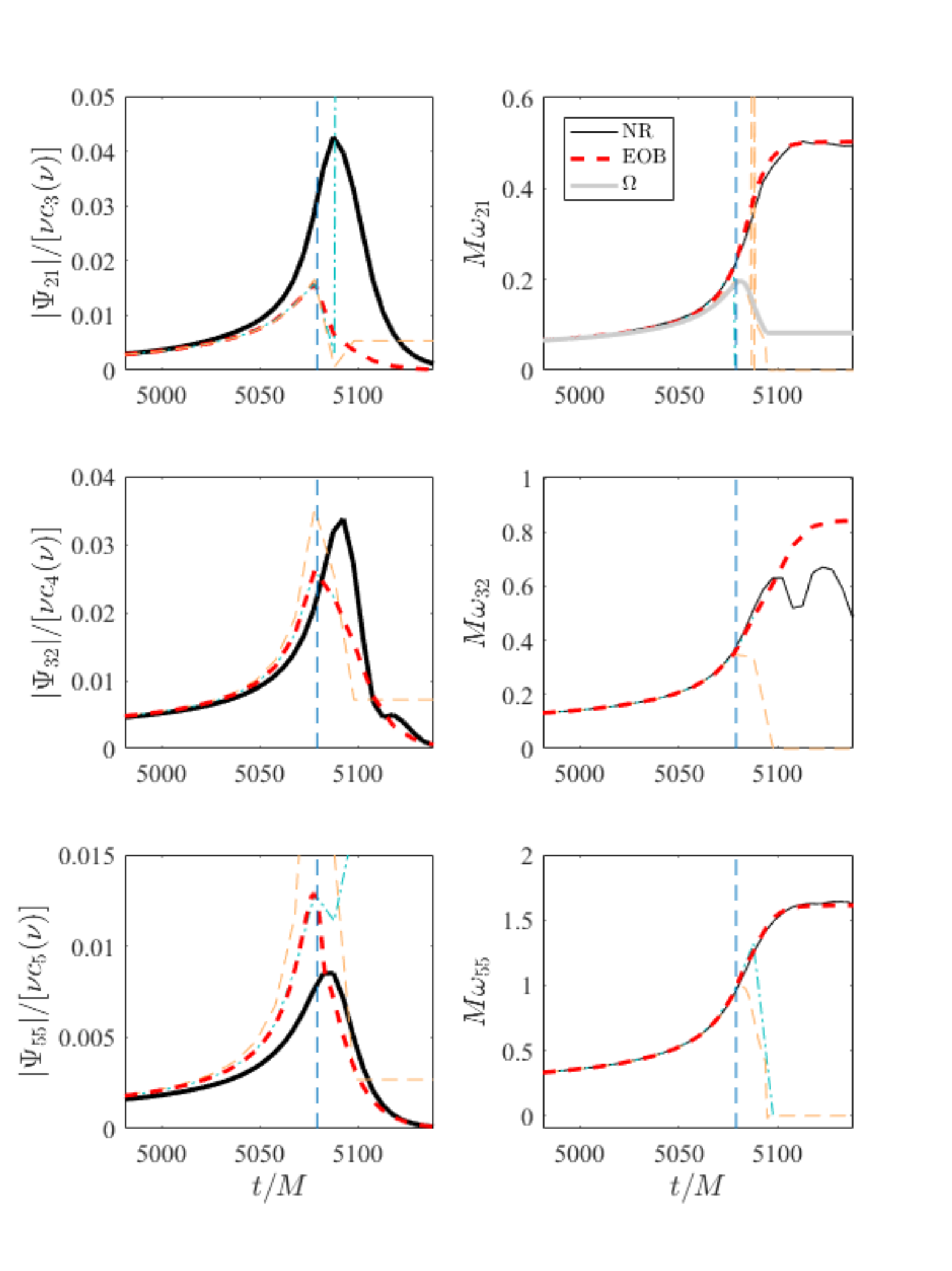} 
  \caption{EOB/NR waveform comparison for {\tt SXS:BBH:1466} for modes $(2,1)$, $(3,2)$ and $(5,5)$.
    The frequency of the $(2,1)$ mode behaves correctly through merger and ringdown, while the EOB
    amplitude largely underestimates the NR one. As in Fig.~\ref{fig:ell_neq_emm}, the orange curve
    is the purely analytical EOB waveform, while the light blue one is the NQC corrected. The vertical
    line marks the merger location}
\label{fig:hlm_q1p9}
\end{center}
\end{figure}
\begin{figure}[t]
\begin{center}
  \includegraphics[width=\columnwidth]{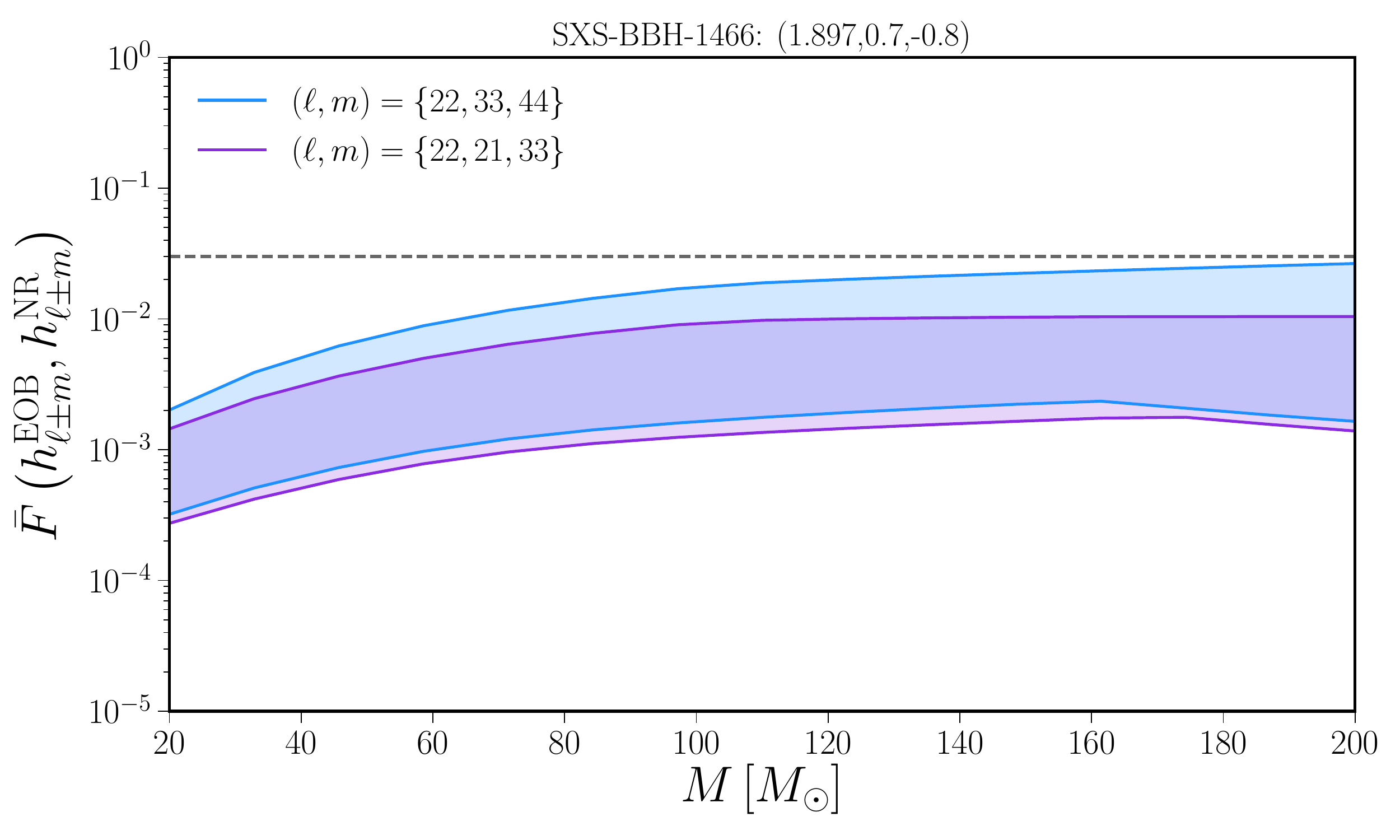}
  \caption{\label{fig:MM_SXS_1466} The minimum and maximum EOB/NR unfaithfulness for
    {\tt SXS:BBH:1466} over the whole sky. The blue curve uses the $(2,2), (3,3)$
    and $(4,4)$ modes. The purple curve uses the $(2,2), (2,1)$ and $(3,3)$ modes.
    Worst case mismatches occur near edge on configurations with the unfaithfulness being below 3\% up to $200 M_{\odot}$.  }
\end{center}
\end{figure}

\begin{figure}[t]
\begin{center}
\includegraphics[width=0.42\textwidth]{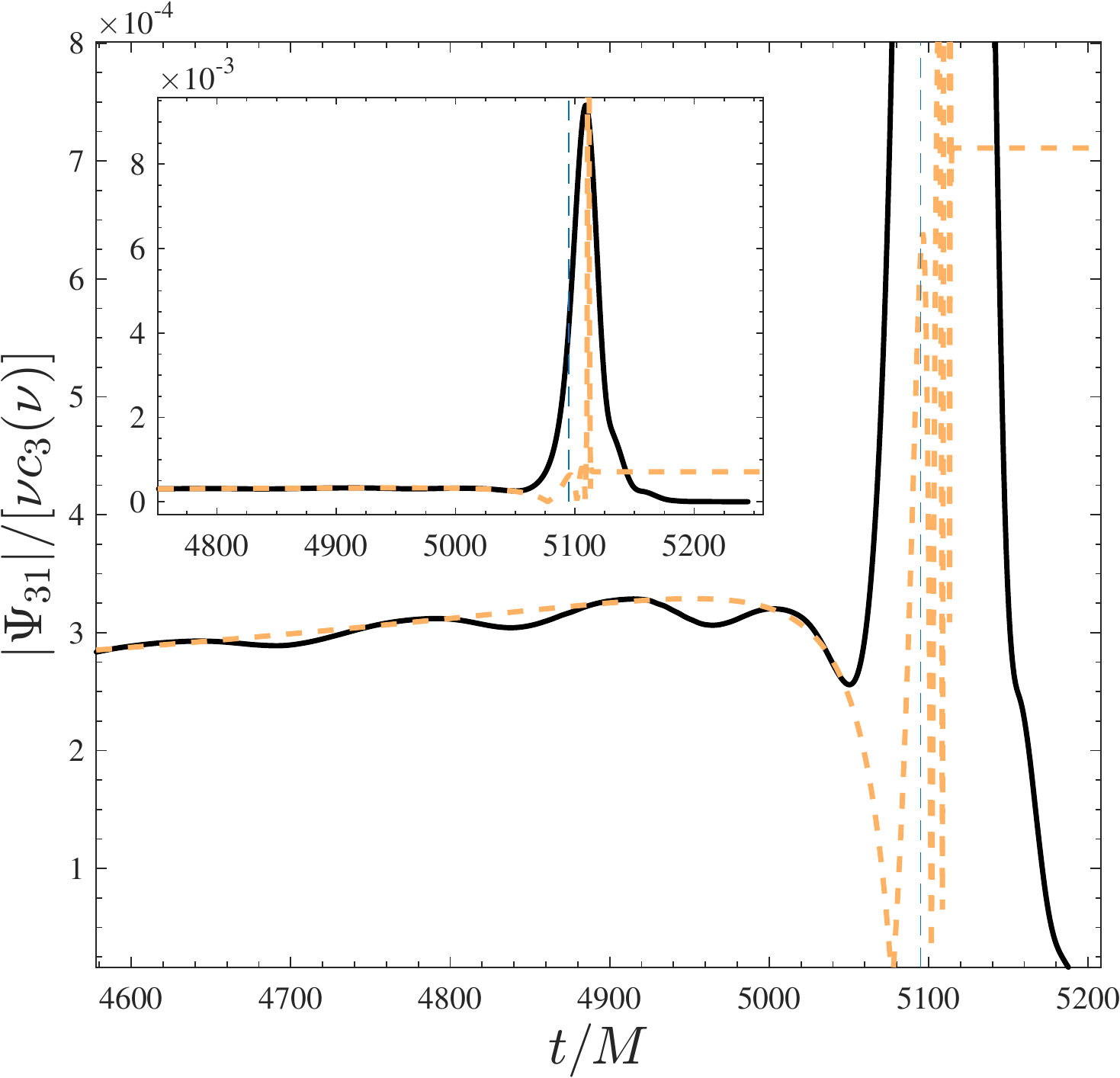}
\caption{Mode $(3,1)$: comparison between the EOB amplitude (orange)
  and the corresponding NR one from dataset {\tt SXS:BBH:1496}, $(1.1584,+0.7997,+0.0285)$.
  While the analytical waveform has a zero because of $\hat{f}_{31}$, 
  the NR one just shows  a glimpse of a global minimum, probably because if insufficient
  numerical resolution. Note however the excellent qualitative and quantitative 
  consistency between the two waveforms up to that point.}
\label{fig:h31_q1p16}
\end{center}
\end{figure}

\subsection{Peculiar behavior of $\bm{m=1}$ waveform amplitudes for $\bm{1\leq q \leq 2}$.}
Reference~\cite{Cotesta:2018fcv} pointed out that a few NR simulations exhibit a minimum
in the $(2,1)$ mode amplitude in the late inspiral phase. Such behavior was found in
4 SXS datasets: {\tt SXS:BBH:0254} $(2,+0.6,-0.6)$; {\tt SXS:BBH:0612} $(1.6,0.5,-0.5)$;
{\tt SXS:BBH:0614} $(2,+0.75,-0.5)$; and {\tt SXS:BBH:1377} $(1.1,-0.4,-0.7)$. 
Only the first among these dataset if public through the SXS catalog.
In addition, Ref.~\cite{Cotesta:2018fcv} noticed that the same feature is present 
in the EOB resummed waveform (both in orbital-factorized and non-orbital factorized form).
An explanation of this phenomenon was suggested on the basis of leading-order
considerations, that were similarly proven using a 3PN-based analysis. 
In addition, Ref.~\cite{Cotesta:2018fcv} compared the PN prediction
for the frequency corresponding to the minimum of the $(2,1)$ mode with
the value extracted from NR simulations. From this PN-based analysis, Ref.~\cite{Cotesta:2018fcv}
suggested that the phenomenon comes from a compensation between the spinning 
and leading-order nonspinning terms entering the $(2,1)$ mode. Notably, the PN 
based analysis aimed at explaining this feature qualitatively as well as 
semi-quantitatively (see Table~I in Ref.~\cite{Cotesta:2018fcv}).

Here we revisit the analysis of Ref.~\cite{Cotesta:2018fcv} and we attempt to
improve it along several directions thanks to the robustness of our factorized and
resummed waveform amplitudes. In brief we can show that: (i) focusing on
the same datasets considered in Ref.~\cite{Cotesta:2018fcv}, we illustrate that
the $(2,1)$, {\it purely analytical} EOB amplitude has a minimum (in fact,
a zero) rather close to the values reported in Table~I of Ref.~\cite{Cotesta:2018fcv},
and definitely much closer than the PN-based prediction; (ii) the phenomenon
is here understood as coming from the compensation, occurring at a given
frequency, between the two (inverse-resummed) macro-terms that compose the
analytically resummed expression of $\hat{f}_{21}^{S}$, one proportional to $X_{12}$
and the other one proportional to $\tilde{a}_{12}$, and that appear with opposite signs;
(iii) guided by this analytical understanding, we investigated whether some
of the currently available simulations in the SXS catalog may develop a zero
(that occurs in fact as a cusp) in the amplitude. Quite remarkably we found that 
it is indeed the case for {\tt SXS:BBH:1466}, $(1.9,+0.70,-0.8)$, that shows a 
clean minimum that is perfectly consistent with the EOB-based analytical prediction;
(iv) since the same structure, with the minus sign, is present also in
other $m=\text{odd}$ modes, we investigated whether the same phenomenon
may show up also in some of the other SXS datasets. Interestingly,
we found that also the $(3,1)$ mode of {\tt SXS:BBH:1496} is consistent 
with the EOB-predicted analytical behavior, suggesting that such features
may occur in several modes.

Let us now discuss in detail the four points listed above. Figure~\ref{fig:h21_m04_m07}
illustrates an EOB analytical waveform for $(1.1,-0.4,+0.7)$, that corresponds to the
dataset SXS:BBH:1377. As mentioned above, this simulation is not public and so
we cannot perform an explicit EOB/NR comparison. The top panel shows the $\ell=m=2$ 
waveform amplitude together with the EOB 
orbital frequency $M\Omega$. The middle panel shows the $(2,1)$ waveform amplitude, 
that develops a zero highlighted by a marker. It turns out that this zero precisely corresponds 
to the zero of the $\hat{f}_{21}^{\rm S}$ function once evaluated at $x=(M\Omega)^{2/3}$. 
This function is shown, versus $x$, in the bottom panel of Fig.~\ref{fig:h21_m04_m07}.
To be more quantitative, the last row of Table~\ref{tab:21freq} lists the 
corresponding frequency, that is identical to the NR-extracted value
reported in the corresponding last column of Table~I of~\cite{Cotesta:2018fcv}.
To check the model further, we explored also the other two cases in the Table,
similarly finding a rather good agreement between the EOB orbital frequency
corresponding to the zero and the NR value\footnote{Note that Ref.~\cite{Cotesta:2018fcv}
does not explain how their $M\Omega_0^{\rm NR}$ is computed. We may imagine that it is
just given by the NR orbital frequency divided by two, which is slightly different
from the EOB orbital frequency we include due to the presence of tail terms and
other effects.}.
Our reasoning relies on our orbital factorized waveform, and in particular
on the definition of $\hat{f}_{21}^{\rm S}$. However, Ref.~\cite{Cotesta:2018fcv}
pointed out that a zero in the amplitude may occur also in the standard, non
orbital-factorized, waveform amplitude. To make some quantitative statement,
we also consider the function
\be
\label{eq:f_taylor}
f_{21}^{\rm orb+S}=X_{12}\left(\rho^{\rm orb}_{21}\right)^2+f_{21}^{\rm S},
\ee
where both $\rho_{21}^{\rm orb}$ and $f_{21}^{\rm S}$ are kept in PN-expanded form.
The orbital term is given in the usual Taylor-expanded
form $\rho_{21}^{\rm orb}= 1 + (\dots)x + (\dots)x^2 + (\dots)x^3 + (\dots)x^4 + (dots)x^5$.
The spin term, at NNLO, reads
\be
f_{21}^{\rm S}=-\dfrac{3}{2}\tilde{a}_{12}x^{1/2}+c_{\rm SO}^{\rm NLO}x^{3/2}+c_{\rm SS}^{\rm LO}x^2 + c_{\rm SO}^{\rm NNLO}x^{5/2},
\ee
where
\begin{align}
  c_{\rm SO}^{\rm NLO} &=\left(\dfrac{110}{21}+\dfrac{79}{84}\nu\right)\tilde{a}_{12}-\dfrac{13}{84}\tilde{a}_0X_{12},\\
  c_{\rm SS}^{\rm LO}  &=-\dfrac{27}{8}\tilde{a}_0\tilde{a}_{12}+\dfrac{3}{8}X_{12}\left(\tilde{a}_1^2+\dfrac{10}{3}\tilde{a}_1\tilde{a}_2+\tilde{a}_2^2\right),\\
  c_{\rm SO}^{\rm NNLO} &= \left(-\dfrac{3331}{1008}-\dfrac{13}{504}\nu+\dfrac{613}{1008}\nu^2\right)\tilde{a}_{12}\nonumber\\
                    &+\left(-\dfrac{443}{252}+\dfrac{1735}{1008}\right)\tilde{a}_0X_{12}+\dfrac{3}{4}\tilde{a}_0^2\tilde{a}_{12}.
\end{align}
For the configuration $(1.1,-0.4,-0.7)$, this function, represented versus $x$, {\it does not}
have a zero, as illustrated by the magenta line in the bottom panel of Fig.~\ref{fig:h21_m04_m07}.

The closeness between the numbers in Table~\ref{tab:21freq} prompted us to
additionally investigate for which values of spin and mass ratio the analytical
$(2,1)$ amplitude develops a zero before merger frequency.
Comparing with the configurations available through the SXS catalog  
(notably those up to February 3, 2019), we found that the parameters
of dataset  {\tt SXS:BBH:1466} are such that the zero in the amplitude
occurs at a frequency that  is {\it smaller} than the merger frequency.
We then explicitly checked the $(2,1)$ mode of this simulation and,
as illustrated in Fig.~\ref{fig:h21_q1p9}, we found that it has a local
minimum, that is very consistent with the cusp in the analytic EOB
waveform modulus.
In addition, Fig.~\ref{fig:hlm_q1p9} illustrates the behavior of the full
waveform completed by NQC corrections and ringdown. As above, we show
together the more difficult modes to model, $(2,1)$ and $(3,2)$, with $(5,5)$.
The figure highlights that the $(2,1)$ frequency is well captured by
the analytical model, although the amplitude is underestimated by more
than a factor two. Consistently, Fig.~\ref{fig:MM_SXS_1466} shows that
the minimum and maximum unfaithfulness over the whole sky is always
below $3\%$. This makes us confident that \TEOBiResumSM{} can give a
reliable representation of the $(2,1)$ mode also in this special
region of the parameter space, since it naturally incorporates
a feature that is absent in {\tt SEOBNRv4HM}~\cite{Cotesta:2018fcv}.
One should however be aware that the $(2,1)$ EOB mode is not as good for
the case $(2,+0.60,-0.60)$, where the corresponding
NR waveform is found to have very a clean minimum much closer
to the merger frequency, as noted in Ref.~\cite{Cotesta:2018fcv}.
This is probably due to lack of additional analytical information
to improve the behavior of the $(2,1)$ mode in the strong-field regime.
It would be interesting to investigate, for future work, whether
higher-order PN terms (e.g. those obtained after hybridization with
test-mass results, similarly to the procedure followed for the $m=$~even)
could be useful to improve the behavior of the $(2,1)$ EOB amplitude for $(2,+0.60,-0.60)$.

As a last exploratory study, we investigated whether some  of the
other $m$-odd multipolar amplitudes can develop a zero at a frequency
smaller than the merger frequency, and we found this happens for several modes.
In the SXS catalog (up to February 3, 2019)
we identified a configuration where, analytically, we may expect a zero in the $(3,1)$ mode.
This is {\tt SXS:BBH:1496}, with parameters $(1.1584,0.7997,0.0285)$. Figure~\ref{fig:h31_q1p16}
compares the analytical EOB waveform amplitude with the NR one. We think it is remarkable
that the NR is consistent with the analytic waveform (modulo some numerical oscillation)
up to $t/M\simeq 5050$. At this time the NR waveform develops a local dip that, 
we conjecture, would eventually lead to an approximate cusp by increasing the
resolution. We hope that these special features of the waveform could be investigated
in more detail by dedicated NR simulations.

\section{Conclusions}
\label{sec:conclusions}
We have introduced \TEOBiResumSM{}, an improved, NR-informed, EOB model for 
nonprecessing, spinning, coalescing black hole binaries. This model incorporates several 
subdominant waveform modes, beyond the quadrupolar one, that are completed
through merger and ringdown. The work presented here generalizes to the spinning
case the nonspinning model {\tt TEOBiResumMultipoles} presented in Paper~I, Ref.~\cite{Nagar:2019wds}.
Generally speaking, we found that modes with $m=\ell$, up to $\ell=5$, are the
most  robust ones all over the parameter space covered by the SXS and BAM 
NR simulations at our disposal. The other modes, and especially
the most relevant $(2,1)$ one, can be nonrobust for medium-to-large value of the spins
anti-aligned with the orbital angular momentum. The waveform modes (and thus the radiation reaction)
rely on a new resummed representation for the waveform multipolar amplitudes,
that improves their robustness and predictive power through late-inspiral
and merger, as well as a new, NR-informed, representation of the ringdown part.

Our results can be summarized as follows:
\begin{enumerate}
\item The new analytical description of the binary relative dynamics due to the orbital-factorized
and resummed radiation reaction entails a new (somehow simpler) determination of 
the EOB flexibility functions $\{a_6^c(\nu),c_3(\nu,S_1,S_2)\}$, that is different from the 
one used in \TEOBResumS{}~\cite{Nagar:2018zoe}. We computed the EOB/NR unfaithfulness 
for the $(2,2)$ mode and found that it is always below $0.5\%$ (except for a single outlier that
grazes $0.85\%$) all over the current release of the SXS NR waveform catalog (595 datasets) 
as  well as on additional data from BAM code spanning up to mass ratio $q=18$. 
We remark that the performance of the model is largely improved, with respect to 
Ref.~\cite{Nagar:2018zoe}, in the large-mass-ratio, large-spin corner, notably for $(8,+0.85,+0.85)$.

\item We provided a prescription for completing higher modes trough merger 
and ringdown. Such prescription is the carbon copy of what previously done in the 
nonspinning case and discussed in Paper~I. No new conceptual modification to the 
procedure were introduced here. The novelty is the introduction of the spin-dependence
in the NR-informed fits of the quantities needed to determine the NQC parameters
and the peak-postpeak (ringdown) behavior. Such fits are done factorizing some
leading-order spin contributions, as well as incorporating test-mass information,
in an attempt to reduce the flexibility in the fits and to improve their
robustness all over the parameter space.

We found that for $\ell=m$ modes, up to $\ell=m=5$, the model is very robust and
reliable. When putting together all $m=\ell$ modes up to $\ell=4$, the maximal
EOB/NR unfaithfulness all over the sky (with Advanced LIGO noise) is always
well below $3\%$ up to total mass $M=120M_\odot$, that is exceeded slightly after
because of lack of accuracy in both the EOB and the NR data itself, especially
in the $(4,4)$ mode. The model peforms similarly well ($\bar{F}\lesssim 3\%$) aso
when the $(2,1)$ mode is included.
We have however pointed out that for large values of the spin, anti-aligned with
the angular momentum, e.g. as $(3,-0.85,-0.85)$, inaccuracies in the postmerger
EOB dynamics prevent one to get accurate $(2,1)$ mode through merger and ringdown.

\item Inspired by previous work, we could confirm that the
  phenomenology of the $(2,1)$ mode is rich, in particular that its amplitude
  can have a zero during the late-inspiral before merger
  for nearly equal-mass binaries. We have presented a quantitative understanding 
  of the phenomenon. We also showed that the EOB waveform, in its orbital-factorized 
  and resummed avatar of Ref.~\cite{Nagar:2016ayt,Messina:2018ghh}, 
  can accurately reproduce NR waveforms with the same phenomenology,
  at least when  the frequency of the zero is sufficiently far from merger. We remark 
  that was achieved {\it without} advocating any additional ad-hoc calibration or 
  tuning of phenomenological parameters entering the waveform amplitude. 
  Quite interestingly, the same phenomenon may occur also in some of the 
  other of the $m=\text{odd}$ modes. In particular, we could find,
  for the  $(3,1)$ mode, a SXS configuration that shows this behavior
  and illustrate how it agrees with the analytical prediction.

\item In general, this work made us aware that the structure of the $(2,1)$ mode 
is very challenging to be modeled properly through peak and ringdown using the 
simple approach developed in Paper~I. Such difficulty is shared by other modes 
with $m\neq \ell$ in certain region of the parameter space, whenever the
peak of such mode is significantly ($\sim 7-8M$) delayed with respect to the 
merger time (e.g. the $(4,3)$ or $(3,1)$). We consider the identification of
this difficulty as one of the most relevant outcomes of this work.
We think that the proper modelization of such $m\neq \ell$ modes in the
transition from the late inspiral up to the waveform peak should not be
done using brute force (e.g. by extending
the effective postmerger fits also {\it before} the peak) but rather that it requires 
a more detailed understanding of the underlying physical elements, in
particular: (i) the structure of the waveform (e.g. with the need of naturally
incorporating the zero in the amplitude also when it is known to exist at
rather high frequencies, e.g. for $(2,+0.60,-0.60)$); and (ii) the behavior 
of the EOB relative dynamics (notably mirrored in the time evolution of 
the orbital frequency $\Omega(t)$) in the extreme region just after the merger, 
corresponding to very small radial separation. We have shown explicitly that
one of the analytical choices adopted in \TEOBResumS{}, i.e. the $u_c$ dependence
in the gyro-gravitomagnetic functions, was (partly) responsible of the problems
we encountered in modeling the $(2,1)$ mode (see Sec.~\ref{sec:no_uc}).
Together with a different choice of gauge, so to incorporate the test-black hole
spin-orbit interaction~\cite{Rettegno:2019tzh}, it might be possible to obtain
an improved EOB dynamics more robust also in the postmerger regime, so to easily
account for more subdominant multipoles via the usual NQC-completion and ringdown
matching procedure.

\item The results discussed in this paper were obtained with the {\tt Matlab} implementation
of the model. However, \TEOBiResumSM{} is freely available via a stand-alone
$C$-implementation~\cite{teobresums}. Tests of the code and evaluation of its performance
in parameter-estimation context are enclosed in the related documentation and will be 
additionally discussed in a forthcoming publication~\cite{Riemenschneider:2020}. 
In particular, instead of iterating on the NQC parameters  $(a_1,a_2)$, the $C$-implementation 
uses suitably designed fits. The performance, in terms of the $\bar{F}_{\rm EOB/NR}$ 
diagnostics, is fully compatible with what discussed here.

\end{enumerate}

\acknowledgments
We are grateful to T.~Damour for discussions.
F.~M., G.~R. and P.~R. thank IHES for hospitality during the development of this work.
We thank S.~Bernuzzi and R.~Gamba for continuous help and assistance in the 
development of the stand-alone $C$-implementation of \TEOBiResumSM{}.

\appendix
\section{Nonspinning limit}
\label{sec:nospin_lim}
\begin{figure*}[t]
	\begin{center}
		\includegraphics[width=0.45\textwidth]{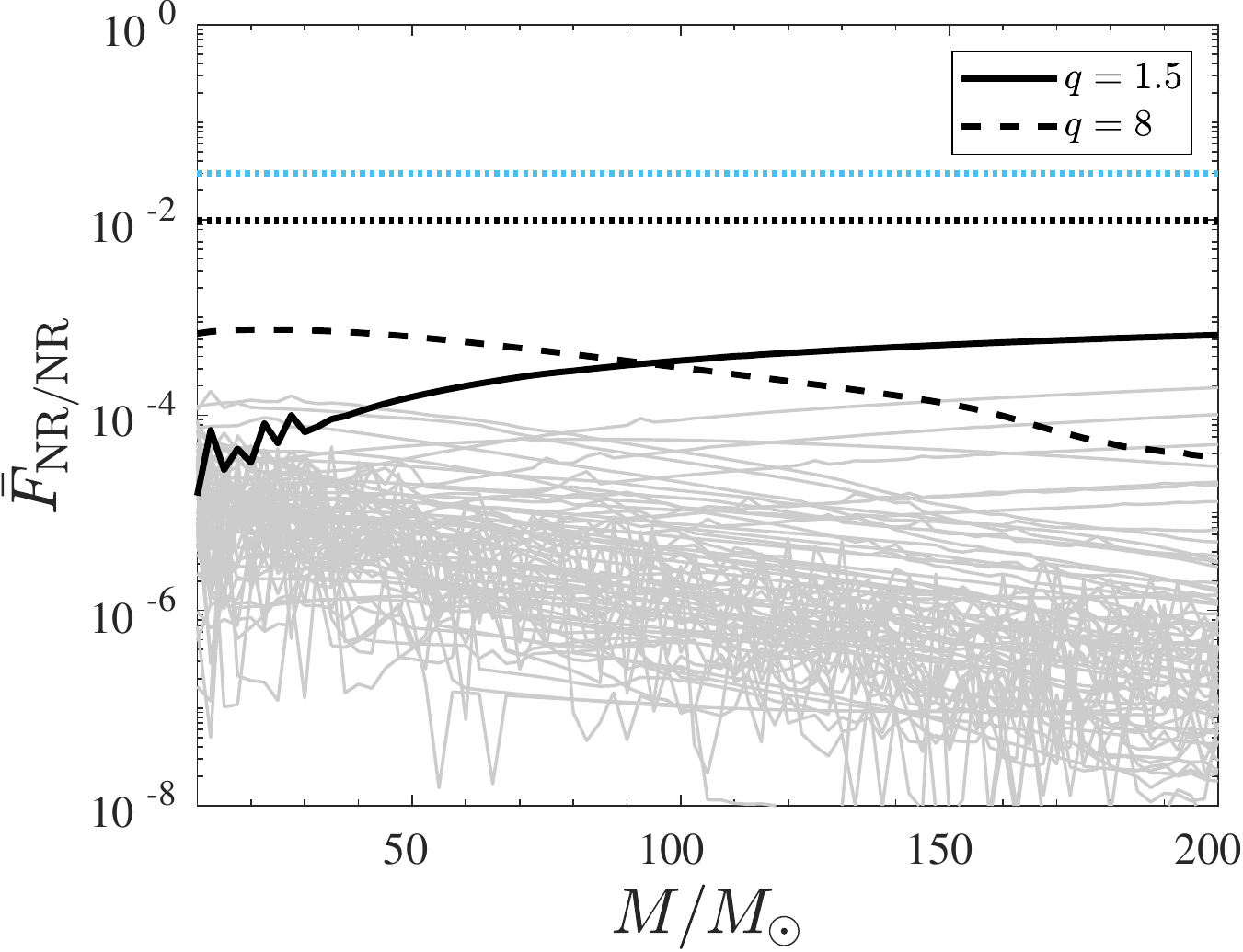}
		\includegraphics[width=0.45\textwidth]{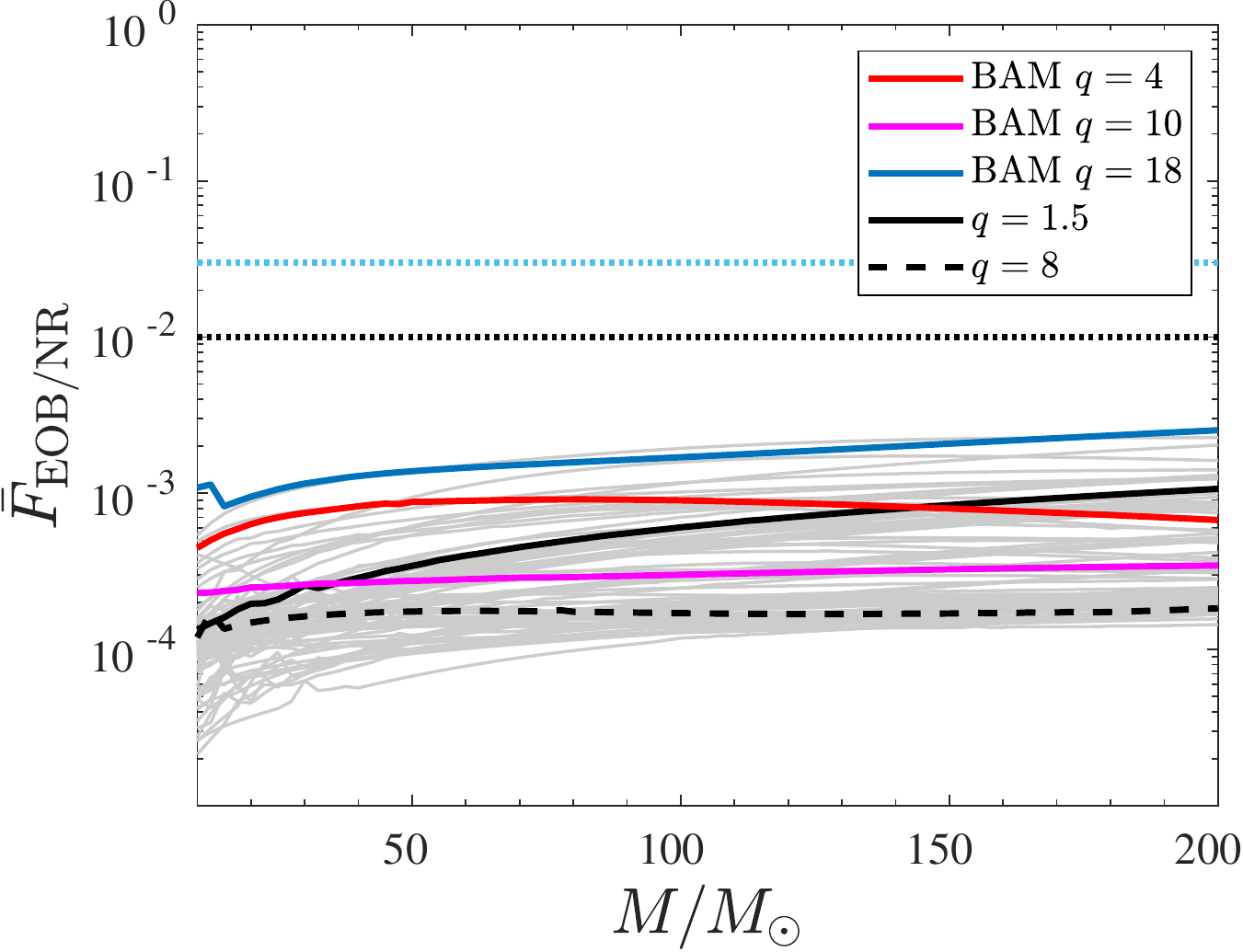}
		\caption{Nonspinning sector. Left panel: the NR/NR unfaithfulness between the highest and 
		                         second highest resolutions available. Right panel: EOB/NR unfaithfulness
					for all available non-spinning datasets. The analytical waveforms are 
					evaluated with $(\chi_1,\chi_2)=(0,10^{-4})$, so as to probe the 
					stability of the model and its robustness in this regime.}
		\label{fig:barF_SXS_nospin}
	\end{center}
\end{figure*}

Here we briefly comment on the performance of the model in the nonspinning limit,
as an addendum to the extensive discussion reported in Ref.~\cite{Nagar:2019wds}.
In total 89 non-spinning NR waveforms are available. These are listed in Tables~\ref{tab:nospin_1}-\ref{tab:nospin_2}.
Of these, 19 \texttt{SXS} and 3 \texttt{BAM} waveforms were used to inform \TEOBiResumM{} for the 
postmerger part, see Ref.~\cite{Nagar:2019wds}.
We compute $\bar{F}$ from Eq.~\eqref{eq:barF}.  Note that the analytical EOB waveforms are obtained with $\chi_1=0$ 
and $\chi_2=10^{-4}$, so to actually probing the spin-dependent dynamics 
in the nonspinning limit. Figure~\ref{fig:barF_SXS_nospin} shows $\bar{F}_{\rm NR/NR}$ (left) 
for the 86 \texttt{SXS} nonspinning waveforms and  $\bar{F}_{\rm EOB/NR}$ (right) for the full set of 89 nonspinning waveforms. 
Only two waveforms show a large $\bar{F}_{\rm NR/NR}$ value: {\tt SXS:BBH:0093} $(q=1.5)$ and {\tt SXS:BBH:0063} $(q=8)$,
though both remain below $8\times 10^{-4}$. 
Consistently with Ref.~\cite{Nagar:2019wds}, $\bar{F}_{\rm EOB/NR}$ is well behaved all over.
The largest unfaithfulness is reached by the {\tt BAM}, $q=18$ waveform at $\max(\bar{F}_{\rm EOB/NR})=0.2533\%$.

\begin{table}
  \caption{\label{tab:equal_SXS} This table summarizes the SXS NR waveform data
    in the \textit{postpeak-calibration} set, with mass-ratio $q=1$. From left to right,
    the columns report: the SXS simulation number, mass ratio and dimensionless spins $\chi_i$,
    and the maximum value of the unfaithfulness $\bar{F}$ between: The two highest resolutions 
    of the NR dataset, if available, see Fig.\ref{fig:barF_SXS_error}, and between EOB and NR, see Fig.~\ref{fig:barF_SXSBAM}.}
\begin{center}
\begin{ruledtabular}
\begin{tabular}{r | l l l l}
\# & {\rm id} & $(q,\chi_1,\chi_2)$ & $\bar{F}^{\rm max}_{\rm NR/NR}[\%]$ & $\bar{F}^{\rm max}_{\rm EOB/NR}[\%]$\\\hline
1 & BBH:0178 & $(1,+0.9942,+0.9942)$ & $0.0066$ & $0.0259$\\
2 & BBH:0177 & $(1,+0.9893,+0.9893)$ & $0.0021$ & $0.0345$\\
3 & BBH:0172 & $(1,+0.9794,+0.98)$ & $0.0022$ & $0.0188$\\
4 & BBH:0157 & $(1,+0.95,+0.95)$ & $0.0027$ & $0.0329$\\
5 & BBH:0160 & $(1,+0.9,+0.9)$ & $0.0118$ & $0.0114$\\
6 & BBH:0153 & $(1,+0.85,+0.85)$ & .. & $0.0249$\\
7 & BBH:0230 & $(1,+0.8,+0.8)$ & $0.0016$ & $0.0737$\\
8 & BBH:0228 & $(1,+0.6,+0.6)$ & $0.0080$ & $0.1458$\\
9 & BBH:0150 & $(1,+0.2,+0.2)$ & $0.0027$ & $0.0723$\\
10 & BBH:0149 & $(1,-0.2,-0.2)$ & $0.0037$ & $0.1369$\\
11 & BBH:0148 & $(1,-0.44,-0.44)$ & $0.0013$ & $0.0688$\\
12 & BBH:0215 & $(1,-0.6,-0.6)$ & $0.0040$ & $0.0903$\\
13 & BBH:0154 & $(1,-0.8,-0.8)$ & $0.0036$ & $0.0836$\\
14 & BBH:0212 & $(1,-0.8,-0.8)$ & $0.0032$ & $0.0610$\\
15 & BBH:0159 & $(1,-0.9,-0.9)$ & $0.0069$ & $0.0295$\\
16 & BBH:0156 & $(1,-0.95,-0.95)$ & $0.0055$ & $0.0798$\\
17 & BBH:0231 & $(1,+0.9,0)$ & $0.0046$ & $0.1094$\\
18 & BBH:0232 & $(1,+0.9,+0.5)$ & $0.0073$ & $0.0430$\\
19 & BBH:0229 & $(1,+0.65,+0.25)$ & $0.0053$ & $0.1411$\\
20 & BBH:0227 & $(1,+0.6,0)$ & $0.0052$ & $0.1776$\\
21 & BBH:0005 & $(1,+0.5,0)$ & $0.0592$ & $0.1396$\\
22 & BBH:0226 & $(1,-0.9,+0.5)$ & $0.0018$ & $0.0679$\\
23 & BBH:0224 & $(1,-0.8,+0.4)$ & $0.0020$ & $0.0842$\\
24 & BBH:0225 & $(1,+0.8,+0.4)$ & $0.0014$ & $0.0784$\\
25 & BBH:0223 & $(1,+0.3,0)$ & $0.1520$ & $0.1071$\\
26 & BBH:0222 & $(1,-0.3,0)$ & $0.1598$ & $0.0616$\\
27 & BBH:0220 & $(1,-0.8,-0.4)$ & $0.0040$ & $0.1042$\\
28 & BBH:0221 & $(1,+0.8,-0.4)$ & $0.0053$ & $0.1238$\\
29 & BBH:0004 & $(1,-0.5,0)$ & $0.0189$ & $0.0998$\\
30 & BBH:0218 & $(1,+0.5,-0.5)$ & $0.2160$ & $0.1794$\\
31 & BBH:0219 & $(1,+0.9,-0.5)$ & $0.0076$ & $0.1173$\\
32 & BBH:0216 & $(1,-0.6,0)$ & $0.0040$ & $0.0797$\\
33 & BBH:0217 & $(1,-0.6,+0.6)$ & $0.0048$ & $0.1103$\\
34 & BBH:0214 & $(1,-0.62,-0.25)$ & $0.0010$ & $0.0621$\\
35 & BBH:0213 & $(1,-0.8,+0.8)$ & $0.0040$ & $0.0938$\\
36 & BBH:0209 & $(1,-0.9,-0.5)$ & $0.0010$ & $0.0610$\\
37 & BBH:0210 & $(1,-0.9,0)$ & $0.0024$ & $0.0708$\\
38 & BBH:0211 & $(1,-0.9,+0.9)$ & $0.0027$ & $0.0918$
\end{tabular}
\end{ruledtabular}
\end{center}
\end{table}

\begin{table}
  \caption{\label{tab:q2_SXS_old}This table summarizes the SXS NR waveform data
    in the \textit{postpeak-calibration} set, with mass-ratio $1<q\leq 2$.
    From left to right, the columns report: the SXS simulation number, mass ratio and dimensionless spins $\chi_i$,
    and the maximum value of the unfaithfulness $\bar{F}$ between: The two highest resolutions 
    of the NR dataset, if available, see Fig.\ref{fig:barF_SXS_error}, and between EOB and NR, see Fig.~\ref{fig:barF_SXSBAM}.}
\begin{center}
\begin{ruledtabular}
\begin{tabular}{r | l l l l}
\# & {\rm id} & $(q,\chi_1,\chi_2)$ & $\bar{F}^{\rm max}_{\rm NR/NR}[\%]$ & $\bar{F}^{\rm max}_{\rm EOB/NR}[\%]$\\\hline
39 & BBH:0306 & $(1.3,+0.96,-0.9)$ & $0.0031$ & $0.2059$\\
40 & BBH:0013 & $(1.5,+0.5,0)$ & .. & $0.1420$\\
41 & BBH:0025 & $(1.5,+0.5,-0.5)$ & $0.0278$ & $0.2446$\\
42 & BBH:0016 & $(1.5,-0.5,0)$ & $0.0009$ & $0.0262$\\
43 & BBH:0019 & $(1.5,-0.5,+0.5)$ & $0.0213$ & $0.0408$\\
44 & BBH:0258 & $(2,+0.87,-0.85)$ & $0.0061$ & $0.2599$\\
45 & BBH:0257 & $(2,+0.85,+0.85)$ & $0.0024$ & $0.4144$\\
46 & BBH:0254 & $(2,+0.6,-0.6)$ & $0.0009$ & $0.2218$\\
47 & BBH:0255 & $(2,+0.6,0)$ & $0.0023$ & $0.1324$\\
48 & BBH:0256 & $(2,+0.6,+0.6)$ & $0.0068$ & $0.0771$\\
49 & BBH:0253 & $(2,+0.5,+0.5)$ & $0.0040$ & $0.0844$\\
50 & BBH:0252 & $(2,+0.37,-0.85)$ & $0.0029$ & $0.1659$\\
51 & BBH:0249 & $(2,+0.3,-0.3)$ & $0.0057$ & $0.0888$\\
52 & BBH:0250 & $(2,+0.3,0)$ & $0.0045$ & $0.0837$\\
53 & BBH:0251 & $(2,+0.3,+0.3)$ & $0.0037$ & $0.0755$\\
54 & BBH:0248 & $(2,+0.13,+0.85)$ & $0.0030$ & $0.0666$\\
55 & BBH:0244 & $(2,0,-0.6)$ & $0.0010$ & $0.0542$\\
56 & BBH:0245 & $(2,0,-0.3)$ & $0.0226$ & $0.0385$\\
57 & BBH:0246 & $(2,0,+0.3)$ & $0.0081$ & $0.0395$\\
58 & BBH:0247 & $(2,0,+0.6)$ & $0.0041$ & $0.0440$\\
59 & BBH:0243 & $(2,-0.13,-0.85)$ & $0.0006$ & $0.0538$\\
60 & BBH:0240 & $(2,-0.3,-0.3)$ & $0.0614$ & $0.0235$\\
61 & BBH:0241 & $(2,-0.3,0)$ & $0.0129$ & $0.0251$\\
62 & BBH:0242 & $(2,-0.3,+0.3)$ & $0.0260$ & $0.0282$\\
63 & BBH:0239 & $(2,-0.37,+0.85)$ & $0.0005$ & $0.0338$\\
64 & BBH:0238 & $(2,-0.5,-0.5)$ & $0.1110$ & $0.0351$\\
65 & BBH:0235 & $(2,-0.6,-0.6)$ & $0.0048$ & $0.0267$\\
66 & BBH:0236 & $(2,-0.6,0)$ & $0.0029$ & $0.0483$\\
67 & BBH:0237 & $(2,-0.6,+0.6)$ & $0.0014$ & $0.0880$\\
68 & BBH:0234 & $(2,-0.85,-0.85)$ & $0.0049$ & $0.0709$\\
69 & BBH:0233 & $(2,-0.87,+0.85)$ & $0.0012$ & $0.1564$
\end{tabular}
\end{ruledtabular}
\end{center}
\end{table}
\begin{table}
  \caption{\label{tab:q3_SXS_old}This table summarizes the SXS NR waveform data in the \textit{postpeak-calibration}
    set, with mass-ratio $3\leq q$. From left to right, the columns report: the SXS simulation number, mass ratio and dimensionless spins spins $\chi_i$,
    and the maximum value of the unfaithfulness $\bar{F}$ between: The two highest resolutions 
    of the NR dataset, if available, see Fig.\ref{fig:barF_SXS_error}, and between EOB and NR, see Fig.~\ref{fig:barF_SXSBAM}.}
\begin{center}
\begin{ruledtabular}
\begin{tabular}{r | l l l l}
\# & {\rm id} & $(q,\chi_1,\chi_2)$ & $\bar{F}^{\rm max}_{\rm NR/NR}[\%]$ & $\bar{F}^{\rm max}_{\rm EOB/NR}[\%]$\\\hline
70 & BBH:0036 & $(3,-0.5,0)$ & $0.0010$ & $0.0405$\\
71 & BBH:0045 & $(3,+0.5,-0.5)$ & .. & $0.1456$\\
72 & BBH:0174 & $(3,+0.5,0)$ & $0.1040$ & $0.1828$\\
73 & BBH:0260 & $(3,-0.85,-0.85)$ & $0.0004$ & $0.0744$\\
74 & BBH:0261 & $(3,-0.73,+0.85)$ & $0.0016$ & $0.1453$\\
75 & BBH:0262 & $(3,-0.6,0)$ & $0.0002$ & $0.0362$\\
76 & BBH:0263 & $(3,-0.6,+0.6)$ & $0.0009$ & $0.0914$\\
77 & BBH:0264 & $(3,-0.6,-0.6)$ & $0.0024$ & $0.0449$\\
78 & BBH:0265 & $(3,-0.6,-0.4)$ & $0.0008$ & $0.0329$\\
79 & BBH:0266 & $(3,-0.6,+0.4)$ & $0.0003$ & $0.0714$\\
80 & BBH:0267 & $(3,-0.5,-0.5)$ & $0.0058$ & $0.0368$\\
81 & BBH:0268 & $(3,-0.4,-0.6)$ & $0.0016$ & $0.0249$\\
82 & BBH:0269 & $(3,-0.4,+0.6)$ & $0.0017$ & $0.0516$\\
83 & BBH:0270 & $(3,-0.3,-0.3)$ & $0.0038$ & $0.0217$\\
84 & BBH:0271 & $(3,-0.3,0)$ & $0.0014$ & $0.0255$\\
85 & BBH:0272 & $(3,-0.3,+0.3)$ & $0.0035$ & $0.0291$\\
86 & BBH:0273 & $(3,-0.27,-0.85)$ & $0.0027$ & $0.0605$\\
87 & BBH:0274 & $(3,-0.23,+0.85)$ & $0.0018$ & $0.0355$\\
88 & BBH:0275 & $(3,0,-0.6)$ & $0.0008$ & $0.0602$\\
89 & BBH:0276 & $(3,0,-0.3)$ & $0.0028$ & $0.0373$\\
90 & BBH:0277 & $(3,0,+0.3)$ & $0.0029$ & $0.0194$\\
91 & BBH:0278 & $(3,0,+0.6)$ & $0.0015$ & $0.0252$\\
92 & BBH:0279 & $(3,+0.23,-0.85)$ & $0.0010$ & $0.0670$\\
93 & BBH:0280 & $(3,+0.27,+0.85)$ & $0.0052$ & $0.0405$\\
94 & BBH:0281 & $(3,+0.3,-0.3)$ & $0.0027$ & $0.0729$\\
95 & BBH:0282 & $(3,+0.3,0)$ & $0.0011$ & $0.0223$\\
96 & BBH:0283 & $(3,+0.3,+0.3)$ & $0.0032$ & $0.0328$\\
97 & BBH:0284 & $(3,+0.4,-0.6)$ & $0.0005$ & $0.1288$\\
98 & BBH:0285 & $(3,+0.4,+0.6)$ & $0.0013$ & $0.0257$\\
99 & BBH:0286 & $(3,+0.5,+0.5)$ & $0.0022$ & $0.0257$\\
100 & BBH:0287 & $(3,+0.6,-0.6)$ & $0.0053$ & $0.1954$\\
101 & BBH:0288 & $(3,+0.6,-0.4)$ & $0.0006$ & $0.0683$\\
102 & BBH:0289 & $(3,+0.6,0)$ & $0.0005$ & $0.0624$\\
103 & BBH:0290 & $(3,+0.6,+0.4)$ & $0.0032$ & $0.0260$\\
104 & BBH:0291 & $(3,+0.6,+0.6)$ & $0.0010$ & $0.0129$\\
105 & BBH:0292 & $(3,+0.73,-0.85)$ & $0.0009$ & $0.3817$\\
106 & BBH:0293 & $(3,+0.85,+0.85)$ & $0.0046$ & $0.4764$\\
107 & BBH:0060 & $(5,-0.5,0)$ & .. & $0.0217$\\
108 & BBH:0110 & $(5,+0.5,0)$ & .. & $0.0383$\\
109 & BBH:0208 & $(5,-0.9,0)$ & $0.0385$ & $0.0667$\\
110 & BBH:0202 & $(7,+0.6,0)$ & $0.0048$ & $0.3976$\\
111 & BBH:0203 & $(7,+0.4,0)$ & $0.0095$ & $0.0556$\\
112 & BBH:0205 & $(7,-0.4,0)$ & $0.0040$ & $0.0484$\\
113 & BBH:0207 & $(7,-0.6,0)$ & $0.0011$ & $0.0613$\\
114 & BBH:0064 & $(8,-0.5,0)$ & $0.0338$ & $0.0325$\\
115 & BBH:0065 & $(8,+0.5,0)$ & $0.0189$ & $0.1440$\\
116 & BBH:1375 & $(8,-0.9,0)$ & .. & $0.1223$
\end{tabular}
\end{ruledtabular}
\end{center}
\end{table}
\begin{table}
  \caption{\label{tab:BAM}This table summarizes the \texttt{BAM} NR waveform data in
    the \textit{postpeak-calibration} set. From left to right, the columns report:
    the simulation number, mass ratio and dimensionless spins spins $\chi_i$,
    and the maximum value of the EOB/NR unfaithfulness $\bar{F}$, see Fig.~\ref{fig:barF_SXSBAM}.
    These waveforms were mostly presented in Refs.~\cite{Husa:2015iqa,Khan:2015jqa,Keitel:2016krm}}
\begin{center}
\begin{ruledtabular}
\begin{tabular}{l r | l l l}
 &\# & $(q,\chi_1,\chi_2)$ & $\bar{F}^{\rm max}_{\rm EOB/NR}[\%]$&\\\hline
&117 & $(2,+0.5,+0.5)$ & $0.3458$&\\
&118 & $(2,+0.75,+0.75)$ & $0.4149$&\\
&119 & $(3,-0.5,-0.5)$ & $0.1895$&\\
&120 & $(4,-0.75,-0.75)$ & $0.2898$&\\
&121 & $(4,-0.5,-0.5)$ & $0.1588$&\\
&122 & $(4,-0.25,-0.25)$ & $0.1096$&\\
&123 & $(4,+0.25,+0.25)$ & $0.0402$&\\
&124 & $(4,+0.5,+0.5)$ & $0.0385$&\\
&125 & $(4,+0.75,+0.75)$ & $0.0378$&\\
&126 & $(8,-0.85,-0.85)$ & $0.0791$&\\
&127 & $(8,+0.8,0)$ & $0.2555$&\\
&128 & $(8,+0.85,+0.85)$ & $0.2530$&\\
&129 & $(18,-0.8,0)$ & $0.1653$&\\
&130 & $(18,-0.4,0)$ & $0.0418$&\\
&131 & $(18,+0.4,0)$ & $0.0232$&\\
&132 & $(18,+0.8,0)$ & $0.1029$&
\end{tabular}
\end{ruledtabular}
\end{center}
\end{table}
\begin{table}
  \caption{\label{tab:SXS_new1}This table summarizes part of the \texttt{SXS} NR waveform data in the \textit{validation} set with $q=1$.
    From left to right, the columns report: the SXS simulation number, mass ratio and dimensionless spins spins $\chi_i$,
    and the maximum value of the unfaithfulness $\bar{F}$ between: The two highest resolutions 
    of the NR dataset, if available, see Fig.\ref{fig:barF_SXS_error}, and between EOB and NR, see Fig.~\ref{fig:barF_SXS_new}.}
\begin{center}
\begin{ruledtabular}
\begin{tabular}{r | l l l l}
\# & {\rm id} & $(q,\chi_1,\chi_2)$ & $\bar{F}^{\rm max}_{\rm NR/NR}[\%]$ & $\bar{F}^{\rm max}_{\rm EOB/NR}[\%]$\\\hline
133 & BBH:1124 & $(1,+0.9980,+0.9980)$ & .. & $0.0800$\\
134 & BBH:0158 & $(1,+0.97,+0.97)$ & $0.0031$ & $0.0510$\\
135 & BBH:0176 & $(1,+0.96,+0.96)$ & $0.0065$ & $0.0264$\\
136 & BBH:0155 & $(1,+0.8,+0.8)$ & $0.0035$ & $0.0749$\\
137 & BBH:1477 & $(1,+0.8,+0.8)$ & $0.0037$ & $0.0743$\\
138 & BBH:0328 & $(1,+0.8,+0.8)$ & $0.0034$ & $0.0733$\\
139 & BBH:2104 & $(1,+0.8,+0.8)$ & $0.0033$ & $0.0742$\\
140 & BBH:1481 & $(1,+0.8,+0.73)$ & $0.0032$ & $0.0839$\\
141 & BBH:0175 & $(1,+0.75,+0.75)$ & $0.0030$ & $0.0964$\\
142 & BBH:2106 & $(1,+0.9,+0.5)$ & $0.0064$ & $0.0441$\\
143 & BBH:1497 & $(1,+0.68,+0.67)$ & $0.0032$ & $0.1165$\\
144 & BBH:1495 & $(1,+0.78,+0.53)$ & $0.0058$ & $0.0698$\\
145 & BBH:0152 & $(1,+0.6,+0.6)$ & $0.0047$ & $0.1483$\\
146 & BBH:2099 & $(1,+0.8,+0.4)$ & $0.0048$ & $0.0790$\\
147 & BBH:2102 & $(1,+0.6,+0.6)$ & $0.0007$ & $0.1469$\\
148 & BBH:1123 & $(1,+0.5,+0.5)$ & $0.0033$ & $0.1899$\\
149 & BBH:0394 & $(1,+0.6,+0.4)$ & $0.0023$ & $0.0936$\\
150 & BBH:2103 & $(1,+0.65,+0.25)$ & $0.0022$ & $0.3458$\\
151 & BBH:2105 & $(1,+0.9,0)$ & $0.0002$ & $0.0652$\\
152 & BBH:1122 & $(1,+0.44,+0.44)$ & $0.0031$ & $0.2051$\\
153 & BBH:1503 & $(1,+0.73,+0.14)$ & $0.0028$ & $0.1642$\\
154 & BBH:1501 & $(1,+0.75,+0.1)$ & $0.0040$ & $0.1363$\\
155 & BBH:0326 & $(1,+0.8,0)$ & $0.0056$ & $0.1324$\\
156 & BBH:1507 & $(1,+0.5,+0.3)$ & $0.0032$ & $0.1965$\\
157 & BBH:1376 & $(1,+0.25,+0.5)$ & $0.0035$ & $0.2068$\\
158 & BBH:2101 & $(1,+0.6,0)$ & $0.0039$ & $0.1777$\\
159 & BBH:0418 & $(1,+0.4,0)$ & $0.0041$ & $0.1332$\\
160 & BBH:2095 & $(1,+0.8,-0.4)$ & $0.0006$ & $0.1245$\\
161 & BBH:2093 & $(1,+0.9,-0.5)$ & $0.0018$ & $0.1189$\\
162 & BBH:2097 & $(1,+0.3,0)$ & $0.0014$ & $0.3091$\\
163 & BBH:1502 & $(1,+0.7,-0.4)$ & $0.0026$ & $0.0786$\\
164 & BBH:0366 & $(1,+0.2,0)$ & $0.0027$ & $0.1055$\\
165 & BBH:1114 & $(1,+0.2,0)$ & .. & $0.1452$\\
166 & BBH:0370 & $(1,+0.4,-0.2)$ & $0.0006$ & $0.4640$\\
167 & BBH:0376 & $(1,+0.6,-0.4)$ & $0.0013$ & $0.4585$\\
168 & BBH:1506 & $(1,+0.46,-0.3)$ & $0.0023$ & $0.1897$\\
169 & BBH:1476 & $(1,-0.8,+0.8)$ & $0.0045$ & $0.0926$\\
170 & BBH:2085 & $(1,-0.9,+0.9)$ & $0.0021$ & $0.0910$\\
171 & BBH:2087 & $(1,-0.8,+0.8)$ & $0.0007$ & $0.0932$\\
172 & BBH:0304 & $(1,-0.5,+0.5)$ & $0.0014$ & $0.2308$\\
173 & BBH:2091 & $(1,-0.6,+0.6)$ & $0.0019$ & $0.3086$\\
174 & BBH:2092 & $(1,+0.5,-0.5)$ & $0.0028$ & $0.3435$\\
175 & BBH:0327 & $(1,-0.8,+0.8)$ & $0.0042$ & $0.0457$\\
176 & BBH:0330 & $(1,-0.8,+0.8)$ & $0.0005$ & $0.0930$\\
177 & BBH:0459 & $(1,-0.4,+0.2)$ & $0.0023$ & $0.0359$\\
178 & BBH:0447 & $(1,-0.6,+0.4)$ & $0.0054$ & $0.2526$\\
179 & BBH:1351 & $(1,-0.23,0)$ & $0.0005$ & $0.0813$\\
180 & BBH:2096 & $(1,-0.3,0)$ & $0.0005$ & $0.1127$\\
181 & BBH:1509 & $(1,-0.24,-0.1)$ & $0.0014$ & $0.2097$\\
182 & BBH:2100 & $(1,-0.9,+0.5)$ & $0.0017$ & $0.0481$\\
183 & BBH:2098 & $(1,-0.8,+0.4)$ & $0.0044$ & $0.0475$
\end{tabular}
\end{ruledtabular}
\end{center}
\end{table}
\begin{table}
  \caption{\label{tab:SXS_new2}This table continues to summarizes the \texttt{SXS} NR waveform data in the \textit{validation} set,
    with mass-ratios $1\leq q\leq 1.2$. From left to right, the columns report: the SXS simulation number, mass ratio and dimensionless spins spins $\chi_i$,
    and the maximum value of the unfaithfulness $\bar{F}$ between: The two highest resolutions 
    of the NR dataset, if available, see Fig.\ref{fig:barF_SXS_error}, and between EOB and NR, see Fig.~\ref{fig:barF_SXS_new}.}
\begin{center}
\begin{ruledtabular}
\begin{tabular}{r | l l l l}
\# & {\rm id} & $(q,\chi_1,\chi_2)$ & $\bar{F}^{\rm max}_{\rm NR/NR}[\%]$ & $\bar{F}^{\rm max}_{\rm EOB/NR}[\%]$\\\hline
184 & BBH:0415 & $(1,-0.4,0)$ & $0.0021$ & $0.1571$\\
185 & BBH:1499 & $(1,-0.75,+0.34)$ & $0.0021$ & $0.0440$\\
186 & BBH:1498 & $(1,+0.22,-0.8)$ & $0.0093$ & $0.0261$\\
187 & BBH:2090 & $(1,-0.6,0)$ & $0.0036$ & $0.0464$\\
188 & BBH:0436 & $(1,-0.4,-0.2)$ & $0.0019$ & $0.1467$\\
189 & BBH:0585 & $(1,-0.6,0)$ & $0.0009$ & $0.0863$\\
190 & BBH:0325 & $(1,-0.8,0)$ & $0.0022$ & $0.1511$\\
191 & BBH:1134 & $(1,-0.44,-0.44)$ & $0.0025$ & $0.0383$\\
192 & BBH:1135 & $(1,-0.44,-0.44)$ & $0.0047$ & $0.0265$\\
193 & BBH:2088 & $(1,-0.62,-0.25)$ & $0.0016$ & $0.1088$\\
194 & BBH:1144 & $(1,-0.44,-0.44)$ & $0.0054$ & $0.0393$\\
195 & BBH:2084 & $(1,-0.9,0)$ & $0.0031$ & $0.1176$\\
196 & BBH:1500 & $(1,-0.77,-0.2)$ & $0.0012$ & $0.0560$\\
197 & BBH:0462 & $(1,-0.6,-0.4)$ & $0.0025$ & $0.0454$\\
198 & BBH:0151 & $(1,-0.6,-0.6)$ & $0.0022$ & $0.1071$\\
199 & BBH:2094 & $(1,-0.8,-0.4)$ & $0.0009$ & $0.0999$\\
200 & BBH:2089 & $(1,-0.6,-0.6)$ & $0.0072$ & $0.0900$\\
201 & BBH:1492 & $(1,-0.8,-0.47)$ & $0.0009$ & $0.0756$\\
202 & BBH:2083 & $(1,-0.9,-0.5)$ & $0.0067$ & $0.0565$\\
203 & BBH:1475 & $(1,-0.8,-0.8)$ & $0.0026$ & $0.0468$\\
204 & BBH:2086 & $(1,-0.8,-0.8)$ & $0.0033$ & $0.0349$\\
205 & BBH:0329 & $(1,-0.8,-0.8)$ & $0.0020$ & $0.0509$\\
206 & BBH:1137 & $(1,-0.97,-0.97)$ & $0.0021$ & $0.1300$\\
207 & BBH:0544 & $(1,0,+0.7)$ & $0.0021$ & $0.2125$\\
208 & BBH:0518 & $(1.1,-0.14,+0.43)$ & $0.0012$ & $0.1204$\\
209 & BBH:1513 & $(1.1,-0.1,0)$ & $0.0051$ & $0.1661$\\
210 & BBH:0409 & $(1.2,+0.4,+0.8)$ & $0.0058$ & $0.1314$\\
211 & BBH:1490 & $(1.2,+0.41,+0.76)$ & $0.0026$ & $0.1835$\\
212 & BBH:1496 & $(1.2,+0.8,+0.03)$ & $0.0008$ & $0.1548$\\
213 & BBH:0311 & $(1.2,+0.42,+0.38)$ & $0.0019$ & $0.2521$\\
214 & BBH:0486 & $(1.2,0,+0.8)$ & $0.0024$ & $0.1712$\\
215 & BBH:0559 & $(1.2,-0.2,+0.8)$ & $0.0035$ & $0.0903$\\
216 & BBH:0475 & $(1.2,-0.4,+0.8)$ & $0.0036$ & $0.0776$\\
217 & BBH:1352 & $(1.2,+0.71,-0.67)$ & $0.0013$ & $0.1645$\\
218 & BBH:0503 & $(1.2,-0.6,+0.8)$ & $0.0017$ & $0.0343$\\
219 & BBH:0312 & $(1.2,+0.4,-0.48)$ & $0.0046$ & $0.1594$\\
220 & BBH:1353 & $(1.2,+0.33,-0.44)$ & $0.0013$ & $0.3409$\\
221 & BBH:0309 & $(1.2,+0.33,-0.44)$ & $0.0103$ & $0.2187$\\
222 & BBH:0305 & $(1.2,+0.33,-0.44)$ & $0.0015$ & $0.1682$\\
223 & BBH:0318 & $(1.2,+0.33,-0.44)$ & $0.0018$ & $0.2182$\\
224 & BBH:0319 & $(1.2,+0.33,-0.44)$ & $0.0096$ & $0.1225$\\
225 & BBH:0313 & $(1.2,+0.38,-0.5)$ & $0.0035$ & $0.2190$\\
226 & BBH:0465 & $(1.2,+0.6,-0.8)$ & $0.0010$ & $0.1743$\\
227 & BBH:0314 & $(1.2,+0.31,-0.46)$ & $0.0021$ & $0.2401$\\
228 & BBH:0307 & $(1.2,+0.32,-0.58)$ & $0.0016$ & $0.1026$\\
229 & BBH:0626 & $(1.2,-0.83,+0.73)$ & $0.0061$ & $0.0217$\\
230 & BBH:0535 & $(1.2,+0.2,-0.8)$ & $0.0023$ & $0.0812$\\
231 & BBH:0523 & $(1.2,-0.2,-0.47)$ & $0.0012$ & $0.0362$\\
232 & BBH:0398 & $(1.2,0,-0.8)$ & $0.0036$ & $0.0475$\\
233 & BBH:0386 & $(1.2,-0.2,-0.8)$ & $0.0012$ & $0.1080$\\
234 & BBH:0438 & $(1.2,-0.6,-0.8)$ & $0.0021$ & $0.0363$
\end{tabular}
\end{ruledtabular}
\end{center}
\end{table}
\begin{table}
  \caption{\label{tab:SXS_new3}This table summarizes the \texttt{SXS} NR waveform data in the \textit{validation} set,
    with mass-ratios $1.3\leq q\leq 1.7$. From left to right, the columns report: the SXS simulation number, mass ratio and dimensionless 
    spins $\chi_i$, and the maximum value of the unfaithfulness $\bar{F}$ between: The two highest resolutions 
    of the NR dataset, if available, see Fig.\ref{fig:barF_SXS_error}, and between EOB and NR, see Fig.~\ref{fig:barF_SXS_new}.}
\begin{center}
\begin{ruledtabular}
\begin{tabular}{r | l l l l}
\# & {\rm id} & $(q,\chi_1,\chi_2)$ & $\bar{F}^{\rm max}_{\rm NR/NR}[\%]$ & $\bar{F}^{\rm max}_{\rm EOB/NR}[\%]$\\\hline
235 & BBH:0507 & $(1.3,+0.8,+0.4)$ & $0.0043$ & $0.1326$\\
236 & BBH:1493 & $(1.3,0,+0.8)$ & $0.0029$ & $0.1639$\\
237 & BBH:0525 & $(1.3,+0.8,-0.4)$ & $0.0008$ & $0.1835$\\
238 & BBH:1505 & $(1.3,-0.1,+0.55)$ & $0.0021$ & $0.1400$\\
239 & BBH:0591 & $(1.2,0,+0.4)$ & $0.0028$ & $0.0926$\\
240 & BBH:1508 & $(1.3,+0.3,-0.07)$ & $0.0057$ & $0.1296$\\
241 & BBH:1474 & $(1.3,+0.72,-0.8)$ & $0.0025$ & $0.2317$\\
242 & BBH:1223 & $(1.2,+0.38,-0.46)$ & $0.0034$ & $0.1685$\\
243 & BBH:0651 & $(1.3,0,+0.03)$ & $0.0051$ & $0.1191$\\
244 & BBH:0650 & $(1.3,0,+0.03)$ & $0.0141$ & $0.3692$\\
245 & BBH:0315 & $(1.3,+0.32,-0.56)$ & $0.0011$ & $0.3037$\\
246 & BBH:0464 & $(1.2,0,-0.4)$ & $0.0013$ & $0.1014$\\
247 & BBH:1487 & $(1.3,-0.8,+0.5)$ & $0.0022$ & $0.0636$\\
248 & BBH:0377 & $(1.3,-0.8,+0.4)$ & $0.0017$ & $0.0739$\\
249 & BBH:0466 & $(1.3,-0.8,-0.4)$ & $0.0006$ & $0.0388$\\
250 & BBH:1471 & $(1.3,-0.78,-0.8)$ & $0.0009$ & $0.0561$\\
251 & BBH:0129 & $(1.4,+0.09,-0.07)$ & .. & $0.1464$\\
252 & BBH:1482 & $(1.4,-0.58,+0.8)$ & $0.0016$ & $0.0301$\\
253 & BBH:0625 & $(1.4,-0.71,+0.22)$ & $0.0046$ & $0.0519$\\
254 & BBH:1146 & $(1.5,+0.95,+0.95)$ & $0.0446$ & $0.8427$\\
255 & BBH:1473 & $(1.4,+0.7,+0.8)$ & $0.0060$ & $0.1843$\\
256 & BBH:0441 & $(1.5,+0.6,+0.8)$ & $0.0034$ & $0.1585$\\
257 & BBH:0385 & $(1.5,+0.8,0)$ & $0.0016$ & $0.1737$\\
258 & BBH:0361 & $(1.5,0,+0.8)$ & $0.0025$ & $0.1085$\\
259 & BBH:0372 & $(1.5,+0.8,-0.4)$ & $0.0025$ & $0.1282$\\
260 & BBH:0499 & $(1.5,+0.01,+0.74)$ & $0.0041$ & $0.1005$\\
261 & BBH:0009 & $(1.5,+0.5,0)$ & .. & $0.2536$\\
262 & BBH:0392 & $(1.5,-0.2,+0.8)$ & $0.0024$ & $0.0555$\\
263 & BBH:0440 & $(1.5,0,+0.4)$ & $0.0014$ & $0.0512$\\
264 & BBH:0369 & $(1.5,+0.6,-0.8)$ & $0.0026$ & $0.3375$\\
265 & BBH:1511 & $(1.5,+0.03,-0.1)$ & $0.0024$ & $0.0952$\\
266 & BBH:0579 & $(1.5,+0.4,-0.8)$ & $0.0022$ & $0.0881$\\
267 & BBH:0012 & $(1.5,-0.5,0)$ & $0.0068$ & $0.0341$\\
268 & BBH:0014 & $(1.5,-0.5,0)$ & $0.0561$ & $0.0425$\\
269 & BBH:0101 & $(1.5,-0.5,0)$ & .. & $0.0284$\\
270 & BBH:0404 & $(1.5,0,-0.8)$ & $0.0032$ & $0.0962$\\
271 & BBH:0437 & $(1.5,-0.2,-0.8)$ & $0.0019$ & $0.0552$\\
272 & BBH:1480 & $(1.5,-0.8,-0.3)$ & $0.0035$ & $0.0304$\\
273 & BBH:0397 & $(1.5,-0.8,-0.4)$ & $0.0017$ & $0.0419$\\
274 & BBH:1470 & $(1.5,-0.73,-0.8)$ & $0.0035$ & $0.0345$\\
275 & BBH:0519 & $(1.6,+0.64,+0.4)$ & $0.0048$ & $0.0812$\\
276 & BBH:1488 & $(1.6,-0.33,+0.75)$ & $0.0061$ & $0.0357$\\
277 & BBH:1479 & $(1.6,-0.56,-0.8)$ & $0.0016$ & $0.0380$\\
278 & BBH:0501 & $(1.7,+0.6,+0.8)$ & $0.0063$ & $0.1260$\\
279 & BBH:0435 & $(1.7,+0.4,+0.8)$ & $0.0029$ & $0.0817$\\
280 & BBH:0566 & $(1.7,+0.2,+0.8)$ & $0.0041$ & $0.0971$\\
281 & BBH:0382 & $(1.7,0,+0.8)$ & $0.0017$ & $0.0701$\\
282 & BBH:0529 & $(1.7,0,+0.53)$ & $0.0035$ & $0.0627$\\
283 & BBH:0550 & $(1.7,-0.2,+0.8)$ & $0.0021$ & $0.0534$\\
284 & BBH:0451 & $(1.7,0,+0.4)$ & $0.0022$ & $0.0526$\\
285 & BBH:0488 & $(1.7,+0.6,-0.8)$ & $0.0040$ & $0.2618$
\end{tabular}
\end{ruledtabular}
\end{center}
\end{table}
\begin{table}
  \caption{\label{tab:SXS_new4}This table continues to summarizes the \texttt{SXS} NR waveform data in the \textit{validation} set,
    with mass-ratios $1.7\leq q\leq 2$. From left to right, the columns report: the SXS simulation number, mass ratio and dimensionless spins $\chi_i$,
    and the maximum value of the unfaithfulness $\bar{F}$ between: The two highest resolutions 
    of the NR dataset, if available, see Fig.\ref{fig:barF_SXS_error}, and between EOB and NR, see Fig.~\ref{fig:barF_SXS_new}.}
\begin{center}
\begin{ruledtabular}
\begin{tabular}{r | l l l l}
\# & {\rm id} & $(q,\chi_1,\chi_2)$ & $\bar{F}^{\rm max}_{\rm NR/NR}[\%]$ & $\bar{F}^{\rm max}_{\rm EOB/NR}[\%]$\\\hline
286 & BBH:0678 & $(1.7,0,+0.03)$ & $0.0073$ & $0.1337$\\
287 & BBH:0677 & $(1.7,0,+0.03)$ & $0.0037$ & $0.1791$\\
288 & BBH:0676 & $(1.7,0,+0.02)$ & $0.0050$ & $0.3450$\\
289 & BBH:0355 & $(1.7,-0.6,+0.8)$ & $0.0015$ & $0.0550$\\
290 & BBH:1491 & $(1.7,+0.2,-0.7)$ & $0.0020$ & $0.0999$\\
291 & BBH:0473 & $(1.7,+0.2,-0.8)$ & $0.0019$ & $0.0876$\\
292 & BBH:1465 & $(1.7,-0.79,+0.77)$ & $0.0038$ & $0.0740$\\
293 & BBH:0510 & $(1.7,-0.02,-0.75)$ & $0.0044$ & $0.0639$\\
294 & BBH:0423 & $(1.7,0,-0.8)$ & $0.0008$ & $0.0577$\\
295 & BBH:0402 & $(1.7,-0.8,+0.4)$ & $0.0017$ & $0.0725$\\
296 & BBH:0414 & $(1.7,-0.4,-0.8)$ & $0.0038$ & $0.0438$\\
297 & BBH:0512 & $(1.7,-0.6,-0.8)$ & $0.0006$ & $0.0476$\\
298 & BBH:0388 & $(1.8,+0.8,+0.4)$ & $0.0034$ & $0.1265$\\
299 & BBH:0552 & $(1.8,+0.8,-0.4)$ & $0.0016$ & $0.2418$\\
300 & BBH:1510 & $(1.8,+0.03,+0.3)$ & $0.0018$ & $0.0549$\\
301 & BBH:0371 & $(1.8,0,-0.4)$ & $0.0060$ & $0.0532$\\
302 & BBH:0545 & $(1.8,0,-0.8)$ & $0.0026$ & $0.0306$\\
303 & BBH:0454 & $(1.8,-0.8,-0.4)$ & $0.0011$ & $0.0203$\\
304 & BBH:1469 & $(1.9,+0.8,+0.67)$ & $0.0042$ & $0.1686$\\
305 & BBH:0530 & $(2,0,+0.54)$ & $0.0015$ & $0.0449$\\
306 & BBH:1466 & $(1.9,+0.7,-0.8)$ & $0.0015$ & $0.2468$\\
307 & BBH:0555 & $(1.9,0,+0.53)$ & $0.0008$ & $0.0483$\\
308 & BBH:0368 & $(2,-0.05,+0.25)$ & $0.0015$ & $0.0351$\\
309 & BBH:0403 & $(1.9,0,-0.05)$ & $0.0048$ & $0.0442$\\
310 & BBH:0580 & $(2,+0.02,-0.8)$ & $0.0020$ & $0.0756$\\
311 & BBH:2131 & $(2,+0.85,+0.85)$ & $0.0011$ & $0.3993$\\
312 & BBH:0333 & $(2,+0.8,+0.8)$ & $0.0115$ & $0.2591$\\
313 & BBH:2130 & $(2,+0.6,+0.6)$ & $0.0032$ & $0.0779$\\
314 & BBH:1478 & $(2,+0.8,+0.13)$ & $0.0047$ & $0.1406$\\
315 & BBH:2127 & $(2,+0.5,+0.5)$ & $0.0102$ & $0.0819$\\
316 & BBH:1148 & $(2,+0.43,+0.5)$ & $0.0048$ & $0.0890$\\
317 & BBH:0410 & $(2,+0.6,0)$ & $0.0047$ & $0.2876$\\
318 & BBH:0574 & $(2,+0.4,+0.4)$ & $0.0016$ & $0.1062$\\
319 & BBH:2129 & $(2,+0.6,0)$ & $0.0016$ & $0.1325$\\
320 & BBH:2122 & $(2,+0.13,+0.85)$ & $0.0022$ & $0.0672$\\
321 & BBH:2125 & $(2,+0.3,+0.3)$ & $0.0017$ & $0.0784$\\
322 & BBH:2132 & $(2,+0.87,-0.85)$ & $0.0039$ & $0.4141$\\
323 & BBH:0399 & $(2,+0.2,+0.4)$ & $0.0013$ & $0.0681$\\
324 & BBH:0332 & $(2,0,+0.8)$ & $0.0021$ & $0.0495$\\
325 & BBH:0513 & $(2,+0.6,-0.4)$ & $0.0030$ & $0.1416$\\
326 & BBH:2128 & $(2,+0.6,-0.6)$ & $0.0051$ & $0.1713$\\
327 & BBH:2121 & $(2,0,+0.6)$ & $0.0017$ & $0.0426$\\
328 & BBH:2124 & $(2,+0.3,0)$ & $0.0014$ & $0.0793$\\
329 & BBH:0903 & $(2,0,+0.6)$ & $0.0009$ & $0.0697$\\
330 & BBH:0893 & $(2,0,+0.58)$ & $0.0020$ & $0.0621$\\
331 & BBH:0885 & $(2,0,+0.57)$ & $0.0024$ & $0.1064$\\
332 & BBH:1504 & $(2,+0.25,+0.08)$ & $0.0013$ & $0.0726$\\
333 & BBH:0448 & $(2,+0.4,-0.4)$ & $0.0031$ & $0.1224$\\
334 & BBH:0407 & $(2,0,+0.4)$ & $0.0020$ & $0.0394$\\
335 & BBH:0599 & $(2,+0.2,0)$ & $0.0027$ & $0.0698$\\
336 & BBH:1147 & $(2,+0.43,-0.5)$ & $0.0094$ & $0.1338$
\end{tabular}
\end{ruledtabular}
\end{center}
\end{table}

\begin{table}
  \caption{\label{tab:SXS_new5}This table continues to summarizes the \texttt{SXS} NR waveform data in the \textit{validation} set,
    with mass-ratios $2\leq q\leq 2.6$. From left to right, the columns report: the SXS simulation number, mass ratio and dimensionless spins $\chi_i$,
    and the maximum value of the unfaithfulness $\bar{F}$ between: The two highest resolutions 
    of the NR dataset, if available, see Fig.\ref{fig:barF_SXS_error}, and between EOB and NR, see Fig.~\ref{fig:barF_SXS_new}.}
\begin{center}
\begin{ruledtabular}
\begin{tabular}{r | l l l l}
\# & {\rm id} & $(q,\chi_1,\chi_2)$ & $\bar{F}^{\rm max}_{\rm NR/NR}[\%]$ & $\bar{F}^{\rm max}_{\rm EOB/NR}[\%]$\\\hline
337 & BBH:2120 & $(2,0,+0.3)$ & $0.0013$ & $0.0368$\\
338 & BBH:2123 & $(2,+0.3,-0.3)$ & $0.0021$ & $0.1055$\\
339 & BBH:2113 & $(2,-0.37,+0.85)$ & $0.0012$ & $0.0344$\\
340 & BBH:0913 & $(2,0,+0.03)$ & $0.0034$ & $0.0802$\\
341 & BBH:0971 & $(2,0,+0.03)$ & $0.0010$ & $0.0460$\\
342 & BBH:0987 & $(2,0,+0.03)$ & $0.0080$ & $0.0745$\\
343 & BBH:0703 & $(2,0,+0.03)$ & $0.0085$ & $0.1011$\\
344 & BBH:0704 & $(2,0,+0.03)$ & $0.0007$ & $0.0949$\\
345 & BBH:0921 & $(2,0,+0.03)$ & $0.0021$ & $0.1244$\\
346 & BBH:0702 & $(2,0,+0.02)$ & $0.0038$ & $0.2062$\\
347 & BBH:0961 & $(2,0,+0.02)$ & $0.0011$ & $0.1312$\\
348 & BBH:0979 & $(2,0,+0.02)$ & $0.0068$ & $0.2057$\\
349 & BBH:0931 & $(2,0,+0.0)$ & $0.0034$ & $0.1406$\\
350 & BBH:0554 & $(2,+0.2,-0.4)$ & $0.0029$ & $0.0954$\\
351 & BBH:0354 & $(2,-0.2,+0.4)$ & $0.0006$ & $0.0234$\\
352 & BBH:2126 & $(2,+0.37,-0.85)$ & $0.0031$ & $0.1498$\\
353 & BBH:0482 & $(2,-0.02,-0.13)$ & $0.0040$ & $0.0409$\\
354 & BBH:2119 & $(2,0,-0.3)$ & $0.0034$ & $0.0466$\\
355 & BBH:2116 & $(2,-0.3,+0.3)$ & $0.0027$ & $0.0208$\\
356 & BBH:1112 & $(2,-0.2,0)$ & .. & $0.0227$\\
357 & BBH:0375 & $(2,0,-0.4)$ & $0.0013$ & $0.0204$\\
358 & BBH:0954 & $(2,0,-0.56)$ & $0.0031$ & $0.0784$\\
359 & BBH:0947 & $(2,0,-0.56)$ & $0.0056$ & $0.1312$\\
360 & BBH:0940 & $(2,0,-0.57)$ & $0.0040$ & $0.1698$\\
361 & BBH:2118 & $(2,0,-0.6)$ & $0.0025$ & $0.0617$\\
362 & BBH:2111 & $(2,-0.6,+0.6)$ & $0.0017$ & $0.0882$\\
363 & BBH:2115 & $(2,-0.3,0)$ & $0.0038$ & $0.0223$\\
364 & BBH:0331 & $(2,0,-0.8)$ & $0.0075$ & $0.0668$\\
365 & BBH:0412 & $(2,-0.2,-0.4)$ & $0.0018$ & $0.0295$\\
366 & BBH:0335 & $(2,-0.8,+0.8)$ & $0.0012$ & $0.1186$\\
367 & BBH:2107 & $(2,-0.87,+0.85)$ & $0.0012$ & $0.1700$\\
368 & BBH:2114 & $(2,-0.3,-0.3)$ & $0.0021$ & $0.0234$\\
369 & BBH:2117 & $(2,-0.13,-0.85)$ & $0.0057$ & $0.0642$\\
370 & BBH:2110 & $(2,-0.6,0)$ & $0.0021$ & $0.0435$\\
371 & BBH:0584 & $(2,-0.4,-0.4)$ & $0.0027$ & $0.0260$\\
372 & BBH:0461 & $(2,-0.6,0)$ & $0.0018$ & $0.0318$\\
373 & BBH:2112 & $(2,-0.5,-0.5)$ & $0.0031$ & $0.0256$\\
374 & BBH:0387 & $(2,-0.6,-0.4)$ & $0.0022$ & $0.0239$\\
375 & BBH:2109 & $(2,-0.6,-0.6)$ & $0.0015$ & $0.0273$\\
376 & BBH:0334 & $(2,-0.8,-0.8)$ & $0.0033$ & $0.0519$\\
377 & BBH:2108 & $(2,-0.85,-0.85)$ & $0.0068$ & $0.0766$\\
378 & BBH:1467 & $(2.2,-0.56,+0.8)$ & $0.0051$ & $0.0538$\\
379 & BBH:1494 & $(2.2,-0.47,-0.4)$ & $0.0025$ & $0.0303$\\
380 & BBH:1459 & $(2.3,+0.76,+0.8)$ & $0.0112$ & $0.1404$\\
381 & BBH:1468 & $(2.3,+0.51,+0.8)$ & $0.0068$ & $0.0635$\\
382 & BBH:0631 & $(2.3,-0.13,-0.36)$ & $0.0036$ & $0.0161$\\
383 & BBH:1453 & $(2.4,+0.8,-0.8)$ & $0.0048$ & $0.1916$\\
384 & BBH:1512 & $(2.4,+0.24,0)$ & $0.0045$ & $0.0639$\\
385 & BBH:1472 & $(2.4,-0.8,-0.1)$ & $0.0022$ & $0.0243$\\
386 & BBH:1454 & $(2.5,-0.8,-0.73)$ & $0.0030$ & $0.0733$\\
387 & BBH:1462 & $(2.6,-0.8,+0.5)$ & $0.0021$ & $0.0714$
\end{tabular}
\end{ruledtabular}
\end{center}
\end{table}
\begin{table}
  \caption{\label{tab:SXS_new6}This table summarizes the \texttt{SXS} NR waveform data in the \textit{validation} set,
    with mass-ratios $2.9\leq q\leq 3$.
    From left to right, the columns report: the SXS simulation number, mass ratio and dimensionless spins $\chi_i$,
    and the maximum value of the unfaithfulness $\bar{F}$ between: The two highest resolutions 
    of the NR dataset, if available, see Fig.\ref{fig:barF_SXS_error}, and between EOB and NR, see Fig.~\ref{fig:barF_SXS_new}.}
\begin{center}
\begin{ruledtabular}
\begin{tabular}{r | l l l l}
\# & {\rm id} & $(q,\chi_1,\chi_2)$ & $\bar{F}^{\rm max}_{\rm NR/NR}[\%]$ & $\bar{F}^{\rm max}_{\rm EOB/NR}[\%]$\\\hline
388 & BBH:1461 & $(2.9,-0.45,-0.8)$ & $0.0044$ & $0.0299$\\
389 & BBH:1484 & $(3,-0.56,+0.3)$ & $0.0019$ & $0.0399$\\
390 & BBH:1456 & $(3,+0.74,+0.7)$ & $0.0123$ & $0.0528$\\
391 & BBH:1150 & $(3,+0.7,+0.6)$ & $0.0088$ & $0.0176$\\
392 & BBH:1151 & $(3,+0.7,+0.6)$ & $0.0093$ & $0.0214$\\
393 & BBH:1152 & $(3,+0.7,+0.6)$ & $0.0079$ & $0.0214$\\
394 & BBH:1382 & $(3,+0.7,+0.6)$ & $0.0072$ & $0.0298$\\
395 & BBH:2163 & $(3,+0.6,+0.6)$ & $0.0029$ & $0.0135$\\
396 & BBH:2162 & $(3,+0.6,+0.4)$ & $0.0020$ & $0.0270$\\
397 & BBH:2158 & $(3,+0.5,+0.5)$ & $0.0089$ & $0.0201$\\
398 & BBH:0047 & $(3,+0.5,+0.5)$ & .. & $0.0212$\\
399 & BBH:2161 & $(3,+0.6,0)$ & $0.0052$ & $0.0673$\\
400 & BBH:2157 & $(3,+0.4,+0.6)$ & $0.0043$ & $0.0256$\\
401 & BBH:2152 & $(3,+0.27,+0.85)$ & $0.0047$ & $0.0392$\\
402 & BBH:0031 & $(3,+0.5,0)$ & $0.0244$ & $0.0679$\\
403 & BBH:0041 & $(3,+0.5,0)$ & .. & $0.0571$\\
404 & BBH:2160 & $(3,+0.6,-0.4)$ & $0.0058$ & $0.1463$\\
405 & BBH:2159 & $(3,+0.6,-0.6)$ & $0.0034$ & $0.1960$\\
406 & BBH:2155 & $(3,+0.3,+0.3)$ & $0.0048$ & $0.0240$\\
407 & BBH:1387 & $(3,+0.47,-0.36)$ & $0.0026$ & $0.1079$\\
408 & BBH:2154 & $(3,+0.3,0)$ & $0.0081$ & $0.0451$\\
409 & BBH:2156 & $(3,+0.4,-0.6)$ & $0.0009$ & $0.1299$\\
410 & BBH:2150 & $(3,0,+0.6)$ & $0.0022$ & $0.0253$\\
411 & BBH:2153 & $(3,+0.3,-0.3)$ & $0.0013$ & $0.0719$\\
412 & BBH:2149 & $(3,0,+0.3)$ & $0.0025$ & $0.0201$\\
413 & BBH:2146 & $(3,-0.23,+0.85)$ & $0.0038$ & $0.0354$\\
414 & BBH:2151 & $(3,+0.23,-0.85)$ & $0.0014$ & $0.0671$\\
415 & BBH:2148 & $(3,0,-0.3)$ & $0.0049$ & $0.0373$\\
416 & BBH:2144 & $(3,-0.3,+0.3)$ & $0.0032$ & $0.0312$\\
417 & BBH:2141 & $(3,-0.4,+0.6)$ & $0.0030$ & $0.0658$\\
418 & BBH:2147 & $(3,0,-0.6)$ & $0.0045$ & $0.0251$\\
419 & BBH:2143 & $(3,-0.3,0)$ & $0.0035$ & $0.0232$\\
420 & BBH:2135 & $(3,-0.6,+0.6)$ & $0.0026$ & $0.1012$\\
421 & BBH:2142 & $(3,-0.3,-0.3)$ & $0.0009$ & $0.0249$\\
422 & BBH:2133 & $(3,-0.73,+0.85)$ & $0.0023$ & $0.1467$\\
423 & BBH:2138 & $(3,-0.6,+0.4)$ & $0.0020$ & $0.0748$\\
424 & BBH:0038 & $(3,-0.5,0)$ & .. & $0.0276$\\
425 & BBH:0039 & $(3,-0.5,0)$ & .. & $0.0271$\\
426 & BBH:0040 & $(3,-0.5,0)$ & .. & $0.0224$\\
427 & BBH:2145 & $(3,-0.27,-0.85)$ & $0.0044$ & $0.0611$\\
428 & BBH:2140 & $(3,-0.4,-0.6)$ & $0.0019$ & $0.0239$\\
429 & BBH:2134 & $(3,-0.6,0)$ & $0.0023$ & $0.0388$\\
430 & BBH:0046 & $(3,-0.5,-0.5)$ & .. & $0.0358$\\
431 & BBH:2139 & $(3,-0.5,-0.5)$ & $0.0029$ & $0.0363$\\
432 & BBH:2137 & $(3,-0.6,-0.4)$ & $0.0019$ & $0.0330$\\
433 & BBH:2136 & $(3,-0.6,-0.6)$ & $0.0020$ & $0.0432$\\
434 & BBH:1172 & $(3,-0.7,-0.6)$ & $0.0021$ & $0.0493$\\
435 & BBH:1170 & $(3,-0.7,-0.6)$ & $0.0021$ & $0.0739$\\
436 & BBH:1171 & $(3,-0.7,-0.6)$ & $0.0024$ & $0.0543$\\
437 & BBH:1173 & $(3,-0.7,-0.6)$ & $0.0013$ & $0.0515$\\
438 & BBH:1174 & $(3,-0.7,-0.6)$ & $0.0004$ & $0.0515$
\end{tabular}
\end{ruledtabular}
\end{center}
\end{table}
\begin{table}
  \caption{\label{tab:SXS_new7}This table summarizes the \texttt{SXS} NR waveform data in the \textit{validation} set, with mass-ratios $3\leq q\geq 5.5$.
 	From left to right, the columns report: the SXS simulation number, mass ratio and dimensionless spins $\chi_i$,
 	and the maximum value of the unfaithfulness $\bar{F}$ between: The two highest resolutions 
 	of the NR dataset, if available, see Fig.\ref{fig:barF_SXS_error}, and between EOB and NR, see Fig.~\ref{fig:barF_SXS_new}.}
\begin{center}
\begin{ruledtabular}
\begin{tabular}{r | l l l l}
\# & {\rm id} & $(q,\chi_1,\chi_2)$ & $\bar{F}^{\rm max}_{\rm NR/NR}[\%]$ & $\bar{F}^{\rm max}_{\rm EOB/NR}[\%]$\\\hline
439 & BBH:1175 & $(3,-0.7,-0.6)$ & $0.0022$ & $0.0514$\\
440 & BBH:1485 & $(3,+0.35,-0.4)$ & $0.0046$ & $0.0889$\\
441 & BBH:1447 & $(3.2,+0.74,+0.8)$ & .. & $0.0558$\\
442 & BBH:1457 & $(3.2,+0.54,+0.8)$ & $0.0095$ & $0.0216$\\
443 & BBH:1483 & $(3.2,+0.56,-0.2)$ & $0.0059$ & $0.0849$\\
444 & BBH:1446 & $(3.2,-0.8,+0.78)$ & $0.0047$ & $0.1088$\\
445 & BBH:0317 & $(3.3,+0.52,-0.45)$ & $0.0040$ & $0.1162$\\
446 & BBH:1489 & $(3.5,+0.3,-0.17)$ & .. & $0.0459$\\
447 & BBH:1452 & $(3.6,+0.8,-0.43)$ & .. & $0.1924$\\
448 & BBH:1486 & $(3.7,+0.43,-0.03)$ & $0.0038$ & $0.0328$\\
449 & BBH:1458 & $(3.8,-0.06,+0.8)$ & $0.0047$ & $0.0271$\\
450 & BBH:2014 & $(4,+0.8,+0.4)$ & .. & $0.1195$\\
451 & BBH:1938 & $(4,+0.4,+0.8)$ & $0.0102$ & $0.0634$\\
452 & BBH:1417 & $(4,+0.4,+0.5)$ & $0.0565$ & $0.0745$\\
453 & BBH:1937 & $(4,+0.4,0)$ & $0.0031$ & $0.0228$\\
454 & BBH:1942 & $(4,+0.4,-0.8)$ & $0.0078$ & $0.1199$\\
455 & BBH:1907 & $(4,0,+0.8)$ & $0.0072$ & $0.0256$\\
456 & BBH:2041 & $(4,0,+0.6)$ & $0.0080$ & $0.0483$\\
457 & BBH:2051 & $(4,0,+0.6)$ & $0.0068$ & $0.0530$\\
458 & BBH:2047 & $(4,0,+0.6)$ & $0.0060$ & $0.0495$\\
459 & BBH:2013 & $(4,0,+0.4)$ & $0.0023$ & $0.0248$\\
460 & BBH:1910 & $(4,0,+0.03)$ & $0.0128$ & $0.0473$\\
461 & BBH:2068 & $(4,0,+0.04)$ & $0.0060$ & $0.0476$\\
462 & BBH:1908 & $(4,0,+0.0)$ & $0.0051$ & $0.0589$\\
463 & BBH:2072 & $(4,0,+0.0)$ & $0.0110$ & $0.0391$\\
464 & BBH:1909 & $(4,0,+0.0)$ & $0.0047$ & $0.0260$\\
465 & BBH:2077 & $(4,0,+0.0)$ & $0.0071$ & $0.0185$\\
466 & BBH:2036 & $(4,0,-0.4)$ & $0.0035$ & $0.0356$\\
467 & BBH:2063 & $(4,0,-0.55)$ & $0.0055$ & $0.0652$\\
468 & BBH:2056 & $(4,0,-0.56)$ & $0.0057$ & $0.0667$\\
469 & BBH:2060 & $(4,0,-0.57)$ & $0.0041$ & $0.0490$\\
470 & BBH:1911 & $(4,0,-0.8)$ & $0.0070$ & $0.0690$\\
471 & BBH:1962 & $(4,-0.4,+0.8)$ & $0.0039$ & $0.0746$\\
472 & BBH:1961 & $(4,-0.4,0)$ & $0.0011$ & $0.0231$\\
473 & BBH:1418 & $(4,-0.4,-0.5)$ & $0.0526$ & $0.0554$\\
474 & BBH:1966 & $(4,-0.4,-0.8)$ & $0.0020$ & $0.0309$\\
475 & BBH:1932 & $(4,-0.8,+0.8)$ & $0.0021$ & $0.1056$\\
476 & BBH:2018 & $(4,-0.8,+0.4)$ & $0.0021$ & $0.0561$\\
477 & BBH:1931 & $(4,-0.8,0)$ & $0.0004$ & $0.0297$\\
478 & BBH:2040 & $(4,-0.8,-0.4)$ & $0.0006$ & $0.0414$\\
479 & BBH:1936 & $(4,-0.8,-0.8)$ & $0.0091$ & $0.1059$\\
480 & BBH:1451 & $(4,+0.31,-0.8)$ & $0.0070$ & $0.0637$\\
481 & BBH:1450 & $(4,-0.28,-0.8)$ & $0.0020$ & $0.0377$\\
482 & BBH:1449 & $(4.2,-0.8,-0.34)$ & $0.0026$ & $0.0392$\\
483 & BBH:1434 & $(4.4,+0.8,+0.8)$ & .. & $0.0347$\\
484 & BBH:1445 & $(4.7,-0.5,+0.8)$ & $0.0058$ & $0.1449$\\
485 & BBH:1463 & $(5,+0.61,+0.24)$ & $0.0032$ & $0.1126$\\
486 & BBH:0061 & $(5,+0.5,0)$ & .. & $0.0403$\\
487 & BBH:0109 & $(5,-0.5,0)$ & .. & $0.0404$\\
488 & BBH:1111 & $(5,-0.9,0)$ & $0.0071$ & $0.0436$\\
489 & BBH:1428 & $(5.5,-0.8,-0.7)$ & $0.0066$ & $0.0691$
\end{tabular}
\end{ruledtabular}
\end{center}
\end{table}

\begin{table}
\caption{\label{tab:SXS_new8}This table summarizes the \texttt{SXS} NR waveform data in the \textit{validation} set, with mass-ratios $5.6\leq q\geq 8$.
		From left to right, the columns report: the SXS simulation number, mass ratio and dimensionless spins $\chi_i$,
		and the maximum value of the unfaithfulness $\bar{F}$ between: The two highest resolutions 
		of the NR dataset, if available, see Fig.\ref{fig:barF_SXS_error}, and between EOB and NR, see Fig.~\ref{fig:barF_SXS_new}.}
\begin{center}
\begin{ruledtabular}
\begin{tabular}{r | l l l l}
\# & {\rm id} & $(q,\chi_1,\chi_2)$ & $\bar{F}^{\rm max}_{\rm NR/NR}[\%]$ & $\bar{F}^{\rm max}_{\rm EOB/NR}[\%]$ \\\hline
490 & BBH:1440 & $(5.6,+0.77,+0.3)$ & $0.0055$ & $0.3153$\\
491 & BBH:1443 & $(5.7,+0.4,-0.74)$ & $0.0064$ & $0.0238$\\
492 & BBH:1432 & $(5.8,+0.66,+0.8)$ & $0.0192$ & $0.3279$\\
493 & BBH:1438 & $(5.9,+0.13,+0.8)$ & $0.0081$ & $0.0467$\\
494 & BBH:1444 & $(6,-0.06,-0.76)$ & $0.0164$ & $0.0329$\\
495 & BBH:1437 & $(6,+0.8,+0.15)$ & $0.0141$ & $0.3749$\\
496 & BBH:1425 & $(6.1,-0.8,+0.67)$ & $0.0098$ & $0.1078$\\
497 & BBH:1436 & $(6.3,0,-0.8)$ & $0.0142$ & $0.0363$\\
498 & BBH:1439 & $(6.5,+0.72,-0.3)$ & .. & $0.4042$\\
499 & BBH:1464 & $(6.5,-0.05,-0.32)$ & $0.0057$ & $0.0196$\\
500 & BBH:1424 & $(6.5,-0.66,-0.8)$ & $0.0134$ & $0.0509$\\
501 & BBH:1442 & $(6.6,-0.7,-0.18)$ & $0.0018$ & $0.0266$\\
502 & BBH:1435 & $(6.6,-0.79,+0.07)$ & $0.0081$ & $0.0284$\\
503 & BBH:1448 & $(7,-0.48,+0.52)$ & .. & $0.0872$\\
504 & BBH:0204 & $(7,+0.4,0)$ & $0.0434$ & $0.0697$\\
505 & BBH:0206 & $(7,-0.4,0)$ & $0.0171$ & $0.0649$\\
506 & BBH:1427 & $(7.4,-0.61,-0.73)$ & $0.0050$ & $0.0389$\\
507 & BBH:1429 & $(7.7,-0.2,-0.78)$ & $0.0049$ & $0.0286$\\
508 & BBH:1421 & $(7.8,-0.6,+0.8)$ & $0.0046$ & $0.1319$\\
509 & BBH:1426 & $(8,+0.48,+0.75)$ & $0.0378$ & $0.1978$\\
510 & BBH:1441 & $(8,+0.6,-0.48)$ & .. & $0.2876$\\
511 & BBH:1430 & $(8,+0.28,-0.75)$ & $0.0302$ & $0.0180$\\
512 & BBH:1431 & $(8,+0.08,-0.78)$ & $0.0153$ & $0.0301$\\
513 & BBH:1455 & $(8,-0.4,0)$ & $0.0023$ & $0.0340$\\
514 & BBH:0114 & $(8,-0.5,0)$ & .. & $0.0345$\\
515 & BBH:1423 & $(8,-0.6,-0.75)$ & $0.0103$ & $0.0388$\\
516 & BBH:1420 & $(8,-0.8,+0.8)$ & $0.0094$ & $0.1288$\\
517 & BBH:1433 & $(8,-0.74,+0.2)$ & $0.0037$ & $0.0500$\\
518 & BBH:1422 & $(8,-0.8,-0.46)$ & $0.0060$ & $0.0493$\\
519 & BBH:1419 & $(8,-0.8,-0.8)$ & .. & $0.0701$
\end{tabular}
\end{ruledtabular}
\end{center}
\end{table}
\begin{table*}[t]
  \caption{\label{tab:SXS_long} This table summarizes the \textit{long-inspiral}
    {\tt SXS} NR waveform data. From left to right, columns report: the {\tt SXS}
    simulation number, mass ratio and dimensionless spins $\chi_{1,2}$, number of orbits $N$,
    eccentricity $\epsilon$ and the maximum value of the unfaithfulness $\bar{F}$
    computed between the highest and second highest resolution available $\bar{F}^{\rm max}_{\rm NR/NR}$
    and between EOB waveform and the NR highest resolution $\bar{F}^{\rm max}_{\rm EOB/NR}$.
    These datasets are part  of the \textit{validation} set,
    while the in depth study of the other waveforms is left for future work.  These waveforms
    are discussed in Section~\ref{sec:longNR}.}
\begin{center}
\begin{ruledtabular}
\begin{tabular}{cc c| c c c l c |ccc}
 &\#  && {\rm id} & $(q,\chi_1,\chi_2)$ & N & $\epsilon$ [$10^{-3}$] & & $\bar{F}^{\rm max}_{\rm NR/NR}[\%]$ & $\bar{F}^{\rm max}_{\rm EOB/NR}[\%]$ &\\\hline
 &520 & & SXS:BBH:1412 & $(1.63,+0.40,-0.30)$ & 145.1 & $0.4450$ & & 0.7295 & 0.1266 &\\
 &521 & & SXS:BBH:1413 & $(1.41,+0.50,+0.40)$ & 145.4 & $<1.0$ & & 1.1856 & 0.1585 &\\
 &522 & & SXS:BBH:1414 & $(1.83,-0.50,+0.40)$ & 143.1 & $<1.6$ & & 0.8919 & 0.1965 &\\
 &523 & & SXS:BBH:1415 & $(1.50,+0.50,+0.50)$ & 147.7 & $<0.043$ & & 1.5238 & 0.1453 &\\
 &524 & & SXS:BBH:1416 & $(1.78,-0.40,-0.40)$ & 139.0 & $<1.7$ & & 0.5986 & 0.0468 &
\end{tabular}
\end{ruledtabular}
\end{center}
\end{table*}

\begin{table}
\caption{\label{tab:nospin_1} The waveforms $1^\prime-22^\prime$ have been used in Ref.~\cite{Nagar:2019wds} to inform the nonspinning sector.
	Waveforms $23^\prime-45^\prime$ are the first part of the validation set and span mass-ratios $1\leq q\leq 1.8$. 
	{\rm EOB} waveforms are computed with $\chi_1=0$ and $\chi_2=10^{-4}$. This comparison demonstrates the robustness of the spinning-sector of
	\TEOBiResumSM{} in the nonspinning limit and the consistency with \TEOBiResumM{} when spins are small.
	From left to right, the columns report: the SXS simulation number, mass ratio and dimensionless spins $\chi_i$,
	and the maximum value of the unfaithfulness $\bar{F}$ between: the two highest resolutions 
	of the NR dataset, if available, see Fig.~\ref{fig:barF_SXS_nospin}(left panel), and between 
	EOB and NR, see Fig.~\ref{fig:barF_SXS_nospin}(right panel).}
\begin{center}
\begin{ruledtabular}
\begin{tabular}{r | l l l l}
\# & {\rm id} & $(q,\chi_1,\chi_2)$ & $\bar{F}^{\rm max}_{\rm NR/NR}[\%]$ & $\bar{F}^{\rm max}_{\rm EOB/NR}[\%]$\\\hline
$1^\prime$ & BBH:0180 & $(1,0,0)$ & $0.0035$ & $0.0873$\\
$2^\prime$ & BBH:0007 & $(1.5,0,0)$ & $0.0020$ & $0.0851$\\
$3^\prime$ & BBH:0169 & $(2,0,0)$ & $0.0032$ & $0.0825$\\
$4^\prime$ & BBH:0259 & $(2.5,0,0)$ & $0.0050$ & $0.0840$\\
$5^\prime$ & BBH:0030 & $(3,0,0)$ & $0.0030$ & $0.0497$\\
$6^\prime$ & BBH:0167 & $(4,0,0)$ & $0.0057$ & $0.0326$\\
$7^\prime$ & BBH:0295 & $(4.5,0,0)$ & $0.0066$ & $0.0247$\\
$8^\prime$ & BBH:0056 & $(5,0,0)$ & $0.0158$ & $0.0197$\\
$9^\prime$ & BBH:0296 & $(5.5,0,0)$ & $0.0177$ & $0.0186$\\
$10^\prime$ & BBH:0166 & $(6,0,0)$ & .. & $0.0176$\\
$11^\prime$ & BBH:0297 & $(6.5,0,0)$ & $0.0069$ & $0.0167$\\
$12^\prime$ & BBH:0298 & $(7,0,0)$ & $0.0023$ & $0.0169$\\
$13^\prime$ & BBH:0299 & $(7.5,0,0)$ & $0.0013$ & $0.0172$\\
$14^\prime$ & BBH:0063 & $(8,0,0)$ & $0.0754$ & $0.0183$\\
$15^\prime$ & BBH:0300 & $(8.5,0,0)$ & $0.0037$ & $0.0200$\\
$16^\prime$ & BBH:0301 & $(9,0,0)$ & $0.0014$ & $0.0203$\\
$17^\prime$ & BBH:0302 & $(9.5,0,0)$ & $0.0039$ & $0.0219$\\
$18^\prime$ & BBH:0185 & $(9.99,0,0)$ & $0.0033$ & $0.0246$\\
$19^\prime$ & BBH:0303 & $(10,0,0)$ & $0.0045$ & $0.0233$\\\hline
$20^\prime$ & BAM & $(4,0,0)$ & .. & $0.0913$\\
$21^\prime$ & BAM & $(10,0,0)$ & .. & $0.0345$\\
$22^\prime$ & BAM & $(18,0,0)$ & .. & $0.2533$\\\hline
$23^\prime$ & BBH:0001 & $(1,0,0)$ & .. & $0.1000$\\
$24^\prime$ & BBH:0066 & $(1,0,0)$ & .. & $0.1252$\\
$25^\prime$ & BBH:0067 & $(1,0,0)$ & .. & $0.1282$\\
$26^\prime$ & BBH:0068 & $(1,0,0)$ & .. & $0.1234$\\
$27^\prime$ & BBH:0070 & $(1,0,0)$ & .. & $0.0946$\\
$28^\prime$ & BBH:0071 & $(1,0,0)$ & .. & $0.1252$\\
$29^\prime$ & BBH:0072 & $(1,0,0)$ & .. & $0.1411$\\
$30^\prime$ & BBH:0073 & $(1,0,0)$ & .. & $0.1306$\\
$31^\prime$ & BBH:0086 & $(1,0,0)$ & .. & $0.1000$\\
$32^\prime$ & BBH:0090 & $(1,0,0)$ & .. & $0.2034$\\
$33^\prime$ & BBH:0389 & $(1,0,0)$ & $0.0028$ & $0.1729$\\
$34^\prime$ & BBH:1132 & $(1,0,0)$ & $0.0192$ & $0.2272$\\
$35^\prime$ & BBH:1153 & $(1,0,0)$ & $0.0051$ & $0.0881$\\
$36^\prime$ & BBH:1154 & $(1,0,0)$ & $0.0071$ & $0.0882$\\
$37^\prime$ & BBH:1155 & $(1,0,0)$ & $0.0077$ & $0.0882$\\
$38^\prime$ & BBH:0198 & $(1.2,0,0)$ & $0.0030$ & $0.0866$\\
$39^\prime$ & BBH:0310 & $(1.2,0,0)$ & $0.0046$ & $0.0640$\\
$40^\prime$ & BBH:1143 & $(1.2,0,0)$ & $0.0062$ & $0.0937$\\
$41^\prime$ & BBH:0008 & $(1.5,0,0)$ & $0.0663$ & $0.1068$\\
$42^\prime$ & BBH:0093 & $(1.5,0,0)$ & .. & $0.0795$\\
$43^\prime$ & BBH:0593 & $(1.5,0,0)$ & $0.0039$ & $0.0519$\\
$44^\prime$ & BBH:0194 & $(1.5,0,0)$ & $0.0042$ & $0.0476$\\
$45^\prime$ & BBH:1354 & $(1.8,0,0)$ & $0.0010$ & $0.0428$
\end{tabular}
\end{ruledtabular}
\end{center}
\end{table}
\begin{table}
\caption{\label{tab:nospin_2} Summary of the second part of the non-spinning SXS data sets available, spanning a mass-ratio $2\leq q \leq 10$.
	{\rm EOB} waveforms are computed with $\chi_1=0$ and $\chi_2=10^{-4}$. This comparison demonstrates the robustness of the spinning 
	sector of \TEOBiResumSM{} in the nonspinning limit and the consistency with \TEOBiResumM{} when spins are small. 
	From left to right, the columns report: the SXS simulation number, mass ratio and dimensionless spins $\chi_i$,
	and the maximum value of the unfaithfulness $\bar{F}$ between: The two highest resolutions 
	of the NR dataset, if available, see Fig.~\ref{fig:barF_SXS_nospin}(left panel), and between EOB and NR, 
	see Fig.~\ref{fig:barF_SXS_nospin}(right panel).}
\begin{center}
\begin{ruledtabular}
\begin{tabular}{r | l l l l}
\# & {\rm id} & $(q,\chi_1,\chi_2)$ & $\bar{F}^{\rm max}_{\rm NR/NR}[\%]$ & $\bar{F}^{\rm max}_{\rm EOB/NR}[\%]$ \\\hline
$46^\prime$ & BBH:1222 & $(2,0,0)$ & $0.0032$ & $0.0359$\\
$47^\prime$ & BBH:0184 & $(2,0,0)$ & $0.0039$ & $0.0830$\\
$48^\prime$ & BBH:1166 & $(2,0,0)$ & $0.0033$ & $0.0795$\\
$49^\prime$ & BBH:0850 & $(2,0,0)$ & $0.0047$ & $0.0804$\\
$50^\prime$ & BBH:0858 & $(2,0,0)$ & $0.0036$ & $0.0521$\\
$51^\prime$ & BBH:1164 & $(2,0,0)$ & $0.0010$ & $0.0226$\\
$52^\prime$ & BBH:1165 & $(2,0,0)$ & $0.0043$ & $0.0226$\\
$53^\prime$ & BBH:1167 & $(2,0,0)$ & $0.0027$ & $0.0223$\\
$54^\prime$ & BBH:0869 & $(2,0,0)$ & $0.0052$ & $0.0499$\\
$55^\prime$ & BBH:0201 & $(2.3,0,0)$ & $0.0028$ & $0.0181$\\
$56^\prime$ & BBH:0191 & $(2.5,0,0)$ & $0.0036$ & $0.0299$\\
$57^\prime$ & BBH:1221 & $(3,0,0)$ & $0.0016$ & $0.0157$\\
$58^\prime$ & BBH:0168 & $(3,0,0)$ & $0.0022$ & $0.0515$\\
$59^\prime$ & BBH:0183 & $(3,0,0)$ & $0.0029$ & $0.0523$\\
$60^\prime$ & BBH:1177 & $(3,0,0)$ & .. & $0.0521$\\
$61^\prime$ & BBH:1178 & $(3,0,0)$ & .. & $0.0517$\\
$62^\prime$ & BBH:1179 & $(3,0,0)$ & $0.0020$ & $0.0517$\\
$63^\prime$ & BBH:2265 & $(3,0,0)$ & $0.0046$ & $0.0553$\\
$64^\prime$ & BBH:0200 & $(3.3,0,0)$ & $0.0013$ & $0.0216$\\
$65^\prime$ & BBH:0193 & $(3.5,0,0)$ & $0.0016$ & $0.0214$\\
$66^\prime$ & BBH:0294 & $(3.5,0,0)$ & $0.0102$ & $0.0420$\\
$67^\prime$ & BBH:1906 & $(4,0,0)$ & $0.0014$ & $0.0189$\\
$68^\prime$ & BBH:0182 & $(4,0,0)$ & $0.0049$ & $0.0327$\\
$69^\prime$ & BBH:2019 & $(4,0,0)$ & $0.0016$ & $0.0221$\\
$70^\prime$ & BBH:2025 & $(4,0,0)$ & $0.0039$ & $0.0248$\\
$71^\prime$ & BBH:2030 & $(4,0,0)$ & $0.0034$ & $0.0324$\\
$72^\prime$ & BBH:1220 & $(4,0,0)$ & $0.0030$ & $0.0208$\\
$73^\prime$ & BBH:0190 & $(4.5,0,0)$ & $0.0012$ & $0.0250$\\
$74^\prime$ & BBH:0054 & $(5,0,0)$ & $0.0024$ & $0.0216$\\
$75^\prime$ & BBH:0055 & $(5,0,0)$ & .. & $0.0208$\\
$76^\prime$ & BBH:0107 & $(5,0,0)$ & $0.0095$ & $0.0193$\\
$77^\prime$ & BBH:0112 & $(5,0,0)$ & .. & $0.0207$\\
$78^\prime$ & BBH:0187 & $(5,0,0)$ & $0.0012$ & $0.0204$\\
$79^\prime$ & BBH:0197 & $(5.5,0,0)$ & $0.0011$ & $0.0179$\\
$80^\prime$ & BBH:0181 & $(6,0,0)$ & $0.0007$ & $0.0181$\\
$81^\prime$ & BBH:0192 & $(6.6,0,0)$ & $0.0020$ & $0.0147$\\
$82^\prime$ & BBH:0188 & $(7.2,0,0)$ & $0.0022$ & $0.0168$\\
$83^\prime$ & BBH:0195 & $(7.8,0,0)$ & $0.0040$ & $0.0186$\\
$84^\prime$ & BBH:0186 & $(8.3,0,0)$ & $0.0014$ & $0.0204$\\
$85^\prime$ & BBH:0199 & $(8.7,0,0)$ & $0.0089$ & $0.0217$\\
$86^\prime$ & BBH:0189 & $(9.2,0,0)$ & $0.0015$ & $0.0282$\\
$87^\prime$ & BBH:1108 & $(9.2,0,0)$ & $0.0032$ & $0.0239$\\
$88^\prime$ & BBH:0196 & $(9.7,0,0)$ & $0.0045$ & $0.0226$\\
$89^\prime$ & BBH:1107 & $(10,0,0)$ & $0.0010$ & $0.0337$
\end{tabular}
\end{ruledtabular}
\end{center}
\end{table}

\section{Numerical Relativity Systematics}
\label{sec:NRsystematics}
\begin{figure*}[t]
\begin{center}
\includegraphics[width=\textwidth]{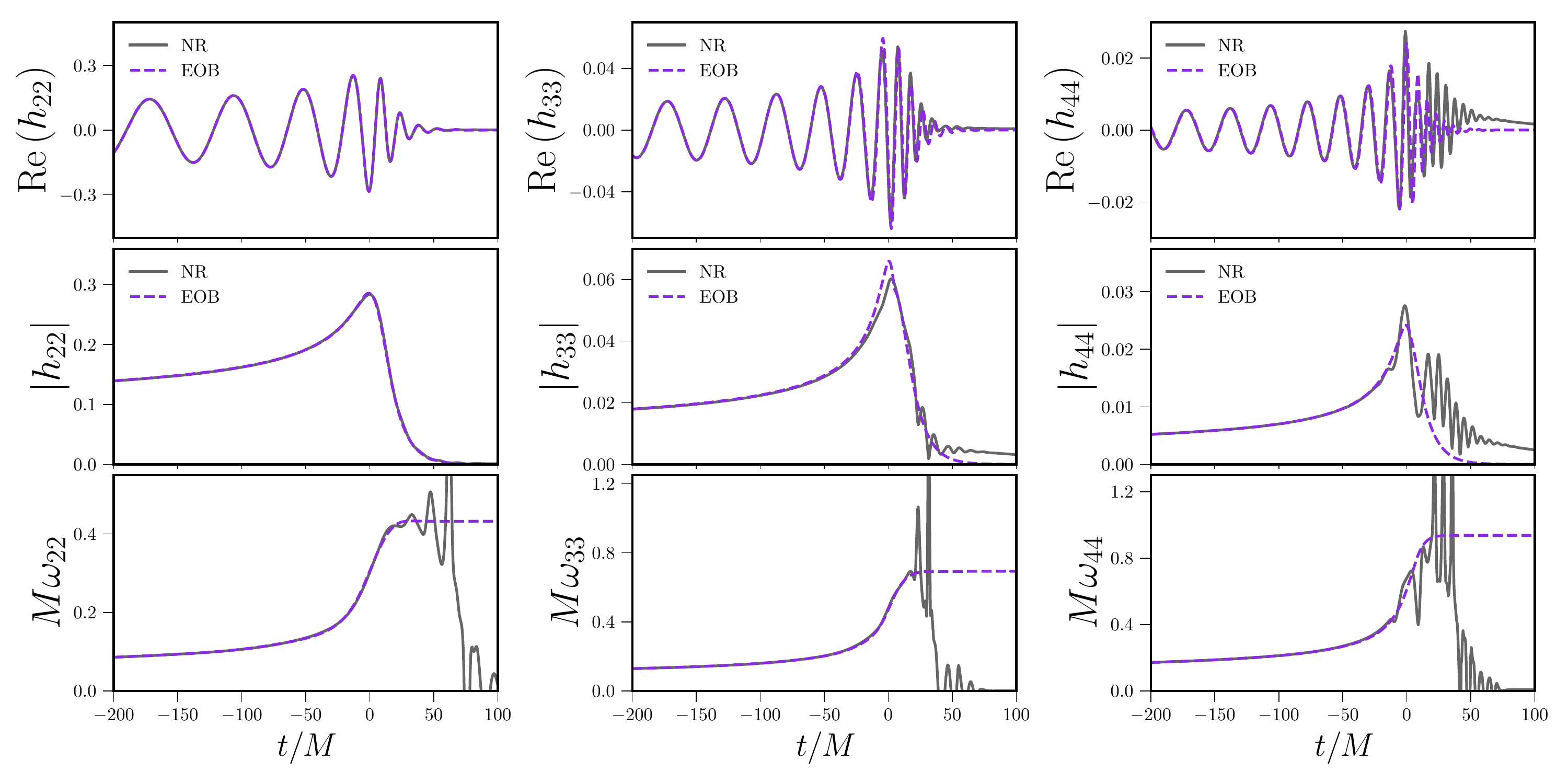}
\caption{Strain, amplitude and instantaenous frequency for {\tt SXS:BBH:0039}.
  Whilst the $(2,2)$ and $(3,3)$ modes are well-behaved, the $(4,4)$ mode
  demonstrates unphysical features, as seen by the non-monotonic behavior
  of the amplitude in the merger-ringdown and the strong, oscillatory
  features in the frequency. This is an example of how NR systematics
  can lead to relatively poor mismatches against the EOB model. }
\label{fig:NR_systematics_1}
\end{center}
\end{figure*}

As was highlighted in~\cite{Nagar:2019wds}, numerical noise and systematics in the NR data can lead to a degradation in the mismatches. 
In Fig.~\ref{fig:mismatch_leqm} we find that the worst mismatches typically come from near edge-on cases, where the power in the $(2,2)$-mode 
is minimized, and for mass ratios near $q \sim 1$, where the amplitude of the odd-$m$ multipoles is suppressed. 
When restricting to the $(\ell, m) = (2,2), (3,3)$ and $(4,4)$ modes, as shown in the bottom panel of Fig.\ref{fig:mismatch_leqm}, the 
mode that contributes the most for the near equal-mass, edge-on configurations is the $(4,4)$ mode. However, as highlighted in 
Fig.~\ref{fig:NR_systematics_1}, the $(4,4)$ mode in the NR datasets can often be particularly problematic, especially through the merger-ringdown. 
In particular, we see strong oscillatory features in the instantaneous frequency and un-physical, non-monotonic behaviour in the amplitude. 
This can result in large mismatches that are relatively uninformative regarding the accuracy of the EOB model against NR. At higher mass ratios, 
where the mode is well-resolved in NR, the mismatches are under control and well below $3$\%. At low total masses, where we compute 
mismatches against a larger portion of the inspiral signal, we see excellent agreement between the EOB model and NR for all modes and 
configurations used in our analysis. In the mismatches shown in Fig.~\ref{fig:mismatch_leqm}, we have removed NR datasets 
that display obvious pathologies, such as those demonstrated in Fig.~\ref{fig:NR_systematics_1}. 

\section{Analytic modeling of the multipolar ringdown waveform}
\label{sec:app_fits}
\subsection{Introduction}
In this Appendix we discuss the fits of the NR data needed
for completing \TEOBiResumSM{} through merger and ringdown.
The fits concern: (i) frequency and amplitude at the peak
of each multipole; (ii) the time delays $\Delta t_{\ell m}$
between the peak of each multipole and the peak of the $(2,2)$
mode; (iii) fits for waveforms quantities at the location at
the NQC extraction point. Technical details are all listed
in Sec.~III~D~and Sec.~V~A~of Ref.~\cite{Nagar:2019wds} and
we address the reader there for complementary information.
In Sec.~\ref{sec:peak} below we report fits of various
waveform quantities at the peak of each multipole, that
is amplitude, frequency and time-delay
$\left(A_{\ell m}^{\rm peak},\omega_{\ell m}^{\rm peak},\Delta t_{\ell m}\right)$.
Following Refs.~\cite{Damour:2014yha,Nagar:2016iwa,Nagar:2018zoe,Nagar:2019wds},
the postpeak waveform needs three additional parameters
$\left(c_3^{{A}_{\ell m}}, c_3^{{\phi_{\ell m}}},c_4^{{\phi_{\ell m}}}\right)$
to be fitted to NR data. This is discussed in Sec.~\ref{sec:fit_evol}
below. We present spin-dependent fits for multipoles
$(2,2),(3,3),(4,4),(5,5)$, although, for robustness,
we prefer to use the nonspinning fits discussed
in~\cite{Nagar:2019wds} except for the $(2,2)$ mode.
This gives a rather accurate representation of the waveform
provided that the other quantities (e.g. the peak ones)
incorporate the complete spin dependence.
The fits of the quasi-normal-mode frequencies and (inverse) damping
timpes entering  $\left(\omega_1^{\ell m},\alpha_1^{\ell m},\alpha_{21}^{\ell m}\right)$
are given in~\cite{Nagar:2019wds}\footnote{The reader
  should note that the fits are done versus the spin
  of the remnant $\hat{a}_f$, which in turn
is obtained from the fits presented in
Ref.~\cite{Jimenez-Forteza:2016oae}.}.
The waveform quantities used to determine the NQC
corrections to the waveform amplitude and phase,
$\left(A_{\ell m}^{\rm NQC},\omega_{\ell m}^{\rm NQC},\dot{A}_{\ell m}^{\rm NQC},\dot{\omega}_{\ell m}^{\rm NQC}\right)$
are usually obtained analytically from the postpeak template
and all details are collected in Sec.~\ref{sec:NQC} below.
For the $(4,4)$ mode, however, this procedure cannot deliver
an accurate time-derivative of the waveform amplitude,
so that a dedicated fit is given. In the case of the $(\ell,m)=(2,2)$
mode fits for all 4 NQC quanties are also given.
Unless otherwise stated all fits are done using \texttt{fitnlm}
of \texttt{matlab} and \texttt{NonLinearModelFit}
of \texttt{MATHEMATICA}. All fits exclusively use the the calibration
set taken from the BAM catalog, test-particle data and
the \textit{calibration set} of SXS waveforms listed in
Appendix~\ref{sec:NR_catalog}. The exception is $\Delta t_{21}$,
which is informed additionally by the \textit{test set}
of SXS waveforms.

\subsection{Modeling the peak of each multipole}
\label{sec:peak}
Firs of all, let us recall some symmetric combinations
of the spin variables that will be useful later on
\begin{align}
\hat{S} &\equiv \frac{S_1 + S_2}{M^2}\ = \frac{1}{2}\left(\tilde{a}_0 + X_{12} \tilde{a}_{12}\right), \\ 
\bar{S} &\equiv \frac{S_1 - S_2}{M^2}\ = \frac{1}{2}\left(X_{12} \tilde{a}_0 + \tilde{a}_{12}\right).
\end{align}
We refer to the multipolar decomposition of the strain 
\be
h \equiv h_+ - i h_\times = \sum_{\ell,m} h_{\ell m}{}_{-2}Y_{\ell m}(\iota, \varphi) \ .
\ee
Here ${}_{-2}Y_{\ell m}$ are the $s=2$ spin-weighted, spherical harmonics. $\iota$ and  $\varphi$
are the polar (with respect to the direction of the orbital angular momentum) and azimuthal
angle in the source frame.
Each multipole is decomposed in amplitude $A_{\ell m}$ and a phase $\phi_{\ell m}$ as 
\be
h_{\ell m} =A_{\ell m} e^{\ii \phi_{\ell m}} \ .
\ee
The instantaneous GW frequency $\omega_{\ell m}$ is defined as
\be
\omega_{\ell m}\equiv -\dot{\phi}_{\ell m} ,
\ee
where the dot indicate the time derivative. Motivated by the leading-order analytical behavior
of each multipole, we introduce the following rescaled multipolar amplitudes $\hat{A}_{\ell m}$:
\begin{align}
\hat{A}_{22}&\equiv  A_{22}/\left[\nu \left(1-\hat{S}\omega_{22} \right)\right],\\
\hat{A}_{21}&\equiv  A_{21}/\nu ,\\
\hat{A}_{33}&\equiv  A_{33}/\nu ,\\
\hat{A}_{32}&\equiv  A_{32}/\left[\nu \left(1-\tilde{a}_0\left(\omega_{32}/2\right)^{1/3} \right)\right] ,\\
\hat{A}_{44}&\equiv  A_{44}/\left[\nu \left(1-\frac{1}{2}\hat{S}\omega_{44} \right)\right] ,\\
\hat{A}_{43}&\equiv  A_{43}/\nu ,\\
\hat{A}_{42}&\equiv  A_{42}/\left[\nu \left(1-\tilde{a}_0\left(\omega_{42}/2\right)^{1/3} \right)\right] .
\end{align}
Then one defines the time where each $(\ell,m)$ mode peaks as
\be
t^{\rm peak}_{\ell m}\equiv t\left({\rm max}\left[ \hat{A}_{\ell m}\right]\right),
\ee
and the merger time, that is defined as the peak of the $(2,2)$ mode, i.e.
\be
t^{\rm mrg} \equiv t^{\rm peak}_{2 2}.
\ee
One then defines the time-delay between merger time and the time where
each mode peak, as
\be
\Delta t_{\ell m}\equiv t^{\rm peak}_{\ell m}-t_{\rm mrg}
\ee
For shortness, we denote quantities calculated at a given time using the corresponding superscript, e.g.
\be
\hat{A}_{\ell m}^{\rm peak}\equiv \hat{A}_{\ell m}\left( t^{\rm peak}_{\ell m}\right).
\ee
Let us now give all details on a mode-by-mode basis.
\subsubsection{$(\ell,m)=(2,2)$ multipole}
\label{sec:fits_merger}
We start by describing the template with which $\omega_{22}^{\rm mrg}$
and $\hat{A}_{22}^{\rm mrg}$ were fitted. The same structure is used both
for the amplitude and frequency at merger.
We here present it explicitly for $\omega_{22}^{\rm mrg}$, w
hile the same for $\hat{A}_{22}^{\rm mrg}$ is obtained by suitably changing the coefficient labels.
The frequency at merger $\omega_{22}^{\rm mrg}$ is factorized as 
\begin{equation}
\label{eq:fttfit}
\omega_{22}^{\rm mrg} 	=  \omega_{22}^{{\rm mrg}_{0}} \omega_{22}^{\rm orb}(\nu)\omega_{22}^{\hat{S}}(\hat{S},X_{12})\ ,
\end{equation}
where $\omega_{22}^{{\rm mrg}_{0}}$ is the value of the merger frequency obtained from a nonspinning
test-particle waveform (see e.g. Table~3 of~\cite{Harms:2014dqa}).
The nonspinning $\nu$-dependence is then introduced by fitting the nonspinning data
with a template of the form
\begin{equation}
\omega_{22}^{\rm orb}(\nu) = 1 + a^\omega_1\nu + a^\omega_2\nu^2\ ,
\end{equation}
where the coefficients $a^\omega_i$ are determined using 19 non-spinning SXS waveforms
with mass ratios $1\leq m_1/m_2\leq 10$.
The spin dependence is introduced in two steps: first one accurately fits the spin-dependence
of equal-mass data. Then, additional flexibility to incorporate the spinning, unequal-mass data
is introduced. More precisely the equal-mass, spin-dependence is obtained with
\begin{equation}
\label{eq:f0}
\omega_{22}^{\hat{S}}(\hat{S},X_{12}=0)=\frac{1+b^{\omega_{m_1=m_2}}_1\hat{S}+b^{\omega_{m_1=m_2}}_2\hat{S}^2}{1+b^{\omega_{m_1=m_2}}_3\hat{S}},
\end{equation}
which is informed by 39 equal-mass, spin-aligned, SXS waveforms. The additional dependence on
mass ratio is incorporated substituting into Eq.~\eqref{eq:f0}
\be
\label{eq:expol2}
b^{\omega_{m_1=m_2}}_i \rightarrow  \frac{b^{\omega_{m_1=m_2}}_i + c^\omega_{i1}X_{12}}{1+c^\omega_{i2}X_{12}} \ ,
\ee
with $i=\left\{ 1,3 \right\}$.
where the additional coefficients $c_{ij}$ are fitted using test-particle data,
77 additional SXS spinning waveforms and 14~additional NR waveforms from BAM.
The coefficients are explicitly given in Table~\ref{tab:mrg_22_fit_coefficients}. 
\begin{table*}
  \caption{\label{tab:mrg_22_fit_coefficients} Explicit coefficients and their errors
    for the merger frequency and amplitude fits of the $(2,2)$ mode. The analytic template
    of the fit is defined in Eqs.~\eqref{eq:fttfit}~--~\eqref{eq:expol2}.}
\begin{ruledtabular}
\begin{tabular}{l c l l  @{\hspace*{25mm}}  l c l l}
$\omega_{22}^{{\rm mrg}_{0}}$ & $=$ & $\;\;\;0.273356$  & 				&  $\hat{A}_{22}^{{\rm mrg}_{=0}}$ 	& $=$ & $\;\;\;1.44959$ & \\
$a^\omega_1$ 					& $=$ & $\;\;\;0.84074$ 	& $\pm 0.014341$  & $a^{\hat{A}}_1$& $=$ 					&$-0.041285$ &$\pm 0.0078878$\\
$a^\omega_2$ 					& $=$ & $\;\;\;1.6976$ 		& $\pm 0.075488$  & $a^{\hat{A}}_2$& $=$ 					&$\;\;\;1.5971$ &$\pm 0.041521$\\ \hline
$b^{\omega_{m_1=m_2}}_1$ 	& $=$ & $-0.42311$ 		& $\pm 0.088583$  & $b^{\hat{A}_{m_1=m_2}}_1$& $=$ 		&$-0.74124$ &$\pm 0.016178$\\
$b^{\omega_{m_1=m_2}}_2$ 	& $=$ & $-0.066699$ 		& $\pm 0.042978$  & $b^{\hat{A}_{m_1=m_2}}_2$& $=$ 		&$-0.088705$&$\pm 0.0081611$\\ 
$b^{\omega_{m_1=m_2}}_3$ 	& $=$ & $-0.83053$ 		& $\pm 0.084516$  & $b^{\hat{A}_{m_1=m_2}}_3$& $=$ 		&$-1.0939$ &$\pm 0.015318$\\ \hline
$c^\omega_{11}$ 					& $=$ & $\;\;\;0.15873$ 	& $\pm 0.1103$  	& $c^{\hat{A}}_{11}$& $=$ 					&$\;\;\;0.44467$ &$\pm 0.037352$\\
$c^\omega_{12}$	 				& $=$ & $-0.43361$ 		& $\pm 0.2393$  	& $c^{\hat{A}}_{12}$& $=$ 					&$-0.32543$ &$\pm 0.081211$\\ 
$c^\omega_{21}$ 					& $=$ & $\;\;\;0.60589$ 	& $\pm 0.076215$  & $c^{\hat{A}}_{31}$& $=$ 					&$\;\;\;0.45828$ &$\pm 0.066062$\\ 
$c^\omega_{22}$ 					& $=$ & $-0.71383$ 		& $\pm 0.096828$  & $c^{\hat{A}}_{32}$& $=$ 					&$-0.21245$ &$\pm 0.080254$ \\ 
\end{tabular}
\end{ruledtabular}
\end{table*}
\subsubsection{$(\ell,m)=(2,1)$ multipole}
The procedure followed for the subdominant modes is similar to what is done for the $(2,2)$.
There are however some differences. First of all, the peak time shift $\Delta t_{\ell m}$
is also fitted to NR simulations. Second, basing ourselves to the analytical behavior
of the multipolar waveform, we have decided to use different factorizations and
different variables to model each mode.
For example, the $(2,1)$ multipole (and every $m$-odd mode) vanishes because
of symmetry in the equal-mass equal-spin case.
This has brought us to consider the following factorization
for $\hat{A}_{21}^{\rm peak}$, which is written as
\be
\hat{A}_{21}^{\rm peak}= \hat{A}_{21}^{{\rm peak}_{0}}X_{12}\hat{A}_{21}^{\rm orb}\left(\nu\right) +  \hat{A}_{21}^{\rm Spin}\left(\bar{S},\nu\right).
\ee
where $\hat{A}_{21}^{{\rm peak}_{0}}$ is the peak amplitude in the test-particle limit.
The factor $\hat{A}_{21}^{\rm orb}$ is informed by non-spinning waveforms
and is fitted with the template
\be
\hat{A}_{21}^{\rm orb}(\nu)= \frac{1+a^{\hat{A}_{21}}_1 \nu+a^{\hat{A}_{21}}_2 \nu^2}{1+a^{\hat{A}_{21}}_3 \nu } \ .
\ee
The spin dependence is first captured in the test-particle limit with the function
\be
\hat{A}_{21}^{\rm Spin}(\bar{S},\nu=0) = \frac{1+b^{\hat{A}^{0}_{21}}_1 \bar{S}+b^{\hat{A}^{0}_{21}}_2 \bar{S}^2}{1+b^{\hat{A}^{0}_{21}}_3 \bar{S}}\ .
\ee
The $\nu$-dependence is then modeled via the replacement
\be
b^{\hat{A}^{0}_{21}}_i\rightarrow b^{\hat{A}^{0}_{21}}_i+   c^{\hat{A}_{21}}_{i1}\nu +   c^{\hat{A}_{21}}_{i2}\nu^2 \ ,
\ee
with $i=\left\{ 1,2,3\right\}$.

The gravitational wave frequency $\omega_{21}$ is instead 
factorized as
\be
\omega^{\rm peak}_{21}=\omega^{{\rm peak}_{0}}_{21}\omega^{\rm orb}_{21}(\nu)\omega^{\rm Spin}_{21}\left(\hat{S},\nu\right),
\ee
where the $\nu$-dependence of the nonspinning part is modeled as
\be
\omega^{\rm orb}_{21}(\nu)=1+a^{\omega_{21}}_1 \nu+a^{\omega_{21}}_2 \nu^2\ .
\ee
The spin dependence is fitted first in the test-particle limit
\be
\omega^{\rm Spin}_{21}\left(\hat{S},\nu = 0\right) = 1+b^{\omega^{0}_{21}}_1 \hat{S}+b^{\omega^{0}_{21}}_2 \hat{S}^2\ ,
\ee
and then extended to a general mass ratio via the replacement 
\be
b^{\omega^{0}_{21}}_i\rightarrow b^{\omega^{0}_{21}}_i+   c^{\omega_{21}}_{i}\nu \ ,
\ee
with $i=\left\{ 1,2 \right\}$.

Finally, to represent analytically the time-delay $\Delta t_{21}$ we use
\be
\Delta t_{21} =\Delta t_{21}^{\rm orb}(\nu)\Delta t_{21}^{\rm spin}\left( \bar{S},X_{12} \right)\ ,
\ee
where the orbital behavior is factorized into two separate parts before fitting with
\bea
\Delta t_{21}^{\rm orb}(\nu) =&\left( \Delta t_{21}^{0}(1-4\nu) +  \Delta t_{21}^{\nu=1/4}4\nu \right) \nonumber\\
&\;\;\; \times\left(1 + a_1^{\Delta t_{21}}\nu\sqrt{1-4\nu} \right)\ .
\eea
The factor $\Delta t_{21}^{\nu=1/4}$ is obtained by fitting a 2nd-order polynomial, in $\hat{a}_0$ to the equal-mass
waveforms. $\Delta t_{21}^{0}$ is the test-particle value.
The equal-mass spin behavior is fitted with
\be
\Delta t_{21}^{\rm spin}\left( \bar{S},X_{12}=0 \right)= 1+ b_1^{\Delta t_{21}^{\nu=1/4}}\hat{a}_0+ b_2^{\Delta t_{21}^{\nu=1/4}}\hat{a}_0^2\ ,
\ee
while the comparable mass case is extrapolated using
\be
b_1^{\Delta t_{21}^{\nu=1/4}} \rightarrow \frac{b_1^{\Delta t_{21}^{0}} + c_{i1}^{\Delta t_{21}}X_{12} }{1+ c_{i2}^{\Delta t_{21}}X_{12}}\ ,
\ee
with $i = \left\{1,2\right\}$. The outcome of the fit, with the explicit values of all coefficients,
id found in Table~\ref{tab:peak_21}.
\begin{table*}
\caption{\label{tab:peak_21}
	Explicit coefficients of the fits of $\hat{A}_{21}^{\rm peak}$, $\omega_{21}^{\rm peak}$ and $\Delta t_{21}$.}
\centering
\begin{ruledtabular}
\begin{tabular}{l c l  @{\hspace*{25mm}}  l c l  @{\hspace*{25mm}}  l c l}
$\hat{A}_{21}^{{\rm peak}_{0}}$ & $=$ &  $0.523878$ 
& $\omega_{21}^{{\rm peak}_{0}}$& $=$ &  $0.290643$
& $\Delta t_{21}^{0}$           & $=$ &  $11.75925$
\\
$a^{\hat{A}_{21}}_1$& $=$ &  $3.33622$ 
& $a^{\omega_{21}}_1$& $=$ &  $-0.563075$ 
&$ \Delta t_{21}^{\nu=1/4}$& $=$ &  $6.6264$
\\
$a^{\hat{A}_{21}}_2$& $=$ &  $3.47085$ 
&$a^{\omega_{21}}_2$& $=$ &  $3.28677$
&$a^{\Delta t_{21}}_1$& $=$ &  $-2.0728$
\\
$a^{\hat{A}_{21}}_3$& $=$ &  $4.76236$ 
& & &
& & &
\\
\hline
$b^{\hat{A}_{21}^{0}}_1$& $=$ &  $-0.428186$ 
& $b^{\omega_{21}^{0}}_1$& $=$ &  $0.179639$
&$b^{\Delta t_{21}^{0}}_1$& $=$ &  $0.0472289$
\\
$b^{\hat{A}_{21}^{0}}_2$& $=$ &  $-0.335659$
& $b^{\omega_{21}^{0}}_2$& $=$ &  $-0.302122$
&$b^{\Delta t_{21}^{0}}_2$& $=$ &  $0.115583$
\\
$b^{\hat{A}_{21}^{0}}_3$& $=$ &  $0.828923$ 
& & & 
& & & 
\\
\hline
$c^{\hat{A}_{21}}_{11}$& $=$ &  $0.891139$ 
&$c^{\omega_{21}}_{1}$& $=$ &  $-1.20684$
&$c^{\Delta t_{21}}_{11}$& $=$ &  $-1976.13$
\\
$c^{\hat{A}_{21}}_{12}$& $=$ &  $-5.191702$ 
& $c^{\omega_{21}}_{2}$& $=$ &  $0.425645$
&$c^{\Delta t_{21}}_{12}$& $=$ &  $3719.88$ 
\\
$c^{\hat{A}_{21}}_{21}$& $=$ &  $3.480139$
& & & 
&$c^{\Delta t_{21}}_{21}$& $=$ &  $-2545.41$
\\
$c^{\hat{A}_{21}}_{22}$& $=$ &  $10.237782$
& & &
& $c^{\Delta t_{21}}_{22}$& $=$ &  $5277.62$
\\
$c^{\hat{A}_{21}}_{31}$& $=$ &  $-13.867475$ 
& & & 
& & & 
\\
$c^{\hat{A}_{21}}_{32}$& $=$ &  $10.525510$
& & &
& & & 
\\
\end{tabular}
\end{ruledtabular}
\end{table*}
\subsubsection{$(\ell,m)=(3,3)$ multipole}
For this mode, the peak amplitude is written as the sum of two terms
\be
\hat{A}_{33}^{\rm peak} = \hat{A}_{33}^{{\rm peak}_{0}} X_{12} \hat{A}_{33}^{\rm orb}(\nu) + \hat{A}_{33}^{\rm Spin}(\tilde{a}_{12},\nu) \ ,
\ee
where $\hat{A}_{33}^{{\rm peak}_{0}} $ is the peak amplitude in the test particle limit. 
The orbital term is modeled as
\be
\hat{A}_{33}^{\rm orb}(\nu)= \frac{1+a^{\hat{A}_{33}}_1 \nu+a^{\hat{A}_{33}}_2 \nu^2}{1+a^{\hat{A}_{33}}_3 \nu } \ .
\ee
The spin dependence is first fitted in the test-particle limit using
\be
\hat{A}_{33}^{\rm Spin}(\tilde{a}_{12},\nu=0) = \frac{b^{\hat{A}^{0}_{33}}_1 \tilde{a}_{12}}{1+b^{\hat{A}^{0}_{33}}_2 \tilde{a}_{12}}\ ,
\ee
and then extended to comparable masses via the replacements
\bea
b^{\hat{A}^{0}_{33}}_1 &\rightarrow &\frac{b^{\hat{A}^{0}_{33}}_1 + c^{\hat{A}_{33}}_{11}\nu }{1+ c^{\hat{A}_{33}}_{12}\nu + c^{\hat{A}_{33}}_{13}\nu^2  }\ ,\\
b^{\hat{A}^{0}_{33}}_2 &\rightarrow &\frac{b^{\hat{A}^{0}_{33}}_2 + c^{\hat{A}_{33}}_{21}\nu }{1+ c^{\hat{A}_{33}}_{22}\nu + c^{\hat{A}_{33}}_{23}\nu^2  }\ .
\eea
The istantaneous frequency $\omega_{33}$ is factorized as
\be
\omega^{\rm peak}_{33}=\omega^{{\rm peak}_{0}}_{33}\omega^{\rm orb}_{33}(\nu)\omega^{\rm Spin}_{33}\left(\hat{S},\nu\right)\ ,
\ee
where
\be
\omega^{\rm orb}_{33}(\nu)= 1+a^{\omega_{33}}_1 \nu+a^{\omega_{33}}_2 \nu^2\ .
\ee
The test-particle spin factor is given by
\be
\omega^{\rm Spin}_{33}\left(\hat{S},\nu = 0\right) = \frac{1+b^{\omega^{0}_{33}}_1 \hat{S}+b^{\omega^{0}_{33}}_2 \hat{S}^2}{1+b^{\omega^{0}_{33}}_3 \hat{S}}\ ,
\ee
while the general spin-dependence stems from the replacement
\be
b^{\omega^{0}_{33}}_i\rightarrow \frac{b^{\omega^{0}_{33}}_i+   c^{\omega_{33}}_{i1}\nu}{1+c^{\omega_{33}}_{i2}\nu}\ ,
\ee
with $i=\left\{ 1,3\right\}$.

To describe $\Delta t_{33}$ we start from the expression
\be
\Delta t_{33} = \Delta t_{33}^{0}\Delta t_{33}^{\rm orb}(\nu)\Delta t_{33}^{\rm spin}\left( \hat{S},\nu \right)\ ,
\ee
with
\begin{align}
\Delta t_{33}^{\rm orb}(\nu) &= 1 + a_1^{\Delta t_{33}}\nu + a_2^{\Delta t_{33}}\nu^2 \ , \\
\Delta t_{33}^{\rm spin}\left( \hat{S},\nu=0 \right) &= \frac{1+ b_1^{\Delta t_{33}^{0}}\hat{S}+ b_2^{\Delta t_{33}^{0}}\hat{S}^2}{1+ b_3^{\Delta t_{33}^{0}}\hat{S}} \ .
\end{align}
The spin-dependence is obtained from the replacement
\be
b_1^{\Delta t_{33}^{0}} \rightarrow \frac{b_1^{\Delta t_{33}^{0}} + c_{i1}^{\Delta t_{33}}\nu }{1+ c_{i2}^{\Delta t_{33}}\nu }\ ,
\ee
with $i = \left\{1,2,3\right\}$.
The explicit values of the fit coefficients are listed in Table~\ref{tab:peak_33}.
\begin{table*}
\caption{\label{tab:peak_33}
Explicit coefficients of the fits of $\hat{A}_{33}^{\rm peak}$, $\omega_{33}^{\rm peak}$ and $\Delta t_{33}$.}
\centering
\begin{ruledtabular}
\begin{tabular}{l c l  @{\hspace*{25mm}}  l c l  @{\hspace*{25mm}}  l c l}
$\hat{A}_{33}^{{\rm peak}_{0}}$ & $=$ &  $0.566017$
&$\omega_{33}^{{\rm peak}_{0}}$ & $=$ &  $0.454128$
& $ \Delta t_{33}^{0}$& $=$ &  $3.42593$
\\
$a^{\hat{A}_{33}}_1$ & $=$ &  $-0.22523$
&$a^{\omega_{33}}_1$ & $=$ &  $1.08224$
&$a^{\Delta t_{33}}_1$& $=$ &  $0.183349$
\\
$a^{\hat{A}_{33}}_2$ & $=$ &  $3.0569$
&$a^{\omega_{33}}_2$ & $=$ &  $2.59333$
&$a^{\Delta t_{33}}_2$& $=$ &  $4.22361$
\\
$a^{\hat{A}_{33}}_3$ & $=$ &  $-0.396851$
& & &
& & & 
\\
\hline
$b^{\hat{A}^{0}_{33}}_1$ & $=$ &  $0.100069$
&$b^{\omega^{0}_{33}}_1$ & $=$ &  $-0.406161$
&$b^{\Delta t_{33}^{0}}_1$& $=$ &  $-0.49791$
\\
$b^{\hat{A}^{0}_{33}}_2$ & $=$ &  $-0.455859$
&$b^{\omega^{0}_{33}}_2$ & $=$ &  $-0.0647944$ 
&$b^{\Delta t_{33}^{0}}_2$& $=$ &  $-0.18754$
\\
  & &
&$b^{\omega^{0}_{33}}_3$ & $=$ &  $-0.748126$
&$b^{\Delta t_{33}^{0}}_3$& $=$ &  $-1.07291$
\\
\hline
$c^{\hat{A}_{33}}_{11}$ & $=$ &  $-0.401156$
&$c^{\omega_{33}}_{11}$ & $=$ &  $0.85777$
&$c^{\Delta t_{33}}_{11}$& $=$ &  $-1.9478$
\\
$c^{\hat{A}_{33}}_{12}$ & $=$ &  $-0.141551$ 
&$c^{\omega_{33}}_{12}$ & $=$ &  $-0.70066$
&$c^{\Delta t_{33}}_{12}$& $=$ &  $13.9828$
\\
$c^{\hat{A}_{33}}_{13}$ & $=$ &  $-15.4949$
&$c^{\omega_{33}}_{31}$ & $=$ &  $2.97025$
&$c^{\Delta t_{33}}_{21}$& $=$ &  $1.25084$
\\
$c^{\hat{A}_{33}}_{21}$ & $=$ &  $1.84962$
&$c^{\omega_{33}}_{32}$ & $=$ &  $-3.96242$
&$c^{\Delta t_{33}}_{22}$& $=$ &  $-3.41811$
\\
$c^{\hat{A}_{33}}_{22}$ & $=$ &  $-2.03512$
& & &
&$c^{\Delta t_{33}}_{31}$& $=$ &  $-1043.15$
\\
$c^{\hat{A}_{33}}_{23}$ & $=$ &  $-4.92334$
& & &
&$c^{\Delta t_{33}}_{32}$& $=$ &  $1033.85$
\\
\end{tabular}
\end{ruledtabular}
\end{table*}
\subsubsection{$(\ell,m)=(3,2)$ multipole}
The peak amplitude of the $(3,2)$ mode is fitted with a factorized template of the form
\be
\hat{A}_{32}^{\rm peak}= \hat{A}_{32}^{{\rm peak}_{0}}\left(1-3\nu \right)\hat{A}_{32}^{\rm orb}\left(\nu\right) \hat{A}_{32}^{\rm Spin}\left(\bar{S},\nu\right) \ ,
\ee
where $\hat{A}_{32}^{{\rm peak}_{0}}$ is the peak amplitude of the mode in the test-particle limit.
The factor $\hat{A}_{32}^{\rm orb}$ is informed by non-spinning waveforms
and is fitted with the template
\be
\hat{A}_{32}^{\rm orb}(\nu)= \frac{1+a^{\hat{A}_{32}}_1 \nu+a^{\hat{A}_{32}}_2 \nu^2}{1+a^{\hat{A}_{32}}_3 \nu } \ .
\ee
The spin dependence is first captured for the test-particle limit with the function
\be
\hat{A}_{32}^{\rm Spin}(\bar{S},\nu=0) = \frac{1+b^{\hat{A}^{0}_{32}}_1 \tilde{a}_0}{1+b^{\hat{A}^{0}_{32}}_2 \tilde{a}_0}\ ,
\ee
while the $\nu$-dependence enters via the replacement
\be
b^{\hat{A}^{0}_{32}}_i\rightarrow \frac{b^{\hat{A}^{0}_{32}}_i+   c^{\hat{A}_{32}}_{i1}\nu +   c^{\hat{A}_{32}}_{i2}\nu^2}{1+   c^{\hat{A}_{32}}_{i3}\nu +   c^{\hat{A}_{32}}_{i4}\nu^2} \ ,
\ee
with $i=\left\{ 1,2 \right\}$.

The instantaneous frequency $\omega_{32}$ mode is factorized as
\be
\omega^{\rm peak}_{32}=\omega^{{\rm peak}_{0}}_{32}\omega^{\rm orb}_{32}(\nu)\omega^{\rm Spin}_{32}\left(\tilde{a}_0,\nu\right) \ .
\ee
The orbital dependence is modeled as
\be
\omega^{\rm orb}_{32}(\nu)=\frac{1+a^{\omega_{32}}_1 \nu+a^{\omega_{32}}_2 \nu^2}{1+a^{\omega_{32}}_3 \nu+a^{\omega_{32}}_4 \nu^2}\ .
\ee
The spin dependence is fitted first for the equal-mass case
\be
\omega^{\rm Spin}_{32}\left(\tilde{a}_0,\nu = 1/4\right)
= \frac{1+b^{\omega^{\nu=1/4}_{32}}_1 \tilde{a}_0+b^{\omega^{\nu=1/4}_{32}}_2 \tilde{a}_0^2}{1+b^{\omega^{\nu=1/4}_{32}}_3 \tilde{a}_0}\ ,
\ee
while the additional dependence on the mass ratio enters via the replacements
\be
b^{\omega^{0}_{32}}_i\rightarrow \frac{b^{\omega^{\nu=1/4}_{32}}_i+   c^{\omega_{32}}_{i1}X_{12}+   c^{\omega_{32}}_{i2}X^2_{12}}{1+   c^{\omega_{32}}_{i3}X_{12}} \ ,
\ee
with $i=\left\{ 1,2\right\}$. The coefficients of $\hat{A}^{\rm peak}_{32}$ and $\omega^{\rm peak}_{32}$
are explicitly listed in Table~\ref{tab:peak_42}.

Moving to $\Delta t_{32}$, it is given by
\be
\Delta t_{32} = \Delta t_{32}^{0}\Delta t_{32}^{\rm orb}(\nu)\Delta t_{32}^{\rm spin}\left( \hat{S},\nu \right)\ ,
\ee
where the orbital behavior is fitted with
\be
\Delta t_{32}^{\rm orb}(\nu) = \frac{1 + a^{\Delta t_{32}}_1\nu + a^{\Delta t_{32}}_2\nu^2}{1 + a^{\Delta t_{32}}_3\nu + a^{\Delta t_{32}}_4\nu^2}\ .
\ee
The spin behavior is more complicated than the corresponding term of other modes. This is separated into two sectors, as
\begin{align}
\Delta t_{32}^{\rm spin}\left( \hat{S},\nu \right) =~& \Delta t_{32}^{{\rm spin}_{\nu>1/5}}\left( \hat{S},\nu \right)\Theta \left(\nu-1/5 \right)\nonumber\\
&+ \Delta t_{32}^{{\rm spin}_{\nu\leq 1/5}}\left( \hat{S},\nu \right)\left[1-\Theta \left(\nu-1/5 \right)\right] \ ,
\end{align}
where $\Theta$ denotes the Heaviside step function. In the $\nu>1/5$ regime the fit is first done to the equal-mass case
\be
\Delta t_{32}^{{\rm spin}_{\nu>1/5}}\left( \hat{S},\nu=1/4 \right)= \frac{1+ b^{\Delta t_{32}^{\nu=1/4}}_1\hat{S}+ b^{\Delta t_{32}^{\nu=1/4}}_2\hat{S}^2}{1+ b^{\Delta t_{32}^{\nu=1/4}}_3\hat{S}}\ .
\ee
Then it is extrapolated following
\be
b^{\Delta t_{32}^{\nu=1/4}}_i \rightarrow \frac{b^{\Delta t_{32}^{\nu=1/4}}_1 + c^{\Delta t_{32}}_{i1}X_{12} + c^{\Delta t_{32}}_{i2}X_{12}^2+ c^{\Delta t_{32}}_{i3}X_{12}^3}{1+ c^{\Delta t_{43}}_{i4}X_{12} + c^{\Delta t_{43}}_{i5}X_{12}^2}\ ,
\ee
with $i=\left\{ 1,2,3\right\}$.

In the $\nu\leq 1/5$ regime the fit is first done to the equal-mass case
\be
\Delta t_{32}^{{\rm spin}_{\nu\leq 1/5}}\left( \hat{S},\nu=0 \right)= \frac{1+ b^{\Delta t_{32}^{0}}_1\hat{S}+ b^{\Delta t_{32}^{0}}_2\hat{S}^2}{1+ b^{\Delta t_{32}^{0}}_3\hat{S}}\ .
\ee
Then it is extrapolated following
\be
b^{\Delta t_{32}^{0}}_i \rightarrow \frac{b^{\Delta t_{32}^{0}}_1 + c^{\Delta t_{32}}_{i1}\nu + c^{\Delta t_{32}}_{i2}\nu^2+ c^{\Delta t_{32}}_{i3}\nu^3}{1+ c^{\Delta t_{32}}_{i4}\nu + c^{\Delta t_{32}}_{i5}\nu^2}\ ,
\ee
with $i=\left\{ 1,2,3\right\}$.
The coefficients appearing in $\Delta t_{32}$ are shown in Table~\ref{tab:Dt_32}.
\begin{table}
\caption{\label{tab:peak_32}
Explicit coefficients of the fits of $\hat{A}_{32}^{\rm peak}$ and $\omega_{32}^{\rm peak}$.}
\centering
\begin{ruledtabular}
\begin{tabular}{l c l @{\hspace*{15mm}} l c l }
$\hat{A}_{32}^{{\rm peak}_{0}}$& $=$ &  $0.199019$
&$\omega_{32}^{{\rm peak}_{0}}$& $=$ &  $0.451607$ 
\\
$a^{\hat{A}_{32}}_1$& $=$ &  $-6.06831$ 
&$a^{\omega_{32}}_1$& $=$ &  $-9.13525$ 
\\
$a^{\hat{A}_{32}}_2$& $=$ &  $10.7505$ 
&$a^{\omega_{32}}_2$& $=$ &  $21.488$ 
\\
$a^{\hat{A}_{32}}_3$& $=$ &  $-3.68883$ 
&$a^{\omega_{32}}_3$& $=$ &  $-8.81384$ 
\\
& &
&$a^{\omega_{32}}_4$& $=$ &  $20.0595$ 
\\
\hline
$b^{\hat{A}_{32}^{0}}_1$& $=$ &  $-0.258378$  &$b^{\omega_{32}^{\nu=1/4}}_1$& $=$ &  $-0.458126$ 
\\
$b^{\hat{A}_{32}^{0}}_2$& $=$ &  $0.679163$
&$b^{\omega_{32}^{\nu=1/4}}_2$& $=$ &  $0.0474616$
\\
& &
&$b^{\omega_{32}^{\nu=1/4}}_3$& $=$ &  $-0.486049$ 
\\
\hline
$c^{\hat{A}_{32}}_{11}$& $=$ &  $4.36263$
&$c^{\omega_{32}}_{11}$& $=$ &  $3.25319$
\\
$c^{\hat{A}_{32}}_{12}$& $=$ &  $-12.5897$ 
&$c^{\omega_{32}}_{12}$& $=$ &  $0.535555$ 
\\
$c^{\hat{A}_{32}}_{13}$& $=$ &  $-7.73233$ 
&$c^{\omega_{32}}_{13}$& $=$ &  $-8.07905$ 
\\
$c^{\hat{A}_{32}}_{14}$& $=$ &  $16.2082$ 
&$c^{\omega_{32}}_{21}$& $=$ &  $1.00066$ 
\\
$c^{\hat{A}_{32}}_{21}$& $=$ &  $3.04724$ 
&$c^{\omega_{32}}_{22}$& $=$ &  $-1.1333$ 
\\
$c^{\hat{A}_{32}}_{22}$& $=$ &  $46.5711$ 
&$c^{\omega_{32}}_{23}$& $=$ &  $0.601572$ 
\\
$c^{\hat{A}_{32}}_{23}$& $=$ &  $2.10475$
& & &
\\
$c^{\hat{A}_{32}}_{24}$& $=$ &  $56.9136$
& & &
\\
\end{tabular}
\end{ruledtabular}
\end{table}

\begin{table*}
\caption{\label{tab:Dt_32}
Explicit coefficients of $\Delta t_{32}$.}
\centering
\begin{ruledtabular}
\begin{tabular}{l c l  @{\hspace*{25mm}}  lcl  @{\hspace*{25mm}}  lcl}
$ \Delta t_{32}^{0}$& $=$ &  $9.16665$ &$c^{\Delta t_{32}^{\nu}}_{11}$& $=$ &  $-0.037634$ &$c^{\Delta t_{32}^{X_{12}}}_{11}$& $=$ &  $2.497188$ \\
$a^{\Delta t_{32}}_1$& $=$ &  $-11.3497$ &$c^{\Delta t_{32}^{\nu}}_{12}$& $=$ &  $12.456704$ &$c^{\Delta t_{32}^{X_{12}}}_{12}$& $=$ &  $-7.532596$ \\
$a^{\Delta t_{32}}_2$& $=$ &  $32.9144$ &$c^{\Delta t_{32}^{\nu}}_{13}$& $=$ &  $2.670868$ &$c^{\Delta t_{32}^{X_{12}}}_{13}$& $=$ &  $4.645986$ \\
$a^{\Delta t_{32}}_3$& $=$ &  $-8.36579$ &$c^{\Delta t_{32}^{\nu}}_{14}$& $=$ &  $-12.255859$ &$c^{\Delta t_{32}^{X_{12}}}_{14}$& $=$ &  $-3.652524$ \\
$a^{\Delta t_{32}}_4$& $=$ &  $20.1017$ &$c^{\Delta t_{32}^{\nu}}_{15}$& $=$ &  $37.843505$ &$c^{\Delta t_{32}^{X_{12}}}_{15}$& $=$ &  $3.398687$ \\
$b^{\Delta t_{32}^{0}}_1$& $=$ &  $-0.34161$ & $c^{\Delta t_{32}^{\nu}}_{21}$& $=$ &  $-25.058475$ &$c^{\Delta t_{32}^{X_{12}}}_{21}$& $=$ &  $7.054185$ \\
$b^{\Delta t_{32}^{0}}_2$& $=$ &  $-0.46107$ &$c^{\Delta t_{32}^{\nu}}_{22}$& $=$ &  $449.470722$ &$c^{\Delta t_{32}^{X_{12}}}_{22}$& $=$ &  $-12.260185$ \\
$b^{\Delta t_{32}^{0}}_3$& $=$ &  $0.34744$ &$c^{\Delta t_{32}^{\nu}}_{23}$& $=$ &  $-1413.508735$&$c^{\Delta t_{32}^{X_{12}}}_{23}$& $=$ &  $5.724802$ \\
$b^{\Delta t_{32}^{\nu=1/4}}_1$& $=$ &  $0.15477$ & $c^{\Delta t_{32}^{\nu}}_{24}$& $=$ &  $-11.852596$ &$c^{\Delta t_{32}^{X_{12}}}_{24}$& $=$ &  $-3.242611$ \\
$b^{\Delta t_{32}^{\nu=1/4}}_2$& $=$ &  $-0.755639$ &$c^{\Delta t_{32}^{\nu}}_{25}$& $=$ &  $41.348059$ &$c^{\Delta t_{32}^{X_{12}}}_{25}$& $=$ &  $2.714232$ \\
$b^{\Delta t_{32}^{\nu=1/4}}_3$& $=$ &  $0.21816$ &$c^{\Delta t_{32}^{\nu}}_{31}$& $=$ &  $-5.650710$&$c^{\Delta t_{32}^{X_{12}}}_{31}$& $=$ &  $2.614565$ \\
& & &$c^{\Delta t_{32}^{\nu}}_{32}$& $=$ &  $-9.567484$&$c^{\Delta t_{32}^{X_{12}}}_{32}$& $=$ &  $-9.507583$ \\
& & &$c^{\Delta t_{32}^{\nu}}_{33}$& $=$ &  $173.182999$&$c^{\Delta t_{32}^{X_{12}}}_{33}$& $=$ &  $7.321586$ \\
& & &$c^{\Delta t_{32}^{\nu}}_{34}$& $=$ &  $-10.938605$&$c^{\Delta t_{32}^{X_{12}}}_{34}$& $=$ &  $-3.937568$ \\
& & &$c^{\Delta t_{32}^{\nu}}_{35}$& $=$ &  $35.670656$&$c^{\Delta t_{32}^{X_{12}}}_{35}$& $=$ &  $4.584970$ \\
\end{tabular}
\end{ruledtabular}
\end{table*}

\subsubsection{$(\ell,m)=(4,4)$ multipole}
The peak amplitude of the $(4,4)$ mode is fitted with 
\be
\hat{A}_{44}^{\rm peak}= \hat{A}_{44}^{{\rm peak}_{0}}\left(1-3\nu\right)\hat{A}_{44}^{\rm orb}\left(\nu\right) \hat{A}_{44}^{\rm Spin}\left(\hat{S},\nu\right) \ ,
\ee
where $\hat{A}_{44}^{{\rm peak}_{0}}$ is the peak amplitude of the mode in the test-particle limit.
The factor $\hat{A}_{44}^{\rm orb}$ is informed by non-spinning waveforms and is fitted with the template
\be
\hat{A}_{44}^{\rm orb}(\nu)= \frac{1+a^{\hat{A}_{44}}_1 \nu+a^{\hat{A}_{44}}_2 \nu^2}{1+a^{\hat{A}_{44}}_3 \nu } \ .
\ee
The spin dependence is first captured for the test-particle limit with the function
\be
\hat{A}_{44}^{\rm Spin}(\hat{S},\nu=0) = \frac{1+b^{\hat{A}^{0}_{44}}_1 \hat{S}+b^{\hat{A}^{0}_{44}}_2 \hat{S}^2}{1+b^{\hat{A}^{0}_{44}}_3 \hat{S}}\ ,
\ee
and then extended in the comparable mass region of the parameter space through 
\be
b^{\hat{A}^{0}_{44}}_i\rightarrow \frac{b^{\hat{A}^{0}_{44}}_i+   c^{\hat{A}_{44}}_{i1}\nu +   c^{\hat{A}_{44}}_{i2}\nu^2 }{1+c^{\hat{A}_{44}}_{i3}\nu+   c^{\hat{A}_{44}}_{i4}\nu^2  }\ ,\ {\rm with} \ i=\left\{ 1,2,3\right\} .
\ee
The peak frequency $\omega_{44}$ is factorized as
\be
\omega^{\rm peak}_{44}=\omega^{{\rm peak}_{0}}_{44}\omega^{\rm orb}_{44}(\nu)\omega^{\rm Spin}_{44}\left(\hat{S},\nu\right) \ .
\ee
The orbital dependence is modeled through
\be
\omega^{\rm orb}_{44}(\nu)=\frac{ 1+a^{\omega_{44}}_1 \nu+a^{\omega_{44}}_2 \nu^2}{1+a^{\omega_{44}}_3 \nu+a^{\omega_{44}}_4 \nu^2}\ .
\ee
The spin dependence is fitted first for the test-particle limit as
\be
\omega^{\rm Spin}_{44}\left(\hat{S},\nu = 0\right) = \frac{1+b^{\omega^{0}_{44}}_1 \hat{S}+b^{\omega^{0}_{44}}_2 \hat{S}^2+b^{\omega^{0}_{44}}_3 \hat{S}^3}{1+b^{\omega^{0}_{33}}_4 \hat{S}}\ .
\ee
The spin dependence in the comparable mass region of the parameter space is modeled through 
\be
b^{\omega^{0}_{44}}_i\rightarrow \frac{b^{\omega^{0}_{44}}_i+   c^{\omega_{44}}_{i1}\nu +   c^{\omega_{44}}_{i2}\nu^2 }{1+c^{\omega_{44}}_{i3}\nu+   c^{\omega_{44}}_{i4}\nu^2  }\ ,
\ee
with $i=\left\{ 1,2,3,4\right\}$.

We fit $\Delta t_{44}$ in a factorized form as
\be
\Delta t_{44} = \Delta t_{44}^{0}\Delta t_{44}^{\rm orb}(\nu)\Delta t_{44}^{\rm spin}\left( \hat{S},X_{12} \right)\ .
\ee
The orbital behavior is fitted with
\be
\Delta t_{44}^{\rm orb}(\nu) =\frac{ 1 + a_1^{\Delta t_{44}}\nu +a_2^{\Delta t_{44}}\nu^2}{1 + a_3^{\Delta t_{44}}\nu +a_4^{\Delta t_{44}}\nu^2}\ ,
\ee
while the spinning one is first fitted to equal mass simulations as
\be
\Delta t_{44}^{\rm spin}\left( \hat{S},X_{12}=0 \right) = \frac{1+ b_1^{\Delta t_{44}^{\nu=1/4}}\hat{S}}{1+ b_2^{\Delta t_{44}^{\nu=1/4}}\hat{S}}\ .
\ee
The general $\nu$-dependence enters via the replacement
\be
b_i^{\Delta t_{44}^{\nu=1/4}} \rightarrow b_i^{\Delta t_{44}^{\nu=1/4}} + c_{i1}^{\Delta t_{44}}X_{12}+ c_{i2}^{\Delta t_{44}}X_{12}^2\ ,
\ee
with $i = \left\{1,2\right\}$.
The explicit values of the fit coefficients can be found in Table~\ref{tab:peak_44}.
\begin{table*}
\caption{\label{tab:peak_44}
Explicit coefficients of the fits of $\hat{A}_{44}^{\rm peak}$, $\omega_{44}^{\rm peak}$ and $\Delta t_{44}$.}
\centering
\begin{ruledtabular}
\begin{tabular}{l c l  @{\hspace*{25mm}}  l c l  @{\hspace*{25mm}}  l c l}
$\hat{A}_{44}^{{\rm peak}_{0}}$ & $=$ &  $0.276618$
&$\omega_{44}^{{\rm peak}_{0}}$ & $=$ &  $0.635659$
&$ \Delta t_{44}^{0}$& $=$ &  $5.27778$
\\
$a^{\hat{A}_{44}}_1$ & $=$ &  $-3.7082$
&$a^{\omega_{44}}_1$ & $=$ &  $-0.964614$
&$a^{\Delta t_{44}}_1$& $=$ &  $-8.35574$
\\
$a^{\hat{A}_{44}}_2$ & $=$ &  $0.280906$
&$a^{\omega_{44}}_2$ & $=$ &  $-11.1828$
&$a^{\Delta t_{44}}_2$& $=$ &  $17.5288$
\\
$a^{\hat{A}_{44}}_3$ & $=$ &  $-3.71276$
&$a^{\omega_{44}}_3$ & $=$ &  $-2.08471$
&$a^{\Delta t_{44}}_3$& $=$ &  $-6.50259$
\\
  & & 
&$a^{\omega_{44}}_4$ & $=$ &  $-6.89287$
&$a^{\Delta t_{44}}_4$& $=$ &  $10.1575$
\\
\hline
$b^{\hat{A}^{0}_{44}}_1$ & $=$ &  $-0.316647$
&$b^{\omega^{0}_{44}}_1$ & $=$ &  $-0.445192$
&$b^{\Delta t_{44}^{\nu=1/4}}_1$& $=$ &  $0.00159701$
\\
$b^{\hat{A}^{0}_{44}}_2$ & $=$ &  $-0.062423$
&$b^{\omega^{0}_{44}}_2$ & $=$ &  $-0.0985658$
&$b^{\Delta t_{44}^{\nu=1/4}}_2$& $=$ &  $-1.14134$
\\
$b^{\hat{A}^{0}_{44}}_3$ & $=$ &  $-0.852876$
&$b^{\omega^{0}_{44}}_3$ & $=$ &  $-0.0307812$
& & &
\\ 
  & &
&$b^{\omega^{0}_{44}}_4$ & $=$ &  $-0.801552$
& & &
\\
\hline
$c^{\hat{A}_{44}}_{11}$ & $=$ &  $1.2436$
&$c^{\omega_{44}}_{11}$ & $=$ &  $-0.92902$
&$c^{\Delta t_{4}}_{11}$& $=$ &  $-2.28656$
\\
$c^{\hat{A}_{44}}_{12}$ & $=$ &  $-1.60555$
&$c^{\omega_{44}}_{12}$ & $=$ &  $10.86310$
&$c^{\Delta t_{44}}_{12}$& $=$ &  $1.66532$
\\
$c^{\hat{A}_{44}}_{13}$ & $=$ &  $-4.05685$
&$c^{\omega_{44}}_{13}$ & $=$ &  $-4.44930$
&$c^{\Delta t_{44}}_{21}$& $=$ &  $-0.589331$
\\
$c^{\hat{A}_{44}}_{14}$ & $=$ &  $1.59143$
&$c^{\omega_{44}}_{14}$ & $=$ &  $3.01808$
&$c^{\Delta t_{44}}_{22}$& $=$ &  $0.708784$
\\
$c^{\hat{A}_{44}}_{21}$ & $=$ &  $0.837418$
&$c^{\omega_{44}}_{22}$ & $=$ &  $1.62523$
& & &
\\
$c^{\hat{A}_{44}}_{22}$ & $=$ &  $-2.93528$
&$c^{\omega_{44}}_{23}$ & $=$ &  $-7.70486$
& & &
\\
$c^{\hat{A}_{44}}_{23}$ & $=$ &  $-11.5591$
&$c^{\omega_{44}}_{23}$ & $=$ &  $15.06517$
& & & 
\\
$c^{\hat{A}_{44}}_{24}$ & $=$ &  $34.1863$
&$c^{\omega_{44}}_{41}$ & $=$ &  $0.93790$
& & &
\\
$c^{\hat{A}_{44}}_{31}$ & $=$ &  $0.950035$ 
&$c^{\omega_{44}}_{42}$ & $=$ &  $8.36038$
& & & 
\\
$c^{\hat{A}_{44}}_{32}$ & $=$ &  $7.95168$ 
&$c^{\omega_{44}}_{43}$ & $=$ &  $-4.85774$
& & & 
\\
$c^{\hat{A}_{44}}_{33}$ & $=$ &  $-1.26899$
&$c^{\omega_{44}}_{44}$ & $=$ &  $4.80446$
& & & 
\\
$c^{\hat{A}_{44}}_{34}$ & $=$ &  $-9.72147$
& & &  
& & & 
\\
\end{tabular}
\end{ruledtabular}
\end{table*}

\subsubsection{$(\ell,m)=(4,3)$ multipole}
\label{sec:43}
The peak amplitude of the $(4,3)$ mode is fitted with 
\be
\hat{A}_{43}^{\rm peak}= \hat{A}_{43}^{{\rm peak}_{0}}X_{12}\left(1-2\nu\right)\hat{A}_{43}^{\rm orb}\left(\nu\right) +  \hat{A}_{43}^{\rm Spin}\left(\tilde{a}_0,\nu\right) \ ,
\ee
where $\hat{A}_{43}^{{\rm peak}_{0}}$ is the peak amplitude of the mode in the test-particle limit.
The factor $\hat{A}_{43}^{\rm orb}$ is informed by non-spinning waveforms and is fitted with the template
\be
\hat{A}_{43}^{\rm orb}(\nu)= \frac{1+a^{\hat{A}_{43}}_1 \nu+a^{\hat{A}_{43}}_2 \nu^2}{1+a^{\hat{A}_{43}}_3 \nu } \ .
\ee
The spin dependence is first captured for the test-particle limit with the function
\be
\hat{A}_{43}^{\rm Spin}(\tilde{a}_0,\nu=0) = \frac{1+b^{\hat{A}^{0}_{43}}_1 \tilde{a}_0+b^{\hat{A}^{0}_{43}}_2 \tilde{a}_0^2}{1+b^{\hat{A}^{0}_{43}}_3 \tilde{a}_0}\ .
\ee
The spin dependence in the comparable mass region of the parameter space is modeled through 
\be
b^{\hat{A}^{0}_{43}}_i\rightarrow \frac{b^{\hat{A}^{0}_{43}}_i+   c^{\hat{A}_{43}}_{i1}\nu}{1 +   c^{\hat{A}_{43}}_{i2}\nu +   c^{\hat{A}_{43}}_{i3}\nu^2}\ ,
\ee
with $i=\left\{ 1,2,3\right\}$.
For the equal mass case however a special fit is made to
accurately capture the correct behavior, i.e.
\be
\hat{A}_{43}^{\rm peak}\left(\tilde{a}_{12},\nu=\frac{1}{4}\right) = \frac{b^{\hat{A}^{\nu=1/4}_{43}}_1 \tilde{a}_{12}+b^{\hat{A}^{\nu=1/4}_{43}}_2 \tilde{a}_{12}^2}{1+b^{\hat{A}^{\nu=1/4}_{43}}_3 \tilde{a}_{12}}\ .
\ee

The istantaneous frequency at peak $\omega_{43}^{\rm peak}$ is factorized as
\be
\omega^{\rm peak}_{43}=\omega^{{\rm peak}_{0}}_{43}\omega^{\rm orb}_{43}(\nu)\omega^{\rm Spin}_{43}\left(\hat{S},\nu\right),
\ee
where the orbital factor is modeled as
\be
\omega^{\rm orb}_{43}(\nu)=\frac{1+a^{\omega_{43}}_1 \nu+a^{\omega_{43}}_2 \nu^2}{1+a^{\omega_{43}}_3 \nu+a^{\omega_{43}}_4 \nu^2}\ .
\ee
The spin dependence is fitted first for the test-particle case
\be
\omega^{\rm Spin}_{43}\left(\hat{S},\nu = 0\right) = \frac{1+b^{\omega^{0}_{43}}_1 \hat{S}+b^{\omega^{0}_{43}}_2 \hat{S}^2}{1+b^{\omega^{0}_{43}}_3 \hat{S}}\ ,
\ee
and then extended to other regions of the parameter space with 
\be
b^{\omega^{0}_{43}}_i\rightarrow \frac{b^{\omega^{0}_{43}}_i+   c^{\omega_{43}}_{i1}\nu+   c^{\omega_{43}}_{i2}\nu^2}{1+   c^{\omega_{43}}_{i3}\nu} \ ,
\ee
where $i = \left\{1,2,3\right\}$.

For what concerns $\Delta t_{43}$, it is represented as
\be
\Delta t_{43} = \Delta t_{43}^{0}\Delta t_{43}^{\rm orb}(\nu)\Delta t_{43}^{\rm spin}\left( \hat{S},\nu \right)\ ,
\ee
with
\begin{align}
\Delta t_{43}^{\rm orb}(\nu) &= \frac{1 + a^{\Delta t_{43}}_1\nu + a^{\Delta t_{43}}_2\nu^2}{1 + a^{\Delta t_{43}}_3\nu + a^{\Delta t_{43}}_4\nu^2}\ ,\\
\Delta t_{43}^{\rm spin}\left( \hat{S},\nu=0 \right) &= \frac{1+ b^{\Delta t_{43}^{0}}_1\hat{S}+ b^{\Delta t_{43}^{0}}_2\hat{S}^2}{1+ b^{\Delta t_{43}^{0}}_3\hat{S}}\ .
\end{align}
We then incorporate the general $\nu$-dependence via the replacement
\be
b^{\Delta t_{43}^{0}}_i \rightarrow \frac{b^{\Delta t_{43}^{0}}_1 + c^{\Delta t_{43}}_{i1}\nu + c^{\Delta t_{43}}_{i2}\nu^2}{1+ c^{\Delta t_{43}}_{i3}\nu + c^{\Delta t_{43}}_{i4}\nu^2}\ ,
\ee
with $i=\left\{ 1,2,3\right\}$. The explicit values of the fit coefficients are listed in Table~\ref{tab:peak_43}.
\begin{table*}
\caption{\label{tab:peak_43}
Explicit coefficients of the fits of $\hat{A}_{43}^{\rm peak}$, $\omega_{43}^{\rm peak}$ and $\Delta t_{43}$.}
\centering
\begin{ruledtabular}
\begin{tabular}{l c l  @{\hspace*{28mm}}  l c l  @{\hspace*{28mm}}  l c l}
$\hat{A}_{43}^{{\rm peak}_{0}}$& $=$ &  $0.0941570$ 
&$\omega_{43}^{{\rm peak}_{0}}$& $=$ &  $0.636130$
&$ \Delta t_{43}^{0}$& $=$ &  $9.53705$
\\
$a^{\hat{A}_{43}}_1$& $=$ &  $-5.74386$
&$a^{\omega_{43}}_1$& $=$ &  $-9.02463$
&$a^{\Delta t_{43}}_1$& $=$ &  $-11.2377$ 
\\
$a^{\hat{A}_{43}}_2$& $=$ &  $12.6016$
&$a^{\omega_{43}}_2$& $=$ &  $21.9802$
&$a^{\Delta t_{43}}_2$& $=$ &  $38.3177$ 
\\
$a^{\hat{A}_{43}}_3$& $=$ &  $-3.27435$
&$a^{\omega_{43}}_3$& $=$ &  $-8.75892$
&$a^{\Delta t_{43}}_3$& $=$ &  $-7.29734$ 
\\
$a^{\omega_{43}}_4$& $=$ &  $20.5624$
& & & 
&$a^{\Delta t_{43}}_4$& $=$ &  $21.4267$ 
\\
\hline
$b^{\hat{A}_{43}^{0}}_1$& $=$ &  $-0.02132252$  
&$b^{\omega_{43}^{0}}_1$& $=$ &  $-0.973324$ 
&$b^{\Delta t_{43}^{0}}_1$& $=$ &  $-1.371832$ 
\\
$b^{\hat{A}_{43}^{0}}_2$& $=$ &  $0.02592749$ 
&$b^{\omega_{43}^{0}}_2$& $=$ &  $-0.109921$
&$b^{\Delta t_{43}^{0}}_2$& $=$ &  $0.362375$ 
\\
$b^{\hat{A}_{43}^{0}}_3$& $=$ &  $-0.826977$
&$b^{\omega_{43}^{0}}_3$& $=$ &  $-1.08036$ 
&$b^{\Delta t_{43}^{0}}_3$& $=$ &  $-1.0808402$ 
\\
$b^{\hat{A}_{43}^{\nu=1/4}}_1$& $=$ &  $-0.00471163$  
& & & 
& & & 
\\
$b^{\hat{A}_{43}^{\nu=1/4}}_2$& $=$ &  $0.0291409$ 
& & & 
& & & 
\\
$b^{\hat{A}_{43}^{\nu=1/4}}_3$& $=$ &  $-0.351031$ 
& & & 
& & & 
\\
\hline
$c^{\hat{A}_{43}}_{11}$& $=$ &  $0.249099$ 
&$c^{\omega_{43}}_{11}$& $=$ &  $11.5224$ 
&$c^{\Delta t_{43}}_{11}$& $=$ &  $3.215984$
\\
$c^{\hat{A}_{43}}_{12}$& $=$ &  $-7.345984$ 
&$c^{\omega_{43}}_{12}$& $=$ &  $-26.8421$
&$c^{\Delta t_{43}}_{12}$& $=$ &  $42.133767$
\\
$c^{\hat{A}_{43}}_{13}$& $=$ &  $108.923746$
&$c^{\omega_{43}}_{13}$& $=$ &  $-2.84285$
&$c^{\Delta t_{43}}_{13}$& $=$ &  $-9.440398$  
\\
$c^{\hat{A}_{43}}_{21}$& $=$ &  $-0.104206$
&$c^{\omega_{43}}_{21}$& $=$ &  $3.51943$ 
&$c^{\Delta t_{43}}_{14}$& $=$ &  $35.160776$
\\
$c^{\hat{A}_{43}}_{22}$& $=$ &  $7.073534$
&$c^{\omega_{43}}_{22}$& $=$ &  $-12.1688$ 
&$c^{\Delta t_{43}}_{21}$& $=$ &  $1.133942$ 
\\
$c^{\hat{A}_{43}}_{23}$& $=$ &  $-44.374738$
&$c^{\omega_{43}}_{23}$& $=$ &  $-3.96385$ 
&$c^{\Delta t_{43}}_{22}$& $=$ &  $-10.356311$ 
\\
$c^{\hat{A}_{43}}_{31}$& $=$ &  $3.545134$ 
&$c^{\omega_{43}}_{31}$& $=$ &  $5.53433$ 
&$c^{\Delta t_{43}}_{23}$& $=$ &  $-6.701429$
\\
$c^{\hat{A}_{43}}_{32}$& $=$ &  $1.341375$ 
&$c^{\omega_{43}}_{32}$& $=$ &  $3.73988$
&$c^{\Delta t_{43}}_{24}$& $=$ &  $10.726960$ 
\\
$c^{\hat{A}_{43}}_{33}$& $=$ &  $-19.552083$
&$c^{\omega_{43}}_{33}$& $=$ &  $4.219$ 
&$c^{\Delta t_{43}}_{31}$& $=$ &  $-6.036207$ 
\\
& & & 
& & & 
$c^{\Delta t_{43}}_{32}$& $=$ &  $67.730599$
\\
& & & 
& & & 
$c^{\Delta t_{43}}_{33}$& $=$ &  $-3.082275$
\\
& & & 
& & & 
$c^{\Delta t_{43}}_{34}$& $=$ &  $11.547917$
\\
\end{tabular}
 \end{ruledtabular}
\end{table*}
\subsubsection{$(\ell,m)=(4,2)$ multipole}
\label{sec:42}
The peak amplitude of the $(4,2)$ mode is fitted with a factorized template of the form
\be
\hat{A}_{42}^{\rm peak}= \hat{A}_{42}^{{\rm peak}_{0}}\left(1-3\nu\right)\hat{A}_{42}^{\rm orb}\left(\nu\right) \hat{A}_{42}^{\rm Spin}\left(\hat{S},\nu\right) \ ,
\ee
where $\hat{A}_{42}^{{\rm peak}_{0}}$ is the peak amplitude of the mode in the test-particle limit.
The factor $\hat{A}_{42}^{\rm orb}$ is informed by non-spinning waveforms and is fitted with the template
\be
\hat{A}_{42}^{\rm orb}(\nu)= 1+a^{\hat{A}_{42}}_1 \nu+a^{\hat{A}_{42}}_2 \nu^2\ .
\ee
The spin dependence is first captured for the test-particle limit with the function
\be
\hat{A}_{42}^{\rm Spin}(\hat{S},\nu=0) = \frac{1+b^{\hat{A}^{0}_{42}}_1 \hat{S}+b^{\hat{A}^{0}_{42}}_2
  \hat{S}^2}{1+b^{\hat{A}^{0}_{42}}_3 \hat{S}+b^{\hat{A}^{0}_{42}}_4 \hat{S}^2}\ .
\ee
The general $\nu$-dependence is then taken into account via the replacement
\be
b^{\hat{A}^{0}_{42}}_i\rightarrow \frac{b^{\hat{A}^{0}_{42}}_i+   c^{\hat{A}_{42}}_{i1}\nu}{1 +   c^{\hat{A}_{42}}_{i2}\nu }\ ,
\ee
with $i=\left\{ 1,2,3,4\right\}$.

The instantaneous frequency $\omega_{42}^{\rm peak}$ is factorized as
\be
\omega^{\rm peak}_{42}=\omega^{{\rm peak}_{0}}_{42}\omega^{\rm orb}_{42}(\nu)\omega^{\rm Spin}_{42}\left(\hat{S},\nu\right)
\ee
The orbital dependence is modeled through
\be
\omega^{\rm orb}_{42}(\nu)=\frac{1+a^{\omega_{42}}_1 \nu+a^{\omega_{42}}_2 \nu^2}{1+a^{\omega_{42}}_3 \nu+a^{\omega_{42}}_4 \nu^2}\ .
\ee
The spin dependence is fitted first for the test-mass case with
\be
\omega^{\rm Spin}_{42}\left(\hat{S},\nu = 0\right) = \frac{1+b^{\omega^{0}_{42}}_1 \hat{S}+b^{\omega^{0}_{42}}_2 \hat{S}^2}{1+b^{\omega^{0}_{42}}_3 \hat{S}+b^{\omega^{0}_{42}}_4 \hat{S}^2}\ ,
\ee
and then the general $\nu$-dependence is taken into account via the replacement
\be
b^{\omega^{0}_{42}}_i\rightarrow \frac{b^{\omega^{0}_{42}}_i+   c^{\omega_{42}}_{i1}\nu}{1+   c^{\omega_{42}}_{i2}\nu+   c^{\omega_{42}}_{i3}\nu^2} \ ,
\ee
with $i = \left\{1,2,3,4\right\}$.
The delay $\Delta t_{42}$ is fitted as
\be
\Delta t_{42} = \Delta t_{42}^{0}\Delta t_{42}^{\rm orb}(\nu)\Delta t_{42}^{\rm spin}\left( \hat{S},\nu \right)\ ,
\ee
where 
\begin{align}
\Delta t_{42}^{\rm orb}(\nu) &= \frac{1 + a^{\Delta t_{42}}_1\nu + a^{\Delta t_{42}}_2\nu^2}{1 + a^{\Delta t_{42}}_3\nu + a^{\Delta t_{42}}_4\nu^2}\ , \\
\Delta t_{42}^{\rm spin}\left( \hat{S},\nu=0 \right) &= \frac{1+ b^{\Delta t_{42}^{0}}_1\hat{S}}{1+ b^{\Delta t_{42}^{0}}_2\hat{S}}\ .
\end{align}
For $\nu<6/25$ the spin factor is approximated by the test-particle fit.
For the other regions, it is extrapolated using
\be
b^{\Delta t_{42}^{0}}_i \rightarrow \frac{b^{\Delta t_{42}^{0}}_1 + c^{\Delta t_{42}}_{i1}\nu}{1+ c^{\Delta t_{42}}_{i2}\nu}\ ,
\ee
with $i=\left\{ 1,2\right\}$. The explicit values of the coefficients of the fits
are listed in Table~\ref{tab:peak_42}.
\begin{table*}
\caption{\label{tab:peak_42}
Explicit coefficients of the fits of $\hat{A}_{42}^{\rm peak}$, $\omega_{42}^{\rm peak}$ and $\Delta t_{42}$.}
\centering
\begin{ruledtabular}
\begin{tabular}{l c l  @{\hspace*{25mm}}  l c l  @{\hspace*{25mm}}  l c l}
$\hat{A}_{42}^{{\rm peak}_{0}}$& $=$ &  $0.0314364$ 
&$\omega_{42}^{{\rm peak}_{0}}$& $=$ &  $0.617533$
&$ \Delta t_{42}^{0}$& $=$ &  $11.66665$
\\
$a^{\hat{A}_{42}}_1$& $=$ &  $-4.56243$ 
&$a^{\omega_{42}}_1$& $=$ &  $-7.44121$ 
&$a^{\Delta t_{42}}_1$& $=$ &  $-9.844617$ 
\\
$a^{\hat{A}_{42}}_2$& $=$ &  $6.4522$ 
&$a^{\omega_{42}}_2$& $=$ &  $14.233$ 
&$a^{\Delta t_{42}}_2$& $=$ &  $23.32294$ 
\\
& & 
&$a^{\omega_{42}}_3$& $=$ &  $-6.61754$ 
&$a^{\Delta t_{42}}_3$& $=$ &  $-5.760481$ 
\\
& &
&$a^{\omega_{42}}_4$& $=$ &  $11.4329$ 
&$a^{\Delta t_{42}}_4$& $=$ &  $7.121793$ 
\\
\hline
$b^{\hat{A}_{42}^{0}}_1$& $=$ &  $-1.63682$ 
&$b^{\omega_{42}^{0}}_1$& $=$ &  $-2.37589$ 
&$b^{\Delta t_{42}^{0}}_1$& $=$ &  $-1.3002045$ 
\\
$b^{\hat{A}_{42}^{0}}_2$& $=$ &  $0.854459$
&$b^{\omega_{42}^{0}}_2$& $=$ &  $1.97249$
&$b^{\Delta t_{42}^{0}}_2$& $=$ &  $-0.9494348$ 
\\
$b^{\hat{A}_{42}^{0}}_3$& $=$ &  $0.120537$
&$b^{\omega_{42}^{0}}_3$& $=$ &  $-2.36107$
& & &
\\
$b^{\hat{A}_{42}^{0}}_4$& $=$ &  $-0.399718$
&$b^{\omega_{42}^{0}}_4$& $=$ &  $2.16383$
& & &
\\
\hline
$c^{\hat{A}_{42}}_{11}$& $=$ &  $6.53943$ 
&$c^{\omega_{42}}_{11}$& $=$ &  $10.1045$ 
&$c^{\Delta t_{42}}_{11}$& $=$ &  $24.604717$ 
\\
$c^{\hat{A}_{42}}_{12}$& $=$ &  $-4.00073$
&$c^{\omega_{42}}_{12}$& $=$ &  $-6.94127$ 
&$c^{\Delta t_{42}}_{12}$& $=$ &  $-0.808279$ 
\\
$c^{\hat{A}_{42}}_{21}$& $=$ &  $-0.638688$ 
&$c^{\omega_{42}}_{13}$& $=$ &  $12.1857$ 
&$c^{\Delta t_{42}}_{21}$& $=$ &  $62.471781$ 
\\
$c^{\hat{A}_{42}}_{22}$& $=$ &  $-3.94066$ 
&$c^{\omega_{42}}_{21}$& $=$ &  $-1.62866$ 
&$c^{\Delta t_{42}}_{22}$& $=$ &  $48.340961$ 
\\
$c^{\hat{A}_{42}}_{31}$& $=$ &  $-0.482148$
&$c^{\omega_{42}}_{22}$& $=$ &  $-2.6756$ 
& & &
\\
$c^{\hat{A}_{42}}_{32}$& $=$ &  $7.668\times 10^{-9} -4$
&$c^{\omega_{42}}_{23}$& $=$ &  $-4.7536$ 
& & &
\\
$c^{\hat{A}_{42}}_{41}$& $=$ &  $1.25617$
&$c^{\omega_{42}}_{31}$& $=$ &  $10.071$ 
& & &
\\
$c^{\hat{A}_{42}}_{42}$& $=$ &  $-4.04848$
&$c^{\omega_{42}}_{32}$& $=$ &  $-6.7299$ 
& & &
\\
& &
&$c^{\omega_{43}}_{33}$& $=$ &  $12.0377$  
& & &
\\
& &
&$c^{\omega_{42}}_{41}$& $=$ &  $-8.56139$ 
& & &
\\
& &
&$c^{\omega_{42}}_{42}$& $=$ &  $-5.27136$ 
& & &
\\
& &
&$c^{\omega_{43}}_{43}$& $=$ &  $5.10653$ 
& & &
\\
\end{tabular}
\end{ruledtabular}
\end{table*}

\subsubsection{$(\ell,m)=(5,5)$ multipole}
For this multipole, the peak amplitude is written as the sum of two terms as
\be
\hat{A}_{55}^{\rm peak} = \hat{A}_{55}^{{\rm peak}_{0}} X_{12}\left(1-2\nu \right) \hat{A}_{55}^{\rm orb}(\nu) + \hat{A}_{55}^{\rm Spin}(\tilde{a}_{12},\nu) \ ,
\ee
where $\hat{A}_{55}^{{\rm peak}_{0}} $ is the peak amplitude in the test particle limit. 
The non-spinning $\nu$-dependence is modeled as
\be
\hat{A}_{55}^{\rm orb}(\nu)= 1+a^{\hat{A}_{55}}_1 \nu+a^{\hat{A}_{55}}_2 \nu^2\ .
\ee
The spin dependence is first fitted to the test-particle limit using
\be
\hat{A}_{55}^{\rm Spin}(\tilde{a}_{12},\nu=0) = \frac{b^{\hat{A}^{0}_{55}}_1 \tilde{a}_{12}}{1+b^{\hat{A}^{0}_{55}}_2 \tilde{a}_{12}}\ ,
\ee
and then extrapolated to the comparable mass region through
\bea
b^{\hat{A}^{0}_{55}}_1 &\rightarrow &\frac{b^{\hat{A}^{0}_{55}}_1 }{1+ c^{\hat{A}_{55}}_{11}\nu + c^{\hat{A}_{55}}_{12}\nu^2  }\ ,\\
b^{\hat{A}^{0}_{55}}_2 &\rightarrow &\frac{b^{\hat{A}^{0}_{55}}_2 }{1+ c^{\hat{A}_{55}}_{21}\nu + c^{\hat{A}_{55}}_{22}\nu^2  }\ .
\eea
The frequency of the $(5,5)$ mode is factorized as
\be
\omega^{\rm peak}_{55}=\omega^{{\rm peak}_{0}}_{55}\omega^{\rm orb}_{55}(\nu)\omega^{\rm Spin}_{55}\left(\hat{S},\nu\right)\ ,
\ee
where
\be
\omega^{\rm orb}_{55}(\nu)= \frac{1+a^{\omega_{55}}_1 \nu+a^{\omega_{55}}_2 \nu^2}{1+a^{\omega_{55}}_3 \nu}\ ,
\ee
and the test-particle spin factor is given by
\be
\omega^{\rm Spin}_{55}\left(\hat{S},\nu = 0\right) = \frac{1+b^{\omega^{0}_{55}}_1 \hat{S}}{1+b^{\omega^{0}_{55}}_2 \hat{S}}\ .
\ee
The spin dependence in the general case is obtained by means of
\be
b^{\omega^{0}_{55}}_i\rightarrow \frac{b^{\omega^{0}_{55}}_i+   c^{\omega_{55}}_{i1}\nu}{1+c^{\omega_{55}}_{i2}\nu}\ ,
\ee
with $i=\left\{ 1,2\right\}$. Note that, in this case, we do not incorporate spin-dependence
in $\Delta_{55}$, but only rely on the nonspinning fit of Ref.~\cite{Nagar:2019wds}.
\begin{table*}
\caption{\label{tab:peak_55}
Explicit coefficients of the fits of $\hat{A}_{55}^{\rm peak}$ and $\omega_{55}^{\rm peak}$}
\centering
\begin{ruledtabular}
\begin{tabular}{@{\hspace*{25mm}} l c l  @{\hspace*{25mm}}  l c l  @{\hspace*{25mm}}  }
$\hat{A}_{55}^{{\rm peak}_{0}}$ & $=$ &  $0.00522697$
&$\omega_{55}^{{\rm peak}_{0}}$ & $=$ &  $0.818117$
\\
$a^{\hat{A}_{55}}_1$ & $=$ &  $-0.29628$
&$a^{\omega_{55}}_1$ & $=$ &  $-2.8918$
\\
$a^{\hat{A}_{55}}_2$ & $=$ &  $6.4207$
&$a^{\omega_{55}}_2$ & $=$ &  $-3.2012$
\\
&  &  
& $a^{\omega_{55}}_3$ & $=$ &  $-3.773$
\\
\hline
$b^{\hat{A}^{0}_{55}}_1$ & $=$ &  $0.04360530$
&$b^{\omega^{0}_{55}}_1$ & $=$ &  $-0.332703$
\\
$b^{\hat{A}^{0}_{55}}_2$ & $=$ &  $-0.5769451$
&$b^{\omega^{0}_{55}}_2$ & $=$ &  $-0.675738$ 
\\
\hline
$c^{\hat{A}_{55}}_{11}$ & $=$ &  $5.720690$
&$c^{\omega_{55}}_{11}$ & $=$ &  $1.487294$
\\
$c^{\hat{A}_{55}}_{12}$ & $=$ &  $44.868515$ 
&$c^{\omega_{55}}_{12}$ & $=$ &  $-2.058537$
\\
$c^{\hat{A}_{55}}_{21}$ & $=$ &  $12.777090$
&$c^{\omega_{55}}_{21}$ & $=$ &  $1.454248$
\\
$c^{\hat{A}_{55}}_{22}$ & $=$ &  $-42.548247$
&$c^{\omega_{55}}_{22}$ & $=$ &  $-1.301284$
\end{tabular}
\end{ruledtabular}
\end{table*}

\subsection{NR-fitting of the postpeak parameters}
\label{sec:fit_evol}
In this Appendix we report the fits of the postpeak parameters
$({c_3^{A_{\ell m}}},{c_3^{\phi_{\ell m}}},{c_4^{\phi_{\ell m}}})$ for
all multipoles multipoles discussed in the main text.
For $(2,2)$, $(3,3)$, $(4,4)$, $(5,5)$ we present fits that
explicitly depend on the spins of the black holes. By contrast,
the same parameters for the other multipoles $(2,1)$, $(3,2)$, $(3,1)$
$(4,3)$, $(4,2)$, are approximated by the spin-independent
fits of Ref.~\cite{Nagar:2019wds}.
Let us note, however, that we prefer to {\it not use} the full
spin-dependent fits of $(c_3^{\phi_{33}},c_4^{\phi_{33}})$
and in $(c_3^{\phi_{44}},c_4^{\phi_{44}})$. Instead the fits of 
Ref.~\cite{Nagar:2019wds} are used to get a
more robust behavior of $\omega_{33}$ and $\omega_{44}$
in all corners of the parameter space, notably when the mass
ratio is between one and two and the spins are large. 
See Appendix~\ref{sec:pm_phase_spin} for a brief discussion.

\subsubsection{The $(\ell,m)=(2,2)$ postpeak}
The data of $({c_3^{A_{22}}},{c_3^{\phi_{22}}},{c_4^{\phi_{22}}})$ were extracted from NR
fitting the NR waveforms in the calibration set over an interval starting at the peak of length
$4\tau^{22}_1$.

The fits are done in three steps, based on the model
\begin{align}
\label{eq:evol_spin}
Y(\nu;\,\hat{S}) = b^Y_0(\nu)+ & b^Y_1\left(X_{12}\right) \hat{S}
  + b^Y_2\left(X_{12}\right) \hat{S}^2 \nonumber\\
& + b^Y_3\left(X_{12}\right) \hat{S}^3 + b^Y_4\left(X_{12}\right) \hat{S}^4.
\end{align}
In the first step $Y(\nu;\,\hat{S}=0$) is fitted to the non-spinning data.
In the second step $b^Y_i\left(X_{12}=0\right)$ are fitted to the equal mass
data. In the third and final step the fits are extrapolated to the comparable mass case
imposing the 1-D fits informed in the previous two steps.
The coefficients of the fit are listed in Table~\ref{tab:rng}.

\begin{table*}
  \caption{\label{tab:rng} The fitted coefficients of $({c_3^{A_{22}}},{c_3^{\phi_{22}}},{c_4^{\phi_{22}}})$
  as defined in Eq.~\eqref{eq:evol_spin}.}
\begin{ruledtabular}
\begin{tabular}{c l c l l | l c l l | l c l l  c}
& \multicolumn{4}{c|}{$Y={c_3^{A_{22}}}$} & \multicolumn{4}{c|}{${Y=c_3^{\phi_{22}}}$} 	& \multicolumn{4}{c}{${Y=c_4^{\phi_{22}}}$} &\\\hline
&  $b^{c_3^A}_0(\nu)$ 	 & $=$ & $-0.5585$ & $0.81196\nu$ 				&  $b^{c_3^\phi}_0(\nu)$ 	 	& $=$ & $\;\;\;3.8436$ 	& $+0.71565\nu$ &  $b^{c_4^\phi}_0(\nu)$ 	 & $=$	& $1.4736$ & $2.2337\nu$ &\\ 
&  $b^{c_3^A}_1(X_{12})$ & $=$ & $-0.398576$ & $+0.1659421X_{12}$ 	&  $b^{c_3^\phi}_1(X_{12})$ 	& $=$& $\;\;\;5.12794$ 	& $-1.323643X_{12}$ &  $b^{c_4^\phi}_1(X_{12})$ & $=$& $8.26539$ & $+0.779683X_{12}$ &\\
&  $b^{c_3^A}_2(X_{12})$ & $=$ & $\;\;\;0.099805$ & $-0.2560047X_{12}$ &  $b^{c_3^\phi}_2(X_{12})$ 	& $=$& $\;\;\;9.9136$ 	& $-3.555007X_{12}$ &  $b^{c_4^\phi}_2(X_{12})$ & $=$& $14.2053$ & $-0.069638X_{12}$ &\\
&  $b^{c_3^A}_3(X_{12})$ & $=$ & $\;\;\;0.72125$ & $-0.9418946X_{12}$ 	&  $b^{c_3^\phi}_3(X_{12})$ 	& $=$& $-4.1075$ 		& $+7.011267X_{12}$ &  $b^{c_4^\phi}_3(X_{12})$ & $=$&\multicolumn{2}{c}{$0$} &\\
&  $b^{c_3^A}_4(X_{12})$ & $=$ & \multicolumn{2}{c|}{$0$} 				&  $b^{c_3^\phi}_4(X_{12})$ 	& $=$& $-31.5562$ 	& $+32.737824X_{12}$ 	&  $b^{c_4^\phi}_4(X_{12})$ & $=$ &\multicolumn{2}{c}{$0$} &\\
\end{tabular}
\end{ruledtabular}
\end{table*}

\subsubsection{The $(\ell,m)=(3,3)$ postpeak}
The data of $({c_3^{A_{33}}},{c_3^{\phi_{33}}},{c_4^{\phi_{33}}})$ were extracted from NR
fitting the NR waveforms in the calibration set over an interval starting at the peak of length
$1\tau^{33}_1$. 
The interpolation is modeled with the template
\begin{align}
\label{eq:evol_spin33}
Y(\nu;\,\hat{S}) = b^Y_0(\nu)+ & b^Y_1\left(X_{12}\right) \hat{S}.
\end{align}
While for the case of ${c_3^{A_{33}}}$ the fit is done versus $\tilde{a}_{12}$.
The fits are done in two hierarchical steps.
(i) $b^Y_0(\nu)$ is fitted to the non-spinning data.
(ii)  $b^Y_1\left(X_{12}\right)$ is fitted with a quadratic polynomial, 
while imposing the fit of $b^Y_0(\nu)$. The fits are given explicitly in Table~\ref{tab:rng33}.
\begin{table*}
  \caption{\label{tab:rng33} The explicit fits of $({c_3^{A_{33}}},{c_3^{\phi_{33}}},{c_4^{\phi_{33}}})$.
		The reader should note that the fits of $({c_3^{\phi_{33}}},{c_4^{\phi_{33}}})$ are not used for any of the results given in the main text. Instead the corresponding fits of Ref.~\cite{Nagar:2019wds} are used. 
		See Appendix~\ref{sec:pm_phase_spin} for a brief discussion.}
\begin{ruledtabular}
\begin{tabular}{ l l c l l  c l l l l l }
&${c_3^{A_{33}}}\left(\nu,X_{12},\tilde{a}_{12}\right) $	&  $=$	&  $-0.5585$ 	& $+0.81196\nu$ 	& $+$	&	$(-0.3502608$ 	&   	 	 $+1.587606X_{12}$ 	& $-1.555325X_{12}^2)$  & $\tilde{a}_{12}$ &\\ 
&${c_3^{\phi_{33}}}\left(\nu,X_{12},\hat{S}\right)$ 		&  $=$	&  $3.0611$ 	& $-6.1597\nu$ 	& $+$	&	$(-0.634377$ 	&   	 	 $+5.983525X_{12}$ 	& $-5.8819X_{12}^2)$  &$\hat{S}$ &\\ 
&${c_4^{\phi_{33}}}\left(\nu,X_{12},\hat{S}\right)$ 		&  $=$	&  $1.789$ 		& $ -5.6684\nu$ 	& $+$	&	$(-3.877528$ 	&   	 	 $+12.0433X_{12}$ 	& $-6.524665X_{12}^2)$  & $\hat{S}$ &\\ 
\end{tabular}
\end{ruledtabular}
\end{table*}
\subsubsection{The $(\ell,m)=(4,4)$ postpeak}
The data of $({c_3^{A_{44}}},{c_3^{\phi_{44}}},{c_4^{\phi_{44}}})$ were extracted from NR
fitting the NR waveforms in the calibration set over an interval starting at the peak of length
$1\tau^{44}_1$. 
The interpolation of  $({c_3^{\phi_{44}}},{c_4^{\phi_{44}}})$ is modeled with the template
\begin{align}
\label{eq:evol_spin44}
Y(\nu;\,\hat{S}) = b^Y_0(\nu)+  b^Y_1\left(X_{12}\right) \hat{S}+  b^Y_2\left(X_{12}\right) \hat{S}^2
\end{align}
in three steps, similar to the the $(2,2)$ mode.
(i) $b^Y_0(\nu)$ is fitted to the non-spinning data.
(ii)  $b^Y_i\left(X_{12}=0\right)$ is fitted to the equal mass data.
(iii) The full dependence of $b^Y_i\left(X_{12}\right)$ on $X_{12}$ is fitted while imposing the one-dimensional 
fits informed in the first two steps.
${c_3^A}_{44}$ is modeled with the template 
\begin{align}
\label{eq:evol_spin44A}
{c_3^{A_{44}}}(\nu;\,\hat{S}) = b^{{c_3^{A_{44}}}}_0(\nu)+  b^{{c_3^{A_{44}}}}_1\nu\hat{S}+  b^{{c_3^{A_{44}}}}_2\nu\hat{S}^2.
\end{align}
The fit is is done in two steps.
(i) $b^{{c_3^{A_{44}}}}_0(\nu)$ is fitted to the non-spinning data.
(ii) The coefficients $b^{{c_3^{A_{44}}}}_i$ are informed using the spinning data, while imposing the non-spinning fit.
The fits are given explicitly in Table~\ref{tab:rng44}.
\begin{table*}
  \caption{\label{tab:rng44} The explicit fits of $({c_3^{A_{44}}},{c_3^{\phi_{44}}},{c_4^{\phi_{44}}})$.
		The reader should note that the fits of $({c_3^{\phi_{44}}},{c_4^{\phi_{44}}})$ are not used for any of the results given in the main text. Instead the corresponding fits of Ref.~\cite{Nagar:2019wds} are used. 
		See Appendix~\ref{sec:pm_phase_spin} for a brief discussion.}
\begin{ruledtabular}
\begin{tabular}{ l l c l l  c l ll c l l ll }
&${c_3^{A_{44}}}\left(\nu,\hat{S}\right) $	&  $=$	&  $ -0.41591 $ 	& $+3.2099\nu$ 	& $-$	&	 	&   	 	 $9.614738\nu$ 	& $\hat{S}$  & $+$ & & $122.461125\nu$& $\hat{S}^2$ &\\ 
&${c_3^{\phi_{44}}}\left(\nu,X_{12},\hat{S}\right)$ 		&  $=$	&   \multicolumn{2}{c}{$\frac{3.6662-30.072\nu +76.371\nu^2}{1-3.5522\nu}$} 	& $+$	&	$(-4.9184$	&   	 	 $+7.911653X_{12})$ 	& $\hat{S}$  & $+$ & $(-15.6772$ & $+21.181688X_{12})$& $\hat{S}^2$ &\\ 
&${c_4^{\phi_{44}}}\left(\nu,X_{12},\hat{S}\right)$ 		&  $=$	&  $0.21595$ 		& $ +23.216\nu$ 	& $+$	&	$(-3.4207$ 	&   	 	 $+11.746452X_{12})$ 	& $\hat{S}$  & $+$ & $(-15.5383$ & $+34.922883X_{12})$& $\hat{S}^2$ &\\ 
\end{tabular}
\end{ruledtabular}
\end{table*}

\subsubsection{The $(\ell,m)=(5,5)$ postpeak}
The data of $({c_3^{A_{55}}},{c_3^{\phi_{55}}},{c_4^{\phi_{55}}})$ were extracted from NR
fitting the NR waveforms in the calibration set over an interval starting at the peak of length
$1\tau^{55}_1$. 
The interpolation is modeled with the template
\begin{align}
\label{eq:evol_spin55}
Y(\nu;\,\hat{S}) = b^Y_0(\nu)+  b^Y_1\left(X_{12}\right) \hat{S} + b^Y_2\left(X_{12}\right) \hat{S}^2.
\end{align}
While for the case of ${c_3^{A_{55}}}$ the fit is done versus $\tilde{a}_{12}$.
The fits are done in two hierarchical steps.
(i) $b^Y_0(\nu)$ is fitted to the non-spinning data.
(ii)  $b^Y_i\left(X_{12}\right)$ are fitted with a linear polynomial, 
while imposing the fit of $b^Y_0(\nu)$. The fits are given explicitly in Table~\ref{tab:rng55}.
\begin{table*}
  \caption{\label{tab:rng55} The explicit fits of $({c_3^{A_{55}}},{c_3^{\phi_{55}}},{c_4^{\phi_{55}}})$.}
\begin{ruledtabular}
\begin{tabular}{ l l c l l  c l l l  c ll l l}
&${c_3^{A_{55}}}\left(\nu,X_{12},\tilde{a}_{12}\right) $	&  $=$	&  $-7.063079$ 	& $+65.464944\nu$ 	& $+$	&	$(-2.055335$ 	&   	 	 $-0.585373X_{12})$  & $\tilde{a}_{12}$ &$+$	&	$(-12.631409$ 	&   	 	 $+19.271346X_{12})$  & $\tilde{a}_{12}^2$ &\\ 
&${c_3^{\phi_{55}}}\left(\nu,X_{12},\hat{S}\right)$ 		&  $=$	&  $-1.510167$ 	& $+30.569461\nu$ 	& $+$	&	$(-2.687133$ 	&   	 	 $+4.873750X_{12})$  & $\hat{S}$ &$+$	&	$( -14.629684$ 	&   	 	 $+19.696954X_{12})$  & $\hat{S}^2$ &\\ 
&${c_4^{\phi_{55}}}\left(\nu,X_{12},\hat{S}\right)$ 		&  $=$	&  $-1.383721$ 	& $+56.871881\nu$ 	& $+$	&	$(+7.198729$ 	&   	 	 $-3.870998X_{12})$  & $\hat{S}$ &$+$	&	$( -25.992190$ 	&   	 	 $+36.882645X_{12})$  & $\hat{S}^2$ &\\ 
\end{tabular}
\end{ruledtabular}
\end{table*}

\subsection{Motivating the choices for the $(3,3)$ and $(4,4)$ postmerger phases}
\label{sec:pm_phase_spin}
\begin{figure*}[t]
  \begin{center}
  \includegraphics[width=0.45\textwidth]{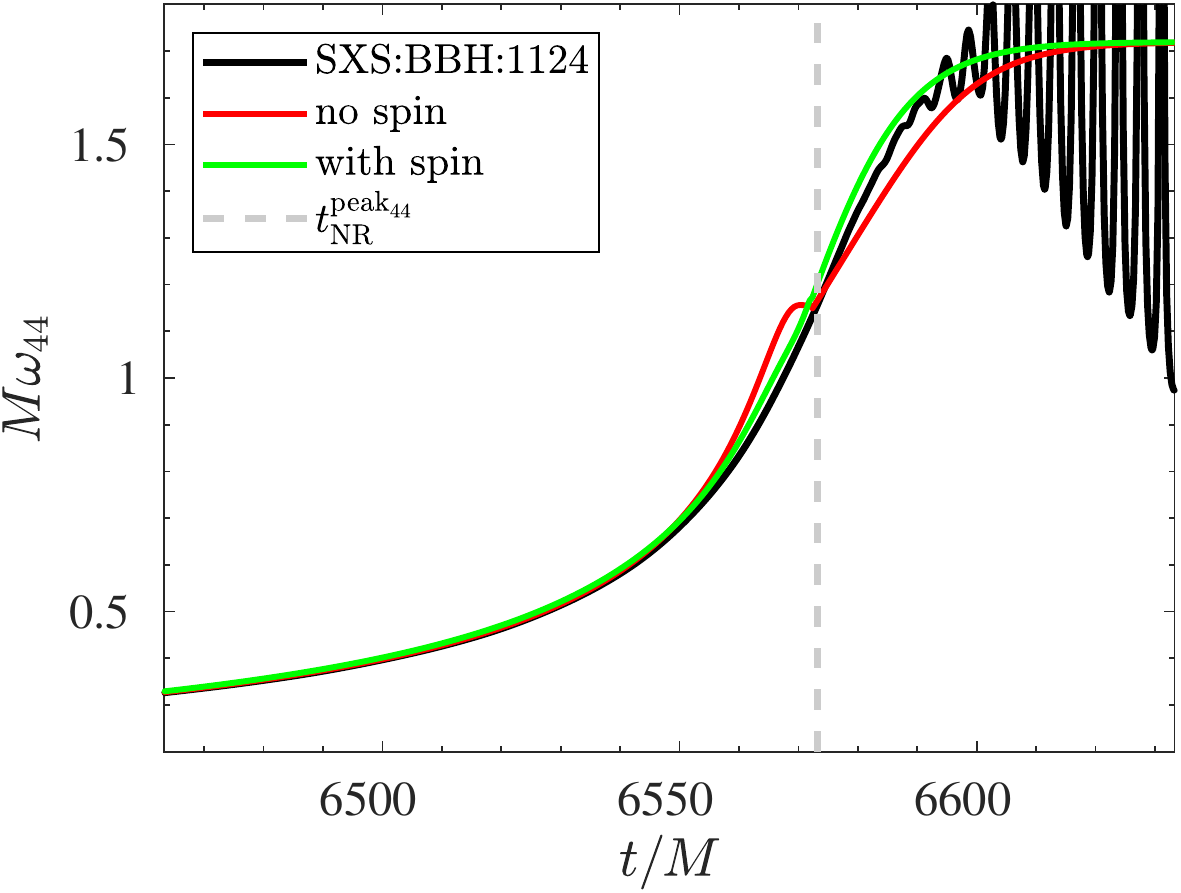}
    \includegraphics[width=0.45\textwidth]{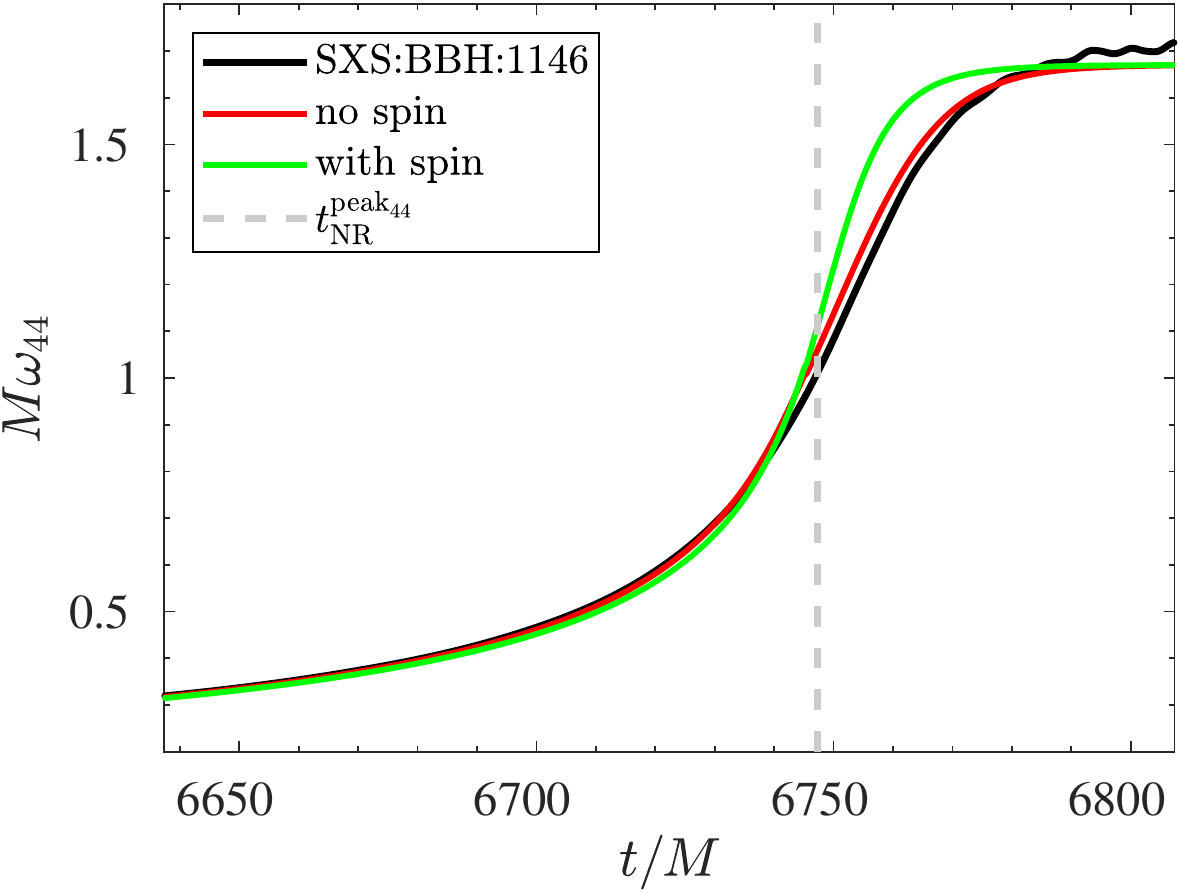}
\caption{In this figure we compare the frequency $M\omega_{44}$ for the two NR waveforms (black) 
			SXS:BBH:1124 $(1,0.998,0.998)$ (left panel) and SXS:BBH:1146 $(1.5,0.95,0.95)$ (right) 
			with the corresponding EOB waveforms, once obtained using the fits of
			Ref.~\cite{Nagar:2019wds} (right panel) and once with the spin-dependent fits presented 
			in  Appenix~\ref{sec:fit_evol} (green).}
\label{fig:pm_spin_v_nospin}
\end{center}
\end{figure*}
As mentioned above, the results presented in the main text {\it do not} rely on the 
the fits of $(c_3^{\phi_{33}},c_4^{\phi_{33}})$ and $(c_3^{\phi_{44}},c_4^{\phi_{44}})$ 
given in Appendix~\ref{sec:fit_evol} with the full spin dependence, but instead use
only their spin-independent part, as already presented  Ref.~\cite{Nagar:2019wds}.
This choice was made so to ensure a  more robust behavior of the frequency
at the beginning of the ringdown when the spins are positive and large.
We illustrate this argument inspecting the behavior of $\omega_{44}$ 
for two highly-spinning configurations. Figure~\ref{fig:pm_spin_v_nospin} 
shows EOB/NR comparisons with two EOB waveforms obtained with either
the nonspinning fits (red online) or those with the full spin dependence (green).
One sees that the spin-dependent fit performs rather well for {\tt SXS:BBH:1124} 
$(1,0.998,0.998)$, consistently with the fact that we used {\tt SXS:BBH:0178}, with 
parameters $(1,0.9942,0.9942)$, to inform the fit. By contrast, one sees that 
the same description applied to a different configuration, $(1.5,0.95,0.95)$,
corresponding to {\tt SXS:BBH:1146},  does not perform equally well, with 
a nonnegiglible gap between the EOB and NR frequencies accumulating right
after the peak. One finds, however, that removing the spin-dependence in
$(c_3^{\phi_{44}},c_4^{\phi_{44}})$ allows one to obtain a much closer EOB/NR
consistency for {\tt SXS:BBH:1146}. For the other case, moving to the nonspinning
description slightly worsens the agreement, both before and after the waveform peak~\footnote{The reader should note
that the postpeak phasing impacts the inspiral waveform through the NQC extraction points obtained from 
the postpeak template. See Appendix~\ref{sec:postpeak_NQC}}. 
On the basis of these results, and especially seen the rather good $\bar{F}$ behavior
illustrated in Fig.~\ref{fig:mismatch_leqm}, we decided to be simple and remove 
the spin dependence in  $(c_3^{\phi_{44}},c_4^{\phi_{44}})$ .  We applied the same
rational also to the $(3,3)$ mode. Clearly, in case of very high-spins, currents fits 
should be improved to some extent, increasing the calibration set so to incorporate
more points in that corner of the parameter space. This will be investigated in future
work.
\subsection{Modeling the NQC extraction points}
\label{sec:NQC}
Let us finally discuss analytic representations of the NR point (amplitude, frequency and derivative) on the multipolar
waveform that is needed for computing the NQC corrections to the waveform multipole by multipole.
For the $(2,2)$ and $(3,3)$ modes we give below dedicated fits. For all other modes, the useful NR quantities 
are obtained analytically from the (fitted) post-peak analytical waveform discussed above.
Let us recall here that, for each mode, the NQC time is always 
\be
t^{\rm NQC}_\lm \equiv t_\lm^{\rm peak}+2.
\ee
All quantities mentioned below with the NQC label are computed at $t=t^{\rm NQC}_\lm$.
\subsubsection{The $(2,2)$ NQC extraction point}
\label{sec:NQC_22}
For the  $(2,2)$ mode the NQC-point quantities 
$\left\lbrace\hat{A}_{22}^{\rm NQC},\dot{A}_{22}^{\rm NQC},\omega_{22}^{\rm NQC},\dot{\omega}_{22}^{\rm NQC}\right\rbrace$ 
are fitted directly. The 3-piece hybrid fit, presented in \cite{Nagar:2017jdw,Nagar:2018zoe} is modified for $q>4$.
The fits of $\left\lbrace\hat{A}_{22}^{\rm NQC}\omega_{22}^{\rm NQC}\right\rbrace$ are done using the template 
discussed already for the peak, see Appendix~\ref{sec:fits_merger}. The reader should note however that the fit of 
$\omega_{22}^{\rm NQC}$ has additional flexibility. The replacement in \eqref{eq:expol2} is also done for $i=2$ for this case.
\begin{table*}[t]
  \caption{\label{tab:NQC} Coefficients of the $(2,2)$ quantities needed to calculate the NQC extraction point.
  From left to right the columns show $\left\lbrace\hat{A}_{22}^{\rm NQC},\dot{\hat{A}}_{22}^{\rm NQC},
  \omega_{22}^{\rm NQC},\dot{\omega}_{22}^{\rm NQC}\right\rbrace$.}
\begin{ruledtabular}
\begin{tabular}{c lll | lll | lll | lll c }
	 &\multicolumn{3}{c|}{$\hat{A}^{\rm NQC}_{22}$} & \multicolumn{3}{c|}{$\dot{A}^{\rm NQC}_{22}$} & \multicolumn{3}{c|}{$\omega^{\rm NQC}_{22}$} & \multicolumn{3}{c}{$\dot{\omega}^{\rm NQC}_{22}$} &\\\hline
  & $\hat{A}_{22}^{{\rm NQC}_{0}}$  & $=$ & $\;\;\; 0.294773$ & $\dot{A}_{22}^{{\rm NQC}_0}/\nu$ & $=$ & $-0.000243654$ 
  & $\omega_{22}^{{\rm NQC}_{0}}$ & $=$ & $\;\;\; 0.285588$ & ${\dot{\omega}_{22}^{{\rm NQC}_{0}}}$ & $=$ & $\;\;\;0.00628027$& \\
  & $a_1^{\hat{A}^{\rm NQC}_{22}}$ & $=$ & $-0.052697$ & $a_1^{\dot{A}^{\rm NQC}_{22}}$ & $=$ & $\;\;\;2.86637$ 
  & $a_1^{\omega^{\rm NQC}_{22}}$ & $=$ & $\;\;\;0.91704$ & $a_1^{\dot{\omega}^{\rm NQC}_{22}}$ & $=$ & $\;\;\;2.4351$& \\
  & $a_2^{\hat{A}^{\rm NQC}_{22}}$ & $=$ & $\;\;\;1.6088$ & $a_2^{\dot{A}^{\rm NQC}_{22}}$ & $=$ & $-1.3667$  
  & $a_2^{\omega^{\rm NQC}_{22}}$ & $=$ & $\;\;\;1.7912$ & $a_2^{\dot{\omega}^{\rm NQC}_{22}}$ & $=$ & $\;\;\;4.4928$& \\\hline
  & $b^{\hat{A}^{{\rm NQC}_{m_1=m_2}}_{22}}_1 $ & $=$ & $-0.705226$ & $b^{\dot{A}^{{\rm NQC}_{m_1=m_2}}_{22}}_1$ & $=$ & $\;\;\;0.02679530$ 
  & $b^{\omega^{{\rm NQC}_{m_1=m_2}}_{22}}_1$ & $=$ & $-0.46550$ & $b^{\dot{\omega}^{{\rm NQC}_{m_1=m_2}}_{22}}_1$ & $=$ & $\;\;\;0.001425242$& \\
  & $b^{\hat{A}^{{\rm NQC}_{m_1=m_2}}_{22}}_2 $ & $=$ & $-0.0953944$ & $b^{\dot{A}^{{\rm NQC}_{m_1=m_2}}_{22}}_2$ & $=$ & $-0.0064409$ 
  & $b^{\omega^{{\rm NQC}_{m_1=m_2}}_{22}}_2$ & $=$ & $-0.078787$ & $b^{\dot{\omega}^{{\rm NQC}_{m_1=m_2}}_{22}}_2$ & $=$ & $-0.00096073$& \\
  & $b^{\hat{A}^{{\rm NQC}_{m_1=m_2}}_{22}}_3 $ & $=$ & $-1.087280$ & & & 
  & $b^{\omega^{{\rm NQC}_{m_1=m_2}}_{22}}_3$ & $=$ & $-0.852284$ & & & \\\hline
  & $c^{\hat{A}^{{\rm NQC}}_{22}}_{11}$ & $=$ & $\;\;\;0.009335$ & $c^{\dot{A}^{{\rm NQC}}_{22}}_{1}$ & $=$ & $-0.015395218$ 
  & $c^{\omega^{{\rm NQC}}_{22}}_{11}$ & $=$ & $-0.338008$ &$c^{\dot{\omega}^{{\rm NQC}}_{22}}_{1}$ & $=$ & $-0.000063766$& \\
  & $c^{\hat{A}^{{\rm NQC}}_{22}}_{12}$& $=$ & $\;\;\;0.582869$ & $c^{\dot{A}^{{\rm NQC}}_{22}}_{2}$ & $=$ & $\;\;\;0.008732589$ 
  & $c^{\omega^{{\rm NQC}}_{22}}_{12}$ & $=$ & $\;\;\;1.077812$ &$c^{\dot{\omega}^{{\rm NQC}}_{22}}_{2}$ & $=$ & $\;\;\;0.000513197$& \\ 
  &  & & & & &
  & $c^{\omega^{{\rm NQC}}_{22}}_{21}$ & $=$ & $\;\;\;0.0555533$ && & &\\
  &  & &  & & &
  & $c^{\omega^{{\rm NQC}}_{22}}_{22}$ & $=$ & $-0.312861$ & & & &\\
  & $c^{\hat{A}^{{\rm NQC}}_{22}}_{31}$ & $=$ & $\;\;\;-0.140747$ & & &
  & $c^{\omega^{{\rm NQC}}_{22}}_{31}$ & $=$ & $\;\;\;0.289185$ && & &\\
  & $c^{\hat{A}^{{\rm NQC}}_{22}}_{32}$ & $=$ & $\;\;\;0.505807$ & & &
  & $c^{\omega^{{\rm NQC}}_{22}}_{32}$ & $=$ & $-0.195838$ & & & &\\
\end{tabular}
\end{ruledtabular}
\end{table*}
In the following the fitting of $\dot{A}_{22}^{\rm NQC}$ and $\dot{\omega}_{22}^{\rm NQC}$. Both rely on the 
same template thus it is only given for the former explicitly. To fit the  time derivative of the amplitude 
at $t_{\rm NQC}$ it was proven useful to not fit it directly, but to fit
$\dot{A}_{22}^{\rm NQC}/\nu\omega_{22}^{\rm NQC}$, starting with the following factorization
\begin{align}
\label{eq:fit_derv_amp}
\frac{\dot{A}_{22}^{\rm NQC}}{\nu\omega_{22}^{\rm NQC}}&= \left[\hat{\dot{A}}_{22}^{{\rm NQC}_{\rm orb}}\left(\nu\right)+\hat{\dot{A}}_{22}^{{\rm NQC}_{\rm Spin}}\left(X_{12},\hat{S}\right)\right].
\end{align}
The nonspinning contribution is fitted as
\begin{align}
\label{eq:fit_derv_amp_orb}
\hat{\dot{A}}_{22}^{{\rm NQC}_{\rm orb}}\left(\nu\right)=1+ a_1^{\dot{A}^{\rm NQC}_{22}}\nu + a_2^{\dot{A}^{\rm NQC}_{22}}\nu^2.
\end{align}
The spin-dependence is represented as
\begin{align}
\label{eq:A_NQC_spin}
\hat{\dot{A}}_{22}^{{\rm NQC}_{\rm Spin}}\left(X_{12}\hat{S}\right)= b^{\dot{A}^{{\rm NQC}_{m_1=m_2}}_{22}}_1\hat{S} +b^{\dot{A}^{{\rm NQC}_{m_1=m_2}}_{22}}_1\hat{S}^2\, .
\end{align}
The extrapolation to the $m_1\neq m_2$ regime is done via the replacement
\be
b^{\dot{A}^{{\rm NQC}_{m_1=m_2}}_{22}}_i \rightarrow b^{\dot{A}^{{\rm NQC}_{m_1=m_2}}_{22}}_i +c^{\dot{A}^{{\rm NQC}_{m_1=m_2}}_{22}}_i X_{12}\, ,
\ee 
with $i = \left\{1,2\right\}$.
All coefficients are listed explicitly in Table~\ref{tab:NQC}.
\subsubsection{The $(3,3)$ NQC extraction point}
\label{sec:NQC_33}
Let us discuss now explicit fits for $\left\lbrace\hat{A}_{33}^{\rm NQC},\dot{A}_{33}^{\rm NQC},\omega_{33}^{\rm NQC},\dot{\omega}_{33}^{\rm NQC}\right\rbrace$.
The amplitude  $\hat{A}_{33}^{\rm NQC}$ is written as two separate terms as
\begin{align}
\label{eq:a33_fact_nqc}
\hat{A}_{33}^{\rm NQC}= \hat{A}_{33}^{{\rm NQC}_0}X_{12}\hat{A}_{33}^{{\rm NQC}_{\rm orb}}(\nu) + \hat{A}_{33}^{{\rm NQC}_{\rm S}}(\nu,\tilde{a}_{12})\ ,
\end{align}
where  $\hat{A}_{33}^{{\rm NQC}_0}$ is the test-particle value. 
The non-spinning sector is fitted  after factorization of 
$\hat{A}_{33}^{{\rm NQC}_0}X_{12}$ with 
\begin{align}
\hat{A}_{33}^{{\rm NQC}_{\rm orb}}(\nu) = 1 + a_1^{\hat{A}_{33}^{{\rm NQC}}}\nu + a_2^{\hat{A}_{33}^{{\rm NQC}}}\nu^2\ .
\end{align}
The spin-dependent factor $\hat{A}_{33}^{{\rm NQC}_{\rm S}}$ is first fitted in the $\nu=0$ limit with
\begin{align}
\label{eq:a33_spin_nqc1}
\hat{A}_{33}^{{\rm NQC}_{\rm S}}= \frac{b_1^{\hat{A}_{33}^{{\rm NQC}}}\tilde{a}_{12}+b_2^{\hat{A}_{33}^{{\rm NQC}}}\tilde{a}_{12}^2}{1+b_3^{\hat{A}_{33}^{{\rm NQC}}}\tilde{a}_{12}} \ .
\end{align}
and then extended to the $\nu\neq 0$ regime through
\begin{align}
\label{eq:a33_spin_nqc}
b_i^{\hat{A}_{33}^{{\rm NQC}}}\rightarrow \frac{b_i^{\hat{A}_{33}^{{\rm NQC}}} + c_{i1}^{\hat{A}_{33}^{{\rm NQC}}}\nu}{1+ c_{i2}^{\hat{A}_{33}^{{\rm NQC}}}\nu}\ {\rm with}\ i=\{1,2,3\}\ .
\end{align}
The time-derivative of the amplitude $\dot{A}_{33}^{{\rm NQC}}$ was fitted in two steps.
In the first step, one is fitting only equal-mass data (but, crucially, including also data with
unequal spins), as
\begin{align}
\frac{\dot{A}_{33}^{{\rm NQC}_{\nu=1/4}}}{10^{5}} =d_0^{\dot{A}_{33}^{{\rm NQC}_{\nu=1/4}}} +& d_1^{\dot{A}_{33}^{{\rm NQC}_{\nu=1/4}}} \tilde{a}_{12}\nonumber\\
& + d_2^{\dot{A}_{33}^{{\rm NQC}_{\nu=1/4}}} \tilde{a}_{12}^2 \ .
\end{align}
The un-equal mass sector is fitted with the same template as $\hat{A}_{33}^{\rm NQC}$ with
3 modifications: (i)  $X_{12}$ is not factorized as in \eqref{eq:a33_fact_nqc}; 
(ii) The spin variable in \eqref{eq:a33_spin_nqc1} is chosen to be $\hat{S}$;  (iii) the transformation 
is only done for $i=2$ in \eqref{eq:a33_spin_nqc},  $c_{1i}^{\hat{A}_{33}^{{\rm NQC}}}=c_{3i}^{\hat{A}_{33}^{{\rm NQC}}}=0$.

Moving now to the NQC frequency $\omega_{33}^{\rm NQC}$, we assume the following factorization
\begin{align}
\omega_{33}^{\rm NQC} =\omega_{33}^{{\rm NQC}_0}\omega_{33}^{{\rm NQC}_{\rm orb}}\omega_{33}^{{\rm NQC}_{\rm S}}\ , 
\end{align}
where $\omega_{33}^{{\rm NQC}_{\rm orb}}$ is fitted to the nonspinning data with a second-order polynomial in $\nu$ as
\begin{align}
\omega_{33}^{{\rm NQC}_{\rm orb}} =  1 + a_1^{\omega_{33}^{{\rm NQC}}}\nu + a_2^{\omega_{33}^{{\rm NQC}}}\nu^2\ .
\end{align}
Then, $\omega_{33}^{{\rm NQC}_{\rm S}}$ if fitted to the test-particle data using
\begin{align}
\omega_{33}^{{\rm NQC}_{\rm S}} = \frac{ 1+ b_1^{\omega_{33}^{{\rm NQC}}}\hat{S}}{1 + b_2^{\omega_{33}^{{\rm NQC}}}\hat{S}}\ .
\end{align}
Finally, the spin-dependence in $\omega_{33}^{{\rm NQC}_{\rm S}}$ incorporates $\nu$-dependent effects as
\begin{align}
b_i^{\omega_{33}^{{\rm NQC}}}\rightarrow b_i^{\omega_{33}^{{\rm NQC}}}+c_i^{\omega_{33}^{{\rm NQC}}}\nu\ {\rm with}\ i=\{1,2\}\ .
\end{align}
Moving finally to the time-derivative of the frequency, $\dot{\omega}_{33}^{\rm NQC}$, it is fitted with the ansatz
\begin{align}
\dot{\omega}_{33}^{\rm NQC} =\dot{\omega}_{33}^{{\rm NQC}_0}\dot{\omega}_{33}^{{\rm NQC}_{\rm orb}}+\dot{\omega}_{33}^{{\rm NQC}_{\rm S}}\ , 
\end{align}
where $\dot{\omega}_{33}^{{\rm NQC}_{\rm orb}}$ is fitted to nonspinning data with 
\begin{align}
\dot{\omega}_{33}^{{\rm NQC}_{\rm orb}} =  1 + a_1^{\dot{\omega}_{33}^{{\rm NQC}}}\nu\ .
\end{align}
$\dot{\omega}_{33}^{{\rm NQC}_{\rm S}}$ if fitted to the test-particle data with
\begin{align}
\dot{\omega}_{33}^{{\rm NQC}_{\rm S}} = b_1^{\dot{\omega}_{33}^{{\rm NQC}}}\hat{S} + b_2^{\dot{\omega}_{33}^{{\rm NQC}}}\hat{S}^2\ .
\end{align}
The spin dependence in $\dot{\omega}_{33}^{{\rm NQC}_{\rm S}}$ is then extrapolated to the comparable mass through
\begin{align}
\label{NQC_33_end}
b_i^{\dot{\omega}_{33}^{{\rm NQC}}}\rightarrow b_i^{\dot{\omega}_{33}^{{\rm NQC}}}+c_i^{\dot{\omega}_{33}^{{\rm NQC}}}\nu\ {\rm with}\ i=\{1,2\}\ .
\end{align}
\begin{table*}[t]
  \caption{\label{tab:NQC_33} Coefficients of the $(3,3)$ quantities needed to calculate the NQC extraction point. 
  From left to right the columns show $\left\lbrace\hat{A}_{33}^{\rm NQC},\dot{\hat{A}}_{33}^{\rm NQC},
  \omega_{33}^{\rm NQC},\dot{\omega}_{33}^{\rm NQC}\right\rbrace$.}
\begin{ruledtabular}
\begin{tabular}{c lll | lll | lll | lll c }
	 &\multicolumn{3}{c|}{$\hat{A}^{\rm NQC}_{33}$} & \multicolumn{3}{c|}{$\dot{A}^{\rm NQC}_{33}$} & \multicolumn{3}{c|}{$\omega^{\rm NQC}_{33}$} & \multicolumn{3}{c}{$\dot{\omega}^{\rm NQC}_{33}$} &\\\hline
  & $\hat{A}_{33}^{{\rm NQC}_{0}}$  & $=$ & $\;\;\; 0.0512928$ & $\dot{A}_{33}^{{\rm NQC}_0}/\nu$ & $=$ & $-3.9568\times 10^{-4}$ 
  & $\omega_{33}^{{\rm NQC}_{0}}$ & $=$ & $\;\;\; 0.476647$ & ${\dot{\omega}_{33}^{{\rm NQC}_{0}}}$ & $=$ & $\;\;\;0.0110394$& \\
  & $a_1^{\hat{A}^{\rm NQC}_{33}}$ & $=$ & $\;\;\;0.09537$ & $a_1^{\dot{A}^{\rm NQC}_{33}}$ & $=$ & $\;\;\; 1.0985$ 
  & $a_1^{\omega^{\rm NQC}_{33}}$ & $=$ & $\;\;\;1.0886$ & $a_1^{\dot{\omega}^{\rm NQC}_{33}}$ & $=$ & $\;\;\; 2.7962$& \\
  & $a_2^{\hat{A}^{\rm NQC}_{33}}$ & $=$ & $\;\;\;3.7217$ & $a_2^{\dot{A}^{\rm NQC}_{33}}$ & $=$ & $-13.458$  
  & $a_2^{\omega^{\rm NQC}_{33}}$ & $=$ & $\;\;\;3.0658$ &  &  & & \\\hline
  & $b^{\hat{A}^{{\rm NQC}_{\nu=0}}_{33}}_1 $ & $=$ & $\;\;\;0.00924494$ & $b^{\dot{A}^{{\rm NQC}_{\nu=0}}_{33}}_1$ & $=$ & $\;\;\;1.41504\times 10^{-4} $ 
  & $b^{\omega^{{\rm NQC}_{\nu=0}}_{33}}_1$ & $=$ & $-0.236271$ & $b^{\dot{\omega}^{{\rm NQC}_{\nu=0}}_{33}}_1$ & $=$ & $-4.5666\times 10^{-4}$& \\
  & $b^{\hat{A}^{{\rm NQC}_{\nu=0}}_{33}}_2 $ & $=$ & $-8.7052\times 10^{-5}$ & $b^{\dot{A}^{{\rm NQC}_{\nu=0}}_{33}}_2$ & $=$ & $\;\;\;1.04680\times 10^{-4}$ 
  & $b^{\omega^{{\rm NQC}_{\nu=0}}_{33}}_2$ & $=$ & $-0.582892$ & $b^{\dot{\omega}^{{\rm NQC}_{\nu=0}}_{33}}_2$ & $=$ & $-0.00388909$& \\
  & $b^{\hat{A}^{{\rm NQC}_{\nu=0}}_{33}}_3 $ & $=$ & $-0.479669$ & $b^{\dot{A}^{{\rm NQC}_{\nu=0}}_{33}}_3$ & $=$ & $-0.422066$ 
  &  &  &  & & & \\\hline
  & $c^{\hat{A}^{{\rm NQC}}_{33}}_{11}$ & $=$ & $\;\;\; 0.0067063$ & $c^{\dot{A}^{{\rm NQC}}_{33}}_{21}$ & $=$ & $-4.671176\times 10^{-4}$ 
  & $c^{\omega^{{\rm NQC}}_{33}}_{1}$ & $=$ & $-0.085544$ &$c^{\dot{\omega}^{{\rm NQC}}_{33}}_{1}$ & $=$ & $\;\;\;0.0290846$& \\
  & $c^{\hat{A}^{{\rm NQC}}_{33}}_{12}$& $=$ & $\;\;\; 4.814781$ & $c^{\dot{A}^{{\rm NQC}}_{33}}_{22}$ & $=$ & $-4.0270198$
  & $c^{\omega^{{\rm NQC}}_{33}}_{2}$ & $=$ & $-0.523365 $ &$c^{\dot{\omega}^{{\rm NQC}}_{33}}_{2}$ & $=$ & $\;\;\;0.0087659$ &\\ \cline{5-7}
  &  $c^{\hat{A}^{{\rm NQC}}_{33}}_{21}$ & $=$ & $\;\;\; 0.0111876$ & $d_0^{\dot{A}_{33}^{{\rm NQC}_{\nu=1/4}}}$ & $=$ & $-0.090676$
& & & &&&& \\
  &  $c^{\hat{A}^{{\rm NQC}}_{33}}_{22}$ & $=$ & $-1.079532 $  & $d_1^{\dot{A}_{33}^{{\rm NQC}_{\nu=1/4}}}$ & $=$ & $-5.1643$
  & && & & & &\\
  & $c^{\hat{A}^{{\rm NQC}}_{33}}_{31}$ & $=$ & $\;\;\; 2.967227$ &  $d_2^{\dot{A}_{33}^{{\rm NQC}_{\nu=1/4}}}$ & $=$ & $-3.2594$
  & && && & &\\
  & $c^{\hat{A}^{{\rm NQC}}_{33}}_{32}$ & $=$ & $-2.571783$ & & &
  & &&& & & &\\
\end{tabular}
\end{ruledtabular}
\end{table*}

\subsubsection{Calculation of NQC quantities from the postpeak analytical waveform}
\label{sec:postpeak_NQC}
Let us finally discuss explicitly the computation of the NQC quantities 
$\left(A_{\ell m}^{\rm NQC},\omega_{\ell m}^{\rm NQC},\dot{A}_{\ell m}^{\rm NQC},\dot{\omega}_{\ell m}^{\rm NQC}\right)$
from the NR-informed analytical description of the postpeak waveform,
as defined in Sec.~V~A~of~\cite{Nagar:2019wds}, to which we refer the reader
for the notation. Although the formulas
have to be intended valid multipole by multipole, in the following we
drop the $(\ell,m)$ indexes for clarity.
The analytical expression for the amplitude and its time derivative read
\begin{widetext}
\begin{align}
A_{h}/\nu = ~&e^{-\alpha_1 \frac{t-t_{\rm peak}}{M_{\rm BH}}} \left[ c_1^A \tanh \left(c_2^A\frac{t-t_{\rm peak}}{M_{\rm BH}} +c_3^A \right) +c_4^A\right],\\
\dot{A}_{h}/\nu=~&\frac{c_1^A c^A_2 e^{-\alpha_1\frac{t-t_{\rm peak}}{M_{\rm BH}}} {\rm sech}^2\left(c^A_2\frac{t-t_{\rm peak}}{M_{\rm BH}}+c^A_3\right)}{M_{\rm BH}} -\frac{\alpha_1 e^{-\alpha_1\frac{t-t_{\rm peak}}{M_{\rm BH}}} \left[c^A_1 \tanh \left(c^A_2\frac{t-t_{\rm peak}}{M_{\rm BH}}+c^A_3\right)+c^A_4\right]}{M_{\rm BH}},
\end{align}
while those for the phase and its derivatives read
\begin{align}
\phi_{h} =&-\omega_1 \frac{t-t_{\rm peak}}{M^2_{\rm BH}}-c_1^{\phi} \mathrm{ln} \left( \frac{1+c_3^{\phi} e^{-c_2^{\phi} \frac{t-t_{\rm peak}}{M_{\rm BH}}} +c_4^\phi e^{-2c_2^\phi \frac{t-t_{\rm peak}}{M_{\rm BH}} }}{1+c_3^{\phi}+c_4^{\phi}}\right),\\
\omega_{h}=&- \dot{\phi}_{h} \hspace{0.15cm}= \frac{\omega_1}{M^2_{\rm BH}}-\frac{c^\phi_1c^\phi_2}{M_{\rm BH}}\frac{c^\phi_3 x(t)+2 c^\phi_4 x^2(t)}{1+c^\phi_3 x(t)+c^\phi_4 x^2(t)},\\
\dot{\omega}_{h}=&-\ddot{\phi}_{h}=\frac{c^\phi_1{c^\phi_2}^2}{M^2_{\rm BH}}\left[\frac{c^\phi_3 x(t)+4 c^\phi_4 x^2(t)}{1+c^\phi_3 x(t)+c^\phi_4 x^2(t)}-\left(\frac{c^\phi_3 x(t)+2 c^\phi_4 x^2(t)}{1+c^\phi_3 x(t)+c^\phi_4 x^2(t)}\right)^2\right],
\end{align}
\end{widetext}
where we introduced
\be
x(t)=e^{-c^\phi_2\frac{t-t_{\rm peak}}{M_{\rm BH}}}.
\ee
The waveform quantities needed to compute the NQC correction to amplitude and phase are simply obtained
by evaluating the above expressions at $t=t^{\rm NQC}_\lm=t_\lm^{\rm peak}+2$ multipole by multipole.
\subsubsection{The fitted derivative of the $(\ell,m)=(4,4)$ amplitude at the NQC extraction point}
Unfortunately, we have realized that the accuracy of the derivative obtained with the above template
does not always have sufficient accuracy. This is due to insufficient flexibility of the fitting template,
that will be modified in future work. To overcome this difficulty, we give here an explicit fit of the amplitude
time-derivative that is then used in the main text.
The derivative of  NQC amplitude is separated in two terms as
\be
\dot{A}_{44}^{\rm NQC} = \nu \dot{A}_{44}^{{\rm NQC}_{0}} \hat{\dot{A}}_{44}^{\rm orb}(\nu) + \hat{\dot{A}}_{44}^{\rm Spin}(\hat{S},\nu) \ ,
\ee
where $\dot{A}_{44}^{{\rm NQC}_{0}} $ is the peak amplitude in the test particle limit. 
The non-spinning behavior is modeled with
\be
\hat{\dot{A}}_{44}^{\rm orb}(\nu)= \frac{1+a^{\dot{A}_{44}^{\rm NQC}}_1 \nu+a^{\dot{A}_{44}^{\rm NQC}}_2 \nu^2}{1+a^{\dot{A}_{44}^{\rm NQC}}_3 \nu+a^{\dot{A}_{44}^{\rm NQC}}_4 \nu^2}\ .
\ee
The spin dependence is first fitted to the test-particle limit using
\be
\hat{\dot{A}}_{44}^{\rm Spin}(\hat{S},\nu=0) = \frac{b^{\dot{A}_{44}^{\rm NQC}}_1 \hat{S}}{1+b^{\dot{A}_{44}^{\rm NQC}}_2 \hat{S}}\ ,
\ee
and then extrapolated to the comparable mass region through
\bea
b^{\dot{A}_{44}^{\rm NQC}}_1 &\rightarrow &\frac{b^{\dot{A}_{44}^{\rm NQC}}_1 + c^{\dot{A}_{44}^{\rm NQC}}_{11}\nu}{1 + c^{\dot{A}_{44}^{\rm NQC}}_{12}\nu  }\ ,\\
b^{\dot{A}_{44}^{\rm NQC}}_2 &\rightarrow &\frac{b^{\dot{A}_{44}^{\rm NQC}}_2+ c^{\dot{A}_{44}^{\rm NQC}}_{21}\nu }{1 + c^{\dot{A}_{44}^{\rm NQC}}_{22}\nu  }\ .
\eea
\begin{table}
\caption{\label{tab:dA_NQC_44}
Explicit coefficients of the fit of $\dot{A}_{44}^{\rm NQC} $.}
\centering
\begin{ruledtabular}
\begin{tabular}{ ll c ll  }
& $\dot{A}_{44}^{{\rm NQC}_{0}}$ & $=$ &  $-1.52614\times 10^{-4}$ & \\
& $a^{\dot{A}_{44}^{\rm NQC}}_1$ & $=$ & $-7.63783$ & \\
& $a^{\dot{A}_{44}^{\rm NQC}}_2$ & $=$ & $\;15.8089$ & \\
& $a^{\dot{A}_{44}^{\rm NQC}}_3$ & $=$ & $-5.88951$ & \\
& $a^{\dot{A}_{44}^{\rm NQC}}_4$ & $=$ & $\;11.1555$ & \\\hline
& $b^{\dot{A}_{44}^{\rm NQC}}_1$ & $=$ & $\;\;\;3.76236\times 10^{-5}$ & \\
& $b^{\dot{A}_{44}^{\rm NQC}}_2$ & $=$ & $-0.819379$ & \\\hline
& $c^{\dot{A}_{44}^{\rm NQC}}_{11}$ & $=$ & $-6.45958\times 10^{-6}$ & \\
& $c^{\dot{A}_{44}^{\rm NQC}}_{12}$ & $=$ & $-2.35613$ & \\
& $c^{\dot{A}_{44}^{\rm NQC}}_{21}$ & $=$ & $-298.678$ & \\
& $c^{\dot{A}_{44}^{\rm NQC}}_{22}$ & $=$ & $-1063.08$ & \\
\end{tabular}
\end{ruledtabular}
\end{table}

\bibliography{refs20200701.bib,local.bib}

\end{document}